\documentclass[10pt]{dmathesis}

\usepackage{fancyhdr}

\usepackage{epsfig}
\usepackage{cite}
\usepackage{graphicx}
\usepackage{amsmath}
\usepackage{color,amscd,subfigure}
\usepackage{theorem}
\usepackage{amssymb}
\usepackage{latexsym}
\usepackage{epic}

\usepackage{mathtools}
\usepackage{feynmp}
\usepackage[makeroom]{cancel}
\usepackage{emptypage} %per non avere titoli in pagine vuote
\usepackage{verbatim}
\usepackage{spverbatim}
\usepackage{fancyvrb}
\usepackage[T1]{fontenc}
\usepackage{multirow}
\usepackage{slashed}
\usepackage[utf8]{inputenc}
\usepackage[english]{babel}
\usepackage[nopanel]{pdfscreen}
\usepackage{datetime}
\usepackage{chngcntr}
\usepackage{verse}
\usepackage{showlabel}
\usepackage{caption}
\usepackage{lipsum}
\usepackage{apalike}
\usepackage[nottoc]{tocbibind}

\makeatletter
\let\@internalcite\cite
\def\cite{\def\citeauthoryear##1##2{##1, ##2}\@internalcite}
\def\shortcite{\def\citeauthoryear##1##2{##2}\@internalcite}
\def\@biblabel#1{\def\citeauthoryear##1##2{##1, ##2}[#1]\hfill}
\makeatother

\newcounter{ind}
\def\eqlabon{
\setcounter{ind}{\value{equation}}\addtocounter{ind}{1}
\setcounter{equation}{0}
\renewcommand{\theequation}{\arabic{chapter}%
         .\arabic{section}.\arabic{ind}\alph{equation}}}
\def\eqlaboff{
\renewcommand{\theequation}{\arabic{chapter}%
         .\arabic{section}.\arabic{equation}}
\setcounter{equation}{\value{ind}}}

\usepackage{hyperref}
\hypersetup{%
    pdfpagemode={UseOutlines},
    bookmarksopen,
    pdfstartview={FitH},
    colorlinks,
    linkcolor={blue},
    citecolor={blue},
    urlcolor={blue}
  }

\definecolor{darkred}{rgb}{0.55, 0.0, 0.0}

\usepackage{afterpage}

\newcommand\blankpage{%
    \null
    \thispagestyle{empty}%
    \addtocounter{page}{-1}%
    \newpage}

\usepackage{xfrac}

\usepackage{tikz}
\usetikzlibrary{shapes.geometric}
\usetikzlibrary{patterns, arrows.meta}
\usetikzlibrary{calc}
\usetikzlibrary{arrows,shapes}
\usetikzlibrary{trees}
\usetikzlibrary{matrix,arrows} 				% For commutative diagram
\usetikzlibrary{positioning}				% For "above of=" commands
\usetikzlibrary{calc,through}				% For coordinates
\usetikzlibrary{decorations.pathreplacing}  % For curly braces
\usepackage{pgffor}							% For repeating patterns

\usetikzlibrary{decorations.pathmorphing}	% For Feynman Diagrams
\usetikzlibrary{decorations.markings}
\tikzset{
    vector/.style={decorate, decoration={snake,mark=at position .55 with {\arrow[draw=black]{>}}}, draw},
    dressed_vector/.style={double, decorate, decoration={snake,mark=at position .55 with {\arrow[draw=black]{>}}}, draw},
	provector/.style={decorate, decoration={snake,amplitude=2.5pt}, draw},
	antivector/.style={decorate, decoration={snake,amplitude=-2.5pt}, draw},
    fermion/.style={draw=black, postaction={decorate},
        decoration={markings,mark=at position .55 with {\arrow[draw=black]{>}}}},
    dressed_fermion/.style={double, draw=black, postaction={decorate},
        decoration={markings,mark=at position .55 with {\arrow[draw=black]{>}}}},
    fermionbar/.style={draw=black, postaction={decorate},
        decoration={markings,mark=at position .55 with {\arrow[draw=black]{<}}}},
    fermionnoarrow/.style={draw=black},
    gluon/.style={decorate, draw=black,
        decoration={coil,amplitude=5pt,segment length=5pt}},
    scalar/.style={dashed,draw=black, postaction={decorate},
        decoration={markings,mark=at position .55 with {\arrow[draw=black]{>}}}},
    scalarbar/.style={dashed,draw=black, postaction={decorate},
        decoration={markings,mark=at position .55 with {\arrow[draw=black]{<}}}},
    scalarnoarrow/.style={dashed,draw=black},
    electron/.style={draw=black, postaction={decorate},
        decoration={markings,mark=at position .55 with {\arrow[draw=black]{>}}}},
	bigvector/.style={decorate, decoration={snake,amplitude=4pt}, draw},
}

\providecommand*{\psibar}{\ensuremath{\overline{\psi}}}
\providecommand*{\Psibar}{\ensuremath{\overline{\Psi}}}

\providecommand*{\ord}{\ensuremath{\text{O}}}

\providecommand*{\Ree}{\ensuremath{\Re\text{e}}}
\providecommand*{\Imm}{\ensuremath{\Im\text{m}}}
\providecommand*{\I}{\ensuremath{\mathbf{1}}}
\providecommand*{\tr}{\ensuremath{\text{Tr}}}
\providecommand*{\e}{\ensuremath{\text{e}}}
\providecommand*{\de}{\ensuremath{\text{d}}}
\providecommand*{\ord}{\ensuremath{\text{O}}}
\providecommand*{\Ree}{\ensuremath{\mathbb{R}\text{e}}}
\providecommand*{\Imm}{\ensuremath{\mathbb{I}\text{m}}}
\providecommand*{\I}{\ensuremath{\mathbf{1}}}
\providecommand*{\tr}{\ensuremath{\text{Tr}}}
\providecommand*{\e}{\ensuremath{\text{e}}}
\providecommand*{\de}{\ensuremath{\text{d}}}
\providecommand*{\De}{\ensuremath{\mathcal{D}}}
\providecommand*{\psibar}{\ensuremath{\overline{\psi}}}

\newcommand{\quotes}[1]{``#1''}
\newcommand{\be}{\begin{equation}}
\newcommand{\ee}{\end{equation}}
\newcommand{\bea}{\begin{eqnarray}}
\newcommand{\eea}{\end{eqnarray}}
\newcommand{\nn}{\nonumber}
\newcommand{\beqn}{\begin{eqnarray}}  
\newcommand{\eeqn}{\end{eqnarray}}   
\newcommand{\beq}{\begin{equation}}   
\newcommand{\eeq}{\end{equation}}

\begin{document}

%% Front page of thesis
%%%%%%%%%%%%%%%%%%%%%%%%%%%%%%%%%%%%%%%%%%%%%%%%%%%%%%%%%%%%%%%%%%%%%%%%%%
%   This is frontpage.tex file needed for the dmathesis.cls file.  You   %
%  have to  put this file in the same directory with your thesis files.  %
%                Written by M. Imran 2001/06/18                          % 
%                 No Copyright for this file                             % 
%                 Save your time and enjoy it                            % 
%                                                                        % 
%%%%%%%%%%%%%%%%%%%%%%%%%%%%%%%%%%%%%%%%%%%%%%%%%%%%%%%%%%%%%%%%%%%%%%%%%%%
%%%%%%%%%%%%%%%%%%%%%%%%%%%%%%%%%%%%%%%%%%%%%%%%%%%%%%%%%%%%%%%%%%%%%%%%%%%
%%%%%%%%%%%%%%%%           The title page           %%%%%%%%%%%%%%%%%%%%%%%  
%%%%%%%%%%%%%%%%%%%%%%%%%%%%%%%%%%%%%%%%%%%%%%%%%%%%%%%%%%%%%%%%%%%%%%%%%%%
\pagenumbering{roman}
%\pagenumbering{arabic}

\setcounter{page}{1}

\newpage
\thispagestyle{empty}

\addtolength{\hoffset}{-0.75cm}

\begin{center}{
\vspace*{\stretch{1}}
\hspace*{\stretch{1}}
\Large
\textup{DOCTORAL THESIS}
}
\vspace*{\stretch{1}}
\hspace*{\stretch{1}}
\end{center}

\begin{center}{
\vspace*{\stretch{1}}
\LARGE 
\hrulefill\\
\bf{High precision applications of lattice gauge theories in the quest for new physics}\\
\hrulefill
}
\vspace*{\stretch{1}}
\end{center}

\hspace{-2cm}
\includegraphics[scale=1.2]{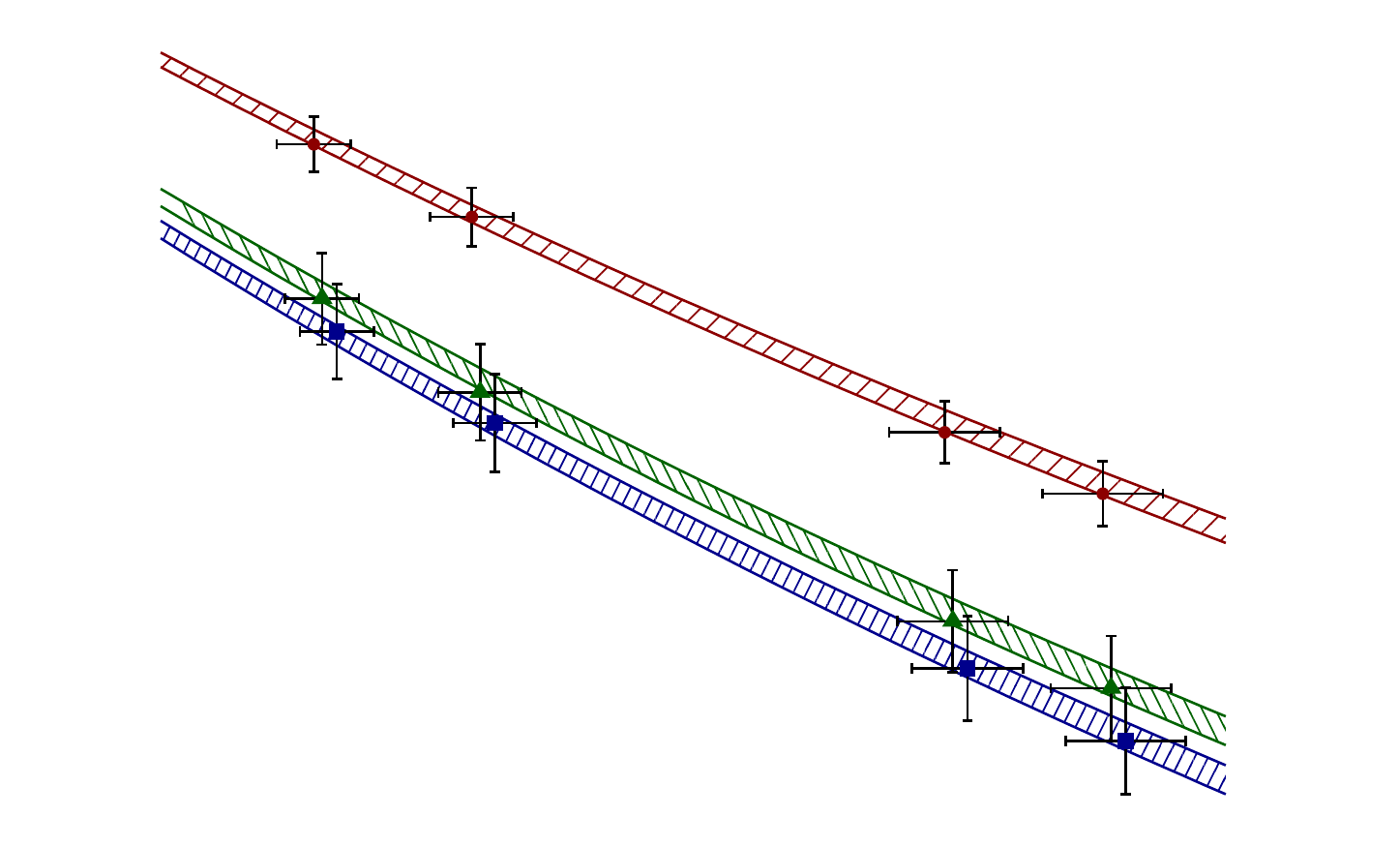}

\newcommand\blfootnote[1]{%
  \begingroup
  \renewcommand\thefootnote{}\footnote{#1}%
  \addtocounter{footnote}{-1}%
  \endgroup
}
%\blfootnote{Now featuring error bars.}

%\begin{center}
%\vspace*{\stretch{1}}
%\includegraphics[scale=1.2]{{figure/logo}.pdf}
%\vspace*{\stretch{1}}
%\end{center}

\vspace{1.cm}
\begin{center}
\vspace*{\stretch{1}}
\LARGE
Andrea Bussone
\vspace*{\stretch{1}}
\end{center}

\newpage
\thispagestyle{empty}

\addtolength{\hoffset}{+0.75cm}

\begin{center}{\LARGE 
\hrulefill\\
\bf{High precision applications of lattice gauge theories in the quest for new physics}\\
\hrulefill
}
\end{center}
\vspace{0.5cm}

\begin{center}\large
\begin{tabular}{ l  l }
  CANDIDATE: &\hspace{1.5 cm}Andrea Bussone \\[0.5ex]
  SUPERVISOR: & \hspace{1.5 cm}Michele Della Morte\\[0.5ex]
  CO-SUPERVISOR: & \hspace{1.5 cm}Claudio Pica
\end{tabular}
\end{center}

\newdate{date}{3}{10}{2017}
\begin{center}{\Large
  % Put your university logo here if you wish.
   \includegraphics[scale=1]{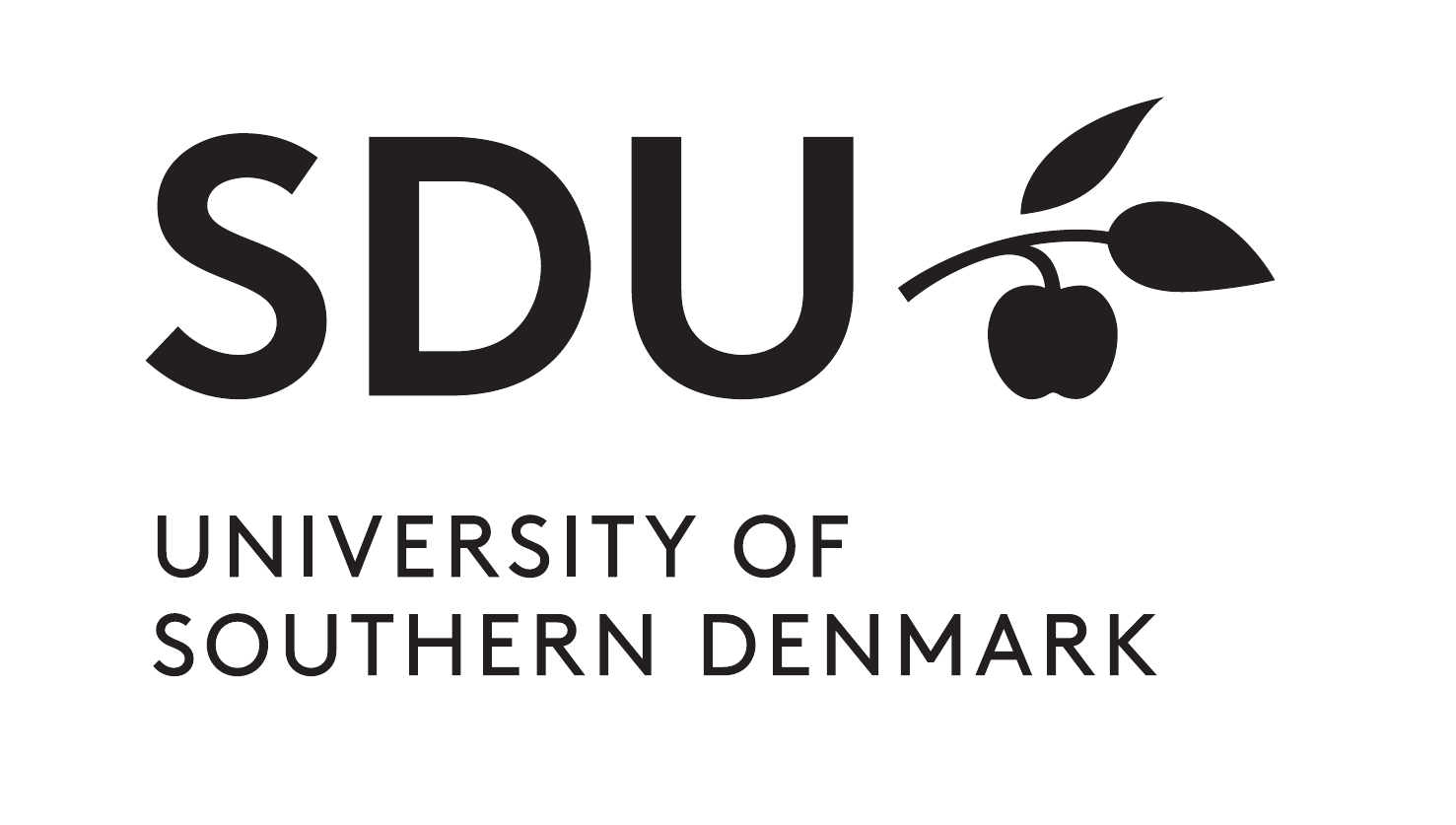}
  \\
   \emph{Dissertation for the degree of Doctor of Philosophy}\\
     \vspace{0.7cm}
\textup{{\textcolor{darkred}{CP$^3$-Origins}\\
Department of Mathematics and Computer Science\\
\vspace{.7 cm}
Submitted to the University of Southern Denmark
\\
\vspace{.2cm}
Corrected version:}}\\
\vspace{.2 cm}
\displaydate{date}}
\end{center}

%%%%%%%%%%%%%%%%%%%%%%%%%%%%%%%%%%%%%%%%%%%%%%%%%%%%%%%%%%%%%%%%%%%%%%%%%%%
%%%%%%%%%%%%%%%%%%           The abstract page         %%%%%%%%%%%%%%%%%%%%  
%%%%%%%%%%%%%%%%%%%%%%%%%%%%%%%%%%%%%%%%%%%%%%%%%%%%%%%%%%%%%%%%%%%%%%%%%%%
\newpage
\thispagestyle{empty}
\addcontentsline{toc}{chapter}{\numberline{}Abstract}
\begin{center}
  \textbf{\large Abstract}
\end{center}
We present some aspects of high precision calculations in the context of Lattice Quantum Field Theory.
This work is a collection of three studies done during my Ph.D.~period.\\

\noindent{First} we present how to use the reweighting technique to compensate for the breaking of unitarity due to
the use of dif\mbox{}ferent boundary conditions in the valence and sea sector.
In particular when twisted boundary conditions are employed, with $\theta$ twisting angle.
In large volume we found that the breaking is negligible, while in rather small volumes an ef\mbox{}fect is present.
The quark mass appears to change with $\theta$ as a cutof\mbox{}f ef\mbox{}fect.\\

\noindent{In} the second part of the dissertation we present an optimization method for Hybrid Monte Carlo performances.
The presented strategy is quite general and can be applied to Beyond Standard Model strongly interacting theories,
for which the need for precision is becoming urgent nowadays.
The work is based on the existence of a shadow Hamiltonian, an exactly conserved quantity along the Molecular Dynamics trajectory.
The optimization method is economic since it only requires the forces to be measured, which are already used for
the evolution from one configuration to the new one.
We found predictions for the cost of the simulations with an accuracy of 10\% and we could estimate the optimal parameters for
the Omelyan integrator with mass-preconditioning and multi time-scale.\\

\noindent{In} the last part of the work we address the calculation of electromagnetic corrections to the 
hadronic contribution to the $(g-2)$ anomaly of the muon.
A long standing discrepancy between theoretical calculations and experimental results is present.
Each new Beyond Standard Model theory aims at solving the discrepancy with the introduction of New Physics.
But before invoking New Physics we need to clear the sight from possible ef\mbox{}fects within the Standard Model.
Firstly we discuss dif\mbox{}ferent implementations of QED on the lattice with periodic boundary conditions.
Then we consider applications of that for the muon anomaly.
In this exploratory study we carefully matched the masses of the charged pions in the theory with and without QED.
In that way we are able to access directly the electromagnetic corrections to the anomaly with smaller statistical errors.
We found a visible ef\mbox{}fect at the percent level although consistent with zero within two sigmas.

\newpage
\thispagestyle{empty}
\addcontentsline{toc}{chapter}{\numberline{}Sammenfatning}
\begin{center}
  \textbf{\large Sammenfatning}
\end{center}

\noindent{Vi} præsenterer nogle vigtige aspekter af højpræcisionsberegninger i forbindelse med lattice kvantefeltteori.
Denne afhandling er en samling af tre studier udført i løbet af min Ph.d.-uddannelse.\\

\noindent{Først} præsenterer vi, hvordan man bruger reweighting-teknikken til at kompensere for unitæritetsbruddet forårsaget af brugen af forskellige randbetingelser i valens- og sea-sektoren. I særdeleshed når der anvendes twistede-randbet
ingelse med twist-vinkel $\theta$. I store volumener fandt vi, at bruddet er ubetydeligt, mens der for små volumener findes en ef\mbox{}fekt. Kvarkmassen ser ud til at ændre sig med vinklen $\theta$ som en såkaldt cut-of-ef\mbox{}fekt.\\

\noindent{I} anden del af afhandlingen præsenterer vi en optimeringsmetode til at forbedre ydelsen af Hybrid Monte Carlo-algoritmen. Den fremstillede strategi er generel og kan anvendes til stærkt vekselvirkende teorier udover standardmodellen
for hvilke behovet for præcision er begyndt at blive nødvendigt.  Strategien er baseret på eksistensen af en shadow Hamiltonian, en eksakt bevaret størrelse langs Molecular Dynamics-banen. Optimeringsmetoden er økonomisk, da den
kun kræver, at kræfterne måles, og disse er allerede nødvendige for udviklingen fra en konfiguration til den næste. Vi fandt forudsigelser for omkostningerne ved simuleringerne med en nøjagtighed på 10 \%, og vi kunne estimere de
optimale parametre for Omelyan-integratoren med mass-preconditioning og multi time-scale.\\

\noindent{Den} sidste del af afhandlingen omhandler beregningen af de elektromagnetiske korrektioner til den hadroniske del af $(g-2)$-anomalien for myonen.  For denne observable størrelse eksisterer der en langvarig uoverensstemmelse mellem
de teoretiske beregninger og de eksperimentelle resultater. Enhver 
ny teori udover stardardmodellen
stiler efter at løse denne uoverensstemmelse ved at introducere ny fysik, men før vi tyer til ny fysik, skal vi sikre os imod
ef\mbox{}fekter genereret af standardmodellen selv. For det første diskuterer vi forskellige implementeringer af QED på gitteret med 
periodiske randbetingelser.
Derefter betragter anvendelsen af disse for myon-anomalien.  I dette
eksplorative studie,  matchede vi nøje masserne af de ladede pioner både i teorien med og uden QED.  På denne måde har vi mulighed for direkte at få adgang til de elektromagnetiske korrektioner til anomalien med mindre statistiske
usikkerheder.  Vi fandt en synlig ef\mbox{}fekt på procentniveauet selvom resultatet er i overensstemmelse med nul inden for to standardafvigelser.

%
%%%%%%%%%%%%%%%%%%%%%%%%%%%%%%%%%%%%%%%%%%%%%%%%%%%%%%%%%%%%%%%%%%%%%%%%%%%%
%%%%%%%%%%%%%%%%%%%          The declaration page         %%%%%%%%%%%%%%%%%%  
%%%%%%%%%%%%%%%%%%%%%%%%%%%%%%%%%%%%%%%%%%%%%%%%%%%%%%%%%%%%%%%%%%%%%%%%%%%%
\chapter*{Publications}
\addcontentsline{toc}{chapter}{\numberline{}Publications}

This dissertation is based on the following published articles done at CP$^3$-Origins
during the three years of my Ph.D.~studies:
\begin{itemize}
\item[article] A.~Bussone, M.~Della Morte, M.~Hansen and C.~Pica,
  ``On reweighting for twisted boundary conditions,''
  Comput.\ Phys.\ Commun.\  (2017)
  doi:10.1016/j.cpc.2017.05.011
  [arXiv:1609.00210 [hep-lat]].
  \cite{Bussone:2016lty}
\item[conf.~proc.] A.~Bussone, M.~Della Morte, M.~Hansen and C.~Pica,
  ``Reweighting twisted boundary conditions,''
  PoS LATTICE {\bf 2015} (2016) 021
  [arXiv:1509.04540 [hep-lat]]. \cite{Bussone:2015yja}
\item[conf.~proc.]\sloppy A.~Bussone, M.~Della Morte, V.~Drach, M.~Hansen, A.~Hietanen, J.~Rantaharju and C.~Pica,
  ``A simple method to optimize HMC performance,''
  PoS LATTICE {\bf 2016} (2016) 260
  [arXiv:1610.02860 [hep-lat]]. \cite{Bussone:2016pmq}
\item[conf.~proc.] A.~Bussone, M.~Della Morte and T.~Janowski,
  ``Electromagnetic corrections to the hadronic vacuum polarization of the photon within QED$_{\rm L}$ and QED$_{\rm M}$,''
  [arXiv:17xx.xxxxx [hep-lat]]. \emph{To appear in EPJ Web of Conferences.}
\end{itemize}

Other pieces of work were published during the same period.
The following articles will be not discussed in the dissertation:
 \begin{itemize}
\item[article]  A.~Bussone {\it et al.} [ETM Collaboration],
  ``Mass of the b quark and B -meson decay constants from N$_f$=2+1+1 twisted-mass lattice QCD,''
  Phys.\ Rev.\ D {\bf 93} (2016) no.11,  114505
  doi:10.1103/PhysRevD.93.114505
  [arXiv:1603.04306 [hep-lat]]. \cite{Bussone:2016iua}
 \end{itemize}
 
All publications can be acquired freely as on-line e-prints at arXiv.org.

%%%%%%%%%%%%%%%%%%%%%%%%%%%%%%%%%%%%%%%%%%%%%%%%%%%%%%%%%%%%%%%%%%%%%%%%%%%
%%%%%%%%%%%%%%%% The dedication page, of you have one  %%%%%%%%%%%%%%%%%%%%  
%%%%%%%%%%%%%%%%%%%%%%%%%%%%%%%%%%%%%%%%%%%%%%%%%%%%%%%%%%%%%%%%%%%%%%%%%%%

\newpage
\thispagestyle{empty}
\settowidth{\versewidth}{Such a monstrous presumption to think}
\begin{verse}[\versewidth]
	Such a monstrous presumption to think that others could benefit from the squalid catalogue of your mistakes!
\end{verse}
\begin{flushright}
    Fellini - 8$\sfrac{1}{2}$
\end{flushright}

%%%%%%%%%%%%%%%%%%%%%%%%%%%%%%%%%%%%%%%%%%%%%%%%%%%%%%%%%%%%%%%%%%%%%%%%%%%%
%%%%%%%%%%%%%%%%%%     The acknowledgements page         %%%%%%%%%%%%%%%%%%  
%%%%%%%%%%%%%%%%%%%%%%%%%%%%%%%%%%%%%%%%%%%%%%%%%%%%%%%%%%%%%%%%%%%%%%%%%%%

\newpage
\thispagestyle{empty}

\chapter*{Acknowledgements}
\addcontentsline{toc}{chapter}{\numberline{}Acknowledgements}

I would like to express my gratitude to
Prof.~Michele Della Morte and Prof.~Claudio Pica.
They guided and helped me in understanding how to do
research during my time at CP$^3$-Origins.\\
I thank all the people that I met at CP$^3$-Origins 
for sharing with me their passion and time.\\

{\noindent I}
 am indebted to Prof.~Roberto Frezzotti 
and Dr.~Petros Dimopoulos for their time
spent giving me valuable advices.\\

{\noindent I}
 would like to thank my parents for the strength they constantly 
give me.

%%%%%%%%%%%%%%%%%%%%%%%%%%%%%%%%%%%%%%%%%%%%%%%%%%%%%%%%%%%%%%%%%%%%%%%%%%%
%%%%%%%%    tableofcontents, listoffigures and listoftables       %%%%%%%%%
%%%%%%%%        Command if you do not have  them                  %%%%%%%%%
%%%%%%%%%%%%%%%%%%%%%%%%%%%%%%%%%%%%%%%%%%%%%%%%%%%%%%%%%%%%%%%%%%%%%%%%%%%
\tableofcontents
\listoffigures
\listoftables
\clearpage

%%%%%%%%%%%%%%%%%%%%%%%%%%%%%%%%%%%%%%%%%%%%%%%%%%%%%%%%%%%%%%%%%%%%%%%%%%%
%%%%%%%%%%%%%%%%%%%%%%   END OF FRONT PAGE %%%%%%%%%%%%%%%%%%%%%%%%%%%%%%%%
%%%%%%%%%%%%%%%%%%%%%%%%%%%%%%%%%%%%%%%%%%%%%%%%%%%%%%%%%%%%%%%%%%%%%%%%%%%

\setcounter{equation}{0}

\chapter*{Introduction}
\addcontentsline{toc}{chapter}{Introduction} \markboth{INTRODUCTION}{}

The current description of interactions among elementary particles, the Standard Model of particle
physics, has passed all experimental tests, and it is a well established theory.
Despite this, it has several shortcomings. 
It is not able to explain what is the nature of dark matter, the cosmological constant,
why neutrinos oscillate and if general relativity can be unified with quantum mechanics, to name some of the main ones.
New Physics signals, beyond the Standard Model, are therefore
extensively searched by particle physicists (both experimentalists and theorists) to gain insights on 
the correct extension of it. That proceeds in two ways, through direct and indirect searches. The lattice approach
can contribute to each of them.
In this thesis we discuss some aspects of high precision calculations necessary to better understand the Standard Model
and its contribution to physical observables.\\

\noindent{\it Indirect Search}: 
new particles can produce sizeable contributions to rare decays,
which are suppressed in the Standard Model because, for example, of its flavor structure. 
For this reason a precise 
determination of the parameters controlling flavor mixing in the Standard Model is needed in the quest for
New Physics. 
Such parameters can be extracted from the comparison
of experimental data with theoretical predictions. 
The latter typically depend
on hadronic matrix elements, which are inherently non-perturbative.
Lattice QCD is the most powerful, first-principle, systematically-improvable
approach to reliably compute those. 
Partially quenched twisted boundary conditions are used 
in lattice calculations for example to compute form factors.
Twisted boundary conditions for fermion fields in spatial directions allow for 
the lattice momenta to be continuously varied.
Usually the sea quarks still obey periodic boundary conditions.
The break of unitarity, already at the perturbative level, can have an ef\mbox{}fect due to the finite volume.
Such an ef\mbox{}fect can have impact when aiming at percent level precision in calculations of physical quantities.
We devoted Chap.~\ref{chap:rtbc} to the discussion of such ef\mbox{}fects in a QCD-like theory.
We used reweighting techniques to change the boundary conditions for the sea fermions and compensate for the 
unitarity violation.\\
Staying within the indirect search, 
one of the most outstanding achievements in experimental physics is the precision measurement of the
muon anomalous magnetic moment $(g-2)_\mu$.
A 3$\sigma$ tension between theoretical calculations and 
experimental results is still present and a better theoretical understanding is needed. 
The main
uncertainty in the theoretical prediction comes from the leading, and next to leading, hadronic
vacuum polarization where the lattice can, and must, significantly improve since the new experiments
will reduce the experimental uncertainty. 
In this context we have started to investigate
electromagnetic corrections to the hadronic leading order vacuum polarization contribution to 
 $(g - 2)_\mu$ by
using QCD configurations with $N_f = 2$ dynamical quarks produced by the Coordinated Lattice Simulations (CLS) initiative.
The accuracy of lattice estimates of some relevant hadronic quantities has reached the percent
level and so electromagnetic corrections cannot be neglected any longer.
In Chap.~\ref{chap:qedlc_ho_muon} we discuss preliminary results on the QED corrections to the 
hadronic contribution to the muon anomaly.\\

\noindent{\it Direct Search}:
Another way of testing possible extensions of the Standard Model is through lattice investigations of Strongly
Interacting gauge theories with dif\mbox{}ferent gauge groups, fermions in representations possibly dif\mbox{}ferent
from the fundamental one, and scalars. 
Such theories may provide viable models for Dynamical Electro-Weak Symmetry Breaking.
An example of that is the Technicolor theory.
Since the field is relatively young on the lattice, existing results so far are at the qualitative level. 
However times are ripe now to develop methods and algorithms in order to introduce some of the 
improvements already exploited in the lattice QCD field.
For this reason any progress to render simulations easier is welcome but sophisticated algorithms and integrators
are quite hard to optimize. 
In this sense we have worked on the optimization for the mass preconditioning 
and multi time-scale integrator in Beyond Standard Model theories, 
for the case of the Omelyan integrator using the measurements 
of the driving
forces and their variances during the molecular dynamics step, in order to estimate Hamiltonian violations.
In this kind of theories the balance between forces can be very
dif\mbox{}ferent from the one in QCD and the need for a general strategy is urgent.
The above mentioned work is presented in Chap.~\ref{chap:oHMCp} were a general strategy to optimize algorithms for a 
class of Beyond Standard Model theories on the lattice is presented.\\

\noindent{ This work is divided in the following way:}
\begin{itemize}
\item In Chap.~\ref{enalgt} we review the main concepts at the heart of Lattice Gauge Theories with focus on the
relevant topics for our studies.
\item In Chap.~\ref{qedothl} we discuss how to implement QED on a finite volume with periodic boundary conditions, were the so called zero mode problem arises.
The main focus there is on the so-called L regularization of the zero mode and the massive approach.
\item In Chap.~\ref{chap:rtbc} we present the use of reweighting techniques to compensate for the breaking of unitarity caused by the use
of dif\mbox{}ferent boundary conditions for valence and sea quarks.
\item In Chap.~\ref{chap:oHMCp} we give a recipe for the optimization of HMC performances in Beyond Standard Model theories making use of the existence of an exactly conserved 
quantity, the Shadow Hamiltonian.
\item In Chap.~\ref{chap:qedlc_ho_muon} we present preliminary results on the electromagnetic corrections on the hadronic vacuum polarization relevant
for the muon anomaly. There we suggest a new strategy to isolate such ef\mbox{}fects.
\item Chap.~\ref{chap:concl_outl} contains the conclusions with the main results given in this work and outlooks.
\end{itemize}

\afterpage{\blankpage}

%% Main text
% set page number starts from 1
\pagenumbering{arabic}
\setcounter{page}{1}

%% To ensure the equation counter works correctly
\eqlabon
\eqlaboff

\setcounter{equation}{0}

\chapter{Elements of non-Abelian lattice gauge theories}
\label{enalgt}

In this chapter we briefly review the relevant concepts for the simulations of \emph{non-Abelian} quantum gauge theories on the lattice.
We specialize to the known example of Quantum Chromodynamics (QCD).
Lattice regularization is a first principle systematically-improvable tool to address non-perturbative physics.
Lattice QCD has been successfully applied to calculations of many properties of hadrons.
Most importantly for determinations of fundamental parameters of the Standard Model, such
as quark masses, strong coupling constant, and calculations of form factors,
needed for the Cabibbo-Kobayashi-Maskawa matrix elements.
Furthermore the lattice gives good control of statistical as well as systematic errors, e.g.~continuum, infinite volume and chiral extrapolations. \\
This chapter does not intend to give a complete description of non-Abelian lattice gauge theories
but more of a review on the main concepts covered by this work.

The chapter is organized as follows; in Section~1 we briefly review the 
fields content in QCD and the definition of a lattice.
In section~2 we review the symmetries associated to a QCD-like theory
and how they can be realized at the quantum level.
There 
we concentrate on \emph{chiral} symmetry, particularly relevant 
for Wilson fermions on the lattice.
In Section~3 we sketch the renormalization procedure on a general ground 
and we review the concept of \emph{asymptotic freedom} and \emph{dimensional transmutation}.
In Section~4 we present the lattice regularization and postulate the 
gauge and fermion action in the Wilson formulation.
We also review the Nielsen-Ninomiya no-go theorem, which predicts the existence of
\emph{doublers}, cured with Wilson fermions.
In Section~5 we show how to systematically reduce the discretization ef\mbox{}fects to $\ord(a^2)$ for
on-shell quantities. We specialize to the  $\ord(a)$ improvement of the action
through the introduction of the so-called \emph{clover} term.
Section~6 contains a discussion on the the critical mass, i.e.~the bare mass
for which the Wilson fermions on the lattice become massless, and how it emerges 
from the explicit \emph{hard} breaking of chiral symmetry.
Section~7 contains a general discussion about the Hybrid Monte Carlo method relevant 
for the generation of gauge configurations on the lattice.

\section{QCD formal theory}

The theory of Quantum Chromodynamics (QCD) is the result of many ideas and experimental results, its aim is to explain the strong force between elementary constituents of matter: quarks and gluons.
	The particles that participate in the strong force are the \emph{hadrons} and they are divided in two classes:
	\begin{itemize}
	\item\emph{Baryons}, e.g.~protons, neutrons, etc.\mbox{}, that have spin 1/2, 3/2, $\dots$ (fermions),
	\item\emph{Mesons}, e.g.~pions, kaons, etc.\mbox{}, that have integer spin 0, 1, $\dots$ (bosons).
	\end{itemize}		
	The concept of quarks arose from the need to have a realization at the Lagrangian level of the $\mathbf{SU}(N_f)$ flavor symmetry  
	that is observed in the low mass spectrum of mesons and baryons
	($N_f$ being the number of fermion species).\\	
	The starting point for the quantization procedure through the Feynman path integral is the classical action, given as 
	the integral over the space-time volume of the Lagrangian density
	\begin{equation}
	\text{S}\left( A, \psi, \overline{\psi}\right)
	=\int \de^4x\,  \mathcal{L} \left( A, \psi, \overline{\psi}\right) .
	\end{equation}
	In QCD the action is a functional of gluon ($A$) and quark fields ($\psibar, \psi$) and 
	is invariant under $\mathbf{SU}(N_c)$ gauge transformations, with $N_c=3$ number of colors in the QCD case.	
	The QCD Lagrangian describes the interaction between spin 1/2 quarks with mass $m$ and massless spin 1 gluons. It is given by
	\begin{equation}
	\mathcal{L}_\text{QCD}=-\frac{1}{4} F^a_{\mu\nu} F^{\mu\nu}_a
	+\sum_{f}^{N_f}\psibar_f \left(i{\not}D-m_f\right) \psi_f,
	\label{classical_qcd_lagrangian}
	\end{equation}
	where natural units are used ($\hbar = c =1$) and $F^a_{\mu\nu}$ is the field strength tensor
	\begin{equation}
	F^a_{\mu\nu}=\partial_\mu A^a_\nu-\partial_\nu A^a_\mu - g f^{abc} A_\mu^b A_\nu^c.
	\label{field_strenght_tensor}
	\end{equation}
	The indexes $a, b$ and $c$
	run over the $N_c^2-1$ color degrees of freedom, $f^{abc}$
	 are the structure constants of 
	 the $\mathbf{SU}(N_c)$ group, 
	 $g$ is the coupling constant of the strong interaction,
	 ${\not}D=\gamma_\mu D^\mu$ is the covariant derivative $D_\mu = \partial_\mu + ig A_\mu$, the $\gamma$ matrices satisfy the anticommutation relation $\{\gamma^\mu,\gamma^\nu\}=2g^{\mu\nu}$, where $g^{\mu\nu}$ is the metric tensor of the Minkowski space-time. 
	In principle nothing forbids to write a $\theta$-term in the action as
	\begin{align}
	\label{eq:axion}
	\theta \epsilon^{\mu\nu\sigma\rho} F^a_{\mu\nu} F^a_{\rho\sigma},
	\end{align}
	 that breaks parity and time-reversal symmetries, with $ \epsilon^{\mu\nu\sigma\rho}$ antisymmetric tensor and $\theta$ real parameter.
	 Here we assume those symmetries to be conserved, as there is no experimental evidence of such a violation in QCD.	 
	 
	The third term of the gluon field strength tensor in Eq.~\ref{field_strenght_tensor} gives rise to a three- and four-gluons vertexes in Eq.~\ref{classical_qcd_lagrangian}. 
	That is a peculiarity of non-Abelian theories, in fact there is no analogous term in QED where the gauge group is Abelian.\\
	The dimensions, in unit of mass, of the fields are:
	\begin{itemize}
	\item fermion spin 1/2 fields $\left[\psi\right]$=3/2,
	\item boson spin 1 fields $\left[A\right]$=1.
	\end{itemize}
	
	QCD is formulated in Minkowski space and by employing a Wick transformation we \emph{rotate} to the Euclidean space. 
	The rotation can be done if the Osterwalder-Schrader reflection positivity condition is fulfilled \cite{Osterwalder:1973dx, Osterwalder:1974tc}.
	The Euclidean continuum space is then replaced by a discretized finite box (lattice)
	\begin{align}
	\nn
	\Lambda = \left\{ x=(x_0, x_1, x_2, x_3): x_\mu = a n_\mu, \; n_\mu \in \left[ 0, L_\mu \right)\subset \mathbb{Z}  \right\},
	\end{align}
	where $a$ is the separation between points on the lattice (\emph{lattice spacing})
	and $L_\mu$ is the lattice extent in direction $\mu$. 
	Boundary conditions have to be chosen and usually periodic ones are employed to retain translational invariance of the theory.\\	
	The quantization procedure is then carried out by means of the Feynman path integral.
	The partition function is defined as 
	\begin{align}
	\mathcal{Z} = \int \De[A,\psi,\psibar] \e^{-{\rm S}_{{\rm QCD}}\left( A, \psi, \overline{\psi}\right)} \, ,
	\end{align}
	where $\De[A,\psi,\psibar]$ is the path integral measure.
	The path integral defined in this way is ill-defined because of gauge ambiguities, we will introduce the gauge fixing procedure in Sect.~\ref{sect:faddeev-popov}.
	The expectation value of an operator $\mathcal{O}$ is then given as
	\begin{align}
	\langle \mathcal{O} \rangle = \frac{1}{\mathcal{Z}} \int \De[A,\psi,\psibar] \mathcal{O}(A, \psi, \psibar)\, \e^{-{\rm S}_{{\rm QCD}}\left( A, \psi, \overline{\psi}\right)}  .
	\end{align}
		
	\section{Symmetries of QCD-like theories}
	\label{sec:symmetriesqcd}
	
	We consider a QCD-like formal theory in Euclidean space
	with a degenerate doublet of quarks\footnote{This choice
	correspond to the so-called \emph{isospin symmetric limit}.} $\Psi=\left(u,~d\right)^t$, i.e.~$m_u=m_d\equiv m$.
	 The Lagrangian for such a theory reads
\begin{equation}
\label{QCD_continuum_lagrangian}
\mathcal{L}[A, \Psi, \Psibar] 
= \mathcal{L}_\text{YM}
+\mathcal{L}_\text{F} 
= \frac{1}{4} F^a_{\mu\nu} F^a_{\mu\nu}+\overline{\Psi}(x)\left({\not}\text{D}+M\right)\Psi(x),
\end{equation}
where $M$ is the mass matrix in \emph{flavor} space.
The above Lagrangian exhibits the following symmetry groups:
\begin{itemize}
\item \emph{Poincaré = Lorentz + space-time translations} group, which is the semi-direct product of the groups
\begin{equation*}
\mathbf{T}\rtimes\mathbf{SO}(4).
\end{equation*}
\item Local (gauge) group of \emph{color}, $N_c=3$
\begin{equation*}
\mathbf{SU}(N_c).
\end{equation*}
\item When the masses vanish, $m=0$, there is the additional global \emph{chiral} symmetry
\begin{align*}
\mathbf{U}(2)_{\rm L}\otimes\mathbf{U}(2)_{\rm R}\sim\mathbf{U}(1)_{\rm V}\otimes\mathbf{U}(1)_{\rm A}\otimes\mathbf{SU}(2)_{\rm L}\otimes\mathbf{SU}(2)_{\rm R}.
\end{align*}
\end{itemize}

Symmetries lead to conservation of \emph{Noether currents}, $\partial_\mu J_\mu=0$, in the classical theory. By integrating over space the temporal component of the current we obtain the conserved \emph{charge} $Q$ along a solution 
of the equation of motion.
	\begin{align}
	& Q \equiv\int_{\mathbb{R}^3} J_0\, \de^3x, \\
	&\int_{\mathbb{R}^3}\partial_\mu J_\mu\,\de^3x
	=\int_{\mathbb{R}^3}\partial_0 J_0\,\de^3x
	+\int_{\mathbb{R}^3}\underline{\nabla}\cdot\underline{ J}\,\de^3x
	=\frac{\partial}{\partial x_0}\int_{\mathbb{R}^3} J_0\,\de^3x 
	= \frac{\partial Q}{\partial x_0} \equiv 0,
	\end{align}
	where we have used that fields are localized in space and they vanish \quotes{quick enough} at infinity\footnote{On the 
	lattice we can think to have chosen periodic boundary conditions}. The charge operator generates symmetry transformations when 
	acting on the fields.

The symmetries of a quantum field theory can be realized in three dif\mbox{}ferent ways:
\begin{itemize}
\item à la Wigner (exact symmetry).
\item à la Nambu-Goldstone (spontaneously broken symmetry).
\item Anomalous (not realized symmetry).
\end{itemize}

\paragraph{Wigner}

The vacuum of the theory $|0\rangle$ is invariant under the correspondent transformation, namely
\begin{align}
\hat{\text{Q}}|0\rangle=0.
\end{align}
In this situation particles form symmetry \emph{multiplets} and
 particles in the same multiplet have same mass.

\paragraph{Nambu-Goldstone}

The vacuum is not invariant under the transformation, i.e.~
\begin{align}
\hat{\text{Q}}|0\rangle\neq0.
\end{align}
In that case particles do not form degenerate multiplets.
Nonetheless \emph{Goldstone theorem} \cite{Goldstone:1961eq}
 states: for each charge $Q$ of the spontaneously broken symmetry exists a massless scalar particle, called \emph{Goldstone boson}.\\

We remark that Wigner and Nambu-Goldstone realizations do not exclude each
 other. For example a symmetry group $\mathbf{G}$ can be partially broken to a proper subgroup $\mathbf{H}\subset\mathbf{G}$. 
The symmetry in $\mathbf{H}$ is then realized à la Wigner while the remnant, the coset
 $\mathbf{G/H}$, is realized à la Nambu-Goldstone.

\paragraph{Anomalous}
	
	A symmetry is \emph{anomalous} if it is an exact symmetry at the level of the action but it is not preserved as a symmetry of the path integral. If the Action is invariant under the symmetry, 
	the invariance can be \quotes{lost} 
	from the integration
	measure or from the specification of boundary conditions. 
	Even though the anomaly raises from a particular choice of the regularization it is independent of that.
	If a global symmetry is anomalous then it contributes finitely to a physical process, as the $\mathbf{U}_{\rm A}(1)$ in QCD. 

\subsection{$\mathbf{SU}(2)_{\rm L}\otimes\mathbf{SU}(2)_{\rm R}$}
	
	We start from the \emph{vector} and \emph{axial} transformations
	\begin{align}
	\label{SU_2_V_transformation}
	\mathbf{SU}(2)_{\rm V}:\quad&\Psi\rightarrow\e^{-i\omega_f\sigma_f/2}\Psi\simeq \left(1-i\omega_f\frac{\sigma_f}{2}\right)\Psi,\quad\overline{\Psi}\rightarrow\overline{\Psi}\e^{i\omega_f\sigma_f/2}\simeq \Psibar\left(1+i\omega_f\frac{\sigma_f}{2}\right),\\
	\label{SU_2_A_transformation}
	\mathbf{SU}(2)_{\rm A}:\quad&\Psi\rightarrow\e^{i\omega_f\sigma_f\gamma_5 /2}\Psi\simeq \left(1+i\omega_f\frac{\sigma_f}{2}\gamma_5\right)\Psi,
	\quad\overline{\Psi}\rightarrow\overline{\Psi}\e^{i\omega_f\sigma_f\gamma_5 /2}\simeq \Psibar\left(1+i\omega_f\frac{\sigma_f}{2}\gamma_5\right),
	\end{align}
	where the $\sigma_f$ are the Pauli matrices in flavor space.
	The associated conserved currents are
	\begin{align}
	\label{eq:noether_currents}
	J_{\mu}^{{\rm V}f}&
	\equiv V_\mu^f=\frac{1}{2}\overline{\Psi}\sigma_f\gamma_\mu\Psi,
	\quad J_{\mu}^{{\rm A}f}\equiv A_\mu^f
	=\frac{1}{2}\overline{\Psi}\sigma_f\gamma_\mu\gamma_5\Psi.
	\end{align}
	From the above currents we can define the charges $\hat{\text{Q}}_{\rm V}^f$ and $\hat{\text{Q}}_{\rm A}^f$, and by considering them as operators we find that they obey to the so-called \emph{charge algebra}
	\begin{equation}
	\left[\hat{\text{Q}}_{\rm V}^f,\hat{\text{Q}}_{\rm V}^g\right]=i\epsilon^{fgh}\hat{\text{Q}}_{\rm V}^h,\quad\left[\hat{\text{Q}}_{\rm A}^f,\hat{\text{Q}}_{\rm V}^g\right]=i\epsilon^{fgh}\hat{\text{Q}}_{\rm A}^h,\quad\left[\hat{\text{Q}}_{\rm A}^f,\hat{\text{Q}}_{\rm A}^g\right]=i\epsilon^{fgh}\hat{\text{Q}}_{\rm V}^h.
	\end{equation}
	The Pauli matrices are the generators of the transformations and they belong to the \emph{Lie algebra} of $\mathbf{SU}(2)$ that we denote as\footnote{The $\mathfrak{su}(2)$ algebra contains \emph{hermitian} and \emph{traceless} 2$\times 2$ matrices.} $\mathfrak{su}(2)$.
	 Even though the $\mathbf{SU}(2)_{\rm A}$ is not closed, as clear from the charge algebra relations, we can factorize the vector and axial algebras by considering the two groups $\mathbf{SU}(2)_{\rm L}$ and $\mathbf{SU}(2)_{\rm R}$.\\	
	In order to write the \emph{left} and \emph{right} transformations we define the correspondent \emph{projectors}
	\begin{equation}
	\text{P}_{{\rm L}/{\rm R}}=\frac{1 {\tiny -/+} \gamma_5}{2},
	\end{equation}
	from which we write down the projected spinors
	\begin{equation}
	\Psi_{{\rm L}/{\rm R}}=\frac{1 {\rm {\tiny -/+}} \gamma_5}{2}\Psi,\quad\overline{\Psi}_{{\rm L}/{\rm R}}=\overline{\Psi}\frac{1 {\tiny +/-} \gamma_5}{2}.
	\end{equation}
	The transformations in the new basis are
	\begin{align}
	\mathbf{SU}(2)_{\rm L}:\quad&\Psi_{\rm L}\rightarrow\e^{i\omega_{{\rm L}f}\sigma_f}\Psi_{\rm L}\simeq \left(1+i\omega_{{\rm L}f}\sigma_f\right)\Psi_{\rm L},\quad\overline{\Psi}_{\rm L}\rightarrow\overline{\Psi}_{\rm L}\e^{-i\omega_{{\rm L}f}\sigma_f}\simeq \Psibar_{\rm L}\left(1-i\omega_{{\rm L}f}\sigma_f\right)\, ,\\
	\mathbf{SU}(2)_{\rm R}:\quad&\Psi_{\rm R}\rightarrow\e^{i\omega_{{\rm R}f}\sigma_f}\Psi_{\rm R}\simeq \left(1+i\omega_{{\rm R}f}\sigma_f\right)\Psi_{\rm R},\quad\overline{\Psi}_{\rm R}\rightarrow\overline{\Psi}_{\rm R}\e^{-i\omega_{{\rm R}f}\sigma_f}\simeq \Psibar_{\rm R}\left(1-i\omega_{{\rm R}f}\sigma_f\right)\, .
	\end{align}
	In this form the two Lie algebras are closed, i.e.~they are factorized and written as $\mathfrak{su}(2)_{\rm L}\otimes\mathfrak{su}(2)_{\rm R}$.
	If we put $\omega_{\rm L}=\omega_{\rm R}$ we obtain again the vector case, indeed the latter is a \emph{sub-algebra} 
	\begin{equation*}
	\mathfrak{su}(2)_{\rm V}\subset\mathfrak{su}(2)_{\rm L}\otimes\mathfrak{su}(2)_{\rm R}.
	\end{equation*}
	
	When the quark masses do not vanish, $m\neq0$, some of the aforementioned symmetries are broken. The group $\mathbf{SU}(2)_{\rm L}\otimes\mathbf{SU}(2)_{\rm R}$ suf\mbox{}fers from two kind of symmetry breaking:
	\begin{enumerate}
	\item \emph{explicit}, but \emph{soft}\footnote{In this context means that the introduced divergences are only logarithmic.}, when $m\neq0$,
	\item \emph{spontaneous}, à la Nambu-Goldstone.
	\end{enumerate}
	
	\paragraph{Explicit symmetry breaking}
	
	This kind of breaking is due to dif\mbox{}ferent and not vanishing quark masses (in QCD only).\\
	When the breaking is due to an operator with dimension smaller than 4 the UV divergences are the same of the theory without its presence.
	The parameter, in front of the operator, is then multiplicatively renormalizable \cite{Weinberg:1951ss}.
	
	\paragraph{Spontaneous symmetry breaking}
	
	The group is spontaneously broken to the vector subgroup
	\begin{equation*}
	\mathbf{SU}(2)_{\rm L}\otimes\mathbf{SU}(2)_{\rm R}\rightarrow\mathbf{SU}(2)_{\rm V}
	\end{equation*}
	which remain an exact symmetry since we are working in the isospin symmetric limit.\\
	The $\mathbf{SU}(2)_{\rm V}$ symmetry is realized à la Wigner while $\mathbf{SU}(2)_{\rm A}$ à la Nambu-Goldstone. This means that the QCD vacuum has to satisfy
	\begin{equation}
	\hat{\text{Q}}_{\rm V}^f|0\rangle=0,\quad\hat{\text{Q}}_{\rm A}^f|0\rangle\neq0.
	\end{equation}
	The $N_f^2-1$ Goldstone bosons are the pions (in $N_f=2$), which have zero mass for vanishing quark masses $m=0$.

	\section{Renormalization}
	
	The Lagrangian in Eq.~\ref{classical_qcd_lagrangian} is the starting point for the quantization of the theory and is called \emph{bare} Lagrangian.
	The procedure of quantization leads to interpret fields as operators that act on the Hilbert space of the physical states. 
	For a given physical process, during the evolution from the initial to the final state, intermediate of\mbox{}f-shell particles also play a r\^ole. 
	These virtual degrees of freedom lead to UV divergences in 
	the pertubative expansion of the Green's functions in the quantum theory. 
	The perturbative expansion is carried out through a loop expansion that is a formal expansion in powers of $\hbar$ of the functional generator of the theory.\\	
	The strategy to give meaning to a quantum field theory is divided in two steps:
	\begin{enumerate}
	\item Regularization.
	The regularization modifies the behavior in the high energy region 
	by introducing, for example, an energy cutof\mbox{}f\footnote{In the case of a lattice this is done by lattice spacing $a\propto 1 /\Lambda$.} $\Lambda$. The regularized theory always leads to finite results.
	\item Renormalization.
	One calculates, in the regularized theory,
	 physical quantities in number equal to the number of free parameters that enter in the bare Lagrangian. 
	 Holding fixed these calculated quantities one performs the limit $\Lambda\rightarrow\infty$, 
	that generates in the couplings a dependence upon $\Lambda$.
	\end{enumerate}
	If at the end of the procedure all the observables stay finite as $\Lambda\rightarrow\infty$ then the theory is renormalizable. 
	One can prove this is the case in $d=4$ space-time dimensions if the Lagrangian density contains only operators\footnote{We can also add operators 
	with $d_{O}>4$ scaled by inverse 
	powers of the UV cutof\mbox{}f ($1/\Lambda^{d_O-4}$)
	as long as we do not include the associated parameters in the set 
	of free parameters to be kept fixed in the renormalization step.
	} with dimension $d_{O}\leq 4$.\\
	The choice of the renormalization conditions (e.g.~the Green's functions on which they are imposed and their precise values) and of the renormalization point $\mu$ (i.e.~the momentum scale at which the conditions are imposed) is conventional. 
	Such choices characterize what is called 
	\emph{renormalization scheme}.
	There are important remarks, which we should have in mind when we build a renormalizable field theory:
	\begin{itemize}
	\item In a given renormalization scheme, renormalized Green's functions (and derived quantities) are independent of the UV-regularization. This leads to the concept of \emph{universality} in Quantum Field Theory.
	\item Physical observables are independent of the renormalization scheme.
	\item Renormalized parameters as well as renormalization conditions can be chosen conventionally (e.g.~mass independent schemes). Conventional renormalized parameters can always be eliminated in favor of (an equal \emph{finite} number of) physical quantities. A theory predicts only relationships among physical quantities.
	\end{itemize}
	
		\subsection{Asymptotic freedom}
		\label{ssec:asyfreed}
		
		The running of the coupling constant is governed by
		the QCD beta function, at one-loop  in perturbation it  is found to be
		\begin{equation}
		\mu\frac{\partial g}{\partial\mu}=\beta(g)=-\frac{g^3}{16\pi^2}\left(\frac{11 N_c-2N_f}{3}\right)+\ord\left(g^5\right)\equiv -\frac{b}{2}g^3,
		\label{beta_function}
		\end{equation}
		where $N_c=3$ and $N_f\leq6$ (\quotes{active} flavors). Upon increasing the energy scale $\mu$, the strong charge decreases and in the limit of high energy scale it vanishes. 
		This is the reason why QCD is an asymptotically free theory, for $b>0$.\\
		By solving the one-loop beta function in Eq.~\ref{beta_function} we can investigate the
		behavior of the strong coupling constant.
		When $g^2(\mu_1)$ and $g^2(\mu_2)$ are both in the perturbative region, we obtain
		\begin{equation}
	g^2(\mu_2)=\frac{g^2(\mu_1)}{1+bg^2(\mu_1)\ln \mu_2 / \mu_1}.
		\label{renormalized_coupling_constant}
		\end{equation}
		From the above equation we see that by fixing $g(\mu_1)$ to be in the peturbative region then for $\mu_2 > \mu_1$ the coupling is decreasing and still in that region.\\
		Usually in the literature the $\Lambda_{\text{QCD}}$ parameter is defined, i.e.~an integration constant of the Renormalization Group Equations.
		At one-loop it is given by
		\begin{equation}
		\label{lambda_qcd_relation}
		g^2(\mu)=\frac{1}{b\ln\frac{\mu}{\Lambda_{\text{QCD}}}}.
		\end{equation}
		From the knowledge of the value $g^2$ at the scale $\mu$ one obtains at which scale the perturbation theory ceases to be valid.
		For $\mu\simeq \Lambda_{\text{QCD}}$ the coupling becomes large. 
		In that region hadrons are the real physical degrees of freedom.		
		A state can propagate freely if and only if it is colorless, according
		to the \emph{confinement hypothesis}, i.e.~in asymptotically free theories only singlet states 
		under gauge force can be asymptotic states.
		\subsection{Hadron masses}
		
		Let be $M_H$ the mass of a given hadron that is a composite state of quarks and gluons in QCD with massless quarks. For dimensional reasons the renormalized mass at the renormalization point will be
		\begin{equation}
		M_H=\mu f_H\left(g(\mu)\right),
		\end{equation}
		where $f_H$ is a typical dimensionless function of the given hadron. The hadron mass cannot depend on the choice of the renormalization point, meaning that
		\begin{equation}
		\frac{\de M_H}{\de\mu}=0=f_H(g)+\beta(g)\frac{\partial f_H}{\partial g}.
		\end{equation}
		By using the one-loop beta function in Eq.~\ref{beta_function} we find the solution
		\begin{equation}
		M_H=c_H\mu\exp\left({-\frac{1}{ b g^2}}\right),
		\end{equation}
		where $c_H$ is an integration constant. 
		If we substitute $\Lambda_{\text{QCD}}$ from Eq.~\ref{lambda_qcd_relation} we may rewrite the solution as
		\begin{equation}
		M_H=c_H\Lambda_{\text{QCD}}.
		\label{eq:qcd_hadron_masses}
		\end{equation}
		The above equation shows the non-perturbative 
		feature of the hadron masses. 
		All the hadronic masses can be expressed in terms of the same fundamental scale $\Lambda_{\text{QCD}}$.

\section{Lattice regularization}

The Euclidean lattice quantum field theory, postulated by Wilson \cite{Wilson:1974sk}, represents an invariant regularization of the theory described by the Lagrangian in Eq.~\ref{QCD_continuum_lagrangian}, where the r\^ole of the cut-of\mbox{}f is played by the inverse of the lattice spacing $a$.

The gauge fields on the lattice are introduced as elements of the gauge group $U_\mu(x)\in\mathbf{SU}(N_c)$. 
They are related to the algebra through the exponentiation
\begin{align}
U_\mu(x) = \exp \left[  i a A^a_\mu(x+a\hat{\mu})T^a \right],
\end{align}
with $T^a$ the $N_c^2-1$ generators.
They live on the links that connect the site $x$ to $x+a\hat{\mu}$ and for this reason they are usually called \emph{links}.
Fermions are described by Grassmann variables, i.e.~anticommuting fields, and they live on the sites of the lattice and carry
Dirac and color indexes. 
The way to deal with fermions is to introduce \emph{pseudofermions}, i.e.~boson variables with the wrong statistics, in the generation of configurations, as we will see in Sect.~\ref{sect:HMCsgt}. 
In addition, for fermionic observables, Wick contractions are employed, meaning that we calculate the exact fermionic contribution on each gauge field background.

Any form of the action on the lattice is allowed as long as it reproduces the correct formal theory Eq.~\ref{QCD_continuum_lagrangian} in the naive continuum limit, i.e.~$a\rightarrow 0$, it is gauge invariant and a Lorentz scalar.
In this sense a lattice action is not unique.
In the lattice community a variety of gauge actions have been employed: 
Wilson plaquette \cite{Wilson:1974sk}, 
Luscher-Weisz \cite{Luscher:1984xn}, 
Iwasaki glue \cite{Iwasaki:1985we}, etc;
and fermion actions: Wilson \cite{Wilson:1974sk},
staggered \cite{Kogut:1974ag},
twisted mass \cite{Frezzotti:1999vv}, 
domain wall \cite{Kaplan:1992bt}, 
overlap, etc.
We present here the Wilson plaquette gauge action and Wilson fermions, relevant for the rest of the work.

\subsection{Wilson plaquette gauge action}

The Wilson plaquette action is given by
\begin{align}
\label{eq:wilson_plaquette}
{\rm S}_{\rm G} = \beta \sum_{\mu < \nu} \left( 1- \frac{1}{N_c}\Ree\tr P_{\mu\nu}  \right),
\end{align}
where $\beta = 2 N_c/g_0^2$, and $P_{\mu\nu}$ the plaquette in the $(\mu, \nu)$ plane,
\begin{align}
P_{\mu\nu} = U_{\mu}(x) U_\nu (x+a\hat{\mu}) U_\mu^\dagger (x+a\hat{\nu}) U_\nu^\dagger (x).
\end{align}
The action is gauge invariant and reproduces the formal Yang-Mills (YM) action up to $\ord(a^2)$ discretization errors.

\subsection{Fermionic action}

We write the action for a fermion as
\begin{align*}
{\rm S}_{\rm F} = \sum_{x, y\in \Lambda} \psibar(x) D(x, y) \psi(y).
\end{align*}
In many Beyond Standard Model (BSM) strongly interacting theories the 
representation for the fermions is varied. 
For the sake of simplicity we consider throughout this introductory chapter the fermion to transform under the fundamental representation of the gauge group.

\subsubsection{Na\"ive fermions}

The na\"ive fermion action is given by the discretized version of the formal one by replacing the derivative with the 
central discretized derivative,
\begin{equation}
\label{eq:naive}
	{\rm S}_{\rm NF} = a^4 \sum_{x\in\Lambda}\sum_\mu	\overline{\psi}(x)\left[\gamma_\mu \overline{\nabla}_\mu  + m_0  \right]\psi(x).
		\end{equation}
Where we have defined the $\nabla^{\pm}$ forward and backward derivatives, and $\overline{\nabla}$ the symmetric derivative
		\begin{align}
		\nn
		\nabla^+_\mu\psi(x)&=\frac{1}{a}\left[U_\mu(x)\psi(x+a\hat{\mu})-\psi(x)\right],\\
		\nn
		\nabla^-_\mu\psi(x)&=\frac{1}{a}\left[\psi(x)-U_\mu^\dagger(x-a\hat{\mu})\psi(x-a\hat{\mu})\right],\\
		\overline{\nabla}_\mu\psi(x)&\equiv\frac{\nabla^+_\mu+\nabla^-_\mu}{2}=\frac{1}{2a}\left[U_\mu(x)\psi(x+a\hat{\mu})-U_\mu^\dagger(x-a\hat{\mu})\psi(x-a\hat{\mu})\right].
		\end{align}
The propagator at tree-level, i.e.~$U_\mu=\I$, is found by transforming in Fourier space and considering the inverse of the bilinear operator in the action in Eq.~\ref{eq:naive},
\begin{equation}
		 \widetilde{D}^{-1}(p)=\frac{m-\frac{i}{a}\gamma^\mu \sin(ap_\mu)}{m^2+\sum_\mu\left[\frac{\sin(ap_\mu)}{a}\right]^2}.
		\end{equation}
The poles of the propagator correspond to on-shell particles that satisfies the correct energy dispersion relation.
In the case of na\"ive fermions we obtain $2^4=16$ fermions instead of one, that means there are 15 unphysical fermions. This is the well known \emph{doublers} problem.

\subsubsection{Nielsen-Ninomiya no-go theorem}

A very nice review on the exact chiral symmetry on the lattice is given in Ref.~\cite{Luscher:1998pqa}. 
%Let us consider the free massless $m_0=0$ Dirac operator.
The Nielsen-Ninomiya no-go theorem states that the following properties cannot hold at the same time:
\begin{enumerate}
\item Ultra-locality\footnote{The interaction range is spread over a finite number of points on the lattice.}. The Dirac operator in momentum space $\widetilde{D}(p)$ is analytic and periodic function of the momenta $p_\mu$ with period $2\pi / a$.
\item Correct continuum behavior. For momenta far below the cutof\mbox{}f $p_\mu \ll \pi / a$ the Dirac operator is $\widetilde{D}(p) = i\gamma_\mu p_\mu + \ord(ap^2)$. 
\item No doublers. $\widetilde{D}(p)$ is invertible $\forall p_\mu \neq 0$ in the first Brillouin zone.
\item Exact chirality. $\{ D, \gamma_5 \} = 0$ in the limit of vanishing mass.
\end{enumerate}

\subsubsection{Wilson fermions}

In order to avoid doublers we introduce the Wilson term, $-\frac{ar}{2}\psibar\nabla_\mu^-\nabla_\mu^+\psi$ an irrelevant operator\footnote{Operators of dimension $d>4$ multiplied by a 
power $d-4$ of the lattice spacing $a^{d-4}$. These are called irrelevant in the sense they do not destroy the renormalizability of the 
theory.}. 
Such an operator breaks in a \emph{hard} way the symmetry $\mathbf{SU}(2)_{\rm L}\otimes\mathbf{SU}(2)_{\rm R}$.
Chiral symmetry is restored in the continuum and massless limit, but on the lattice the limit $m_0\rightarrow0$ does no longer 
correspond to the chiral limit. 
See Sect.~\ref{subsect:pcac} for further details.\\
The Wilson fermionic Lagrangian on the lattice is written as
\begin{equation}
\mathcal{L}_{\rm F}=\overline{\psi}(x)\left[\gamma_\mu\overline{\nabla}_\mu-\frac{ar}{2}\nabla_\mu^-\nabla_\mu^+ +m_0\right]\psi(x),
\end{equation}
with $r$ the Wilson parameter, usually set to one, and we have introduced the Laplacian
\begin{align}
\nn
\nabla^-_\mu\nabla^+_\mu\psi(x)&=\frac{1}{a^2}\left[U_\mu(x)\psi(x+a\hat{\mu})-U_\mu^\dagger(x-a\hat{\mu})U_\mu(x)\psi(x)-\psi(x)+U_\mu^\dagger(x-a\hat{\mu})\psi(x-a\hat{\mu})\right]\\
		&=\frac{1}{a^2}\left[U_\mu(x)\psi(x+a\hat{\mu})+U_\mu^\dagger(x-a\hat{\mu})\psi(x-a\hat{\mu})-2\psi(x)\right].
		\end{align}
The Lagrangian on the lattice reads		
		\begin{align}
		\nn
\mathcal{L}_{\rm F}&=\overline{\psi}(x)\left(m_0+\frac{4r}{a}\right)\psi(x)+\\
&\phantom{=\overline{\psi}(x)} 
+\frac{1}{2a}\sum_\mu\left[\,\overline{\psi}(x)U_\mu(x)(\gamma_\mu-r)\psi(x+a\hat{\mu})-\overline{\psi}(x)U^\dagger_\mu(x-a\hat{\mu})(\gamma_\mu+r)\psi(x-a\hat{\mu})\right]\\
\nn
&=\overline{\psi}(x)\left(m_0+\frac{4r}{a}\right)\psi(x)\\
\label{eq:wilson_fermions}
&\phantom{=\overline{\psi}(x)}
+\frac{1}{2a}\sum_\mu\left[\,\overline{\psi}(x)U_\mu(x)(\gamma_\mu-r)\psi(x+a\hat{\mu})-\overline{\psi}(x+a\hat{\mu})U^\dagger_\mu(x)(\gamma_\mu+r)\psi(x)\right],
\end{align}
in the last step we have shifted the space-time of the second term since when we consider the action, 
that is a sum over the whole space-time, a shift does not af\mbox{}fect the result.\\
The Wilson term gives an additional mass proportional to $2 r / a$ to all the doublers. 
Hence in the continuum limit they get infinite mass and they decouple from the theory.
It should be mentioned that the Wilson fermion action reproduces the formal fermionic one up to $\ord(a)$ ef\mbox{}fects.
The Osterwalder-Schrader reflection positivity condition was proven
to hold for Wilson fermions at finite lattice spacing \cite{Luscher:1976ms}.

\section{Symanzik improvement programme}

As we saw in the previous section discretization errors are typically of $\ord(a^2)$ for the gauge action and of $\ord(a)$ for the fermionic part.
The Symanzik improvement programme \cite{Symanzik:1983dc, Symanzik:1983gh} gives a systematical reduction of the discretization ef\mbox{}fects.
The central idea is to add irrelevant terms that eliminate the unwanted ef\mbox{}fects. 
To do so we write an ef\mbox{}fective action (and operators) that describes the behavior of the lattice theory at finite lattice spacing toward the continuum limit.
Let us assume that the lattice action ${\rm S}$ generates $\ord(a)$ discretization errors and it depends on the field $\phi$ and a coupling $g_0$.
The ef\mbox{}fective action can be written as 
\begin{align}
{\rm S}_{\rm ef\mbox{}f} &= {\rm S}_0 + a \int\de^4 x \mathcal{L}_1 (x) + \ord(a^2),
\end{align}
where ${\rm S_0}$ is the formal-continuum action we would like to reproduce as $a\rightarrow 0$, and $\mathcal{L}_1$
is given by linear combination of operators of dimension 5 that have the same symmetry of the lattice action ${\rm S}$.
Similarly we write for the field
\begin{align}
\phi_{\rm ef\mbox{}f} (x) = \phi_0 (x) + a \phi_1(x) + \ord(a^2),
\end{align}
where $\phi_0$ is the continuum counterpart and $\phi_1$ is a $d+1$ dimensional operator, if $d=[\phi_0]$, with same quantum numbers as $\phi$.\\
The strategy is to look at expectation values of composite ef\mbox{}fective operators and expand it in $a$ through the formulae given above,
\begin{align}
\nn
\langle \phi_{\rm ef\mbox{}f}(x_1) \phi_{\rm ef\mbox{}f}(x_2 ) \dots \phi_{\rm ef\mbox{}f}(x_n) \rangle 
= & \langle \phi_0(x_1) \phi_0(x_2 ) \dots \phi_0(x_n) \rangle  - a \int \de^4 z \langle \phi_0(x_1) \phi_0(x_2 ) \dots \phi_0(x_n)  \mathcal{L}_1(z) \rangle\\
& + a \sum_{k = 1}^n \langle \phi_0(x_1) \phi_0(x_2 ) \dots  \phi_1(x_k)   \dots \phi_0(x_n) \rangle + \ord(a^2),
\end{align}
where the expectation values on the r.h.s.~are taken in the formal theory with action ${\rm S}_0$.
Then we modify the lattice action ${\rm S}$ and the field $\phi$ by higher dimensional operators such that they cancel the $\ord(a)$ terms in the equation.
Contact terms can arise from the integration over the space-time, i.e.~when $z = x_k$ for some $k$, and a prescription should be given to treat them. 
They can be included in a redefinition of the field $\phi_1$ and we do not worry about them in the following.

\subsection{$\ord(a)$ improved action}

Here we briefly review the improvement for the Wilson fermion action. The final result has $\ord(a^2)$ cutof\mbox{}f ef\mbox{}fects.
The Lagrangian $\mathcal{L}_1$ can be written as linear combination of the following fields
\begin{align}
\label{eq:one}
\mathcal{O}_1 &= \psibar i \sigma_{\mu\nu} F_{\mu\nu} \psi,\\
\mathcal{O}_2 &= \psibar D_\mu D_\mu \psi + \psibar \overleftarrow{D}_\mu \overleftarrow{D}_\mu \psi,\\
\label{eq:three}
\mathcal{O}_3 &= m \tr\left[ F_{\mu\nu} F_{\mu\nu}   \right],\\
\mathcal{O}_4 &= m \left( \psibar \gamma_\mu D_\mu \psi - \psibar \overleftarrow{D}_\mu \gamma_\mu \psi   \right),\\
\label{eq:five}
\mathcal{O}_5 &= m^2 \psibar\psi.
\end{align}
We can eliminate redundant terms if we restrict ourself to improvement of \emph{on-shell quantities}, for which the equations of motion can be used. 
After this step we are left with the operators in Eqs.~\ref{eq:one},\ref{eq:three} and \ref{eq:five}. 
Another reduction can be made by noticing that the operators $\mathcal{O}_3$ and $\mathcal{O}_5$ are already present in the original Lagrangian. 
Hence their insertion amount in a rescaling of the bare coupling $g_0$ and mass $m_0$.
Finally for the improved action we obtain \cite{Sheikholeslami:1985ij}
\begin{align}
{\rm S}_{\rm IMP}  =  {\rm S}_{\rm F} + a^5 \sum_{x \in \Lambda} c_{\rm SW} \psibar(x) \frac{i}{4} r \sigma_{\mu\nu} \widehat{F}_{\mu\nu} (x) \psi(x),
\end{align}
where $\sigma_{\mu\nu} = \frac{1}{2}\left[\gamma_\mu,\gamma_\nu\right]$ and $\widehat{F}_{\mu\nu} = \frac{1}{8i} \left[ Q_{\mu\nu}(x) - Q_{\nu\mu}(x) \right]$, with the definition
\begin{align}
\nn
Q_{\mu\nu}(x) =& \,U_\mu(x)\, U_\nu(x+a\hat{\mu})\, U^\dagger_\mu(x+a\hat{\nu})\, U^\dagger_\nu(x)
+ U_\nu(x)\, U^\dagger_\mu(x-a\hat{\mu}+a\hat{\nu})\, U^\dagger_\nu(x-a\hat{\mu})\, U_\mu(x-a\hat{\mu})\\
\nn
& + U^\dagger_\mu(x-a\hat{\mu})\, U^\dagger_\nu(x-a\hat{\mu}-a\hat{\nu})\, U_\mu(x-a\hat{\mu}-a\hat{\nu})\, U_\nu(x-a\hat{\nu})\\
& + U^\dagger_\nu(x-a\hat{\nu})\, U_\mu(x-a\hat{\nu})\, U_\nu(x+a\hat{\mu}-a\hat{\nu})\, U^\dagger_\mu(x).
\end{align}

\section{Ward-Takahashi identities}

The Ward-Takahashi identities (WTIs) are obtained by imposing that expectation
values are invariant under transformations of fermionic variables
that are symmetries. For simplicity, in
the following we omit the Yang-Mills part of the action.
The Ward-Takahashi identities are the true content of a symmetry, as the
Noether theorem is in the classical theory. At the quantum level, thanks to these identities,
we have an infinite set of relations among quantities.
The concepts presented here will be used in Chap.~\ref{chap:rtbc},
where we investigate the change of the critical mass in small volumes due to twisted boundary conditions.

\subsection{Formal theory}

In this section we derive the na\"ive WTI for the $\mathbf{SU}(2)_{\rm V}$ and $\mathbf{SU}(2)_{\rm A}$ transformations. For a formal derivation of the WTI structure see App.~\ref{app:wti}, in the following we use a local operator $\mathcal{O}(y)$ for the sake of simplicity.
We read the fermionic action from Eq.~\ref{QCD_continuum_lagrangian}, and we assume that the mass matrix is proportional to the identity in flavor space $M=m\I$.

	\subsubsection{$\mathbf{SU}(2)_V$}
	
	The infinitesimal transformations are written in Eq.~\ref{SU_2_V_transformation}. 
	The associated WTI is
	\begin{align}
	\partial_\mu^x\big\langle V^f_\mu(x)\mathcal{O}(y)\big\rangle=\delta^4(x-y)\langle\delta\mathcal{O}(y)\rangle,
	\end{align}
	where $V^f_\mu$ is given in Eq.~\ref{eq:noether_currents}.
	
	\subsubsection{$\mathbf{SU}(2)_A$}
	
	The infinitesimal transformations are given in Eq.~\ref{SU_2_A_transformation}. 
	The associated WTI is
	\begin{align}
%	\label{PCAC}
\nn
	\partial_\mu^x\big\langle A^f_\mu(x)\mathcal{O}(y)\big\rangle
	&= 2m\big\langle P^f(x)\mathcal{O}(y)\big\rangle 
	+ \delta^4(x-y)\langle\delta\mathcal{O}(y)\rangle\\
	&\stackrel{m=0}{=} \delta^4(x-y)\langle\delta\mathcal{O}(y)\rangle,
	\end{align}
	where $A^f_\mu$ is given in Eq.~\ref{eq:noether_currents} and 
	\begin{align}
	P^f &\equiv \overline{\Psi}\frac{\sigma_f}{2}\gamma_5\Psi.
	\end{align}
We see that the axial current $A_\mu^f$ is not conserved if $m\neq0$, since there is an explicit symmetry breaking. 
The above equation in the case of $m\neq0$ is called \emph{partially conserved axial current} (PCAC) \emph{relation} and the mass term defined in this way is called $m_\text{PCAC}$.\\
	By assuming the fields in $\mathcal{O}$ to be localized outside a region $R$, in which the parameter $\omega_f(x)$ lives, 
	the variation of the operator vanishes $\delta\mathcal{O} = 0$ and we obtain
	\begin{align}
		\label{eq:axial}
	\partial_\mu^x\big\langle A^f_\mu(x)\mathcal{O}(y)\big\rangle=2m\big\langle P^f(x)\mathcal{O}(y)\big\rangle.
	\end{align}
	
\subsection{Critical mass on the lattice}
\label{subsect:pcac}

For the derivation of the vector and axial current on the lattice with Wilson fermions see App.~\ref{app:wti}.
The vector current on the lattice satisfies the na\"ive WTI, up to cutof\mbox{}f ef\mbox{}fects, but for the axial case the relation is not straightforward \cite{Bochicchio:1985xa}. 
A very nice review on broken symmetries is given in Ref.~\cite{Testa:1998ez}.
The main issue there is that the variation under axial transformations of the Wilson term, $\chi_{\rm A}^f$, cannot be recast as a total derivative.
That means we are left with an \emph{unwanted piece}.
We report here the bare PCAC relation on the lattice
\begin{align}
\label{eq:bare_pcac}
\partial_\mu^-\langle A_\mu^f(x)\mathcal{O}(y)\rangle = 2m_0\langle P^f(x)\mathcal{O}(y)\rangle + \langle\chi_{\rm A}^f(x)\mathcal{O}(y)\rangle .
\end{align}
We know that in the formal classical limit $\chi_{\rm A}^f$ has to vanish, 
its general form is an operator of dimension 5 times the lattice spacing $\chi_{\rm A}^f = a\mathcal{O}_5$,
and the said operator may contain power divergences that compensate the factor $a$. \\
Hence the task is to build a finite operator out of $\mathcal{O}_5$ by renormalization procedure.
We know from renormalization that composite operators mix under renormalization with operators of the same and lower dimension.
There are no operators of dimension 5 entering in the mixing \cite{Curci:1985se}, which would appear with logarithmically
divergent coef\mbox{}ficients, hence we can write
%To simplify the discussion we do not consider mix with operators of dimension 5, which appear with logarithmically divergent coef\mbox{}ficients, then
\begin{align}
\overline{\mathcal{O}}_5 = Z_5 \left[ \mathcal{O}_5 + \frac{2\overline{m}}{a} P^f + \frac{Z_A-1}{a}\partial_\mu^- A_\mu^f \right].
\end{align}
The coef\mbox{}ficients $\overline{m}$ and $Z_A$ do not run with the scale, and $Z_5$ is a logarithmically divergent coef\mbox{}ficient. 
By substituting back in the bare PCAC relation the above equation we obtain
\begin{align}
\label{eq:ren_pcac}
\partial_\mu^-\langle \hat{ A}_\mu^f(x)\mathcal{O}(y)\rangle = 2\left(m_0 - \overline{m}\right)\langle P^f(x)\mathcal{O}(y)\rangle + \langle\overline{\chi}_{\rm A}^f(x)\mathcal{O}(y)\rangle,
\end{align}
where $\hat{ A}_\mu^f\equiv Z_A A_\mu^f$ and $\overline{\chi}_{\rm A}^f\equiv a\overline{\mathcal{O}}_5 / Z_5$.\\
Now we enforce the condition $\langle\overline{\chi}_{\rm A}^f(x)\mathcal{O}(y)\rangle\rightarrow 0$ as $a\rightarrow 0$.
By choosing $\mathcal{O}(y) = P^g_R(y)$ and renormalizing the fundamental fields we obtain
\begin{align}
\label{eq:m_awi}
\partial_\mu^-\langle A_{\mu R}^f(x) P_R^g(y)\rangle = 2m_{\rm AWI} \langle P^f(x) P_R^g(y)\rangle,
\end{align}
where $m_{\rm AWI} = m_0 - \overline{m}$. 
It should be stressed that the $\overline{m}$ is itself a function of $m_0$ and the other parameters of the theory, i.e.~$\overline{m} = \overline{m}(m_0, g_0) = f(g_0, am_0)/a$.\\
The unrenormalized PCAC mass is
\begin{align}
m_{\rm PCAC} \equiv \frac{m_{\rm AWI}}{Z_A} =\frac{\partial_\mu^-\langle A^f(x) P^g(y)\rangle}{2 \langle P^f(x) P^g(y)\rangle} = \frac{Z_P}{Z_A} m_{{\rm PCAC}, R}.
\end{align}
The renormalization constants and the renormalized mass do not depend on the kinematical parameters such as the time $x_0$ we insert the current. 
Changing the kinematical parameters accounts for probing the PCAC relation in a dif\mbox{}ferent way. 
If we consider two dif\mbox{}ferent kinematical configurations at the same $(g_0, a m_0)$ point in the bare space and we measure the associated unrenormalized current masses then the dif\mbox{}ference will be of order $a$.
This is exactly what we studied in Chap.~\ref{chap:rtbc} testing dif\mbox{}ferent kinematical configurations by injecting momentum through twisted boundary conditions.

\section{Hybrid Monte Carlo simulations of gauge theories}
\label{sect:HMCsgt}

The Hybrid Monte Carlo is the most widely used algorithm to generate lattice QCD configurations including dynamical fermions, see Ref.~\cite{Kennedy:2006ax} for a review on the topic.
In this section we present the HMC method for simulations of gauge theories on the lattice. The following discussion is a standard-textbook one, nice reviews are found in Refs.~\cite{Duane:1987de, Gattringer:2010zz, Lippert:1997qx, Luscher:2010ae}.\\
Suppose we want to compute the expectation value of an operator $\Omega[U]$, with $U_\mu$ links on the lattice governed by the action $\text{S}_{\rm G}[U]$, in QFT this is given by
\begin{align}
\langle \Omega \rangle = \frac{1}{\mathcal{Z}} \int \De[U] \e^{-{\rm S}_{\rm G}[U]} \Omega[U].
\end{align}
In the Monte Carlo method one samples the distribution $P_{\rm S}$ given by
\begin{align}
P_{\rm S} = \frac{\e^{-{\rm S}_{\rm G}[U]}}{\mathcal{Z}} ,\,\text{ and evaluates }\,\overline{\Omega} = \frac{1}{N_{\rm cnfs}} \sum_{j=1}^{N_{\rm cnfs}} \Omega[U_j],
\end{align}
on a sequence $\{U_j\}$ of configurations. By taking $N_{\rm cnfs}\gg 1$ we have, thanks to the central limit theorem, 
\begin{align}
\overline{\Omega} = \langle \Omega \rangle + \ord\left(\frac{1}{\sqrt{N_{\rm cnfs}}}\right).
\end{align}
The way to generate such a sequence of configurations is to build a \emph{Markov process}.\\
The concepts illustrated in this Section will be used as building blocks for the study in Chap.~\ref{chap:oHMCp} where we optimize the performance 
of the HMC algorithm in presence of multilevel integrators and mass-preconditioning for the fermionic forces.

\subsection{Markov chain}

For the sake of simplicity let us consider a generic field $\phi$ with associated action ${\rm S}(\phi)$.
A Markov process is a stochastic procedure that generates new configuration $\phi'$ starting from a previous $\phi$
with a probability $P_{\rm M} (\phi \rightarrow \phi')$, called \emph{transition probability}.\\
The transition probability has to satisfy the following properties:
\begin{itemize}
\item Aperiodicity, $P_{\rm M} (\phi \rightarrow \phi') >0$ $\forall\phi$, i.e.~the process does not get trapped in cycles.
\item $P_{\rm M} (\phi \rightarrow \phi') \geq 0$ $\forall\phi,\, \phi'$, and $\sum_{\phi'} P_{\rm M} (\phi \rightarrow \phi') = 1$ $\forall\phi$. It guarantees that $P_{\rm M}$ is a probability distribution in $\phi'$ for each $\phi$ and the corresponding Markov chain is ergodic\footnote{Any point of the configuration space is reached.}.
\end{itemize}
Since a Markov process does not have sinks or sources in the probability space, we need to satisfy the \emph{balance equation}
\begin{align}
\sum_{\phi} P_{\rm S} (\phi) P_{\rm M} (\phi \rightarrow \phi')
= \sum_{\phi} P_{\rm S} (\phi') P_{\rm M} (\phi' \rightarrow \phi),
\end{align}
i.e.~the total probability to end up in a configuration $\phi'$ has to be the same to the total probability to hop out of a configuration $\phi'$.
On the r.h.s.~$P_{\rm S} (\phi')$ can be factorized out of the sum, which then is equal to one, and the above equation becomes
\begin{align}
\sum_{\phi} P_{\rm S} (\phi) P_{\rm M} (\phi \rightarrow \phi')
= P_{\rm S} (\phi').
\end{align}
The balance equation states that $P_{\rm S} (\phi')$ is a fixed point of the Markov process, i.e.~the equilibrium distribution is preserved by the update process.\\
Any Markov chain converges to a unique fixed point distribution $P_{\rm S} (\phi)$ provided that satisfies
the \emph{detailed balance condition}
\begin{align}
P_{\rm S} (\phi) P_{\rm M} (\phi \rightarrow \phi')
=  P_{\rm S} (\phi') P_{\rm M} (\phi' \rightarrow \phi).
\end{align}

\subsection{How to build a Markov process}

Here we describe how to build a Markov process, to this end we divide the strategy in two parts:
\begin{enumerate}
\item Suggest a new configuration $\phi'$ with probability $P_{\rm C} (\phi \rightarrow \phi')$.
\item Accept the suggestion with probability $P_{\rm A} (\phi \rightarrow \phi')$, and if rejected stay with the old configuration $\phi$.
\end{enumerate}
A choice for $P_{\rm A}$ that satisfies the detailed balance for any choice of $P_{\rm C}$ is the Metropolis accept/reject algorithm,
\begin{align}
P_{\rm A}(\phi\rightarrow \phi') = {\rm min} \left(  1, \frac{P_{\rm S} (\phi') P_{\rm C} (\phi' \rightarrow \phi)}
{P_{\rm S} (\phi) P_{\rm C} (\phi \rightarrow \phi')}   \right).
\end{align}
An ideal choice for $P_{\rm C} (\phi \rightarrow \phi')$ should give large acceptance rate which does not 
depend too strongly on the size of the system we want to simulate and 
should minimize the autocorrelation between successive configurations.
The standard choice for $P_{\rm C}$ is given by the Hybrid Molecular Dynamics algorithm.

\subsection{Hybrid Monte Carlo simulation}

The HMC algorithm is a Markov process and it consists of
\begin{itemize}
\item Molecular dynamics (MD) trajectories,
\item Metropolis test.
\end{itemize}
We introduce a fictitious time $\tau$, \emph{Markov time}, and a set of conjugate momenta\footnote{In $\mathbf{SU}(N_c)$ gauge theories the conjugate momenta are $\pi_\mu(x) = i \pi_\mu^a T^a$, with $T^a$ generators.} $\pi(\tau)$,
for each dynamical degree of freedom, with Gaussian distribution. 
The Hamiltonian of the system is then given by, in matrix/vector notation,
\begin{align}
H(\pi, \phi)  = \frac{1}{2} \pi^2 + {\rm S}(\phi) = {\rm T}(\pi) + {\rm S}(\phi),
\end{align}
where $\pi^2 = \sum_{x\in\Lambda} \pi^2(x)$.
The HMC forms a Markov chain with fixed point $\e^{-H(\pi, \phi)}$, which is the desired distribution,  
\begin{align}
\langle \Omega \rangle = \langle \Omega \rangle_H = \frac{1}{\mathcal{Z}_H} \int \De[\pi, \phi] \e^{-H(\pi, \phi)} \Omega[\phi],
\end{align}
since $\Omega$ does not depend on the momenta the functional integration in $\pi$ is canceled out by the partition function.
The procedure consists in
\begin{enumerate}
\item Select initial random momenta $\pi$ from a Gaussian distribution $P_{\rm G}(\pi)$ of zero mean and unit variance,
\item Evolve the system in the $(\phi, \pi)$ space according to the Hamilton equations
\begin{align}
\frac{\de \pi}{\de \tau} = -\frac{\partial {\rm S}}{\partial \phi}, \quad \frac{\de \phi}{\de \tau} = \pi .
\end{align}
$\Delta H$ is the violation in the energy conservation caused by the integrator at the end of the trajectory, 
and the probability is denoted as $P_{\rm H} \left[ (\phi, \pi) \rightarrow (\phi', \pi')  \right]$.
\item Accept/reject the suggested configuration with probability $P_{\rm A}\left[ (\phi, \pi) \rightarrow (\phi', \pi')  \right]=$\\ $ {\rm min} \left(1, \e^{-\Delta H}\right)$.
\end{enumerate}
The transition probability for the $\phi$ field is given by
\begin{align}
P_{\rm M} (\phi' \rightarrow \phi )
= \int \mathcal{D}[\pi, \pi'] 
P_{\rm G}(\pi) P_{\rm H} \left[ (\phi, \pi) \rightarrow (\phi', \pi')  \right]
P_{\rm A}\left[ (\phi, \pi) \rightarrow (\phi', \pi')  \right],
\end{align}
that has to satisfy the detailed balance condition.

\paragraph{Detailed balance condition}

To satisfy the detailed balance condition we need
\begin{itemize}
\item Area preserving of the integration measure $\mathcal{D}[\phi, \pi]$,
\item Reversibility of the trajectory.
\end{itemize}
The dynamic has to be reversible, hence
\begin{align}
P_{\rm H} \left[ (\phi, \pi) \rightarrow (\phi', \pi')  \right] = 
P_{\rm H} \left[ (\phi', -\pi') \rightarrow (\phi, -\pi)  \right].
\end{align}
First we write the identity
\begin{align}
\nn
P_{\rm G}(\pi) P_{\rm S} (\phi)
P_{\rm A}\left[ (\phi, \pi) \rightarrow (\phi', \pi')  \right] &=
\e^{- H (\phi, \pi)} {\rm min} \left(1, \e^{-\Delta H}\right) \\
&= 
{\rm min} \left(\e^{- H (\phi, \pi)}, \e^{- H (\phi', \pi')}\right) =
\e^{- H (\phi', \pi')} {\rm min} \left(1, \e^{\Delta H}\right) ,
\end{align}
and we notice that $H(\phi, \pi) = H(\phi, -\pi)$ from which we infer 
\begin{align}
P_{\rm S}(\phi) P_{\rm G} (\pi) \propto P_{\rm H} (\phi, \pi) = \e^{- H (\phi, \pi)} = \e^{- H (\phi, -\pi)} = P_{\rm H} (\phi, -\pi) \propto P_{\rm S}(\phi) P_{\rm G} (-\pi).
\end{align}
Combining the two equations we found earlier integrating over $\mathcal{D}[\pi, \pi']$ recalling that $\mathcal{D}[\pi, \pi'] = \mathcal{D}[-\pi, -\pi']$, we obtain
\begin{align}
\nn
P_{\rm S} (\phi)  P_{\rm M} (\phi\rightarrow\phi') &=
P_{\rm S} (\phi) 
\int \mathcal{D}[\pi,\pi'] 
P_{\rm G}(\pi) 
P_{\rm H} \left[ (\phi, \pi) \rightarrow (\phi', \pi')\right]
P_{\rm A}\left[ (\phi, \pi) \rightarrow (\phi', \pi')  \right] 
\\
\nn
&= 
P_{\rm S} (\phi') 
\int \mathcal{D}[\pi,\pi'] 
P_{\rm G}(-\pi) 
P_{\rm H} \left[ (\phi', -\pi') \rightarrow (\phi, -\pi)\right]
P_{\rm A}\left[ (\phi', -\pi') \rightarrow (\phi, -\pi)  \right] \\
&= P_{\rm S} (\phi')  P_{\rm M} (\phi' \rightarrow\phi) ,
\end{align}
that is the detailed balance condition.

\subsection{Dynamical fermions}

We would like to include in our simulations fermions $\psi$ governed by the action ${\rm S}_{\rm F} = \psibar M[\phi] \psi$, with $M(\phi)$ a generic local operator.
To do so we replace the Grassmann fields $\psibar, \psi$ with bosonic fields $\chi^*, \chi$. 
The simulated theory is obtained by integrating out the fermion fields 
and dynamical fermion loops are included through a stochastic representation of the fermionic determinant.
Hence the distribution to consider is 
\begin{align}
\nn
P_{\rm S} & \propto \int \De [\psibar, \psi]  \exp\left(-{\rm S}(\phi) - \psibar M(\phi) \psi \right) \propto \e^{-{\rm S} (\phi)}  \det M(\phi)  \\
\label{eq:distribution}
& \propto \int \De [\chi^*, \chi]  \exp\left\{-{\rm S}(\phi) -\chi^* M^{-1} \chi\right\}.
\end{align}
In the relevant case of the $\gamma_5$-version of the Dirac-Wilson operator\footnote{Notice it is hermitian, but it can have negative eigenvalues.}
 $M(\phi) = \gamma_5 D_W(U)$,
to ensure the convergence of the bosonic Gaussian integral we double the number of simulated fermions, $\det M\rightarrow (\det M)^2$, and the probability becomes
\begin{align}
\nn
P_{\rm S} &\propto \e^{-{\rm S} (\phi)} (\det M(\phi))^2 
\propto \e^{-{\rm S} (\phi)} \det (M(\phi) M^\dagger(\phi)) \\
\label{eq:probability}
&\propto \int \De [\chi^*, \chi]  \exp\left\{-{\rm S}(\phi) -\chi^*(M^\dagger M)^{-1} \chi\right\}.
\end{align}
The Hamilton equations for the $\phi$ fields in this framework become
\begin{align}
\nn
\frac{\de \pi}{\de \tau} &= -\frac{\partial {\rm S}}{\partial \phi} - \chi^* \frac{\partial(M^\dagger M)^{-1}}{\partial \phi}\chi\\
&= -\frac{\partial {\rm S}}{\partial \phi} + \chi^* (M^\dagger M)^{-1} \left[M^\dagger\frac{\partial M}{\partial \phi} + \frac{\partial M^\dagger}{\partial \phi} M \right] (M^\dagger M)^{-1} \chi \equiv F_\phi + F_{\rm Ferm},\\
\frac{\de \phi}{\de \tau} &= \pi.
\end{align}
It is easy to see that a number of conjugate gradient inversions, $\eta = (M^\dagger M)^{-1}\chi$, are required in order to compute the HMC part of the force, $F_{\rm Ferm}$.
The fields $\chi$ and $\chi^*$ are held fixed during the molecular dynamics steps and updated in between.\\
It should be mentioned that the sample of the distribution in Eq.~\ref{eq:distribution} can be performed if the measure of the path integral is real and positive, only in this way we can interpret it as a probability distribution.
That interpretation breaks down, for example, if the term in Eq.~\ref{eq:axion} with $\theta \neq 0$ is included in the action.

\afterpage{\blankpage}

\setcounter{equation}{0}

\chapter{Quantum electrodynamics on the lattice}
\label{qedothl}

The main goal of lattice QCD is to calculate in a non-perturbative manner observables needed in phenomenological
applications. Most of the relevant hadronic quantities are computed in the so called isosymmetric limit, i.e.~neglecting
the up and down quark mass dif\mbox{}ference and the QED interactions.
Nowadays, for some observables, the accuracy of lattice estimates
 has reached the percent level and the above mentioned ef\mbox{}fects
cannot be neglected any longer, a recent review on that is found in Ref.~\cite{Tantalo:2013maa} .\\
The inclusion of the quark mass dif\mbox{}ference is quite straightforward from the 
theoretical point of view and it requires only additional computational ef\mbox{}fort, e.g.~reweighting technique \cite{Finkenrath:2012cz} or direct simulations. 
Since the inclusion of QED on a
lattice presents some subtleties that are ultimately related to the long-range nature of electromagnetic interactions, we have reserved to them this introductory chapter.

In QED simulations the so-called \emph{zero mode problem} is present, 
which appears by looking at unphysical contributions to a given process or amplitude. 
The dif\mbox{}ferent contributions are divergent and this is well understood in the view of the \emph{Bloch-Nordsieck theorem}, 
which states that physical quantities in QED are IR divergent free \cite{Bloch:1937pw}. 
Furthermore in a periodic lattice Gauss's law forbids charged states to propagate \cite{Hayakawa:2008an}.\\
The most popular way to deal with this problem is to subtract the zero mode and remove it from the dynamics. This is done in several ways:
\begin{itemize}
\item QED$_{\rm TL}$, the zero mode is set to zero on each configuration.
\item QED$_{\rm L}$, the spatial zero modes are set to zero, on each time-slice on each configuration.
\item QED$_{\rm C}$, the zero mode is absent due to a special choice of boundary conditions.
\item QED$_{\rm M}$, a photon mass term is introduced, i.e.~the zero mode is generated with a Gaussian distribution with zero mean value.
\end{itemize}
It is well understood that QED$_{\rm TL}$ does not possess reflection positivity, and consequently a positive definite Hamiltonian, while QED$_{\rm L}$ does have it.
In Ref.~\cite{Borsanyi:2014jba} finite volume ef\mbox{}fects were computed only for masses in QED$_{\rm TL}$ and QED$_{\rm L}$, a similar work was done for QED$_{\rm C}$ \cite{Lucini:2015hfa}. 
In QED$_{\rm M}$ the spatial zero modes are regularized, as in perturbation theory, with a Gaussian weight.
Here we are forced to produce configurations with dif\mbox{}ferent photon masses, since eventually we will extrapolate the results to vanishing mass \cite{Endres:2015gda}.

The implementation of QED has been studied since long time with pioneering work in \cite{Duncan:1996xy} in the framework of quenched QED+QCD (qQ(C+E$_{\rm TL}$)D).\\
All the studies, until now, are mostly focused on the hadron spectrum. 
In the last years the field was very active and recent studies have been performed qQED+dynamical QCD (qQED+QCD) \cite{Blum:2010ym, Basak:2014vca}.
The best result, at the present, in the fully dynamical Q(C+E$_{\rm L}$)D simulations has been achieved by the BMW collaboration \cite{Borsanyi:2014jba} where the proton-neutron mass
splitting with about $5\sigma$ statistical significance is given.
In Ref.~\cite{Endres:2015gda} a first hadron spectrum determination in qQED$_{\rm M}$+QCD was presented.
It should be mentioned that another quenched QED strategy is given in \cite{deDivitiis:2013xla} in which the path integral and corresponding observables are expanded to order $\alpha_{\rm em}$.
There the QED corrections are extracted directly from the observables as $\ord(1)$ ef\mbox{}fects rather than $\ord(\alpha_{\rm em})$ ones.\\
The introduction of a photon mass
is particularly interesting since it changes the finite volume ef\mbox{}fects from power-like to exponential-like, avoiding the
infinite volume extrapolation but introducing an extrapolation in the photon mass. 
The C$^\star$ approach is also an elegant solution to the zero mode problem and it comes at the cost of flavor violations. 
No further discussions are given here since we did not explored that approach in our projects.

The chapter is organized as follows; in Section~1 we briefly discuss the global symmetries in the framework of Q(C+E)D,
in Section~2 we present the problem of IR divergences on the lattice QED in non-compact formulation. 
Section~3 is devoted to general considerations on the gauge symmetry, as gauge fixing and  
completeness in finite volume.
The fixing of eventual residual symmetries and zero modes is discussed in Section~4.
Section~5 contains the recipe for the generation of quenched QED configurations in Feynman gauge.
In Section~6 we illustrate the impossibility of building charged states in a finite volume with periodic boundary conditions
and how this is avoided in QED$_{\rm TL}$ and QED$_{\rm L}$.
In Section~7 we present the other solution to the Gauss law problem, the \emph{massive IR regularization} on the lattice for QED.
%,and in Section~8 the generation of QED$_{\rm M}$ is presented.
Section~8 contains the comparison of Wilson loops in QED$_{\rm TL}$, QED$_{\rm L}$ and QED$_{\rm M}$ with the corresponding 
predictions in the infinite volume.

\section{Global symmetries}

The electromagnetic formal theory with fermions in Euclidean space is given by the following Lagrangian
\begin{align}
\mathcal{L} [A_\mu, \Psi, \Psibar] = \mathcal{L}_{\rm QED} + \mathcal{L}_{\text{F}} = \frac{1}{4} F_{\mu\nu}F_{\mu\nu} + \Psibar(x) \left( eQ{\not}D + M\right) \Psi(x),
\end{align}
where $F_{\mu\nu} = \partial_\mu A_\mu - \partial_\nu A_\mu$ is the field-strength tensor for the gauge degrees of freedom, $\Psi$ is a fermionic flavor multiplet, $M$ is the mass and $Q$ the charge matrices.
$A_\mu$ is a real vector field with dimension [1]. 
A $\theta$-term is not forbidden, that is parity and time reversal violating term
\begin{align}
\theta \epsilon^{\mu\nu\rho\sigma} F_{\mu\nu}F_{\rho\sigma} = \theta F_{\mu\nu}\widetilde{F}_{\mu\nu},
\end{align}
with $\theta$ real parameter and $\epsilon^{\mu\nu\rho\sigma}$ antisymmetric tensor, but QED conserves those symmetries. 
The $\theta$-term is related to \emph{instantonic contributions} and it is known that in an Abelian $\mathbf{U}(1)$ 
theory it does not have any physical ef\mbox{}fect, in contrast to non-Abelian theories, hence we can safely neglect it.\\
The following discussion can be found in Ref.~\cite{Blum:2007cy}.
We know that QCD exhibits chiral symmetry, as we saw in 
Sect.~\ref{sec:symmetriesqcd}, given 
by the following group, in presence of three fermion species, $N_f=3$
\begin{align}
\mathbf{G}_\text{QCD}(N_f=3) = \mathbf{SU}(N_f)_L \otimes \mathbf{SU}(N_f)_R \otimes \mathbf{U}(1)_V,
\end{align}
where we have omitted the $\mathbf{U}(1)_A$ since it is \emph{anomalous}.\\
In Q(C+E)D we find that some global symmetries, present in QCD, are explicitly broken by 
the electromagnetic interactions.
Notice we are considering two electromagnetic charges to be the same.
The explicit breaking can be seen through the Ward-Takahashi identity for the flavor non-singlet axial current $A_\mu^f = \Psibar \gamma_\mu \gamma_5 T^f \Psi$, 
with $T^f$ one of the generators of $\mathbf{SU}(N_f)$ 
(see Eq.~\ref{eq:axial} for the case $N_f=2$ in QCD),
\begin{align}
\partial_\mu A_\mu^f = ieA_\mu \Psibar \left[T^f, Q\right] \gamma_\mu \gamma_5\Psi - \frac{\alpha}{2\pi} \tr\left[Q^2T^f\right] F_{\mu\nu}\widetilde{F}_{\mu\nu}.
\end{align}
For $N_f=3$ we have $N_f^2 -1 = 8$ generators and the first term in the above equation vanishes when $f=3, 6, 7, 8$ while the second does it for $f=1,2,4,5,6,7$ and a combination of $f=3$ and 8, which is
\begin{align}
\frac{1}{2} T'^3 = \frac{\sqrt{3}}{2} T^8 - \frac{1}{2} T^3.
\end{align}
Thus the inclusion of QED ef\mbox{}fects leads to the symmetry group
\begin{align}
\mathbf{G}_\text{QCD+QED}(N_f=3) = \mathbf{SU}'(2)_L \otimes \mathbf{SU}'(2)_R \otimes \mathbf{U}(1)_V
\end{align}
where the prime are subgroups $\mathbf{SU}'(2)\subset \mathbf{SU}(3)$ generated by $T^6, T^7, T'^3$. 
Then the QCD dynamics break spontaneously the group to 
\begin{align}
\label{eq:sub_qcd_qed}
\mathbf{H}_\text{QCD+QED}(N_f=3) = \mathbf{SU}'(2)_V \otimes \mathbf{U}(1)_V.
\end{align}

\subsection*{Landau Pole}

In QED the known \emph{Landau pole} is present. 
The one-loop $\beta$-function in QED is given by
\begin{align}
\mu\frac{\partial e}{\partial \mu} = \frac{e^3}{12\pi^2},
\end{align}
and by solving the dif\mbox{}ferential equation we find 
\begin{align}
e^2(\mu) = \frac{e^2(\mu_0)}{1 - \frac{e^2(\mu_0)}{6\pi^2} \ln \frac{\mu}{\mu_0}}.
\end{align}
It is clear that there is a singularity at $\mu = \mu_0 \exp\left[ \frac{6\pi^2}{e^2(\mu_0)}  \right]$, the so-called Landau pole.\\
By plugging in physical number we find that the pole is beyond the Planck scale, and practically unreachable in present lattices.

\section{Infrared divergences}

On the lattice we introduce the electromagnetic 
field in a dynamical way by employing the 
\emph{non-compact} formulation. 
QED with compact formulation suf\mbox{}fers from a phase transition 
around $\beta \approx 1.2$ as first seen in Ref.~\cite{Creutz:1979zg}.
For large unphysical electromagnetic coupling a transition
from the Coulomb phase to a confining phase is present\footnote{The phase transition 
is due to the introduction of lattice artifacts, 
like photon self-interaction.}. 
In the following we will deal only with fields in the non-compact formulation to avoid the above mentioned unphysical phase transition.
Gauge invariance on the lattice still requires the coupling 
between fermions and compact variables but 
we have the freedom to choose the non-compact formulation 
for the pure QED part of the action. 
But in this framework we are forced to fix the gauge, in order to make finite the path integral.

One of the main dif\mbox{}ference with QCD is that the physical states can have energy arbitrary close to zero because the theory does not have a mass gap. For this reason finite volume ef\mbox{}fects are expected to be important\footnote{QCD finite volume ef\mbox{}fects go as $\e^{-m_\pi L}$.}, since the interaction is long-ranged.
Furthermore QED can lead to an IR divergent theory \cite{Portelli:2013jla}, if one considers dif\mbox{}ferent 
contributions (diagrams) to a physical quantity. 
To elucidate this point we look at a one photon loop 
contribution to a correlation function in the formal theory
\begin{align}
I = \int\frac{\de^4 k}{(2\pi)^4} \frac{f(k, p_1, p_2, \dots, p_n)}{k^4},
\end{align}
where $k$ is the momentum flowing in the loop, $p_1, p_2, \dots, p_n$ are the external momenta, and we assume that $f(k, p_1, p_2, \dots, p_n) \underset{k\rightarrow 0}{\propto} k^{-\epsilon}$. 
In the situation where $\epsilon>0$ the integrand is divergent in the limit $k\rightarrow 0$, and it is an IR divergence if the singularity is not integrable. 
However it should be noted that the IR divergences do not af\mbox{}fect physical results when we consider all the possible contributing diagrams.
This statement is the essence of the Bloch-Nordsieck theorem \cite{Bloch:1937pw} and the more general \emph{Kinoshita-Lee-Nauenberg theorem} \cite{Kinoshita:1962ur, Lee:1964is}.\\
On the lattice the integral becomes a discrete sum over momenta, which depend on the volume and the type of boundary conditions. By choosing periodic boundary conditions in the gauge sector the allowed momenta are $p_\mu = \frac{2\pi}{L_\mu} k_\mu$, where $L_\mu$ is the lattice extent in direction $\mu$ and $k_\mu\in\left(-\frac{L_\mu}{2}, \frac{L_\mu}{2}\right]\subset\mathbb{Z}$. The vanishing momentum $p^2 = 0$ is part of the reciprocal lattice and the discretized version of the integral is hence \emph{ill-defined}.
Thus we need an IR regularization of the theory on the lattice if we are interested in potentially IR divergent contributions.

\section{Gauge symmetry: general considerations}

In the following we will show how to generate QED configurations in the quenched approximation\footnote{Sea and valence quarks are treated in a dif\mbox{}ferent way, i.e.~sea quarks are not charged while valence ones are.}. The crucial point is the gauge fixing procedure and the removal of zero mode (the latter prescription makes the theory IR \quotes{divergent free} in the sense explained in the previous Section).

\subsection{Gauge fixing}
\label{sect:faddeev-popov}

In the non-compact formalism we are forced to fix the gauge in order to make the path integral well defined. The ill-definiteness arises from the fact that we cannot invert the gauge operator and calculate the photon propagator. This is a direct consequence of gauge invariance: there exist gauge equivalent transformations that do not af\mbox{}fect the physics, called \emph{orbits}.

In the Faddeev-Popov approach \cite{Faddeev:1967fc} one considers a constraint in the path integral in a way that each orbit is intersected only once\footnote{In other words we want a representative configuration for each orbit and the other elements in the orbit are related to it by a smooth gauge transformation.
No Gribov ambiguities are present.}, then the path integral can be re-written as
\begin{align}
\nn
\int \mathcal{D}[A_\mu] \e^{-\text{S}(A)} 
&\propto \int \mathcal{D}[A_\mu] \det\left(\frac{\partial G}{\partial \Omega}\right)\bigg|_{G=0} \delta\left(G(A)\right) \e^{-\text{S}(A)}\\
\label{eq:Faddeev-Popov}
& \propto \int \mathcal{D}[A_\mu] \det\left(\frac{\partial G}{\partial \Omega}\right)\bigg|_{G=0} \exp{\left(-\text{S}(A) -\frac{1}{2\alpha}\int\de^4x\, G(A)^2 \right)}
\end{align}
where $G$ is the gauge condition to impose, i.e.~$G(A)=0$, 
%$\Omega$ is an element of the gauge group, 
and $\alpha$ is a free parameter used to approximate the Dirac delta. 
The derivative of the gauge condition with respect to $\Omega$ has to be interpreted as the functional derivative of the gauge transformed field, i.e.~$G(A^\Omega)$, with respect to a gauge transformation, given by the function $\Omega$.
The determinant is called \emph{Faddeev-Popov determinant} and can be expressed as a Gaussian integral by introducing fictitious fermionic variable called \emph{ghosts}.
The ghost term contributes to the overall normalization and it does not couple with the gauge fields, i.e.~QED is ghosts free.

\subsection{Gauge symmetry in finite volume}

When we introduce fermions in our system we should care about the zero mode of the photon field, and removing it leads to an IR \quotes{divergent free} theory on the lattice. 
The following discussion is based on Ref.~\cite{Blum:2007cy}.
In the rest of the work we do not explicitly say when the lattice spacing $a$ is set to 1. 
It is easy to restore the proper powers of lattice spacing in the expressions using dimensional analysis.
We introduce only one fermion $\psi$, with charge one, and we couple it to the photon through the term in the Lagrangian
\begin{align}
\mathcal{L}_{\rm F}[A_\mu, \psibar, \psi] = \frac{1}{2}\sum_\mu \psibar(x)\, \gamma_\mu\left(\nabla_\mu^+ + \nabla_\mu^-\right) \psi(x),
\end{align}
where the variables $A_\mu$ are placed in the mid-point of the link $x, x+\hat{\mu}$
and the forward and backward derivatives are
\begin{align}
\nn
\nabla^+_\mu\psi(x)& = U_\mu(x)\psi(x+\hat{\mu})-\psi(x)
= \e^{ieqA_\mu(x+\hat{\mu}/2)}\psi(x+\hat{\mu})-\psi(x),\\
\nabla^-_\mu\psi(x)& = \psi(x)-U^\dagger_\mu(x-\hat{\mu})\psi(x-\hat{\mu})
= \psi(x)-\e^{-ieqA_\mu(x-\hat{\mu}/2)}\psi(x-\hat{\mu}).
\end{align}
The Lagrangian is invariant under a gauge transformation $\Lambda$, which we recall to act on the fields as
\begin{align}
\nn
A'_\mu(x+\hat{\mu}/2) &= A_\mu(x+\hat{\mu}/2) + \partial_\mu^+\Lambda(x),\\
\nn
\psi'(x) &= \e^{i\Lambda(x)} \psi(x),\\
\psibar\,'(x) &= \psibar(x) \e^{-i\Lambda(x)}.
\end{align}
By imposing periodic boundary condition on $A'_\mu$ and $\psi'$ we discover that the function $\Lambda$ does not have to respect periodic boundary condition but instead a weaker condition:
\begin{align}
\label{eq:quant_cond}
\Lambda(x+L_\mu\hat{\mu}) = \Lambda(x) + 2\pi r_\mu,
\end{align}
where the $r_\mu\in\mathbb{Z}$, are \emph{quantized} as a consequence of the periodicity of the fermions\footnote{This makes clear how the allowed gauge transformations depend on the fermionic boundary conditions. One may modify the latter to get rid of unwanted terms~\cite{Lucini:2015hfa}.}.\\
A general form for a gauge transformation that respect the above condition is
\begin{align}
\label{eq:lambda_split}
\Lambda(x) = \Lambda^{(0)}(x) + 2\pi \sum_{\mu=0}^3 r_\mu \frac{x_\mu}{L_\mu},
\end{align}
where $\Lambda^{(0)}(x)$ satisfies periodic boundary conditions.\\
By Fourier transforming the gauge transformed gauge field we learn how it is expressed in momentum space
\begin{align}
\label{eq:momenta_gauge_transf}
\widetilde{A}'_\mu(p) = \widetilde{A}_\mu(p) + \sqrt{V} \frac{2\pi r_\mu}{L_\mu} \delta(p) + i \hat{p}_\mu \widetilde{\Lambda}^{(0)}(p),
\end{align}
where we have defined $\hat{p}_\mu = \frac{2}{a} \sin\left(ap_\mu /2\right)$.
We would like to eliminate all the $\mathbf{U}(1)$ redundancies. 
For this reason we need to gauge fix through a condition and we shall use the Coulomb gauge. The matter is to determine whether the Coulomb gauge is suf\mbox{}ficient to do it, 
or in other words we need to establish if it is a \emph{complete gauge fixing}.

\paragraph{Intermezzo: Non-completeness of Coulomb gauge}

We want to find a particular gauge transformation that transforms the fields satisfying the Coulomb condition into another one satisfying the same condition. For the sake of simplicity let us assume to be in a finite box with vanishing lattice spacing.\\
We are searching for a function $\phi$ that does the job. We split $\phi$ according to the relation Eq.~\ref{eq:lambda_split} and the gauge transformation on the gauge field becomes
\begin{align}
A'_\mu(x) = A_\mu(x) + \frac{2\pi r_\mu}{L_\mu} + \partial_\mu \phi^{(0)}(x).
\end{align}
By imposing the Coulomb condition on both fields before and after
the transformation we find an equation for $\phi^{(0)}$,
\begin{align}
\sum_j \partial_j A'_j(x) = 0 = \cancel{\sum_j \partial_j A_j(x)} + \sum_j \partial_j\partial_j \phi^{(0)}(x) = \nabla^2 \phi^{(0)}(x).
\end{align}
Since the dif\mbox{}ferential equation does not give informations about the time behavior we can write the solution in the following way 
\begin{align}
\phi^{(0)}(t,\underline{x}) = T(t)\widehat{\phi}^{(0)}(\underline{x}), \quad\text{ with } T(t)\neq 0,
\end{align}
and the dif\mbox{}ferential equation is restated as 
\begin{align}
\label{eq:laplacian_phi}
\nabla^2 \widehat{\phi}^{(0)}(\underline{x}) = \left( \frac{\partial^2}{\partial x^2} + \frac{\partial^2}{\partial y^2} + \frac{\partial^2}{\partial z^2} \right) \widehat{\phi}^{(0)}(\underline{x}) = 0.
\end{align}
In rectangular coordinates the equation is separable and the solution is written in terms of the three ordinary dif\mbox{}ferential equation by assuming \cite{Jackson:1998nia}
\begin{align}
\widehat{\phi}^{(0)}(x, y, z) = X(x) Y(y) Z(z).
\end{align}
Substituting back in Eq.~\ref{eq:laplacian_phi} and dividing the result by $\widehat{\phi}^{(0)}(x,y,z)$ we obtain
\begin{align}
\frac{1}{X(x)}\frac{\partial^2 X}{\partial x^2} + \frac{1}{Y(y)}\frac{\partial^2 Y}{\partial y^2} + \frac{1}{Z(z)}\frac{\partial^2 Z}{\partial z^2} = 0.
\end{align}
The solution of such an equation is given by
\begin{align}
\displaystyle{
\begin{cases}
\frac{1}{X(x)}\frac{\partial^2 X}{\partial x^2} &= -\alpha^2,\\
\frac{1}{Y(y)}\frac{\partial^2 Y}{\partial y^2} &= -\beta^2,\\
\frac{1}{Z(z)}\frac{\partial^2 Z}{\partial z^2} &= \gamma^2.
\end{cases}}
\longrightarrow
\begin{cases}
X(x) &= A\cos(\alpha x) + B\sin(\alpha x),\\
Y(y) &= C\cos(\beta y) + D\sin(\beta y),\\
Z(z) &= E\cosh(\gamma z) + F\sinh(\gamma z).
\end{cases}
\end{align}
where we have the relation $\alpha^2 + \beta^2 = \gamma^2$.\\
The solution $\phi^{(0)}$ is just a product of the ones we found and the unknown function $T(t)$ (but periodic in $t$). 
In order to determine the constants we impose 
the boundary conditions on $ X,  Y,  Z$ 
which follow from the periodic boundary 
conditions for $\phi^{(0)}$.\\
At this stage it is clear that if we try to impose periodic boundary conditions on $Z$ the only solution is $Z=\text{const}$ ($\gamma = 0$ implies $\alpha, \beta=0$) which makes $\widehat{\phi}^{(0)}=\text{const}$.\\
We conclude that there is no gauge transformation left in the spatial direction satisfying Eq.~\ref{eq:laplacian_phi}, apart for the constant. We still have the redundancy $T(t)$, which means that the Coulomb gauge is not complete in a finite volume.

\section{QED$_{\rm TL}$ \& QED$_{\rm L}$}
\label{sec:drstrangelove}

In this section we present the two IR regularization know in the
literature as QED$_{\rm TL}$ and QED$_{\rm L}$. 
Both are inspired by the gauge fixing procedure.

\subsection{Fix of the residual gauge symmetry and zero mode}

In the previous section we have shown how the redundancies related to $\widetilde{\Lambda}(p_0, \underline{p}\neq\underline{0})$ are eliminated by the Coulomb gauge fixing. 
Based on the previous Section, we need to fix the residual symmetry generated by:
\begin{itemize}
\item $\widetilde{\Lambda}(p_0\neq 0, \underline{0})$, spatially uniform gauge transformations, i.e.~$\phi^{(0)}(x) = T(t)\times{\rm const}$,
\item $r_\mu$, space-time uniform gauge transformations, i.e.~$\phi(x) = {\rm const}$.
\end{itemize}

\subsection*{$\widetilde{\Lambda}(p_0\neq 0, \underline{0})$ redundancy}

We can see from Eq.~\ref{eq:momenta_gauge_transf} that $\Lambda(p_0\neq 0, \underline{0})$ is acting on $\widetilde{A}_0$,
 while does not change the spatial components $\widetilde{A}_i$.
 The transformation of the temporal one is
\begin{align}
\widetilde{A}'_0(p_0\neq 0, \underline{0}) = \widetilde{A}_0(p_0\neq 0, \underline{0}) + i\hat{p}_0 \widetilde{\Lambda}^{(0)}(p_0\neq 0, \underline{0}).
\end{align}
Then we can always gauge transform $\widetilde{A}_0$ to zero and in this way fix the the redundancy
\begin{align}
\label{eq:supplement}
\widetilde{A}'_0(p_0\neq 0, \underline{0}) = 0 \Longrightarrow \widetilde{\Lambda}^{(0)}(p_0\neq 0, \underline{0}) = i\frac{1}{\hat{p}_0} \widetilde{A}_0(p_0\neq 0, \underline{0}).
\end{align}
The Coulomb gauge condition with the supplement of the Eq.~\ref{eq:supplement} is usually referred to as \emph{extended Coulomb gauge}.

\subsection*{$r_\mu$ redundancy}

Eq.~\ref{eq:momenta_gauge_transf} for space-time uniform gauge transformations becomes
\begin{align}
\widetilde{A}'_\mu(0) = \widetilde{A}_\mu(0) + \sqrt{V} \frac{2\pi r_\mu}{L_\mu}
\end{align}
Here we cannot always gauge transform to zero, since the $r_\mu$ are integers, but we can in principle \quotes{partially} eliminate $\widetilde{A}_\mu$, which means to transform it in the interval
\begin{align}
0 \leq \widetilde{A}'_\mu(0) <  \frac{2\pi\sqrt{V}}{L_\mu}\, ,
\end{align}
by choosing a suitable $r_\mu$. In this way we have eliminated all the redundancies in the Coulomb gauge in finite volume\footnote{This prescription is also known as 
QED$_{\rm SF}$ and it was argued in Ref.~\cite{Patella:2017fgk} that this theory unlikely possesses a transfer matrix.}.

\subsection*{Zero modes}

We already saw that the zero modes $\widetilde{A}_\mu(0), \widetilde{A}_0(p_0\neq0, \underline{0})$ enter in the dynamics only through the coupling of the photon to matter fields.\\
A \emph{quenching} of the zero mode, used in the literature, can be the following 
\begin{align}
\label{eq:zero_mode_fix}
{\rm QED}_{\rm TL}:\; \widetilde{A}_\mu(0) &= 0.
\end{align}
That clearly violates the \quotes{quantization condition} on the $r_\mu$, see Eq.~\ref{eq:quant_cond}, i.e.~they must be integers. This is equivalent to violate the periodic boundary conditions on the fermion field, hence it is a quantifiable finite volume ef\mbox{}fect.
Another way to quench those modes is to add the following constraints
\begin{align}
{\rm QED}_{\rm L}:\; \begin{cases}
\widetilde{A}_\mu(0)= 0\\
\widetilde{A}_j(p_0\neq0, \underline{0})=0
\end{cases}.
\end{align}
This choice is used in simulation with dynamical fermion 
and the error consists, again, in a quantifiable finite volume ef\mbox{}fect \cite{Borsanyi:2014jba}.

\section{Generation of quenched QED configurations}

In this section we summarize how to generate QED configurations
in Feynman gauge.
For the generation in Coulomb gauge see App.~\ref{app:ecg_gen}.

\subsection{Generation in Feynman gauge}
\label{subsect:feynm_gauge_gen}

The Feynman gauge is given by fixing $\alpha=1$ in the Faddeev-Popov approach in Eq.~\ref{eq:Faddeev-Popov}
 and using the Lorenz gauge condition as constraint, i.e.
\begin{align}
G(A) = \partial_\mu A_\mu .
\end{align}
As we noticed earlier the ghosts are factorized and they do not modify the form of the action, so the Lagrangian becomes
\begin{align}
\mathcal{L}_\text{QED}[A_\mu] = \frac{1}{4}\sum_{\mu, \nu} \left(\partial_\mu A_\nu - \partial_\nu A_\mu\right)^2 + \frac{1}{2}\big( \sum_\mu \partial_\mu A_\mu\big)^2.
\end{align}
By recalling that on the lattice the integration by parts reads
\begin{align}
\sum_{x\in\Lambda} g(x) \partial_\mu^+ f(x) = -\sum_{x\in\Lambda} f(x) \partial_\mu^- g(x),
\end{align}
the total action can be rewritten as
\begin{align}
\nn
\text{S} =& \frac{1}{4}\sum_{x\in\Lambda} \left\{ \sum_{\mu, \nu} \left[\partial_\mu^+ A_\nu - \partial_\nu^+ A_\mu\right]^2 + \frac{1}{2}\big[ \sum_\mu \partial_\mu^+ A_\mu\big]^2 \right\} \\
\nn
=& \frac{2}{4}\sum_{x\in\Lambda}\sum_{\mu, \nu} \bigg\{ \partial_\mu^+ A_\nu\partial_\mu^+ A_\nu - \partial_\mu^+ A_\nu\partial_\nu^+ A_\mu + \partial_\nu^+ A_\nu\partial_\mu^+ A_\mu \bigg\}\\
\nn
=& \frac{1}{2}\sum_{x\in\Lambda}\sum_{\mu, \nu} \bigg\{ -A_\nu \partial_\mu^-\partial_\mu^+ A_\nu + \cancel{A_\nu \partial_\mu^-\partial_\nu^+ A_\mu} - \cancel{ A_\nu \partial_\nu^-\partial_\mu^+ A_\mu} \bigg\}\\
=& -\frac{1}{2}\sum_{x\in\Lambda}\sum_{\mu, \nu} A_\nu(x+\hat{\nu}/2) \partial_\mu^-\partial_\mu^+ A_\nu(x+\hat{\nu}/2).
\end{align}
By Fourier transform the fields, Eq.~\ref{app:fourier_transf}, we find that the gauge operator is diagonal in momentum space, for the computations see the App.~\ref{gen_feyn_gauge}, and the action is
\begin{align}
\text{S} = \sum_{p\in\widetilde{\Lambda}} \frac{\hat{p}^2}{2} \sum_\nu \big|\widetilde{A}_\nu(p)\big|^2,
\end{align}
where $\hat{p}^2 = \sum_\mu \hat{p}_\mu^2$.\\
The generation of the electromagnetic field in Feynman gauge amount in generate the 
Fourier components according to a Gaussian distribution and then transform back to obtain the field in coordinate space.

\section{Gauss law problem}

In this section we present the Gauss (and Ampere) law problem in finite
volume and how it is solved 
by fixing the zero mode \cite{Hayakawa:2008an, Davoudi:2014qua}.
The problem consists in the impossibility to build states with non-vanishing charge.\\
The Gauss law is given by
\begin{align}
\sum_j \partial_j E_j(x) = \rho(x),
\end{align}
where $\underline{E}$ is the electric field and $\rho$ the charge density distribution in space.
In a finite volume with periodic boundary conditions on the fields the charge is always vanishing, indeed
\begin{align}
Q = \int_{V_3} \de^3 x\, \rho(x) =  \int_{V_3} \de^3 x\, \sum_j \partial_j E_j(x) = \int_{\partial {V_3}} \de\Sigma\, \underline{E}(x)\cdot\hat{n} = 0,
\end{align}
where $Q$ is the charge, $V_3$ the spatial volume, $\partial {V_3}$ its boundary, $\Sigma$ the surface element and $\hat{n}$ its normal. 
The above result can be understood as follow: the lines of the field that are going outside of the volume are the same coming inside after a complete revolution over the torus, 
because of the periodic boundary conditions. 
This simple argument shows how a single charge is forbidden in a finite volume.

By fixing the zero mode to be constant\footnote{Zero is not strictly necessary.} one can show that we allow for non-vanishing charges in the finite volume.
The constraint $\int\de^4x\,A_\mu = c_\mu$, with $c_\mu$ constant (in configuration space), leads to the \emph{non-local} Lagrangian, see the App.~\ref{app:ELELC} for details,
with $j_\mu$ external current
\begin{align}
\mathcal{L}[A_\mu, j_\mu] = \frac{1}{4} \sum_{\mu, \nu} F_{\mu\nu}^2 + \sum_{\mu}(j_\mu - \frac{c_\mu}{\alpha}) A_\mu + \frac{1}{2\alpha}\sum_\mu A_\mu \int\de^4y\, A_\mu(y),
\end{align}
which violates reflection positivity because the last term connects fields at arbitrary positive and negative times, as noted in Ref.~\cite{Borsanyi:2014jba}.\\
The Euler-Lagrange equation of motion are, Eq.~\ref{app:modified_euler_lagrange}
\begin{align}
\sum_{\nu} \partial_\nu F_{\nu\mu}(x) = j_\mu(x) + b_\mu,
\end{align}
with $b_\mu$ given by Eq.~\ref{app:b_nu},
\begin{align}
b_\mu = \frac{1}{\alpha}\left(\int\de^4y\, A_\mu(y) - c_\mu \right).
\end{align}
Ef\mbox{}fectively, by constraining the zero mode to a constant, we introduced a uniform time independent background current in the system.
By integrating in the spatial volume the time component to obtain the charge we find
\begin{align}
\int_{V_3} \de^3 x\, \sum_\mu \partial_\mu F_{\mu 0}(x) = \int_{V_3} \de^3 x\, \sum_j \partial_j E_j(x) \equiv 0 = \int_{V_3} \de^3x\, j_0(x) + \int_{V_3} \de^3x\, b_0 = Q + b_0 V_3,
\end{align}
which implies $ b_0 = - Q/V_3$. Intuitively we are spreading an opposite charge over the volume, in this way we solve locally the problem of a charge in a finite volume but not globally, because the total net charge is still vanishing. 
Analogous arguments hold for the Ampere law.\\
Furthermore one can recover the imposed constraint by looking at $b_0$
\begin{align}
b_0 = \frac{1}{\alpha}\left(\int\de^4x\, A_0(x) -c_0\right) =- Q/V_3 \Longrightarrow \int\de^4x\, A_0(x) = - \alpha Q/V_3 + c_0 \underset{\alpha\rightarrow 0}{\longrightarrow} c_0.
\end{align}

\section{Massive QED}

As we saw in the previous sections one way to correct the Gauss law, and therefore allow for a net charge on a finite volume, is to fix the zero mode to be constant, e.g.~$\tilde{A}_\mu(p=0)=0$, but the resulting Lagrangian violates reflection positivity. 
The other solution, QED$_{\rm L}$, possesses reflection positivity but the constraint is still non-local and many properties of local quantum field theory are not automatically guaranteed\footnote{For example: renormalizabilty, volume-independence of renormalization constants, Symanzik improvement programme, etc.}.\\
The most economical and straightforward choice that retains locality is the massive approach on the lattice, first presented in Ref.~\cite{Endres:2015gda}. As in perturbation theory one can add a mass term for the photon, which act as an IR regulator, and at the end take the limit of vanishing photon mass, $m_\gamma \rightarrow 0$. We know that under some condition, which will be clarified in the following, the theory is still renormalizable.

In the following we will investigate massive QED with dif\mbox{}ferent gauge fixing terms.

\subsection{Renormalizability: the need for Feynman gauge}

The first issue we encounter with massive QED is the renormalizability, nice discussions on the 
subject are found in Refs.~\cite{ZinnJustin:2002ru, Ruegg:2003ps}. We start with the Proca action
\begin{align}
\text{S}_{\text{Proca}} = \int\de^4x \left[\frac{1}{4}F_{\mu\nu}^2 + \frac{1}{2}m^2_\gamma A_\mu^2\right].
\end{align}
The propagator in Fourier space is
\begin{align}
\Delta_{\mu\nu}(k) = \frac{\delta_{\mu\nu} + k_\mu k_\nu /m^2_\gamma }{k^2 + m^2_\gamma } \xrightarrow{\: k \to \infty \: }\frac{1}{m^2_\gamma }  ,
\end{align}
that in the infinite momentum limit goes to a constant, hence the resulting theory seems to be non-renormalizable by power-counting \cite{Reisz:1987da}.
We can overcome the non-renormalizability by adding a \emph{free massive scalar} field $\chi$ to the action 
\begin{align}
\text{S}_{\text{Stueck}} = \int\de^4x \left[\frac{1}{4}F_{\mu\nu}^2 + \frac{1}{2}m^2_\gamma A_\mu^2 - \frac{1}{2} \left(\partial_\mu\chi\right)^2 + m_\chi^2\chi^2 \right] = \text{S}_{\text{Proca}} + \int\de^4x \left[- \frac{1}{2} \left(\partial_\mu\chi\right)^2 + m_\chi^2\chi^2 \right] .
\end{align}
Now by operating the following change of variables
\begin{align}
\label{eq:stueck_transf}
A_\mu = A'_\mu + \frac{1}{m_\gamma}\partial_\mu \chi,
\end{align}
which has the look of a gauge transformation, the field-strength is left invariant, while the mass term changes as
\begin{align}
\frac{1}{2}m^2_\gamma A_\mu^2 = \frac{1}{2}m^2_\gamma (A'_\mu)^2 + m_\gamma A'_\mu\partial_\mu\chi + \frac{1}{2}\left(\partial_\mu\chi\right)^2.
\end{align}
The resulting action is given by
\begin{align}
\text{S} = \int\de^4x \left[\frac{1}{4}(F'_{\mu\nu})^2 + \frac{1}{2}m^2_\gamma (A'_\mu)^2 + m_\gamma A'_\mu\partial_\mu\chi - \frac{1}{2} m_\chi^2\chi^2\right],
\end{align}
and the integration over $\chi$, done by completing the square, leads to
\begin{align}
\nn
\text{S} &= \int\de^4x \left[\frac{1}{4}(F'_{\mu\nu})^2 + \frac{1}{2}m^2_\gamma (A'_\mu)^2 + \frac{m^2_\gamma }{2m_\chi^2} \left(\partial_\mu A'_\mu\right)^2\right] \\
\label{eq:mass_feyn}
&\equiv \int\de^4x \left[\frac{1}{4}(F'_{\mu\nu})^2 + \frac{1}{2}m^2_\gamma (A'_\mu)^2 + \frac{1}{2\xi} \left(\partial_\mu A'_\mu\right)^2\right],
\end{align}
that is the Proca action with the addition of a Lorenz gauge fixing term with parameter $\xi \equiv m_\chi^2 / m_\gamma^2$. 
The propagator in momentum space now looks
\begin{align}
\Delta_{\mu\nu}(k) = \frac{\delta_{\mu\nu}}{k^2+m^2_\gamma } + \frac{(\xi-1)k_\mu k_\nu}{(k^2+m^2_\gamma )(k^2+\xi m^2_\gamma )}\xrightarrow{\: k \to \infty \: } 0,
\end{align}
which as the correct UV behavior, and the theory is renormalizable by power counting.

\subsection{Solution of the Gauss law problem}

We want to investigate the possibility of having a charge in a box in the case of massive QED in the Feynman gauge ($\xi=1$), see Eq.~\ref{eq:mass_feyn}.\\
The action in the presence of a coupling with an external current $j_\mu$ is 
\begin{align}
\int\de^4x \left[\frac{1}{4}(F'_{\mu\nu})^2 + \frac{1}{2}m^2_\gamma (A'_\mu)^2 + \frac{1}{2} \left(\partial_\mu A'_\mu\right)^2 + j_\mu A'_\mu\right]
\end{align}
and the Euler-Lagrange equation of motion is given by
\begin{align}
\label{eq:eu_mass_feyn}
\sum_\nu \partial_\nu F'_{\nu\mu}(x) = j_\mu(x) + (m^2_\gamma -\Box) A'_\mu.
\end{align}
By integrating in the spatial volume the time component, to find the charge, we have
\begin{align}
{\rm l.h.s. :}\, \int_{V_3} \de^3x \sum_\nu \partial_\nu F'_{\nu 0}(x) = \int_{V_3} \de^3x \sum_j \partial_j E'_j(x)\equiv 0
\end{align}
that is still vanishing due to periodic boundary condition imposed on the fields, while the r.h.s.~of Eq.~\ref{eq:eu_mass_feyn} becomes 
\begin{align}
{\rm r.h.s. :}\, \int_{V_3} \de^3x \,j_0(x) + \int_{V_3} \de^3x (m^2_\gamma -\Box) A'_0 = Q + \int_{V_3} \de^3x (m^2_\gamma -\Box) A'_0.
\end{align}
By combining the results we get that the condition on the charge is 
\begin{align}
Q = \int_{V_3} \de^3x (\Box-m^2_\gamma ) A'_0,
\end{align}
from which we conclude that the charge is not restricted to be zero, in contrast to massless QED in a finite box (with the same periodic boundary conditions).

\paragraph{Coulomb gauge and mass term}

We can ask ourself what happens if we use a dif\mbox{}ferent gauge condition on top of the addition of the mass term. The Lagrangian in the case of Coulomb gauge fixing is given by
\begin{align}
\mathcal{L} = \frac{1}{4} (F_{\mu\nu})^2 + \frac{1}{2} m_\gamma^2 A_\mu^2 + \frac{1}{2} \sigma \left(\partial_i A_i\right)^2,
\end{align}
where $\sigma$ is a dimensionless parameter and it ensures that in the massless limit the fields satisfies the Coulomb gauge condition.\\
By following Ref.~\cite{Mueck:2013wba} we find that the propagator for the above Lagrangian is given by
\begin{align}
\Delta_{\mu\nu}(k) \propto \frac{1}{k^2+m^2_\gamma }\left[\delta_{\mu\nu} + \frac{ (k^2+m^2_\gamma  +\sigma |\underline{k}|^2)k_\mu k_\nu + 2 \sigma |\underline{k}|^2 k_i (\delta_{i\mu}k_\nu + \delta_{i\nu} k_\mu) + \sigma m^2_\gamma  \delta_{i\mu}\delta_{j\nu}k_i k_j  }{m^2_\gamma (k^2+m^2_\gamma  +\sigma |\underline{k}|^2) + \sigma  |\underline{k}|^2}\right],
\end{align}
which has a smooth massless limit but in the infinite momentum limit goes to a constant, hence the theory is not renormalizable by power counting. 
It should be noted that the second term, which is responsible for the non renormalizability of the theory, becomes a pure gauge term, in the massless limit, when the photon is coupled to a conserved current, i.e.~$\partial_\mu J_\mu =0 \leftrightarrow k_\mu \widetilde{J}_\mu=0$, and it does not play any role\footnote{On the lattice we may encounter
a noise cancellations problem, since the term has to disappear on average.}.

The problem of the charge it is also solved in this particular setting and the conclusion made for the previous section can be applied also here. 
Although one has to remember to take first the infinite volume limit, such that the zero mode problem is resolved, then the massless limit, such that we recover 
a renormalizable theory (at least by perturbation theory power counting) and finally the continuum limit, which strictly speaking does not exists
because of the Landau pole.
That because there is no mechanism similar to the Stueckelberg one at the moment for the massive QED with Coulomb gauge condition.

\section{Wilson loops}

We tested the code written for the generation of quenched QED configurations by calculating the exact value of the Wilson loops in an infinite lattice, see appendix \ref{app:wils_loop}. 
The expectation value of a square Wilson loop in the plane $(\mu, \nu)$ is given by
\begin{align}
w_{\mu\nu}(I,I) & = \exp\left( -2e^2Q^2\left[C_\mu(I,0) - C_\nu(I,I\hat{\nu})\right]  \right),
\end{align}
where $C_\mu(I,x)$ is the sum of propagators in the $\mu$ direction, which is one side of the Wilson loop,
\begin{align}
C_\mu(I,x) = a^2 I D(x) + a^2\sum_{\tau=1}^{I-1} (I-\tau) D(x+a\tau\hat{\mu}),
\end{align}
and $D$ is the massless/massive scalar propagator in coordinate space on the lattice.
The propagator is calculated in the infinite volume through the L\"uscher-Weisz algorithm \cite{Luscher:1995zz}, see App.~\ref{app:CSM} for the algorithm details, 
while the Borasoy-Krebs algorithm in Ref.~\cite{Borasoy:2005ha} is employed for the massive propagator.
For Lorenz invariance $w_{\mu\nu}(I,I)$ takes always the same value for each choice $\mu,\nu$,
so we write directly $w(I, I)$.

\subsection{Massless case}

\begin{figure}[!ht]
\begin{center}
\subfigure[\emph{Comparison between the Wilson loop values in Feynman (red segments) and extended Coulomb gauge (blue stars) and the infinite lattice predictions (LW) (dashed line).}]{\includegraphics[scale=0.8,angle=-0]{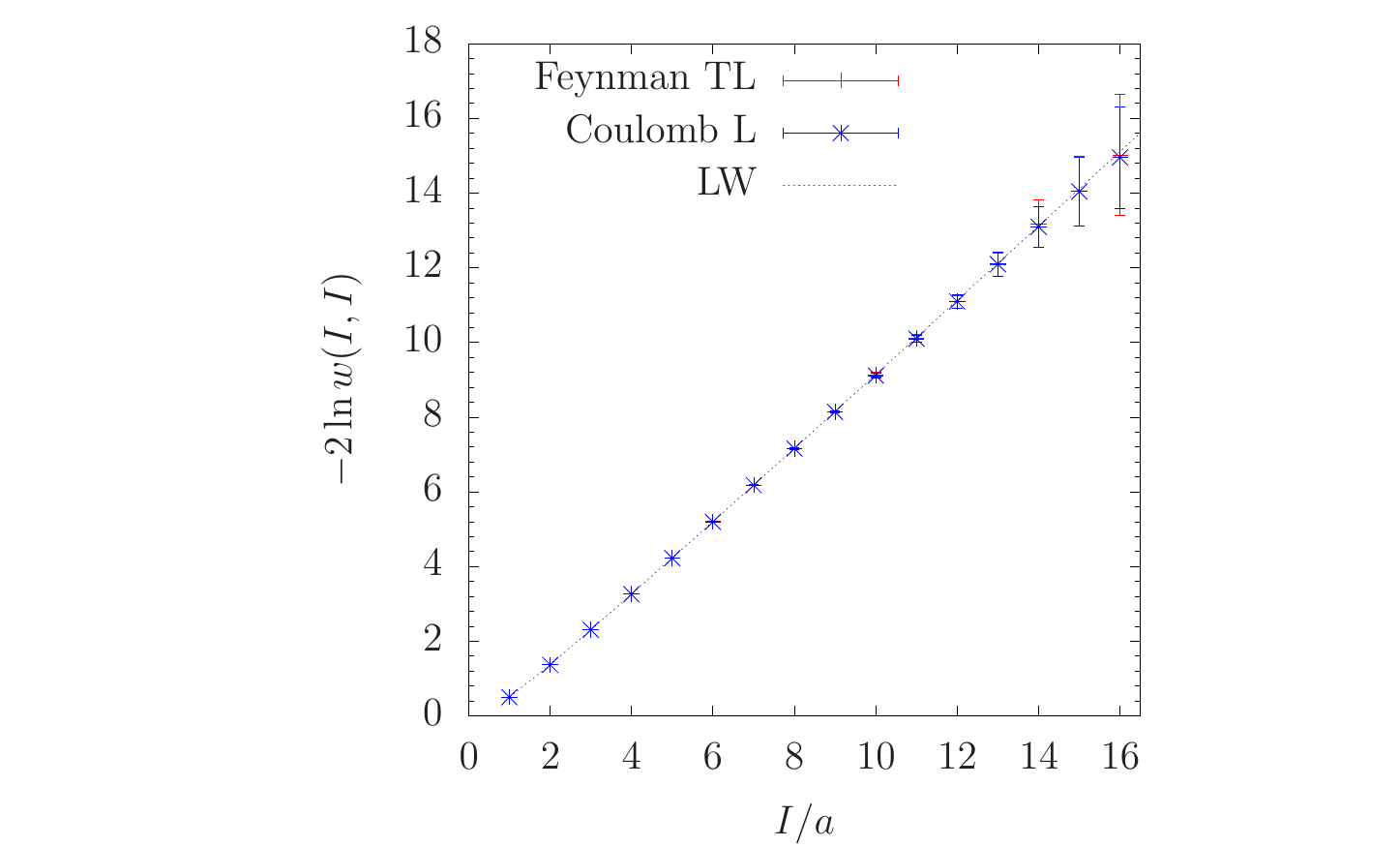}}
\subfigure[\emph{Dif\mbox{}ference between the Wilson loop values calculated in Feynman and extended Coulomb gauge and the infinite lattice predictions.}]{\includegraphics[scale=0.8,angle=-0]{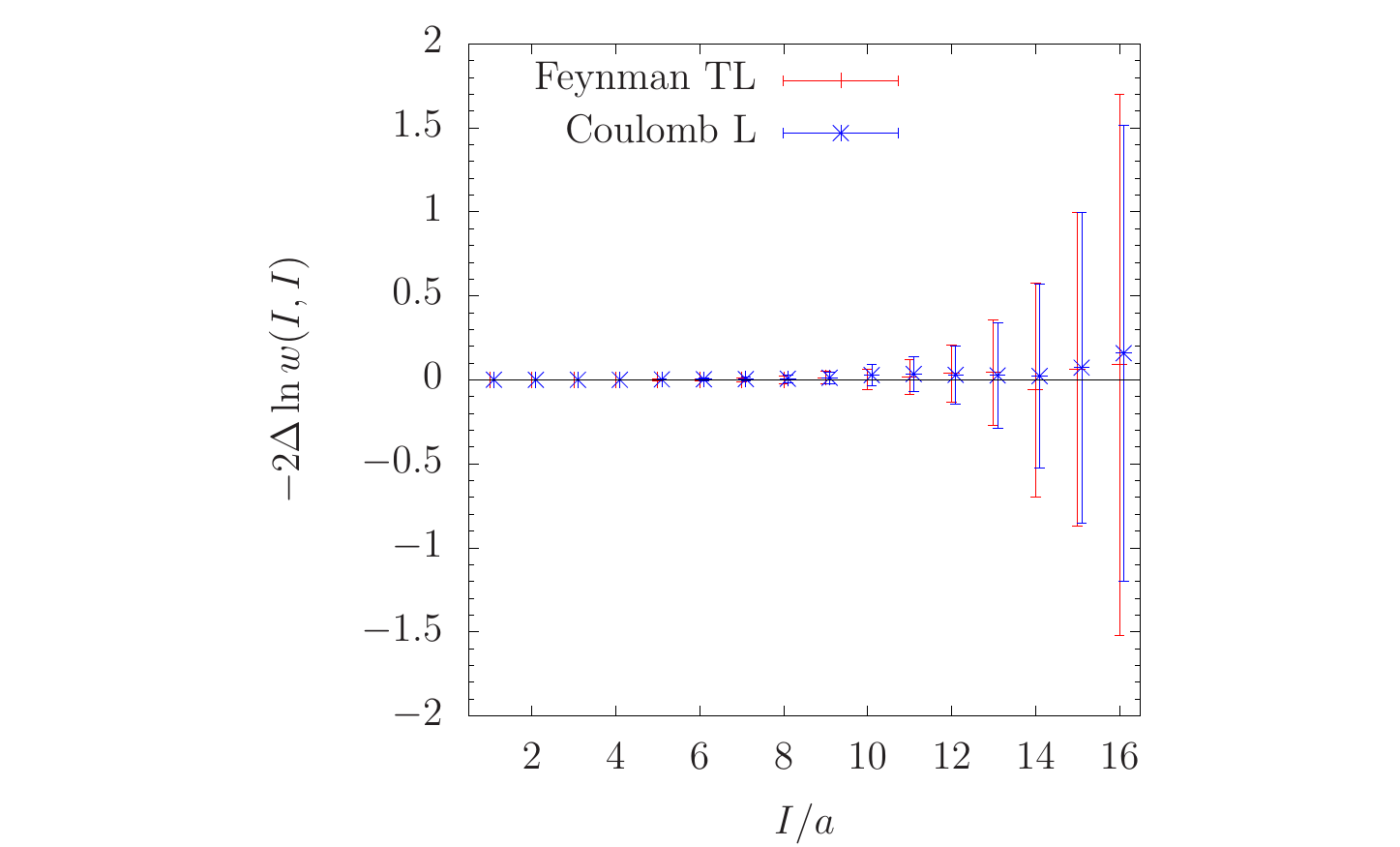}}
\caption{\emph{Wilson loop values in a volume $V=32^4$ and their comparison with the infinite volume predictions calculated through the L\"uscher-Weisz algorithm.}}
\label{fig:tests}
\end{center}
\end{figure}

In Fig.~\ref{fig:tests} we show the comparison between the generation in Feynman and (extended) Coulomb gauge with the infinite lattice predictions (calculated in App.~\ref{app:wils_loop}). 
The quantity plotted is the log of the average of the Wilson loop over 100 configurations in a $32^4$ volume, with $eQ=1$. 
The errors are calculated with the standard deviations since the configurations are not correlated at all.

%%%%%%%%%%%%%%%%%%%%%%%%%%%%%%%%%%%%%%%%%%%%%%%%%%%%%%%%%%%%%%%%%%%%%%%%%
%%%%%%%%%%%%%%%%%%%%%%%%%%%%%%%%%%%%%%%%%%%%%%%%%%%%%%%%%%%%%%%%%%%%%%%%%%

\subsection{Massive case}

\begin{figure}[h!t]
\begin{center}
\subfigure[\emph{Massive Feynman gauge. Comparison with the massless result (dashed line) using the Luscher-Weisz algorithm and massive one (solid line) using the Borasoy-Krebs algorithm.}]{\includegraphics[scale=0.7]{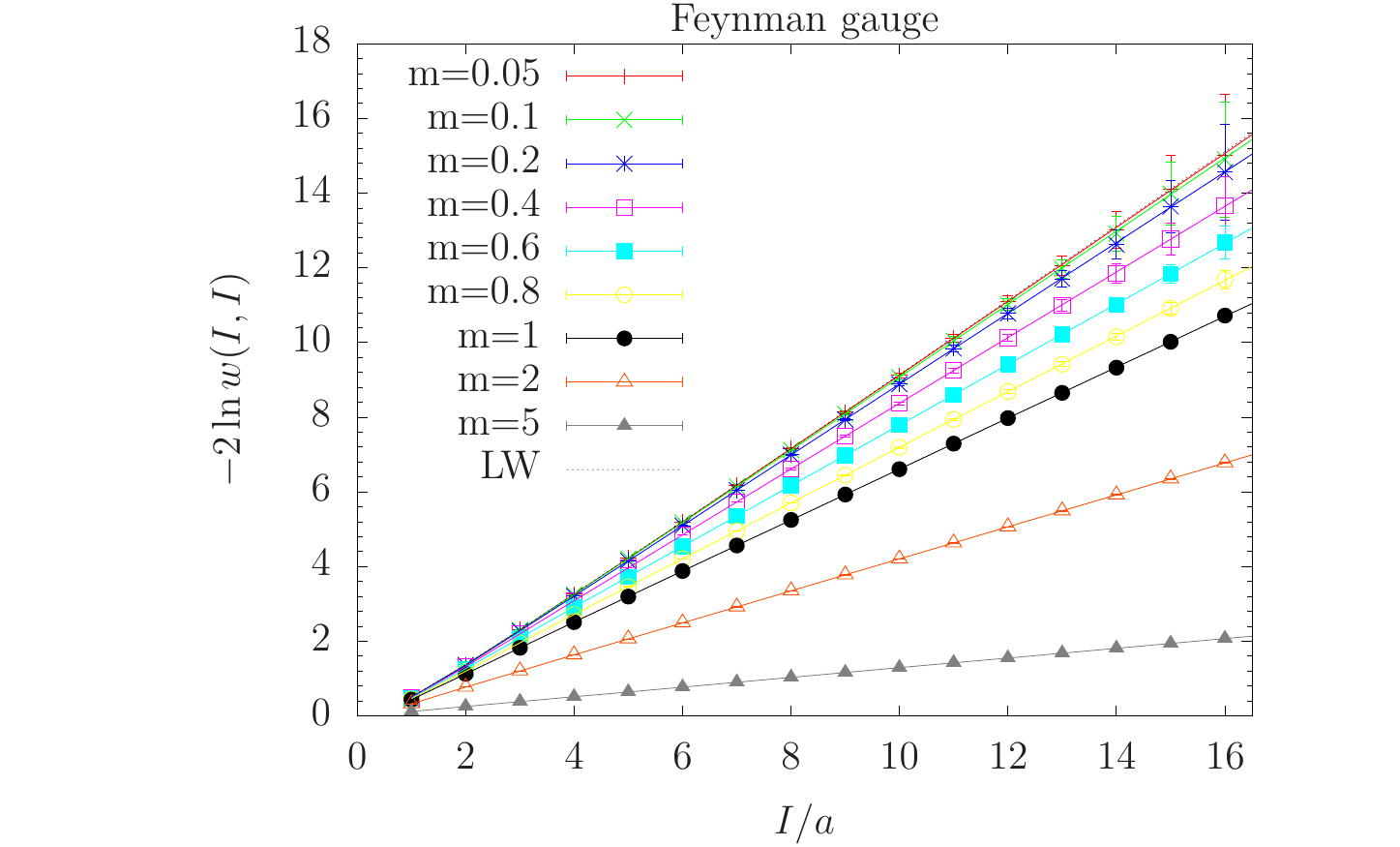}\label{fig:feynman}}
\subfigure[\emph{Massive Coulomb gauge. Comparison with the massless result (dashed line) Luscher-Weisz.}]{\includegraphics[scale=0.7]{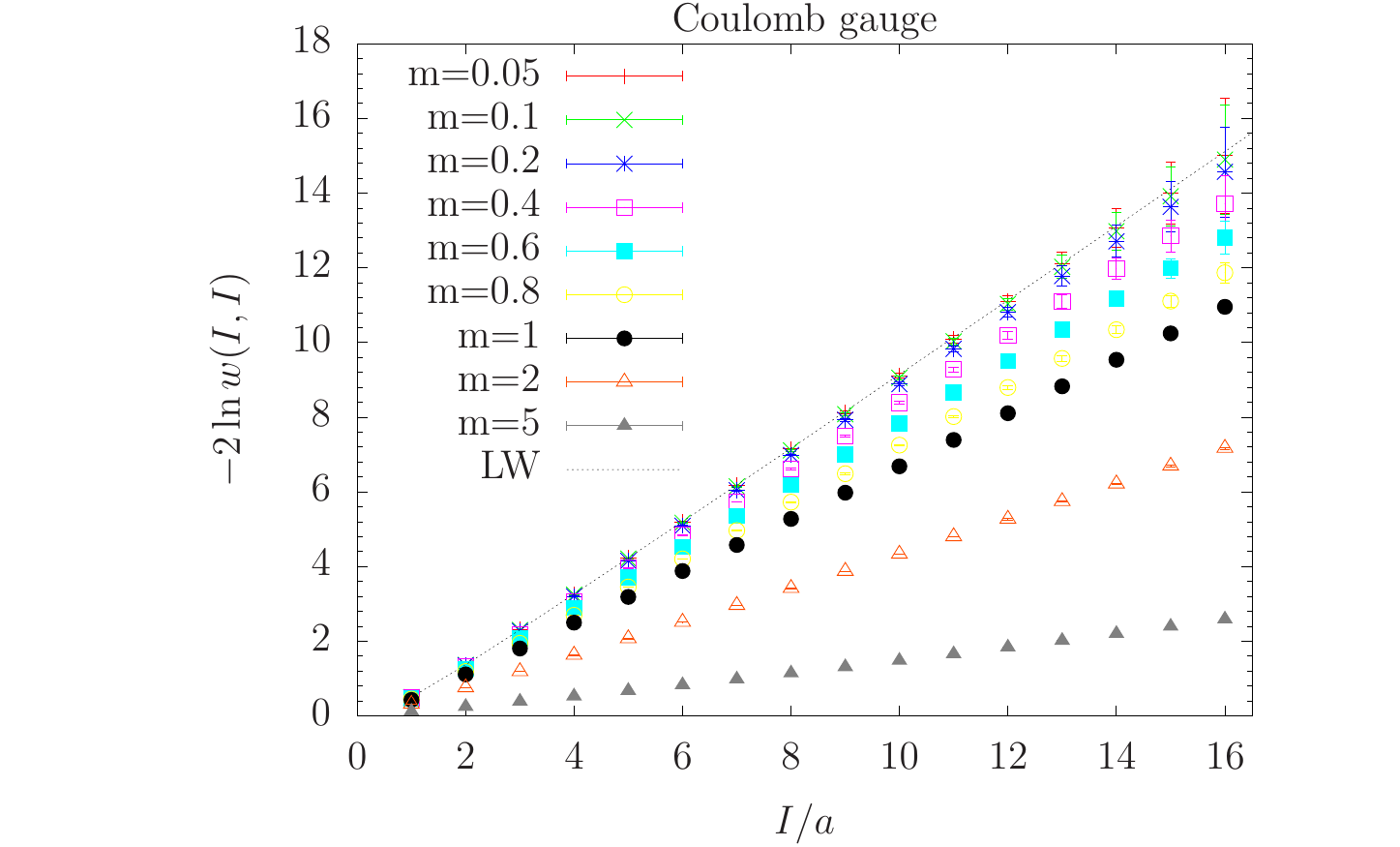}\label{fig:coulomb}}
\caption{\emph{Comparison of Wilson loop values from simulations and infinite volume predictions calculated through the Borasoy-Krebs algorithm.}}
\label{fig:plaquette}
\end{center}
\end{figure}

In Fig.~\ref{fig:plaquette} we show the comparison between the generation in Feynman and Coulomb gauge, $eQ=1$,
with the infinite lattice predictions, calculated as explained above. The quantity plotted is the log of 
the average of the Wilson loop over 100 configurations in a $32^4$ volume, with $eQ=1$.
The errors are calculated through the standard deviations.
%%%%%%%%%%%%%%%%%%%%%%%%%%%%%%%%%%%%%%%%%%%%%%%%%%%%%%%%%%%%%%%%%%%%%%%%%%
%%%%%%%%%%%%%%%%%%%%%%%%%%%%%%%%%%%%%%%%%%%%%%%%%%%%%%%%%%%%%%%%%%%%%%%%%

\afterpage{\blankpage}

\setcounter{equation}{0}

\chapter{Reweighting twisted boundary conditions }
\label{chap:rtbc}

Simulations of Euclidean quantum field theories are necessarily performed 
in a finite space-time volume, hence imposition of boundary conditions on the fields 
has to be made.
Although dif\mbox{}ferent choices for boundary conditions have the same correct infinite volume 
limit, its approach depends on the choice.
Recent examples, in lattice QED, of dif\mbox{}ferent finite volume ef\mbox{}fects can be found in \cite{Lucini:2015hfa} 
(${\rm C}^\star$ boundary conditions) and in \cite{Borsanyi:2014jba} (periodic boundary conditions along with exclusion of zero modes).
Algorithmic ef\mbox{}ficiency and sampling properties of the simulations are also af\mbox{}fected by the boundary 
conditions.
Examples of that can be found in \cite{Luscher:2011kk} for the use of open boundary conditions in lattice QCD
to bypass the freezing of topology (\cite{Mages:2015scv} for the same with C$^\star$ boundary conditions), and in \cite{DellaMorte:2010yp} for the use of generalized boundary conditions in order to exponentially
improve the signal-to-noise ratio in glueball correlation functions computed in the pure gauge theory. 

Usually in lattice QCD (anti)-periodic boundary conditions in (time)-space are imposed on the fermionic fields. 
That leads to a quantization of the spatial momenta in units of $2\pi/L_\mu$.
In a two body hadron decay the energies of the resulting particles cannot assume physical values unless their masses are consistent with the momentum 
quantization rule.
In addition, for several phenomenological applications a fine resolution of small momenta is needed, examples include form factors 
(as $K \to \pi \ell \nu$ transition), charge radius of the pion, and the hadronic vacuum polarization of the photon (relevant for $g-2$ anomaly).
This still represents a severe limitation, in order to have the lowest non-zero momentum around 100 MeV 
we need lattices of about 12 fm. Since we also want to keep discretization ef\mbox{}fects under control lattice spacings $a$ around 0.1 fm and 
below are necessary, then simulations are still expensive in computer time.

Twisted boundary conditions (TBCs) were first introduced in \cite{deDivitiis:2004kq, Bedaque:2004kc} and they serve as a solution to the problem of 
quantized momenta on the lattice in QCD. TBCs correspond to periodic ones up to a phase, the twisting angle $\theta$, 
for fermion fields in spatial directions only and they allow the lattice momenta to be continuously varied.
TBCs represent a standard tool in the lattice community. 
They were used in the computation of quantities above mentioned (see \cite{Boyle:2007wg,Brandt:2013dua,DellaMorte:2011aa} for a selection of them)
and also for RI-MOM scheme \cite{Arthur:2010ht} and matching between Heavy Quark Ef\mbox{}fective Theory and QCD \cite{DellaMorte:2013ega}.

In all the simulations so far the twisting is only applied in the valence sector while the sea fermions obey periodic boundary conditions.
This kind of set-up is usually referred to as \emph{partial twisting}. 
That introduces a unitarity breaking as a boundary ef\mbox{}fect, which has to disappear in the infinite volume limit.
In \cite{Sachrajda:2004mi} this was explicitly checked by means of Chiral Perturbation Theory for meson masses, decay constants and semileptonic form factors.
This also suggests that reweighting techniques might be used to change the boundary conditions for the sea fermions and compensate for the 
unitarity violation.

In the following we will see that resulting reweighting factors are ratios of fermionic determinants that go to one in the infinite volume limit.
In that regime is also dif\mbox{}ficult to give a reliable estimate of the reweighting factors, since they are extensive quantities.
Unitarity violations, if present, are expected in rather small volumes where the reweighting factors can be reliably calculated.

The chapter is organized as follows; in Section~1 we review twisted boundary conditions,
in Section~2 we recall the reweighting method and specialize it to the twisted boundary conditions case,
in Section~3 we introduce the multi-step method based on tree-level exact studies 
on reweighting factors and their variances.
Simulation parameters and Monte-Carlo results in
small and large volumes 
are presented in Section~4.
Section~5 contains pion and quark mass dependence on the twisting angle.
Section~6 contains our conclusions.
The chapter reflects the work presented in Refs.~\cite{Bussone:2016lty, Bussone:2015yja}.

\section{Twisted boundary conditions}

Generic boundary conditions for fermion fields in lattice QCD on a torus are discussed in \cite{Sachrajda:2004mi}.
There it is pointed out that fields do not need to be single valued on the torus, while the action does. 
Hence periodicity conditions on the fermions can be of the form
\begin{equation}
	\Psi\left(x+L_\mu\hat{\mu}\right) = V_\mu \Psi\left(x\right)\;, \quad \mu=1,2,3 \,,
\end{equation}
where $\Psi$ is a flavor multiplet and $V_\mu$ represents a unitary transformation associated to a symmetry of the action.
Similarly, for the $\overline{\Psi}$ one requires
\begin{equation}
	\overline{\Psi}\left(x+L_\mu\hat{\mu}\right) = \overline{\Psi}\left(x\right) V_\mu^\dagger \;, \quad \mu=1,2,3 \,.
\end{equation}
If we consider generic values of the diagonal quark mass matrix, one concludes that $V_\mu$ also has to be diagonal in flavor space, i.e.~
\begin{equation}
	\psi\left(x+L_\mu\hat{\mu}\right) = e^{i\theta_\mu} \psi\left(x\right)\,,
\label{theperiodicity}
\end{equation}
where the twisting angles $\theta_\mu$ have been introduced for each flavor ($\psi$ is a component of the $\Psi$ flavor multiplet) and 
\begin{equation}
\theta_\mu=\left(0,\, \underline{\theta}\right),\quad\text{where }\theta_j\in\left[0,2\pi\right)\text{ are fixed}.
\end{equation}
By Fourier transforming the equation Eq.~\ref{theperiodicity} we obtain
\begin{equation}
\int\de^4k\, 
\e^{ik\cdot (x+L_\mu\hat{\mu})}
\widetilde{\psi}(k)
=\e^{i\theta_\mu}
\int\de^4k\,
\e^{ik\cdot x}\widetilde{\psi}(k).
\end{equation}
Since the fermionic field is an arbitrary function of $k$, in order to satisfy the above equality the exponentials have to be identical
\begin{align}
\label{eq:theta_momenta}
\exp\left[i\left(k_\mu-\frac{\theta_\mu}{L_\mu}\right)L_\mu\right] = 1
\Longrightarrow \begin{cases}
k_0=\frac{2\pi}{T}z_0\\
k_j=\frac{2\pi}{L}z_j+\frac{\theta_j}{L}, \quad\text{ when } j=1,2,3
\end{cases},
\end{align}
where the $z$'s are integer numbers, $z_\mu\in \left( -L_\mu /2 , L_\mu/2 \right] \subset \mathbb{Z}$.
Eq.~\ref{eq:theta_momenta} shows the equivalence between TBCs and momenta shift.
Note that in the infinite volume limit the $\theta$-term disappear, for fixed $\theta$, in Eq.~\ref{eq:theta_momenta}.
On the contrary, when the gap between momenta is not negligible, thanks to $\theta$ we have direct access to values of momenta that are not allowed with PBCs.

An equivalent way to impose TBCs is due to \cite{Bedaque:2004kc}. We can fix the fermionic fields to be periodic and introduce a
\emph{constant}  $\mathbf{U}(1)$ interaction with constant background magnetic potential $\underline{A}=\underline{\theta}/L$,
vanishing electric, magnetic fields and vanishing electric potential.
This interaction can be implemented by transforming the standard $\mathbf{SU}(N_c)$ gauge links $U_\mu(x)$ in the following way
\begin{equation}
\label{eq:modified_links}
\mathcal{U}_\mu(x)=\exp\left[i\frac{\theta_\mu}{L_\mu}\right] U_\mu(n)
= \begin{cases}
U_0(x)\\
\exp\left[i\frac{\theta_j}{L_j} \right] U_j(x), \quad\text{ when } j=1,2,3,
\end{cases}
\end{equation}
note that the exponential factor is independent of $x$. 
The fermions $\psi$ are kept with PBCs and in order to see the equivalence between the two formulations it is enough to observe that the phase 
can be re-absorbed by re-writing the $\psi$ fields in terms of
\begin{equation}
\label{eq:modified_fermion}
					\chi(x)=
						\e^{i\, \underline{\theta} \,\cdot\, \underline{x} / L}\psi(x)\;,
\end{equation}
and to notice that the $\chi$ are indeed periodic up to a phase, as for Eq.~\ref{theperiodicity}.\\
In practice, $\mathbf{SU}(N_c)$ gauge configurations are typically produced 
for one specific choice of $\theta$, and the angle is then varied only when computing the
quark propagators, which is cheaper in terms of CPU-time with respect to the generation
of configurations. As a consequence, the quark propagators in the sea and valence sectors dif\mbox{}fer,
which causes a breaking of unitarity already at the perturbative level.

\section{Reweighting technique}

In the following we use the un-improved Wilson action in Eq.~\ref{eq:wilson_fermions},
 with the links $U_\mu(x)$ replaced by the $\mathcal{U}_\mu(x)$ 
as in Eq.~\ref{eq:modified_links}, and Wilson plaquette action in Eq.~\ref{eq:wilson_plaquette} for the gauge links. 
The link redefinition does not change the plaquette (see Eq.~\ref{eqapp:plaquette}), and therefore the pure gauge term in the action.
The covariant derivatives and the Wilson term are, on the contrary, modified. 
For this purpose we denote the Dirac operator with gauge link $U_\mu$ and $\theta$-angle as  $D[U,\theta]$.
Once the fermionic degrees of freedom are integrated out on each 
$\mathbf{SU}(N_c)$ gauge background, the $\theta$-dependence from the sea sector is completely absorbed in the fermionic determinant.

Let us imagine we want to compute the value of some observables
for one choice of bare parameters  $B = \{\beta', m'_1, m'_2, \dots, m'_{n_f}, \theta'_\mu, \dots\}$ using
the configurations produced at a slightly dif\mbox{}ferent set of parameters  $A = \{\beta, m_1, m_2, \dots, m_{n_f}, \theta_\mu, \dots\}$.
To do so we need to compute on each configuration of the $A$-ensemble the reweighting factor $W_{A B} = P_B /P_A$, 
which is the ratio of the two probability distributions and it is an extensive quantity, 
$P_A[U] = \e^{-\text{S}_{\text{G}}[\beta, U]} \prod_{i=1}^{N_f}\det\left(D[U,\theta]+m_i\right)$ ($N_f$ being the number of fermion species).
The expectation values on the $B$-ensemble can then be expressed as
\begin{equation}
\langle\mathcal{O}\rangle_B = \frac{\langle\widetilde{\mathcal{O}} W_{A B}\rangle_A}{\langle W_{A B}\rangle_A}\, ,
\end{equation}
with $\widetilde{\mathcal{O}}$ being the observable defined after Wick contractions, 
and $\langle\dots\rangle_{A}$ indicates that expectation values have to be taken on the  $A$-ensemble.
By specializing ourself to the case where
only the fermionic $\theta$-angles are changed from one bare set to the other, 
we obtain the following expression of the reweighting factor 
\begin{align}
\label{eq2:rew_fact}
		W_\theta = \det\left(D_W[U,\theta]D_W^{-1}[U,0]\right) = \det\left(D_W[\mathcal{U},0]D_W^{-1}[U,0]\right),
\end{align}
where we have also chosen $D$ to be $D_W$, i.e., the massive Dirac-Wilson operator.
Following \cite{Finkenrath:2013soa} we see that ratios of determinants as those above can be estimated stochastically.
For a normal matrix $M$, whose spectrum is $\lambda(M)$, the following representation of the determinant holds
\begin{equation}
\label{eq:appl_cond}
\frac{1}{\det M}  = \int \De\left[\eta\right] \exp\left(-\eta^\dagger M \eta\right)<\infty\, \Longleftrightarrow\, \mathbb{R}\text{e}\lambda\left( M\right)>0.
\end{equation}
The positivity condition ensures the absolute convergence of the integral and the integral can be evaluated stochastically.
We can generate an ensemble of complex random vectors, say $\{\eta_k,\, k=1,2,\dots, N_\eta\}$ according to a probability distribution $p(\eta)$.
The distribution $p(\eta)$ of the vectors $\eta$ is usually taken to be Gaussian, $p(\eta) = \exp\left(-\eta^\dagger\eta\right)$, 
and in that case, the determinant (or its inverse) can be written as 
\begin{align}
\frac{1}{\det M}  = \bigg\langle \frac{\e^{-\eta^\dagger M \eta}}{p(\eta)} \bigg\rangle_{p(\eta)} = \frac{1}{N_\eta} \sum_{k=0}^{N_\eta} \e^{-\eta_k^\dagger (M-\mathbf{1}) \eta_k} + \ord\left(\frac{1}{\sqrt{N_\eta}}\right).
\end{align}
It is straightforward to generalize the positivity condition above in order to ensure the convergence of the stochastic estimates of all Gaussian moments.
In the case of an hermitian matrix one obtains
\begin{align}
\nn
\bigg\langle \frac{ \e^{-2\eta^\dagger M \eta} }{p(\eta)^2} \bigg\rangle_{p(\eta)} & = \int \De\left[\eta\right] \exp\left[-\eta^\dagger (2M-\mathbf{1}) \eta\right] < \infty\, \Longleftrightarrow\, \lambda\left( M\right)>\frac{1}{2},\\
\nn
\vdots \hspace{1.4cm}&\\
\bigg\langle \frac{ \e^{- N \eta^\dagger M \eta} }{p(\eta)^N} \bigg\rangle_{p(\eta)} & = \int \De\left[\eta\right] \exp\left[-\eta^\dagger [NM-(N-1)\mathbf{1}] \eta\right] <\infty \Longleftrightarrow\, \! \lambda\left( M\right) \! >\! \frac{N-1}{N}\! \underset{N\rightarrow\infty}{\longrightarrow} \!1.
	\end{align}
All eigenvalues should therefore be larger than unity. In particular, in the numerical studies presented here we will
always consider the square of the hermitian version of the Dirac-Wilson operator, $Q= \gamma_5 D_W$, which is to say
we consider the case of two degenerate flavors.

\section{Tree-level studies}

\begin{figure}[h!t]
\begin{center}
\subfigure[\emph{Mean of the reweighting factor at tree-level for $\theta = 0.1$ as a function of $L$.}]{\includegraphics[scale=0.7]{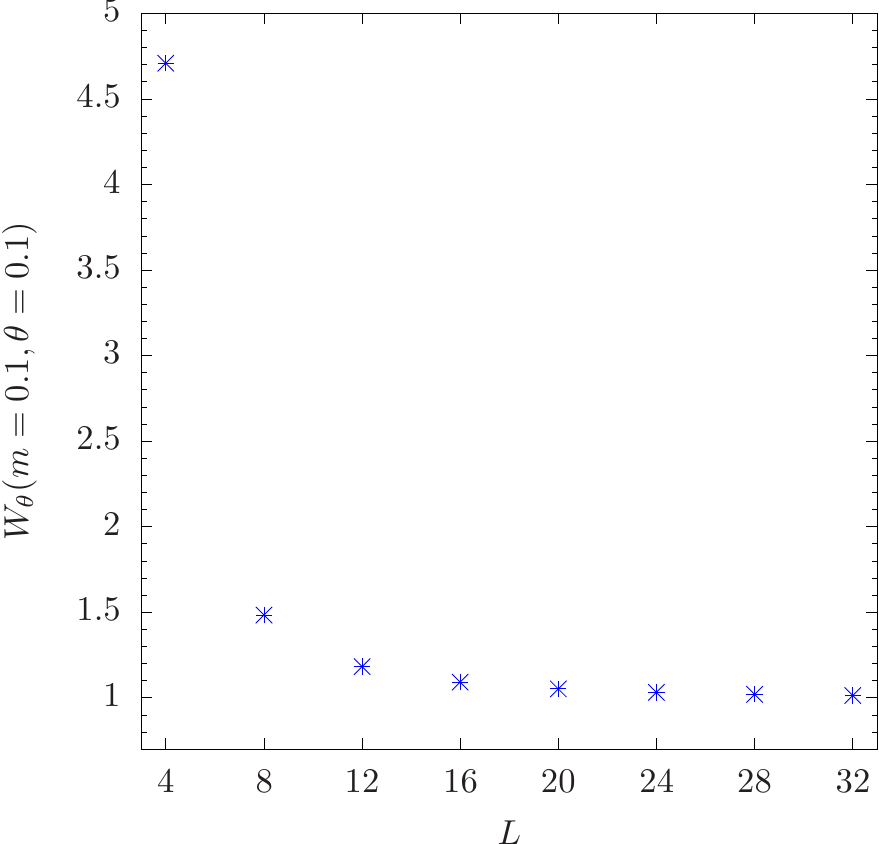}\label{fig:tree_level_test1}}
\hspace{0.5cm}
\subfigure[\emph{Relative variance of the reweighting factor at tree-level, vs $L$, for $\theta = 0.1$.}]{\includegraphics[scale=0.7]{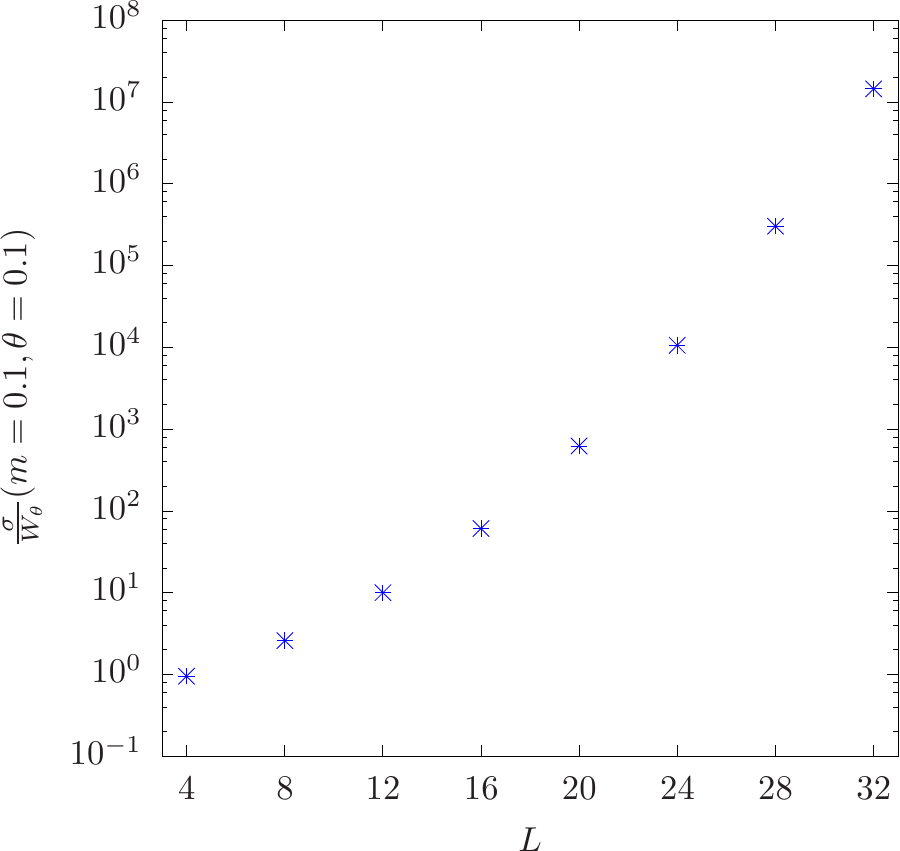}\label{fig:tree_level_test2}}
%\hspace{0.2cm}
\subfigure[\emph{Relative variance of the reweighting factor at tree-level, vs $\theta$, for $L=8$.}]{\includegraphics[scale=0.7]{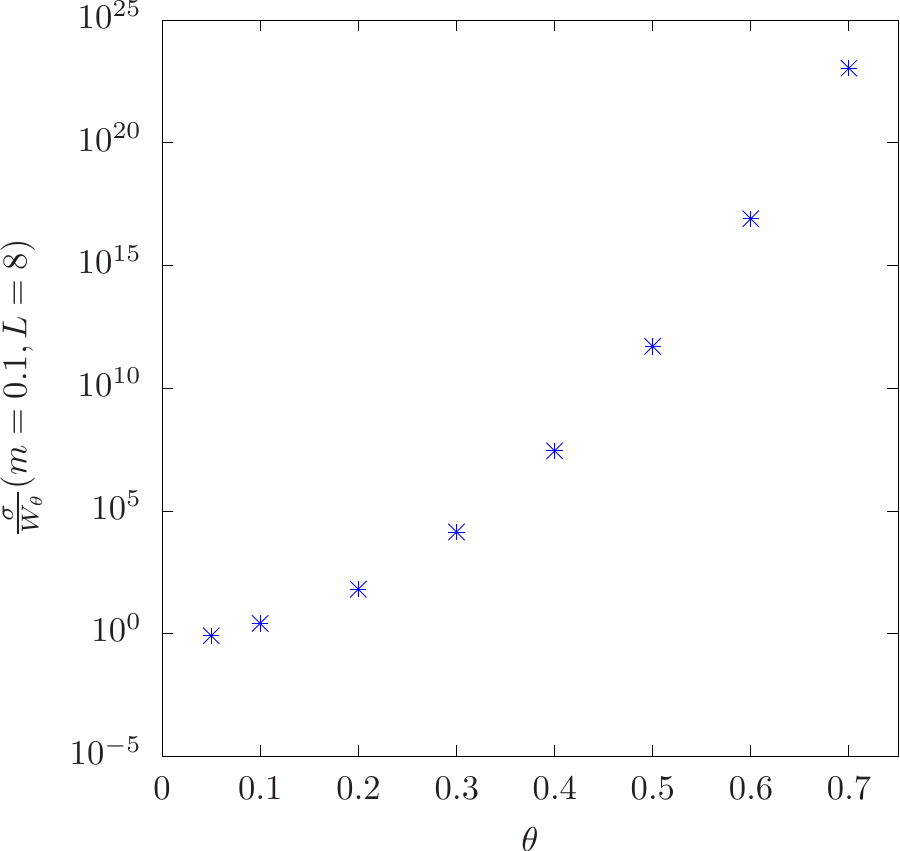}\label{fig:tree_level_test3}}
\caption{\emph{Results for the reweighting factor mean and variance employing the exact formulae in the tree-level case. Each point corresponds to a cubic lattice of the form $L^4$ with $N_f=2$, $N_c=2$ and $m=0.1$.}}
%\label{fig:tree_level_test}
\end{center}
\end{figure}

In the App.~\ref{app:tree-level_rew_fact} we give results for the spectrum of the free Dirac-Wilson operator.
This provides some insights on the large volume asymptotic scaling of the reweighting factors and their variances.
Given the finite-volume nature of twisting, 
a rather obvious expectation is that the reweighting factors approach the value 1 at fixed $\theta$ and in large volumes.
That is confirmed at tree-level, as shown in Fig.~\ref{fig:tree_level_test1}.
On the other hand, the reweighting factor remains an  extensive quantity, and
the corresponding variance grows in the same limit, which is therefore very dif\mbox{}ficult to be reached numerically.
It is clear from Fig.~\ref{fig:tree_level_test2} that the noise to signal ratio grows at least exponentially as the volume
is increased at fixed $\theta$. The same exponential growth is observed as $\theta$ is made larger at fixed $L$ (see Fig.~\ref{fig:tree_level_test3}).
A factorization of the observable has been proven to be ef\mbox{}fective in this case in various 
instances \cite{DellaMorte:2010yp, Luscher:2001up}
and in analogy to what has been done in~\cite{Finkenrath:2013soa} for the case of mass reweighting,
we will pursue a similar approach here.
%%%%%%%%%%%%%%%%%%%%%%%%%%%%%%%%%%%%%%%%%%%%%%%%%%%%%%%%%%%%%%%%%%%%%%%%%%
%%%%%%%%%%%%%%%%%%%%%%%%%%%%%%%%%%%%%%%%%%%%%%%%%%%%%%%%%%%%%%%%%%%%%%%%%%
A natural choice is to split the determinants ratio in the following telescopic way
\begin{align}
\nn
		&A = D_W(\theta)D_W^{-1}(0) = \prod_{\ell=0}^{N-1}A_\ell,\,\text{ with }A_\ell = D_W(\theta_{\ell + 1})D_W^{-1}(\theta_\ell) \simeq
		\mathbf{1}+\ord\left(\delta\theta\right)\;, \\
		&\theta_\ell = \delta \theta \cdot \ell \text{ and }  \delta\theta = \frac{\theta}{N}\;,
	\end{align}
where $A_\ell$ are now matrices deviating from the identity 
by small amounts of O$(\delta\theta)$.
This also reduce the relative fluctuation of the reweighting factor, as can be seen in Fig.~\ref{fig:tree_level_test2}.
The inverse determinant and its error $\varepsilon_{|A^{-1}|}$ are then reconstructed in terms of the $N$ corresponding estimators $1/\det A_\ell$
and $\varepsilon_{|A_\ell^{-1}|}$,
one for each factor $A_\ell$ in the equation above, as 
\begin{equation}
				\frac{1}{\det A} =\prod_{\ell=0}^{N-1}\frac{1}{\det A_\ell}= \prod_{\ell=0}^{N-1}\bigg\langle \frac{\exp\left(-\eta^{(\ell),\dagger} A_\ell \eta^{(\ell)}\right)}{p\left(\eta^{(\ell)}\right)} \bigg\rangle_{p\left(\eta^{(\ell)}\right)},
\end{equation}
and
\begin{equation}
				\varepsilon^2_{|A^{-1}|}  = \sum_{\ell=0}^{N-1}\left[\varepsilon^2_{|A_\ell^{-1}|}\prod_{k\neq\ell}\det \left(A_k\right)^{-2}\right].
\end{equation}
Based on the tree-level results we have presented,
the expression on the r.h.s.~of the equation above 
is given by a sum of $N$ terms, each one depending exponentially on $\delta \theta$ 
(for values of $\delta \theta$ such that the variance in Fig.~\ref{fig:tree_level_test3} is approximately growing exponentially
with the twisting angle)
and therefore, at
fixed $\delta \theta$, the squared error of the telescopic product is expected to grow linearly with $\theta$. 
That is to be compared to the exponential growth of the error one would obtain by
attempting to compute the ratio of determinants for large shifts in $\theta$ in one single step.

\section{Simulations and results}

For the numerical computations, we have used gauge configurations produced for
the $\mathbf{SU}(2)$ gauge theory with two fermions in the fundamental representation.
Dynamical configurations have been generated using un-improved Wilson fermions and the Wilson plaquette gauge action.
This model is a QCD-like theory, featuring chiral symmetry breaking and confinement. 
The outcome of the present study should hence remain qualitatively unchanged for the case of lattice QCD.
Indeed, at tree-level the reweighting factors scale with a power of $N_c$, see Eq.~\ref{eq:rew}.

In Table~\ref{tab:simpar} we collect details about  the ensembles used in this work, for completeness, the value $m_c$ 
of the bare mass parameter yielding massless fermions
is estimated to be $-0.77(2)$ at $\beta = 2.2$ \cite{Lewis:2011zb, Hietanen:2014xca, Arthur:2016dir}.
%%%%%%%%%%%%%%%%%%%%%%%%%%%%%%%%%%%%%%%%%%%%%%%%%%%%%%%%%%%%%%%%%%%%%%%%%%%%
\begin{table}[htb]
\begin{center}
	\begin{tabular}{|c|c|c|c|c|}
	\hline
	$V$ & $\beta$ & $m_0$ & $N_\text{cnf}$ & traj. sep.\\
	\hline
	$8^3 \times 16$ & 2.2 & -0.6 & $980$ & 10 \\
	\hline
	$24^3 \times 32$ & 2.2 & -0.65 & $374$ & 20 \\
	$24^3 \times 32$ & 2.2 & -0.72 & $360$ & 10 \\
	\hline
	\end{tabular}
\caption{\emph{Ensembles used and simulation parameters in the reweighting of TBCs.}}
\label{tab:simpar}
\end{center}
\end{table}
%%%%%%%%%%%%%%%%%%%%%%%%%%%%%%%%%%%%%%%%%%%%%%%%%%%%%%%%%%%%%%%%%%%%%%%%%%%%
We have considered both small and large volumes, which we will discuss separately. 

At fixed value of $\theta$ the ef\mbox{}fect of twisting and therefore of reweighting is at its largest in 
small volumes, based on the tree-level studies. This is the region where 
the stochastic methods should provide reliable estimates.
We restricted ourselves to the case of reweighting to spatially isotropic\footnote{Notice that for non-isotropic $\theta$ parity is
explicitly  broken.}
 $\theta$ angles and 
adopted the hermitian $\gamma_5$-version of the Dirac-Wilson operator with two flavors, since that automatically fulfills the applicability condition\footnote{Notice that $Q[U,0]$ and $Q[\mathcal{U},0]$ commute.} in 
Eq.~\ref{eq:appl_cond}.

\begin{figure}[t!]
\begin{center}
\subfigure{\includegraphics[scale=0.72]{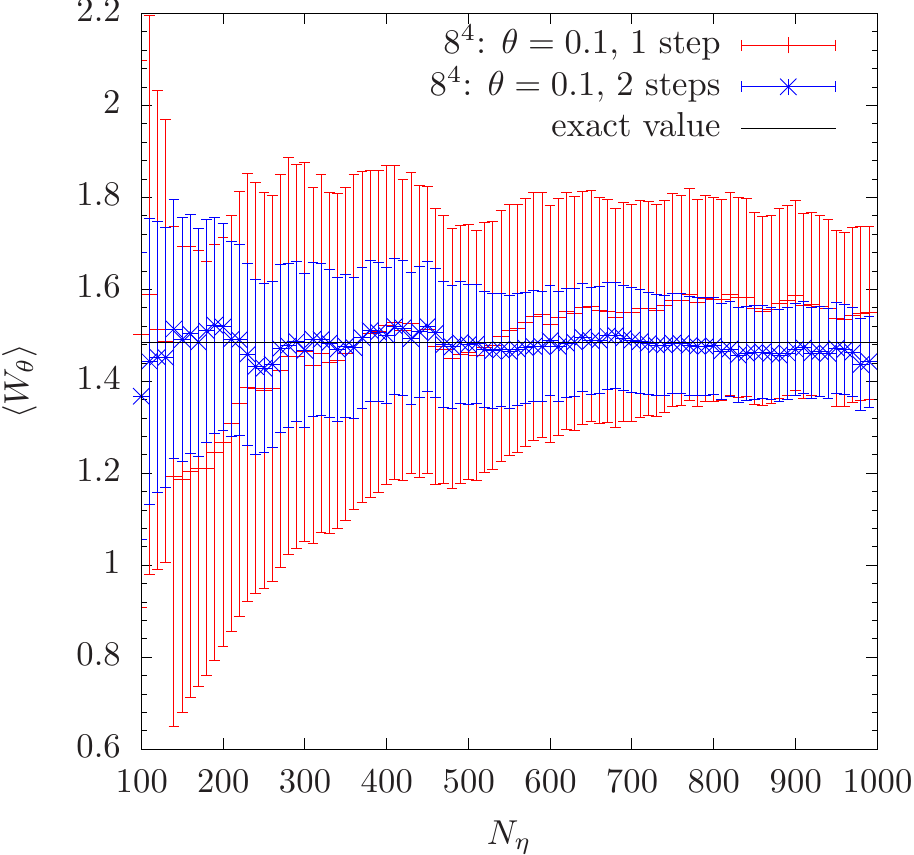}}
\hspace{0.4cm}
\subfigure{\includegraphics[scale=0.72]{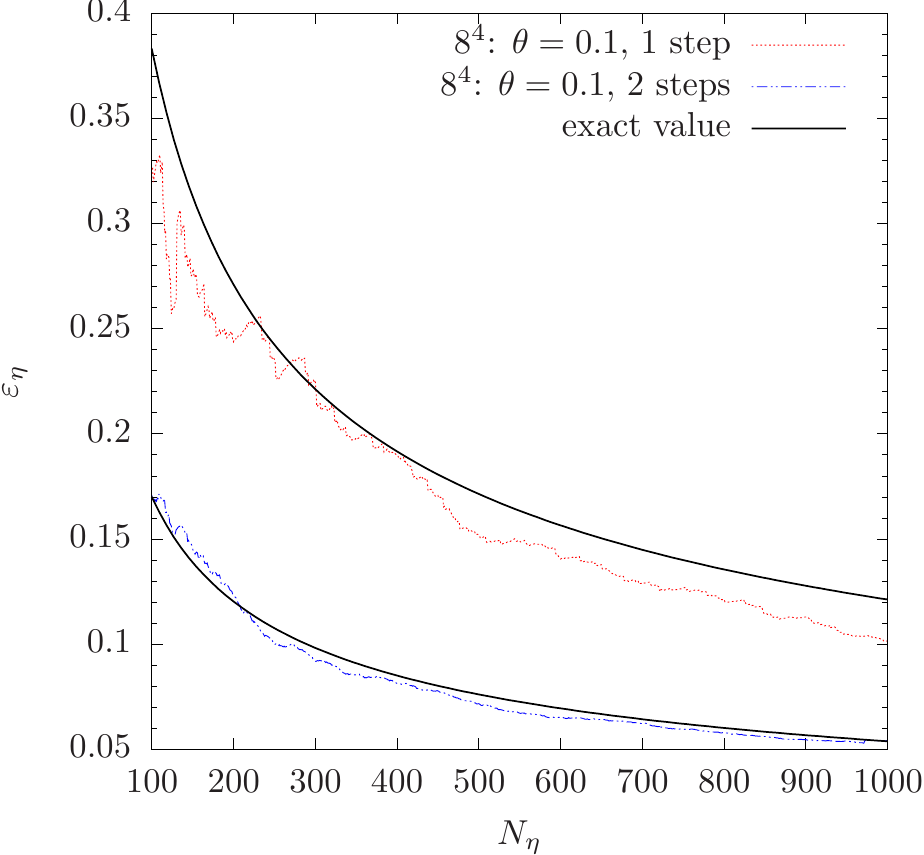}}
\caption{\emph{Comparison between the stochastic and exact estimates of the reweighting factor and its variance on the trivial gauge configuration for $\theta=0.1$, $L=8$ using
one and two levels of factorization in the numerical case. Both plotted against the number $N_\eta$ of Gaussian vectors in each level.} } 
\label{fig:treecomp}
\end{center}
\end{figure}

In order to get an insight on the number $N_\eta$ of Gaussian vectors necessary to obtain a reliable estimate of
the reweighting factor, we computed it on the trivial ($U_\mu(x)=\I$) gauge configuration and compared it to the analytical tree-level prediction.
The results are shown in Fig.~\ref{fig:treecomp}.
%
%%%%%%%%%%%%%%%%%%%%%%%%%%%%%%%%%%%%%%%%%%%%%%%%%%%%%%%%%%%%%%%%%%%%%%%%%%
%%%%%%%%%%%%%%%%%%%%%%%%%%%%%%%%%%%%%%%%%%%%%%%%%%%%%%%%%%%%%%%%%%%%%%%%%%
There, by looking at the statistical error, one sees that about 300 Gaussian vectors are necessary for $\epsilon_\eta$ to reach the correct scaling with $N_\eta$.

\begin{figure}[t!]
\begin{center}
\subfigure{\includegraphics[scale=0.72]{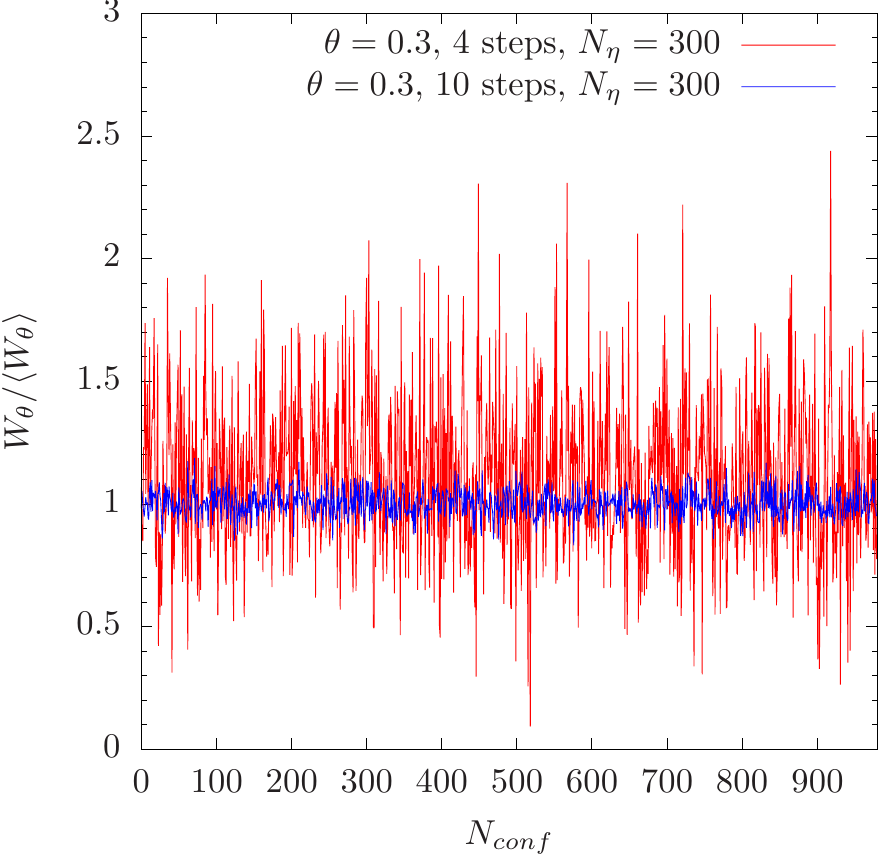}}
\hspace{0.4cm}
\subfigure{\includegraphics[scale=0.72]{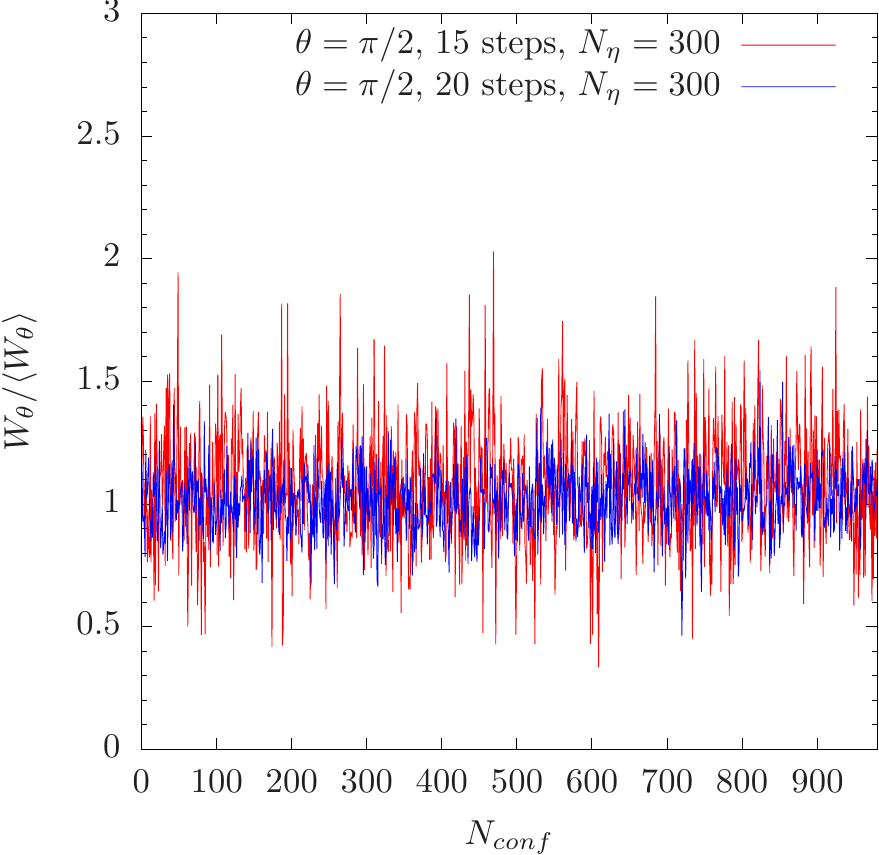}}
\caption{\emph{Monte Carlo history of the reweighting factor relative to its mean for the $8^3 \times 16$ ensemble and with $N_\eta$=300.}}
\label{fig:maybe}
\end{center}
\end{figure}

\begin{figure}[h!t]
\begin{center}
\subfigure[\emph{Mean of the reweighting factor for $\theta = 0.3$ as a function of $N_\eta$.}]{\includegraphics[scale=0.72]{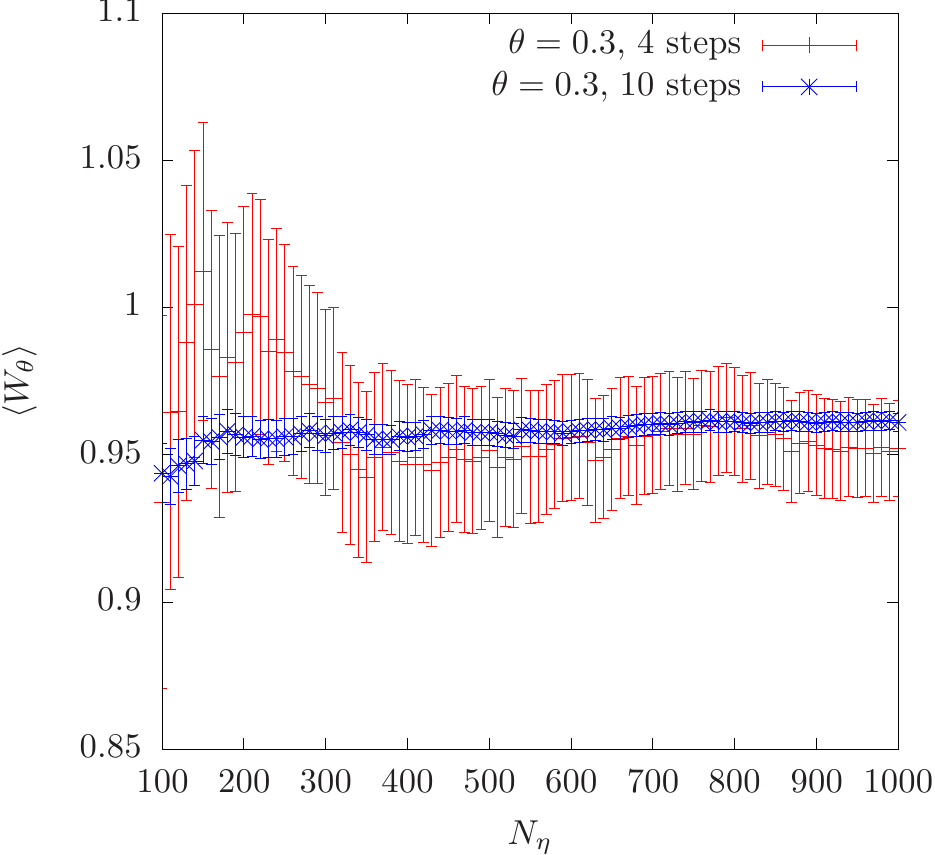}\label{fig:mc_hist_1}}
\hspace{0.4cm}
\subfigure[\emph{Mean of the reweighting factor for $\theta = \pi /2$ as a function of $N_\eta$.}]{\includegraphics[scale=0.72]{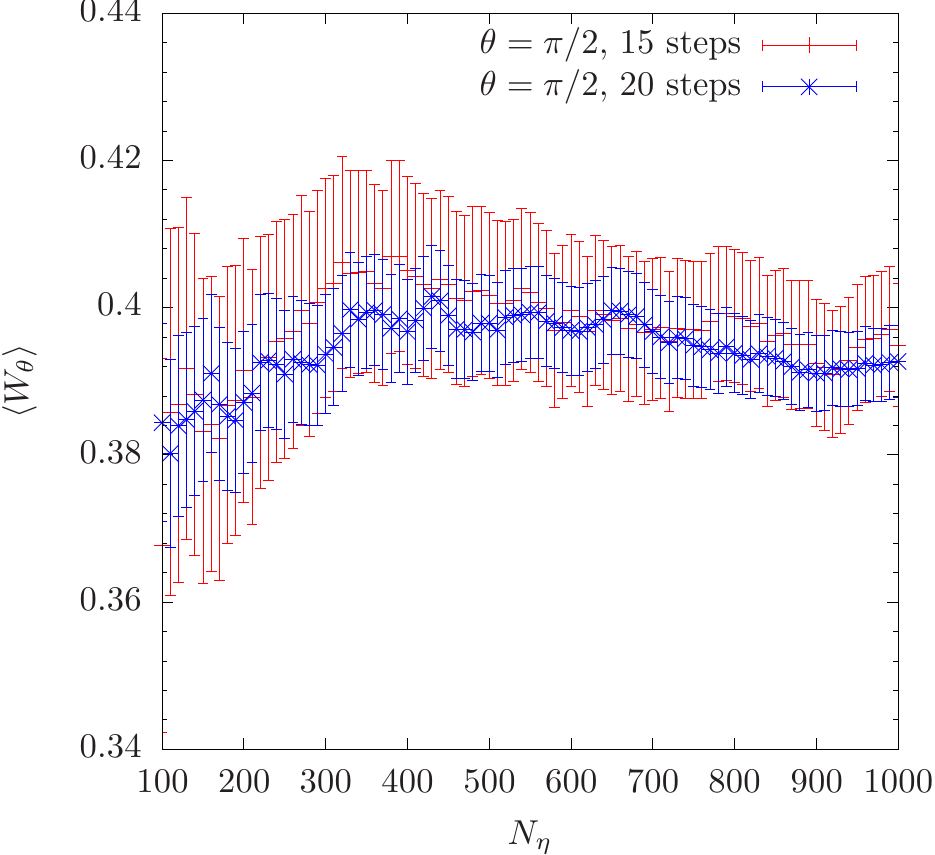}\label{fig:mc_hist_2}}
\caption{\emph{Monte Carlo average of the reweighting factor, averaged over the entire number of configurations vs $N_\eta$. The figures correspond to a volume $V = 8^3 \times 16$.}}

\end{center}
\end{figure}

Turning now to actual Monte Carlo data, in particular from the $8^3 \times 16$ ensemble, we found that under a similar  condition ($N_\eta \gtrsim 300$)
the reweighting factor on each configuration deviates by more than a factor 10 in magnitude from its gauge average in a few cases only (see Fig.~\ref{fig:maybe}).
%%%%%%%%%%%%%%%%%%%%%%%%%%%%%%%%%%%%%%%%%%%%%%%%%%%%%%%%%%%%%%%%%%%%%%%%%%
%%%%%%%%%%%%%%%%%%%%%%%%%%%%%%%%%%%%%%%%%%%%%%%%%%%%%%%%%%%%%%%%%%%%%%%%%%
This prevents the averages to be dominated by large fluctuations (``spikes''), which would cause large statistical errors, and sets a lower limit on $N_\eta$. 
In Figs.~\ref{fig:mc_hist_1} and~\ref{fig:mc_hist_2} we show the mean of the reweighting factor, with each point resulting from an average over $10^3$ configurations, as a function of $N_\eta$.
A good scaling of the error is  visible, according to $N_\eta^{-1/2}$, up to $N_\eta \approx 500$.
At that point the statistical noise saturates the gauge noise, and therefore the error on the gauge average does not decrease any further by increasing $N_\eta$.
This sets an upper limit to about 600 for the number of Gaussian vectors to be used in the stochastic evaluation of the determinants ratio. In addition, it implies that for $N_\eta \gtrsim  600$ one can safely consider the gauge noise only in the error analysis. In the following, we do that by a standard (single-elimination) jackknife plus binning procedure.

%%%%%%%%%%%%%%%%%%%%%%%%%%%%%%%%%%%%%%%%%%%%%%%%%%%%%%%%%%%%%%%%%%%%%%%%%%
%%%%%%%%%%%%%%%%%%%%%%%%%%%%%%%%%%%%%%%%%%%%%%%%%%%%%%%%%%%%%%%%%%%%%%%%%%

\subsection{Small volumes}

In the small volume regime the ef\mbox{}fect of partial twisting and the associated breaking of unitarity may be large, and we want to 
compensate for it by reweighting the observable at hand.
In order to isolate the contribution due to the determinants ratio, we looked first at the plaquette, for which
the entire dependence on $\theta$ comes from the quarks in the sea only, see App.~\ref{app:obs_tbcs}.
Here and in the following we neglect autocorrelations since measurements are separated by 10 to 20 molecular dynamics units. 
In any case, a binning procedure, using bins of length up to $10$, provides entirely consistent results.

\begin{figure}[h!t]
\begin{center}
\subfigure[\emph{Average plaquette after reweighting only one flavor.}]{\includegraphics[scale=0.72]{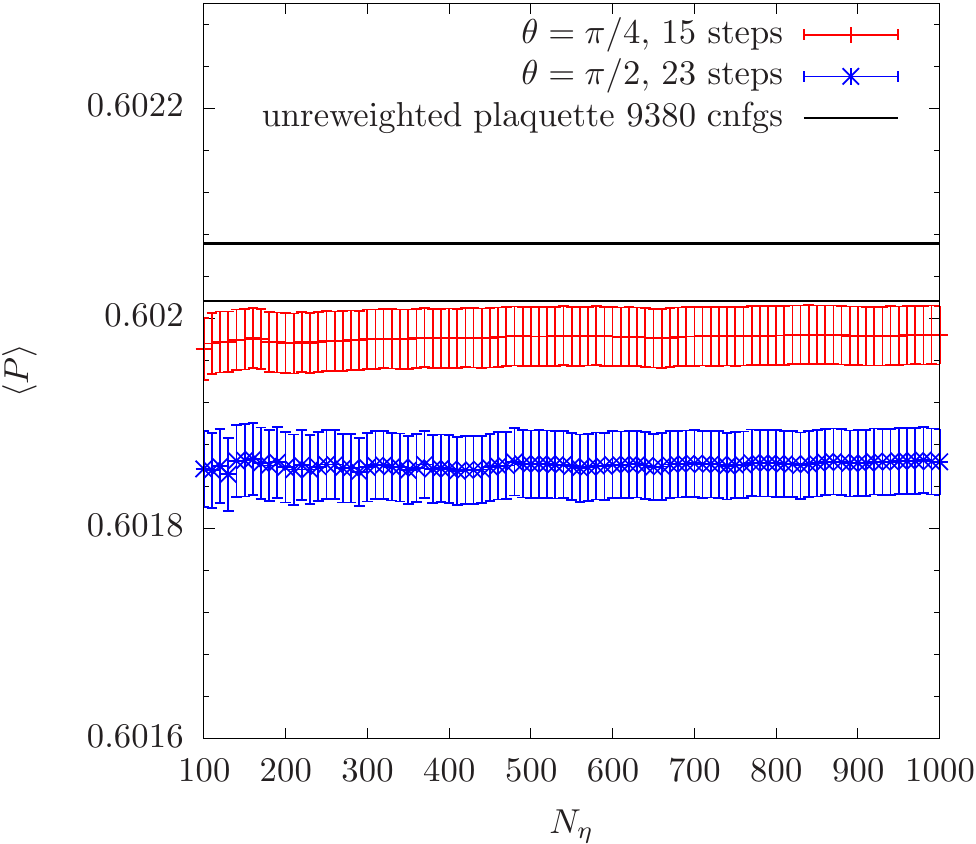}\label{fig:plaquette_small_1}}
\hspace{0.4cm}
\subfigure[\emph{Average plaquette with both flavors reweighted.}]{\includegraphics[scale=0.72]{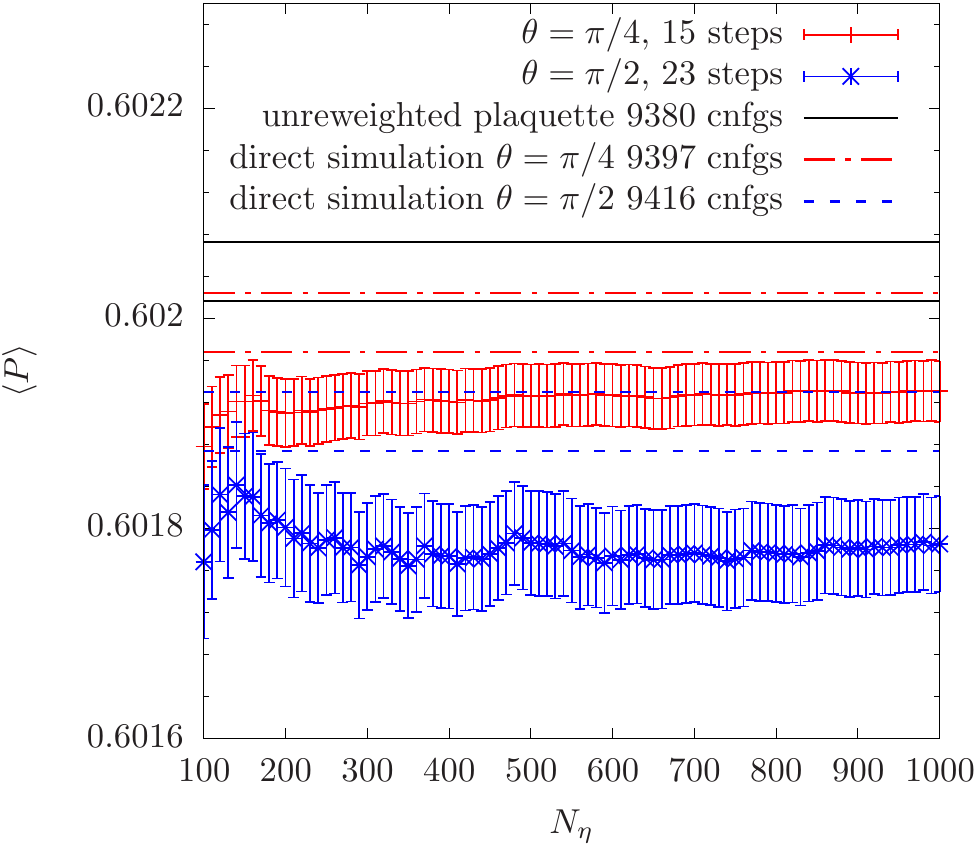}\label{fig:plaquette_small_2}}
\caption{\emph{Monte Carlo history of the reweighted plaquette. The figures correspond to a volume $8^3 \times 16$.}}
\end{center}
\end{figure}

In Fig.~\ref{fig:plaquette_small_1} we show the results after reweighting only one flavor, i.e., by taking the square root of the stochastic
evaluation of the determinants ratio estimated for the $Q^2[U, \theta]$ operators~\cite{Aoki:2012st,Finkenrath:2012cz}.
Notice that here and in the following, whenever the root-trick above is used, we consider rather heavy quarks and pions ($am_\pi \geq 0.45$) and 
we therefore do not expect ambiguities in the sign of the one-flavor determinant.
Ef\mbox{}fects are visible within statistical errors for large values of $\theta$ only.
Those are more pronounced when both flavors are reweighted, as depicted 
in Fig.~\ref{fig:plaquette_small_2}. In addition, in this case, the reweighted 
results can be checked by a direct HMC simulation\footnote{For the
one-flavor case one would have to consider the RHMC algorithm.} at $\theta \neq 0$. Notice that we are discussing permil shifts, which we access by 
using very large statistics ($\approx$ $10^4$ configurations).
We interpret this slight tension as signalling the limit of validity of the reweighting method for the application discussed here.
In the following, we therefore restrict the values of the twisting angle to the interval $\left[0,\pi / 2\right]$.

%%%%%%%%%%%%%%%%%%%%%%%%%%%%%%%%%%%%%%%%%%%%%%%%%%%%%%%%%%%%%%%%%%%%%%%%%%
%%%%%%%%%%%%%%%%%%%%%%%%%%%%%%%%%%%%%%%%%%%%%%%%%%%%%%%%%%%%%%%%%%%%%%%%%%

\begin{figure}[h!t]
\begin{center}
\includegraphics[scale=0.72]{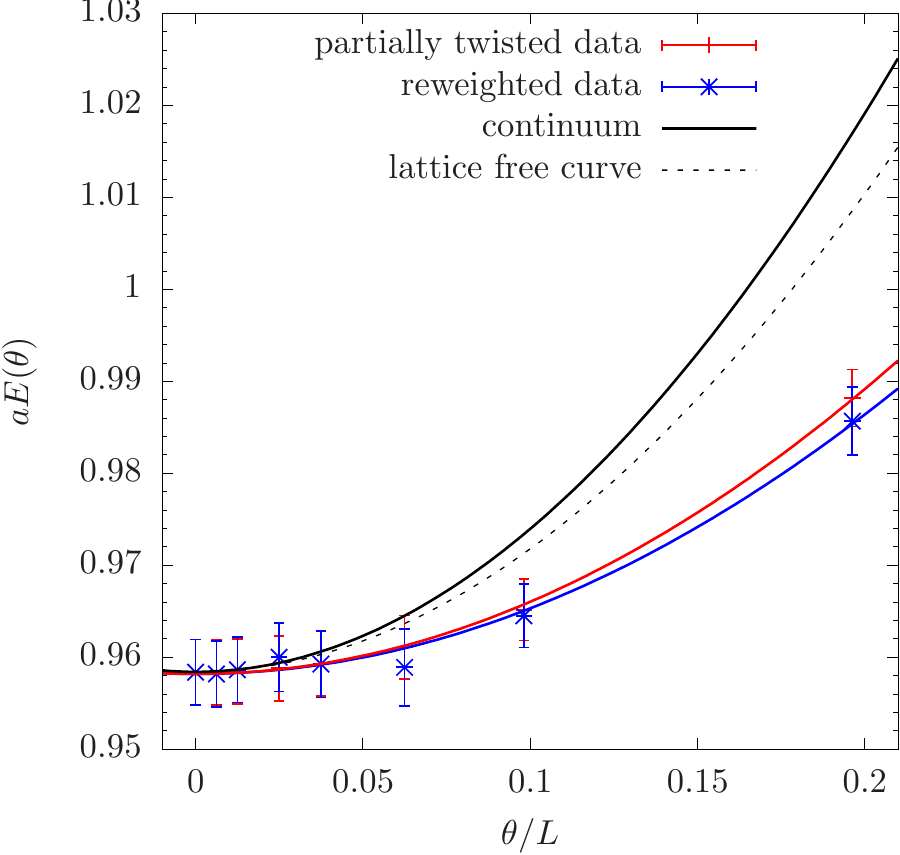}
\caption{\emph{Pion dispersion relation for $V = 8^3 \times 16$. Each point is obtained with a dif\mbox{}ferent (growing with $\theta$) number of independent steps in the determination of the reweighting factor and $N_\eta \gtrsim 600$ in each step.}}
\label{fig:disp_rel_small}
\end{center}
\end{figure}

The other quantity  we have analyzed is the pion dispersion relation.
After twisting only one flavor in the valence, the lowest energy state coupled
to a spatially summed interpolating field
is expected to become a ``quenched'' pion with momentum $\vec{p} = \pm\vec{\theta} / L$, see App.~\ref{app:obs_tbcs}. 
In order to remove this quenching ef\mbox{}fect we have reweighted the relevant correlators 
for the twisting of one flavor in the sea and we have extracted the ef\mbox{}fective energies
from their time-symmetrized versions.
In Fig.~\ref{fig:disp_rel_small} we display the results for the dispersion relation and we compare them to 
the un-reweighted, partially twisted, data (i.e., with twisting in the valence only), to the continuum prediction $(a E)^2 = (a m_\pi)^2 + 3 (a\theta)^2 / L^2$ 
(we use $\vec{\theta}=\theta(1,1,1)$)
and to the lattice free boson theory prediction $\cosh(a E) = 3 + \cosh(a m_\pi) + 3 \cos( a\theta / L)$.
Over the entire range of  $\theta$ values explored there is no significant ef\mbox{}fect within errors.
Let us remark that, following~\cite{Boyle:2008rh}, all two-point functions 
have been computed using $Z_2 \times Z_2$ single time-slice stochastic sources.
%%%%%%%%%%%%%%%%%%%%%%%%%%%%%%%%%%%%%%%%%%%%%%%%%%%%%%%%%%%%%%%%%%%%%%%%%%
%%%%%%%%%%%%%%%%%%%%%%%%%%%%%%%%%%%%%%%%%%%%%%%%%%%%%%%%%%%%%%%%%%%%%%%%%%
	\subsection{Large volumes}
\begin{figure}[h!t]
\begin{center}
\subfigure[\emph{``Heavy'' case of $m_0\simeq -0.65$.}]{\includegraphics[scale=0.72]{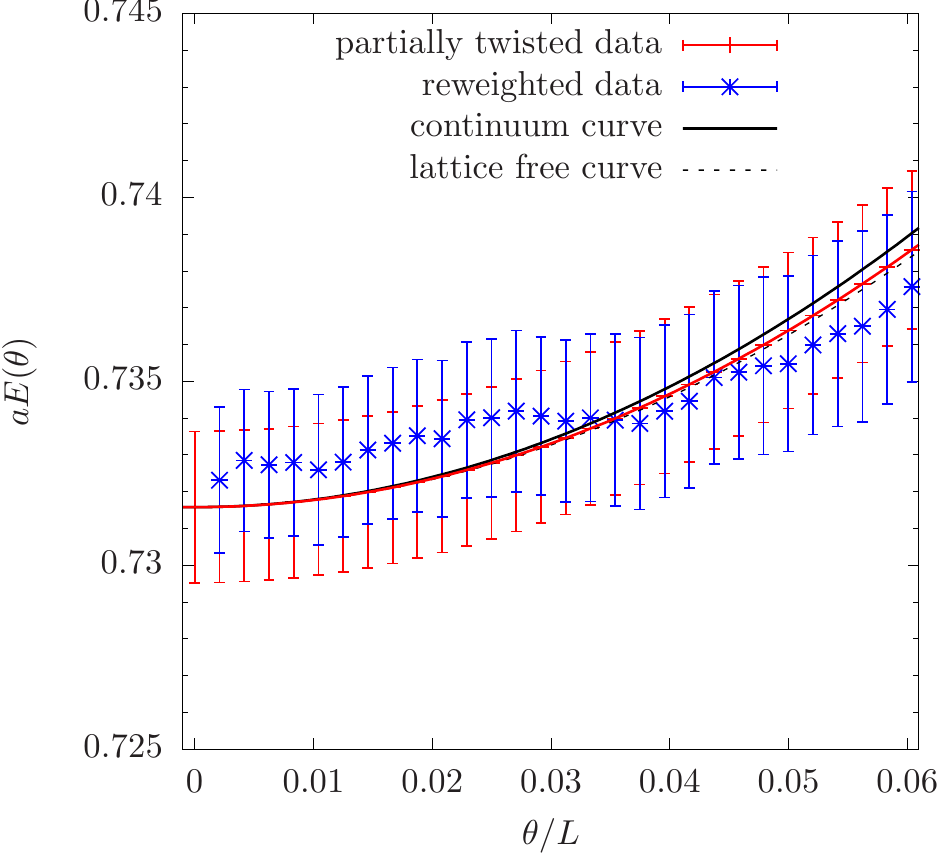}}
\hspace{0.4cm}
\subfigure[\emph{``Light'' case of $m_0\simeq -0.72$.}]{\includegraphics[scale=0.72]{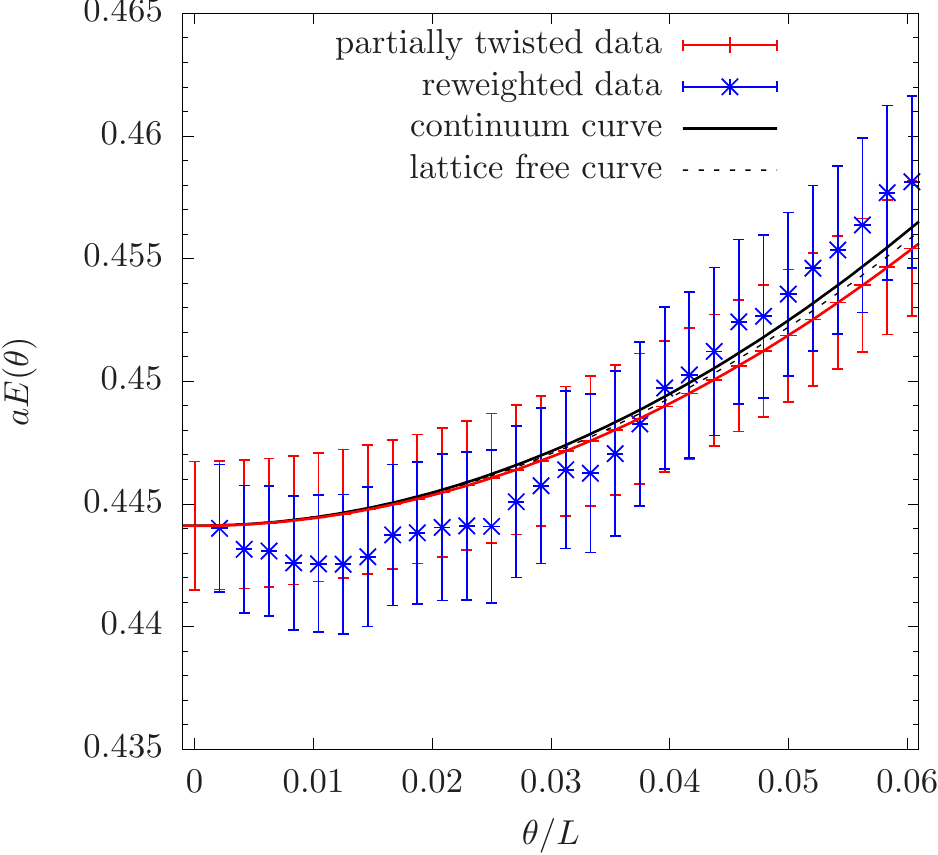}}
\caption{\emph{Pion dispersion relation for $V = 24^3 \times 32$. The reweighting factor at a given $\theta$ is obtained by a telescopic product involving all the previous ones, each one estimated using $N_\eta \gtrsim 600$.}}
\label{fig:disp_rel_large}
\end{center}
\end{figure}

As suggested by the tree-level studies, in large volumes, the accuracy in the determination of the reweighting factors and the overall ef\mbox{}fect of twisting
are very much reduced, compared to the previous case.
We have looked at the pion dispersion relation for two dif\mbox{}ferent values of $m_\pi$ as 
we expect to detect possibly sizeable ef\mbox{}fects for rather light quarks.
However, at the volume considered ($V=24^3 \times 32$), it appears as one can safely neglect any breaking of unitarity.
The results are shown in Fig.~\ref{fig:disp_rel_large}.
Indeed, reweighting does not seem to yield any significant ef\mbox{}fect within the half a percent statistical errors. 
%%%%%%%%%%%%%%%%%%%%%%%%%%%%%%%%%%%%%%%%%%%%%%%%%%%%%%%%%%%%%%%%%%%%%%%%%%
%%%%%%%%%%%%%%%%%%%%%%%%%%%%%%%%%%%%%%%%%%%%%%%%%%%%%%%%%%%%%%%%%%%%%%%%%%

\section{Quark mass dependence on the twisting angle}

As mentioned above, non-periodic boundary conditions ($\theta\neq 0$) for fermionic fields are equivalent to introducing a constant $\mathbf{U}(1)$ interaction, 
through an external field $B_\mu(x)=B_\mu$ coupled to periodic fermions. 
That amounts to replacing the links $U_\mu$ with the links $\mathcal{U}_\mu$ given by
\begin{align}
\label{eq:links_u1}
\mathcal{U}_\mu(x) = \e^{iaB_\mu}U_\mu(x)\;,
\end{align}
where $B_\mu = \theta_\mu / L_\mu$.
In this case the PCAC relation remains unaf\mbox{}fected since the new links $\mathcal{U}_\mu(x)$, being proportional to the 
identity in flavor, still commute with the Pauli matrices\footnote{We are working with two degenerate massive flavors, i.e.~the isosymmetric limit.}. That implies that the vector transformations are still exact symmetries at finite lattice spacing with
Wilson fermions, and that the PCAC relation remains formally the same.
Cutof\mbox{}f ef\mbox{}fects on the other hand depend on the choice of boundary conditions, that has actually been exploited in order to compute improvement 
coef\mbox{}ficients (see, for example, Refs.~\cite{Luscher:1996ug,Durr:2003nc}), within the Symanzik improvement programme for Wilson fermions in QCD.
In some instances, even after improvement, the quark mass defined through the PCAC relation in rather small volumes, turned out
to have a quite pronounced residual (i.e., O$(a^2)$) dependence on the boundary conditions~\cite{Sommer:2003ne}. That is expected
to be even larger here, with un-improved Wilson fermions. For completeness, the bare PCAC quark mass $m_{\rm PCAC}$ can be defined through the spatially integrated axial Ward identity as:
\begin{equation}
\label{eq:pcac_bla}
m_{\rm PCAC}= \frac{\partial_0\langle A_0(x_0) O(0)\rangle}{2\langle P(x_0) O(0) \rangle}\;,
\end{equation}
with $A_\mu$ the axial current, $P$ the pseudoscalar density (both spatially summed over the $x_0^{\rm th}$ time-slice) 
and $O$ an interpolating field, which in our case will be simply given by the pseudoscalar density localized at the origin.

\hrulefill

\paragraph{Intermezzo: PCAC with TBCs}

In the PCAC relation we want the variation of the Wilson term to vanish in the na\"ive continuum limit, $a\rightarrow 0$ (see Sect.~\ref{subsect:pcac}).
Now we have another dimension $[1]$ field to take into account with respect to case with no constant $\mathbf{U}(1)$ interaction.
This means that we can write additional operators that mix with $\mathcal{O}_5$. In particular we find, following App.~\ref{app:lattice_PCAC} in the case of degenerate flavors,
\begin{align}
\overline{\mathcal{O}}_5 
= Z_5 \left[ \mathcal{O}_5 
+ \frac{2\overline{m}}{a} P^f 
+ \frac{Z_A-1}{a}\partial_\mu A_\mu^f 
+ 2\left(\overline{B}_\mu\right)^2 P^f 
+ \widetilde{B}_\mu\partial_\mu^- P^f\right],
\end{align}
where $f$ is a flavor index. 
By substituting back the above expression in Eq.~\ref{eq:bare_pcac} we obtain
\begin{align}
\partial_\mu^-\langle \hat{A}_\mu^f(x)\mathcal{O}(y)\rangle =
2\left(m_0 - \overline{m} - a\overline{B}_\mu\overline{B}_\mu\right)
 \langle P^f(x)\mathcal{O}(y)\rangle 
 - a\widetilde{B}_\mu \partial_\mu^- \langle P^f(x)\mathcal{O}(y)\rangle 
 + \langle\overline{\chi}_{\rm A}^f(x)\mathcal{O}(y)\rangle.
\end{align}
One of the two extra term, compared to Eq.~\ref{eq:ren_pcac},
is vanishing when we consider the integrated WTI. 
In particular by projecting to zero momentum we find that the following integral is vanishing ($B_0=0$)
\begin{align}
a\widetilde{B}_j \int\de^3 x \, \partial_j^- \langle P^f(x)\mathcal{O}(y)\rangle = 0,
\end{align}
because we have chosen PBCs for our fields.\\
In this case $m_{\rm AWI}$ in Eq.~\ref{eq:m_awi} is given by
\begin{align}
m_{\rm AWI} = m_0 - \overline{m} - a\overline{B}_j^2 = m_0 - f(g_0, am_0, a^2 B_j^2)  / a,
\end{align}
hence the zero quark mass is attained when the bare mass is equal to the critical value
\begin{align}
m_c =  f(g_0, am_c, a^2 B_j^2) / a \, ,
\end{align}
which is a function of $g_0$ and $B_j^2 = \theta_j^2 / L^2$ bare parameters. We should note that the function $f$ depends on quadratically on $B_j$ due to Lorentz invariance.

\hrulefill

For the present discussion, it is useful to introduce $\theta_{\rm v}$ as the spatially isotropic twisting angle used for all fermions
in the valence and  $\theta_{\rm s}$ as the corresponding one for all fermions in the sea. We produced small volume configurations
for dif\mbox{}ferent values of $\theta_{\rm s}$ and on those we measured quark and pion masses while changing $\theta_{\rm v}$.
That can actually be used to get an idea of what the ef\mbox{}fect of reweighting should be, and we will
see that the observations above are confirmed. In particular, at fixed  $\theta_{\rm v}$, results depend 
at most at the percent level only on  $\theta_{\rm s}$.

\begin{figure}[h!t]
\begin{center}
\subfigure{\includegraphics[scale=0.72]{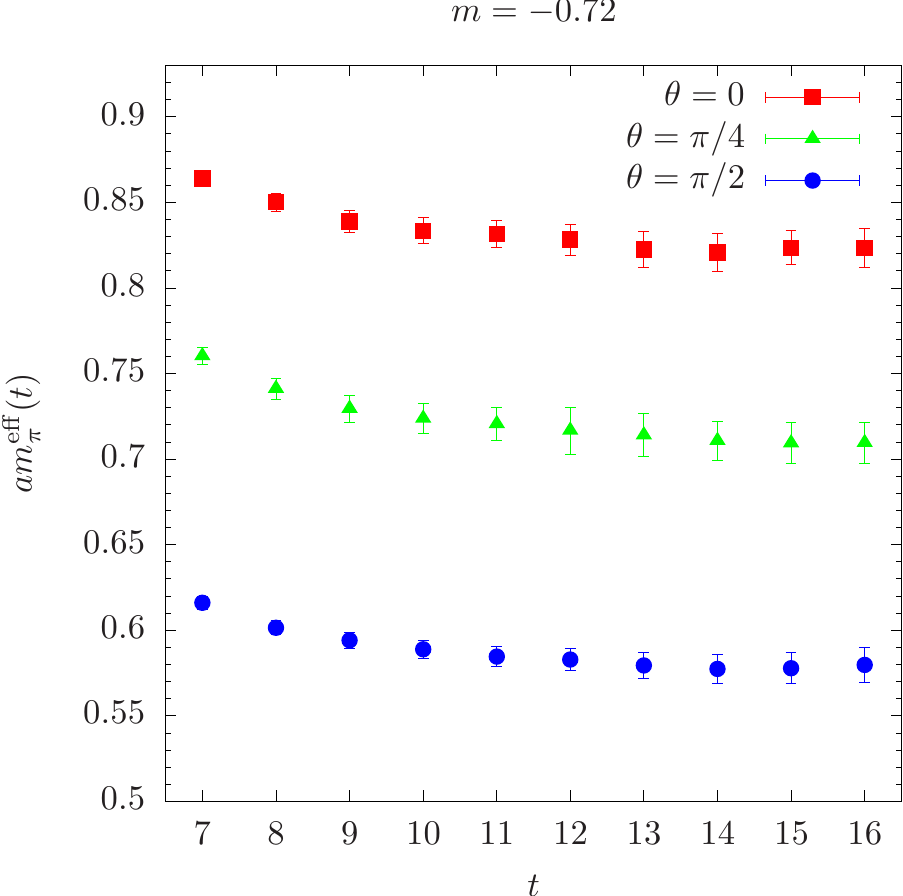}}
\hspace{0.4cm}
\subfigure{\includegraphics[scale=0.72]{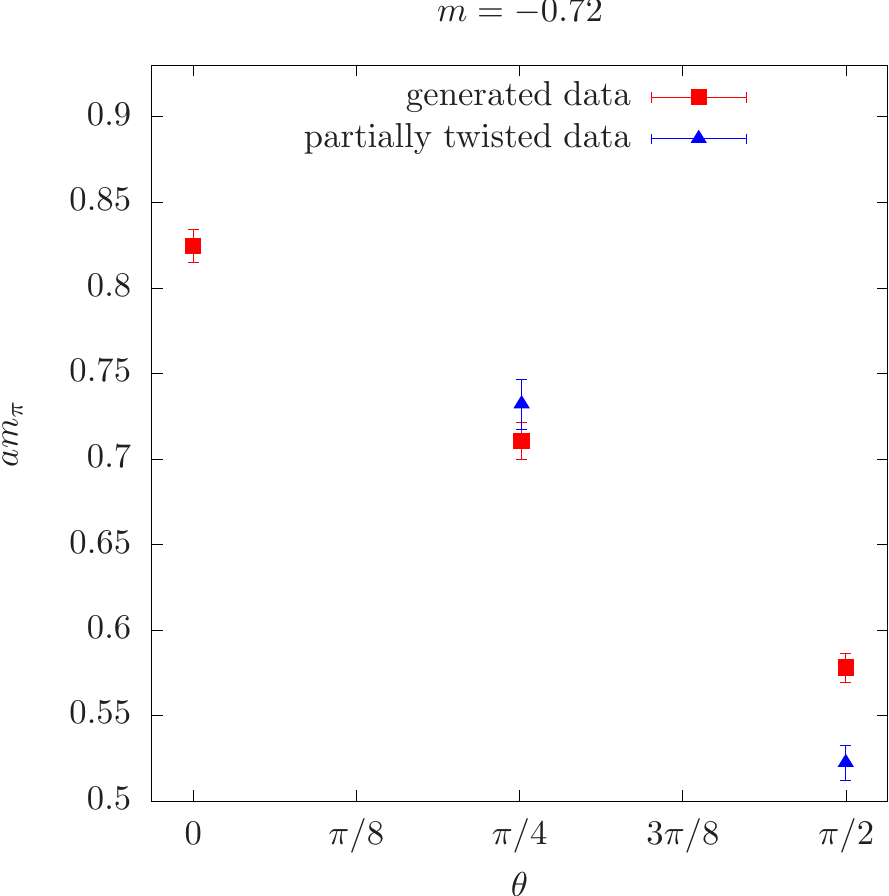}}
\caption{\emph{Pion ef\mbox{}fective masses for $\theta_{\rm s}=\theta_{\rm v}=\theta$ on a $8^3 \times 32$ lattice at $\beta=2.2$, $m_0=-0.72$ (left panel) and comparison
of the corresponding plateau masses (red) with the result from $\theta_{\rm s}=0$ as a function of $\theta=\theta_{\rm v}$ (right panel).}
}
\label{fig:thetaseandval}
\end{center}
\end{figure}

In Fig.~\ref{fig:thetaseandval} we show the pion ef\mbox{}fective masses computed on about 4000 independent configurations generated for lattices 
of size $8^3 \times 32$ at $\beta=2.2$ and $m_0=-0.72$. All the results in the left panel refer to unitary points, i.e. $\theta_{\rm s}=\theta_{\rm v}$.
Naively one would expect the results to lie on top of each other as they all correspond to zero-momentum pions at the same bare parameters.
We ascribe the dif\mbox{}ference to the dependence of the critical bare mass
$m_c$ on both $\theta_{\rm s}$ and $\theta_{\rm v}$, as explained above. 
Since $m_c$ is obtained from the PCAC operator identity the dependence
on the twisting angles is a boundary, therefore finite volume, cutof\mbox{}f ef\mbox{}fect.
This explains why the continuum dispersion relation is rather poorly reproduced in small volumes and for large values
of the twisting angle, as shown in Fig.~\ref{fig:disp_rel_small}. Upon twisting, not only the pions get boosted, but also at least
one of the quark  masses decreases, such that the two ef\mbox{}fects partly compensate. 
In the right panel of Fig.~\ref{fig:thetaseandval} the final estimates of the masses are shown as a function of $\theta=\theta_{\rm v}$.
The square points are the unitary ones, corresponding to the plateaux in the left panel, whereas the
triangle ones are obtained by using configurations produced at $\theta_{\rm s}=0$ on which
two-point functions are computed for dif\mbox{}ferent values of $\theta_{\rm v}=\theta$. 
It is clear that by reweighting the sea twisting angle to the valence one at most a percent ef\mbox{}fect could have been produced here 
in the case  $\theta_{\rm s}=0$, $\theta_{\rm v}=\pi/2$.
%%%%%%%%%%%%%%%%%%%%%%%%%%%%%%%%%%%%%%%%%%%%%%%%%%%%%%%%%%%%%%%%%%%%%%%%%%
%%%%%%%%%%%%%%%%%%%%%%%%%%%%%%%%%%%%%%%%%%%%%%%%%%%%%%%%%%%%%%%%%%%%%%%%%%

\begin{figure}[h!t]
\begin{center}
\includegraphics[scale=0.9]{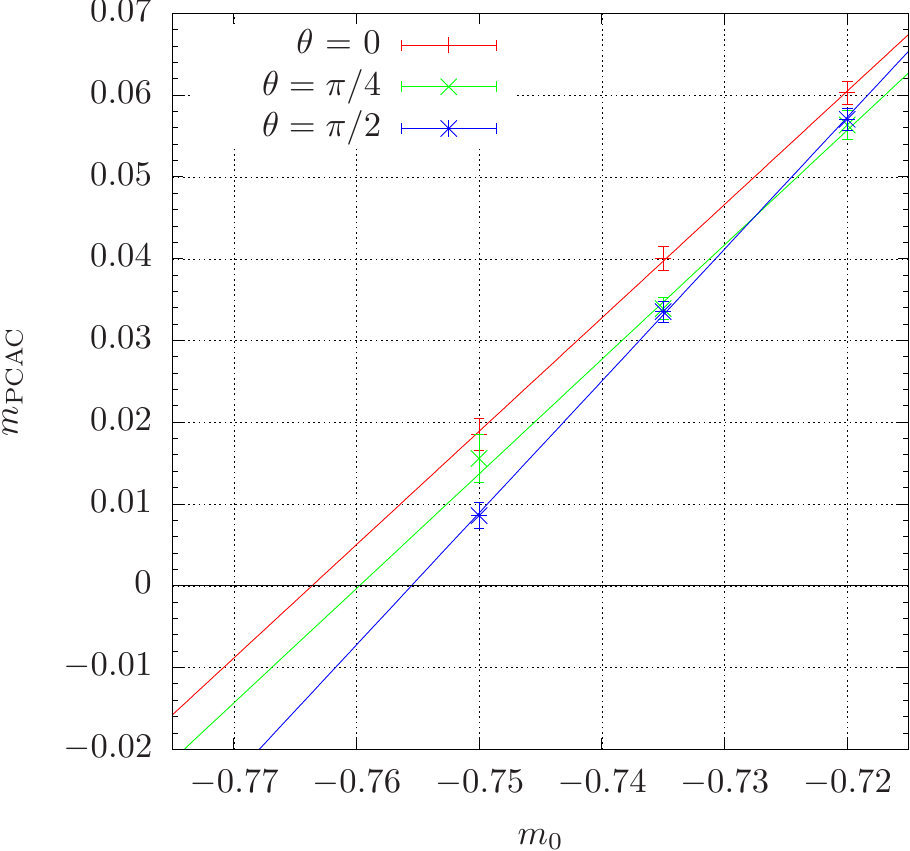}
\caption{\emph{$m_{\rm PCAC}$ as a function of $m_0$ computed for $m_0=-0.72$, $-0.735$ and $-0.75$ at $\beta=2.2$ and $V=8^3 \times 16$.
The curves correspond to three dif\mbox{}ferent values of $\theta=\theta_{\rm v}=\theta_{\rm s}$.}}
\label{fig:mpcac-vs-m0}
\end{center}
\end{figure}

The bare PCAC quark mass, in the un-improved theory, can be related to the bare parameter $m_0$ as
\begin{equation} 
m_{\rm PCAC}(\beta, \theta)=Z(\beta,\theta)\left( m_0-m_c(\beta,\theta) \right)\; ,
\end{equation}
where $Z$ is a normalization factor and $m_c$ is the value of the bare mass parameter defining the massless limit. 
As a consequence of the breaking of chiral symmetry with Wilson fermions $m_c$ is dif\mbox{}ferent from zero, as opposite
to the case of Ginsparg-Wilson or staggered fermions, where at least some axial transformations are preserved 
and that is enough to rule out an additive renormalization of the bare mass.
We are here restricting the attention to the case $\theta_{\rm s}=\theta_{\rm v}=\theta$ and we are working at fixed $\beta$.
By looking at $m_{\rm PCAC}$ as a function of $m_0$, for $m_0$ slightly larger than $m_c$, one obtains a family of linear
curves parameterized by $\theta$. The slope of the curves is given by $Z$, while the value of $m_0$ where the curves
intercept the horizontal axis corresponds to $m_c$. That is exactly what is shown in Fig.~\ref{fig:mpcac-vs-m0} for 
$m_0=-0.72$, $-0.735$ and $-0.75$ (with $\beta$ fixed to $2.2$ and $V=8^3 \times 32$). Whereas the dependence 
of $Z$ on $\theta$ is not significant, $m_c$, as anticipated, changes substantially with the twisting angle.
At $\theta =  \pi/2$ and for $m_0=-0.75$ the PCAC mass is roughly one half of the value at $\theta = 0$. 
We find that corresponding ratio for pion masses is within $1.5$ and $1.7$, which, given the small volumes we have 
considered, is quite consistent with the scaling in a chirally broken theory.

%%%%%%%%%%%%%%%%%%%%%%%%%%%%%%%%%%%%%%%%%%%%%%%%%%%%%%%%%%%%%%%%%%%%%%%%%%
%%%%%%%%%%%%%%%%%%%%%%%%%%%%%%%%%%%%%%%%%%%%%%%%%%%%%%%%%%%%%%%%%%%%%%%%%%

Finally, let us remark that
similar shifts in the quark mass have been discussed for the case of constant external magnetic fields~\cite{Bali:2015vua}.
We emphasize that here we are showing that such ef\mbox{}fects are present also in the case of vanishing magnetic field and constant (vector) potential.

\section{Conclusions}
We explored in detail an application of reweighting techniques to the case of modifications in the spatial periodicity of fermions.
We have thoroughly studied the approach at tree-level and we have provided constraints for the convergence 
of the stochastic estimates of all the Gaussian moments of the reweighting factors. We have also
established the large volume scaling of the average and the variance at tree-level.

At the numerical level, we have performed a complete and detailed study of purely gluonic as well as fermionic quantities in both large
and small volumes. We considered the plaquette and the pion dispersion relation, the latter in the two regimes.
In both cases we found the ef\mbox{}fects of reweighting to be at the sub-percent level for values of $\theta$ up to $\pi/2$.
In our implementation, for the large volumes considered, we found that the reweighting method, with $N_\eta \approx 600$ is a factor 4-5 cheaper compared to generating new configurations for dif\mbox{}ferent
values of the twisting angle, assuming that about 20 molecular dynamics units are needed to decorrelate subsequent configurations.
We performed this comparison in units of matrix-vector multiplications.

Perhaps the most important observation, which, as far as we know, has so far not been investigated in a dedicated way in the literature,
is on the dependence of the critical mass for Wilson fermions on the periodicity phases in the boundary conditions.
Although a cutof\mbox{}f ef\mbox{}fect, it could be rather large in a theory with O$(a)$ discretization ef\mbox{}fects. 
Since the result is that the hadron masses, and not only their momentum, change as $\theta$ changes, the 
corresponding dispersion relations may look very dif\mbox{}ferent at finite and coarse lattice spacings compared to the
continuum predictions. 

\afterpage{\blankpage}

\setcounter{equation}{0}

\chapter{Optimization of HMC performance}
\label{chap:oHMCp}

Gauge theories formulated on Euclidean lattices can be treated as 
statistical system and are amenable to numerical 
simulations.
When matter fields (scalars or fermions)
are dynamically present, the gauge-field configurations are typically
generated using Molecular Dynamics (MD) algorithms,
which come with dif\mbox{}ferent variantions of the
Hybrid Monte Carlo (HMC) \cite{Duane:1987de}.
Such algorithms depend on a large number of parameters,
whose optimal choices depend on the model and the 
regime (e.g.~concerning masses and volumes) considered.
The generation of configurations in large volumes at light quark masses 
is still very demanding.
For larger volumes the simulations become obviously more expensive.
For lighter quarks the condition number of the Dirac operator increases, 
hence the number of iterations needed by the solver 
(to invert the operator) increases.

The case of QCD has been extensively studied because of its obvious 
phenomenological relevance.
Roughly speaking HMC algorithms can be classified according to the
factorization of the quark determinant adopted
(tipically either mass-preconditioning \cite{Hasenbusch:2001ne},
or even/odd preconditioning \cite{DeGrand:1990dk},
or domain-decomposition \cite{Luscher:2005rx},
or rational factorization \cite{Clark:2006fx})
and the symplectic integrator(s) used,
the simplest being the leapfrog integrator,
and the most popular being the second order minimum norm integrator 
or Omelyan integrator \cite{Omelyan:2003, Takaishi:2005tz}.
The factorization of the determinant translates into 
a splitting of fermionic forces in the Molecular Dynamics,
and depending on the hierarchy of such forces various nested integrators can 
be used, where each force is integrated along a trajectory with
a dif\mbox{}ferent time-step \cite{Sexton:1992nu, Urbach:2005ji}.
No complete studies are present for Beyond Standard Model (BSM) strongly interacting models,
where the hierarchy of the forces can be completely dif\mbox{}ferent from the QCD case.

In dynamical simulations the Molecular Dynamics step 
is the most time consuming part, hence
it is of paramount importance 
to speed it up as much as possible.
For symplectic integrators a shadow Hamiltonian exists, that is conserved,
and can be used to optimize the parameters of a simulation.
In Ref.~\cite{Kennedy:2007cb} the shadow Hamiltonian was introduced to make
simulations with dynamical fermions cheaper
by means of symplectic integrators that conserve in a better way
the energy without decreasing the integration step-size too much.
The idea in the present work is to minimize the fluctuations of the
Hamiltonian deviations from the shadow Hamiltonian along the trajectory.
The strategy we will follow is general and it is also suitable 
for BSM models.\\
In the following we restrict the attention to the case of one particular integrator, Omelyan (with $\alpha=1/6$),
because it is a rather common choice, requires less computations and 
existing data (as long as the norm of the forces have been stored) can be readily used for the tuning.
We employ a three-time scale integrator with mass-preconditioning
 and we
show that once the forces and their variances are measured, assuming
they only depend on the mass-preconditioning parameter $\mu$ 
(and bare mass parameter $m_0 - m_{\rm c}$), we are able
to find the optimal integrator parameters.
Thanks to choice of $\alpha = 1/6$ in the integrator, the implementation 
of the Poisson brackets is not needed since their r\^ole is played 
by the forces.
It should be noted that the tuning of the integrator parameters also depends on the choice
of the inverter, the precision of the force and the Hamiltonian measurement. 
The strategy we present is applicable also for a dif\mbox{}ferent set of the aforementioned choices.
Our minimization method gives predictions accurate at the 10\% level, which we
consider quite satisfactory.
For completeness we consider an $\mathbf{SU}(2)$ gauge group with a doublet of
un-improved Wilson fermions in the fundamental representation and Wilson
plaquette gauge action, further details on the simulations set-up can be found in App.~\ref{app:ncp_op}.

The chapter is organized as follow; in Section~1 we review symplectic integrators and the 
existence of a shadow Hamiltonian associated to them for the case
of position and momentum variables in $\mathbb{R}^2$.
In Section~2 we give our set-up 
for the integrator, preconditioning and multi-scale integration and we show the explicit form 
of the shadow Hamiltonian and how it depends on the forces.
In Section~3 we test the goodness of our measurement of Poisson brackets and we show that the
shadow Hamiltonian is indeed conserved along the trajectory.
In Section~4 we relate the acceptance rate in simulations to the variance of the shadow Hamiltonian
and ultimately to the force variances; the cost of a simulation is also defined.
There we also present the functional form of the fits employed in $\mu$ (and the bare mass $m_0$).
Section~5 contains the comparison of the cost of actual simulation with the one found through 
our minimization method.
Section~6 contains our conclusions.
The work presented in this chapter is presented in \cite{Bussone:2016pmq}.

\section{Symplectic integrators: a toy model}

To each symplectic integrator corresponds a constant of motion 
related to the original Hamiltonian of the system we want to simulate.
The constant of motion is referred to as \emph{shadow} Hamiltonian.\\
%The following discussion follows Ref.~\cite{Yoshida:1992}.
See Ref.~\cite{Yoshida:1992} for a review.
Let us examine the case of the 1D harmonic oscillator.
The Hamiltonian of such a system is
\begin{align}
H = \frac{p^2}{2} + \frac{q^2}{2},
\end{align}
and the exact solution is given by the following map
\begin{align}
\begin{pmatrix}
p'\\ q'
\end{pmatrix} = \begin{pmatrix}
\cos\tau & -\sin\tau\\
\sin\tau & \cos\tau
\end{pmatrix}\begin{pmatrix}
p(0)\\q(0)
\end{pmatrix},
\end{align}
where $(q(0), p(0))$ is the initial condition and $(q', p')$ is the $\tau$-time evolved point in the phase space $(q, p)$.
Without loss of generality let us assume $\tau >1$.

\subsection*{Euler}

The easiest way to solve the dif\mbox{}ferential equation numerically is to employ the Euler method.
In that case the map is given by
\begin{align}
\label{eq:euler_map}
\begin{pmatrix}
p'\\ q'
\end{pmatrix} = \begin{pmatrix}
1 & -\tau\\
\tau & 1
\end{pmatrix}\begin{pmatrix}
p(0)\\q(0)
\end{pmatrix}.
\end{align}
The map is not symplectic and the time evolved Hamiltonian 
is found to be
\begin{align}
H(\tau) = H(0) + \frac{\tau^2}{2} \left[ p^2(\tau) + q^2(\tau) \right]\neq H(0).
\end{align}
The above equation tells us that at each integration time step 
the energy of the system is increasing and eventually 
diverges from $H$, i.e.~the error is \emph{unbounded}.

\subsection*{Symplectic Euler}

One can make the Euler map in Eq.~\ref{eq:euler_map} symplectic by modifying it as
\begin{align}
\label{eq:exact}
\begin{pmatrix}
p'\\ q'
\end{pmatrix} = \begin{pmatrix}
1-\tau^2 & -\tau\\
\tau & 1
\end{pmatrix}\begin{pmatrix}
p(0)\\q(0)
\end{pmatrix}.
\end{align}
The associated Hamiltonian at the time $\tau$ is given by
\begin{align}
H(\tau) = H(0) + \frac{\tau^2}{2} \left[3p^2(0) + q^2(0)\right]\neq H(0),
\end{align}
and is not conserved along the trajectory. 
Nonetheless we do have a constant of motion $H'$ given by
\begin{align}
\label{eq:new_const}
H' = \frac{1}{2} (p^2+q^2) + \frac{\tau}{2} pq = H + \frac{\tau}{2} pq.
\end{align}
The parameter $\tau$ is fixed from the beginning in the integration,
and we can see from direct inspection that indeed 
$H'(p(t=\tau), q(t=\tau) ; \tau) = H'( p(t=0), q(t=0); \tau)$.
The error in the Hamiltonian is now bounded and cannot grow. 
Thus if we start from a point $(p(0), q(0) ) = (0,1)$ after few iterations the time evolved point will be on the ellipse\footnote{The 
exact solution is parametrized by a circle $p^2+q^2 = 1$ in the phase space.} given by $p^2 + q^2 + \tau pq = 1$
and therefore the error is of order $\tau$.\\
One can prove that the symplectic map in Eq.~\ref{eq:exact} describes the exact $\tau$-time evolution of a Hamiltonian system, with $\widetilde{H}$ that has the expression in power of $\tau$
\begin{align}
\nn
\widetilde{H} &= H + \tau H_1 + \tau^2 H_2 + \tau^3 H_3 + \dots,\\
\nn
H_1 & = \frac{1}{2} H_p H_q = \frac{1}{2} pq,\\
\nn
H_2 & = \frac{1}{12} \left( H_{pp} H_q^2 + H_{qq} H_p^2\right) = \frac{1}{12} (p^2 + q^2) = \frac{1}{6} H,\\
H_3 & = \frac{1}{12} H_{pp}H_{qq}H_q H_p = \frac{1}{12} pq = \frac{1}{6} H_1,\\
\nn
\vdots
\end{align}
and it is exactly conserved during the trajectory.
Through a \quotes{brute force} calculation to the 17-th order in $\tau$, using MATHEMATICA along with Ref.~\cite{Weyrauch2009},
 we found that the shadow Hamiltonian is given by
\begin{align}
\nn
\widetilde{H} &=
\frac{1}{2}\left( p^2 + q^2 + \tau p q \right)
\left[  1 + \frac{\tau^2}{6} + \frac{\tau^4}{30} + \frac{\tau^6}{140} + \frac{\tau^8}{630} + \frac{\tau^{10}}{2772} + \frac{\tau^{12}}{12012} + \frac{\tau^{14}}{51480} + \frac{\tau^{16}}{218790}  \right]\\
& \equiv \frac{1}{2}\left( p^2 + q^2 + \tau p q \right) P(\tau) = H'\, P(\tau).
\end{align}
There is a clear factorization of a polynomial in $\tau$, $P(\tau)$, and since $\widetilde{H}$ is conserved then also $H'$ is conserved.

\subsection{Shadow hamiltonian}

The Hamilton equations in compact form are given by
\begin{align}
\frac{\de z}{\de t} = \left\{ z, H(z)\right\} \equiv D_H z.
\end{align}
where we introduced $z\in\mathbb{R}^{2n} = \{ (p, q)\}$, and the 
Poisson brackets are given by
\begin{align}
\{ \star, H\} = 
\frac{\partial \star}{\partial q} \frac{\partial H}{\partial p} 
- \frac{\partial \star}{\partial p} \frac{\partial H}{\partial q}.
\end{align}
The formal solution of the dif\mbox{}ferential equation takes the form $z(\tau) = \e^{\tau D_H} z(0)$. 
By choosing a separable Hamiltonian, $H(q, p) = T(p) + V(q)$, we have the operator identification $D_H = D_T + D_V$ and the 
solution can be formally written through the evolution operator as follows
\begin{align}
z(\tau) = \e^{\tau(D_T+D_V)} z(0).
\end{align}
More explicitly
\begin{align}
\nn
\text{ momenta evolution}\quad\e^{\tau D_V}:\quad &f(z(0))=f( p, q) \rightarrow f(z(\tau)) = f( p - \tau V_q, q),\\
\text{ positions evolution}\quad\e^{\tau D_T}:\quad &f(z(0))=f( p, q) \rightarrow f(z(\tau)) = f( p, q + \tau T_p),
\end{align}
where $f$ is an arbitrary function of $z$.\\
Let us specialize to the Omelyan integrator \cite{Omelyan:2003}.
The evolution operator in this case is
\begin{align}
\label{eq:omelyan}
%\exp\left[ \tau D_H \right] = 
\exp[\alpha \tau D_V] 
\exp\left[\frac{\tau}{2} D_T\right] 
\exp\left[\tau (1-2\alpha) D_V\right] 
\exp\left[\frac{\tau}{2} D_T\right] 
\exp[\alpha \tau D_V],
\end{align}
with $\alpha$ free parameter. 
By using the Baker-Campbell-Haussdorf\mbox{}f (BCH) formula 
we obtain that the shadow Hamiltonian $\widetilde{H}$ is related to the one we want to simulate, $H$, as follows
\begin{align}
\widetilde{H} = H + \tau^2 \left\{ \frac{6\alpha^2 - 6\alpha +1}{12} D_V\left[ D_V\left(T\right)\right] + \frac{1 - 6\alpha}{24} D_T\left[ D_V\left(T\right)\right] \right\} + \ord(\tau^4).
\end{align}
By putting $\alpha= 1/6$ the second term in curly brackets vanishes.
The remainder is proportional to $D_V\left[ D_V\left(T\right)\right]$.
In the case of $V=V(q)$ and $T=p^2/2$ this operator is related to the force,
\begin{align}
\nn
D_V(T) &= \left\{ T, V\right\} = {\cancel {V_p T_q} } - V_q T_p = -p V_q,\\
D_V\left[ D_V(T)\right] &= \left\{ - p V_q, V\right\} =  \left\{ V, p V_q\right\} = V_q (p V_q)_p - {\cancel {V_p (pV_q)_q} } = V_q ( V_q + {\cancel {p V_{qp}}} ) = V_q^2 \equiv F^2.
\end{align}
The shadow Hamiltonian in the case of Omelyan integrator with $\alpha = 1/ 6$ is given by
\begin{align}
\label{eq:shadow_harmonic}
\widetilde{H} = H + \frac{(\tau F)^2}{72} + \ord(\tau^4).
\end{align}

\subsection*{Time reversible integrators}

Let us assume we have a time reversible integrator given by
\begin{align}
E(\tau) = \exp\left(\tau A\right) \exp\left(\tau B \right),
\end{align}
if we expand it according to BCH formula we get
\begin{align}
E(\tau) = \exp\left(\tau \gamma_1 + \tau^2\gamma_2 + \tau^3\gamma_3 + \dots \right).
\end{align}
By imposing the time reversibility condition $E(\tau)E(-\tau)=\I$ we get, at the second order in $\tau$
\begin{align}
E(\tau)E(-\tau) = \exp\left(\tau \gamma_1 + \tau^2\gamma_2\right) \exp\left(-\tau \gamma_1 + \tau^2\gamma_2\right) = \exp\left( 2\tau^2\gamma_2 +\ord(\tau^3)\right) = \I, 
\end{align}
which tells us that $\gamma_2=0$.
By iterating the same steps at the order $\tau^4, \tau^6, \dots$ we obtain $\gamma_4=\gamma_6=\dots = 0$, \cite{Yoshida:1990zz}.

\section{Mass preconditioning \& multi time-scale}
\label{sec:multiscale}

We turn now to real HMC simulations of QFT.
A way to reduce the fluctuations of the force is to employ the \emph{mass preconditioning} of the quark
determinant \cite{Hasenbusch:2002ai}. The definitions for the massive and hermitian Dirac operators are the following
\begin{align}
D_m = D+m_0,\, \quad Q=\gamma_5 D_m.
\end{align}
The Dirac matrix is modified as follows
\begin{align}
\nn
D_m &= D_{\rm HMC}\times D_{\rm Hase},\, \quad D_{\rm HMC} = D_m + \mu,\, \quad D_{\rm Hase } = \left( D_m +\mu\right)^{-1} D_m ,\\
Q^2 &= Q^2_{\rm HMC}\times Q^2_{\rm Hase},\, \quad Q^2_{\rm HMC} \equiv \left(D^\dagger_m + \mu\right)  \left(D_m + \mu\right)  ,\, \quad Q^2_{\rm Hase } \equiv Q \left(D^\dagger_m + \mu\right)^{-1}  \left(D_m + \mu\right)^{-1} Q,
\end{align}
where we introduced a new mass term $\mu$, the \emph{Hasenbusch mass}.
The probability distribution in Eq.~\ref{eq:probability} depends on the determinant of the Dirac operator
that now becomes the product of two determinants
\begin{align}
\nn
\det D_m & = 
\det D_{\rm HMC}\times \det D_{\rm Hase}, \\
\label{eq:mass_prec}
\det Q^2 & = 
\det Q^2_{\rm HMC}\times \det Q^2_{\rm Hase}.
\end{align}
In particular, employing the $\gamma_5$ version of the Dirac-Wilson operator, the 
distribution with which the configurations are generated is given by
\begin{align}
P_S \propto \int \De[\phi_1,\phi_1^\dagger, \phi_2, \phi_2^\dagger] \exp\bigg( -{\rm S}_{\rm G}[U] 
- \phi_1^\dagger \left(Q^2_{\rm Hase}\right)^{-1}\phi_1 
- \phi_2^\dagger \left(Q^2_{\rm HMC}\right)^{-1}\phi_2 \bigg),
\end{align}
where we introduced a set of pseudo-fermions for each determinant. 
In our simulation we will have three (\emph{driving}) forces: Gauge, HMC, Hasenbusch.
The preconditioned HMC operator should be a reasonable approximation of the massive Dirac operator, cheaper in time to invert, and its determinant should be positive.
The \emph{condition number} of the HMC and the Hasenbusch operator are reduced compared to the original one, $Q^2$.
This is best understood in the twisted mass-preconditioning as explained in Ref.~\cite{Urbach:2005ji}.
Suppose that the $\lambda_{\rm max}$ and $\lambda_{\rm min}$ are respectively the maximal and minimal eigenvalues of $Q^2$. 
Then its condition number is given by $\lambda_{\rm max} / \lambda_{\rm min}$.
The condition number of $D_m D_m^\dagger + \mu^2$ is about $\lambda_{\rm max} / \mu^2$ and the one of $(D_m D_m^\dagger + \mu^2) / Q^2$ around $\mu^2 / \lambda_{\rm min} $.
Now by taking $\mu^2 \approx \sqrt{\lambda_{\rm max} / \lambda_{\rm min}}$ the Hasenbusch and the HMC term will be better 
conditioned with respect to $Q^2$.

A further acceleration can be achieved by considering \emph{multiple time-step integrators} \cite{Urbach:2005ji}.
It consists of taking dif\mbox{}ferent integration step sizes for dif\mbox{}ferent forces. 
Multiple time-step also helps to maintain under control large driving forces. 
As pointed out in 
Ref.~\cite{Joo:2000dh} large driving forces lead to \emph{instabilities} in HMC simulations.
To avoid them the time step must be small if the driving force are large.
Multiple time-step integrators are a valuable tool if the forces dif\mbox{}fer by orders of magnitude
in absolute value.\\
We employ a three time-scale integrator and assume that in the
outermost level there is the evolution for ${\rm S}_1$ with time step $\delta\tau = \tau /n$, in the middle the integrator for
${\rm S}_2$ with $m$ steps and the innermost is for ${\rm S}_3$ with $k$ steps.

The shadow Hamiltonian associated to the Omelyan integrator with three time-scales and mass
preconditioning is a quite lengthy expression (see App.~\ref{app:mts_osh}) but once we set the parameter $\alpha = 1/6$ it is given by
\begin{align}
\widetilde{H} = H + \frac{\delta\tau^2}{72}\left[ [\hat{ S }_1, [\hat{S}_1, \hat{T}]] + \frac{1}{4m^2} [\hat{S}_2, [\hat{S}_2, \hat{T}]]  +\frac{1}{16m^2k^2} [\hat{S}_3, [\hat{S}_3, \hat{T}]]  \right] + \ord(\delta\tau^4).
\end{align}
We use the conventions adopted in \cite{Kennedy:2012gk, Kennedy:2007cb} and the above formula becomes
\begin{align}
\widetilde{H} 
= H + \frac{\delta\tau^2}{72}\left[ \mathbf{e}_i\left( {\rm S}_1\right) \mathbf{e}^i\left( {\rm S}_1\right)
+ \frac{\mathbf{e}_i\left( {\rm S}_2\right) \mathbf{e}^i\left( {\rm S}_2\right)}{4m^2} 
+ \frac{\mathbf{e}_i\left( {\rm S}_3\right) \mathbf{e}^i\left( {\rm S}_3\right)}{16m^2k^2} \right]
+ \ord(\delta\tau^4),
\end{align}
where $\mathbf{e}_i$ is the $\mathbf{SU}(N_c)$ derivative and $i$ a collective index for space-time, direction and number of generators. 
$\mathbf{e}_i({\rm S})\mathbf{e}^i({\rm S})$ gives the force associated to the action ${\rm S}$ summed 
over all space-time point, directions, color and eventually spin, as expected, since the Hamiltonian is an extensive quantity.
The shadow Hamiltonian reduces to 
\begin{align}
\nn
\widetilde{H} &= H + \frac{\delta\tau^2}{72} \sum_{x,\mu,a}  \left[ T_{R,1}\left(F_1^{a\mu}(x)\right)^2  + T_{R,2}\frac{\left(F_2^{a\mu}(x)\right)^2}{4m^2} + T_{R,3}\frac{\left(F_3^{a\mu}(x)\right)^2}{16m^2k^2}  \right] + \ord(\delta\tau^4)\\
\label{eq:var}
&\equiv H + \frac{\delta\tau^2}{72} \left( |\mathcal{F}_1|^2 + \frac{|\mathcal{F}_2|^2}{4m^2}  + \frac{|\mathcal{F}_3|^2}{16m^2k^2}  \right) + \ord(\delta\tau^4) \equiv H + \delta H + \ord(\delta\tau^4),
\end{align}
that is a function only of the forces used during the simulations.\\
For the derivation of the various forces see the results in App.~\ref{app:mts_osh}.
We immediately see that the shadow Hamiltonian is related to the dif\mbox{}ferent parts of the force
weighted by the corresponding normalization for the generators $T_R$ in the given representation $R$.
Note that the gauge is always in the fundamental $T_R = T_f = 1/2$, while the fermionic parts may be in dif\mbox{}ferent representation.
Already at this point one can
see what drove our assignments for the dif\mbox{}ferent levels.
We want to suppress the contribution of the \emph{largest}
force (the gauge one) and hence that will go in the innermost level, followed by the HMC force and the Hasenbusch one at the outermost level.

In the following we specialize to the case of the Omelyan integrator with $\alpha = 1/6$. 
We consider an $\mathbf{SU}(2)$ gauge group with a doublet of un-improved Wilson fermions in the fundamental representation
and Wilson plaquette gauge action. 
For completeness the value $m_c$ 
of the bare mass parameter yielding massless fermions
is estimated to be $-0.77(2)$ at $\beta = 2.2$ \cite{Lewis:2011zb, Hietanen:2014xca, Arthur:2016dir}.

\section{Benchmarks in small volumes}

\begin{figure}[h!t]
\begin{center}
\begin{minipage}[t]{0.4\linewidth}
%\centering
\subfigure[\emph{Scaling of $|\Delta H|$ and $|\Delta(\delta H) + \Delta H|$ with $\delta \tau$ in $V=8^4$ at fixed inner levels steps $m=10$ and $k=10$.}]
{\includegraphics[scale=0.72]{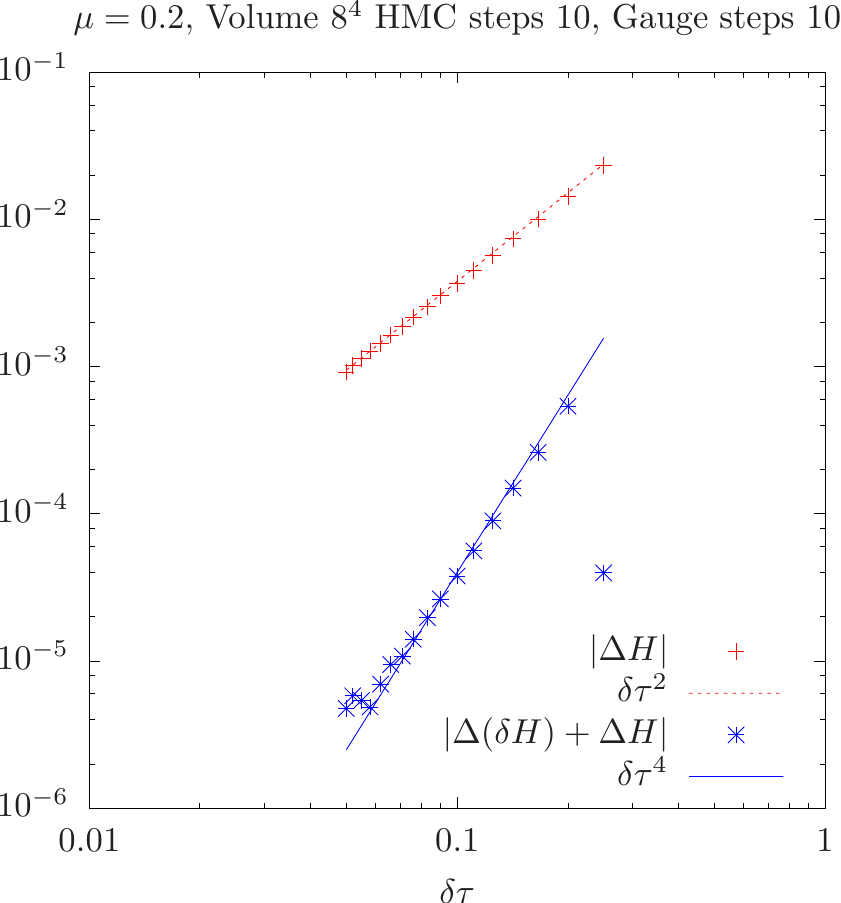}\label{fig:benchmarks11}}
\end{minipage}
\hspace{0.2cm}
\begin{minipage}[t]{0.5\linewidth}
\subfigure[\emph{Scaling of $|\Delta(\delta H) + \Delta H|$ with $\delta \tau$ in $V=4^4$ for two dif\mbox{}ferent choices of the inner level steps.}]
{\includegraphics[scale=0.68]{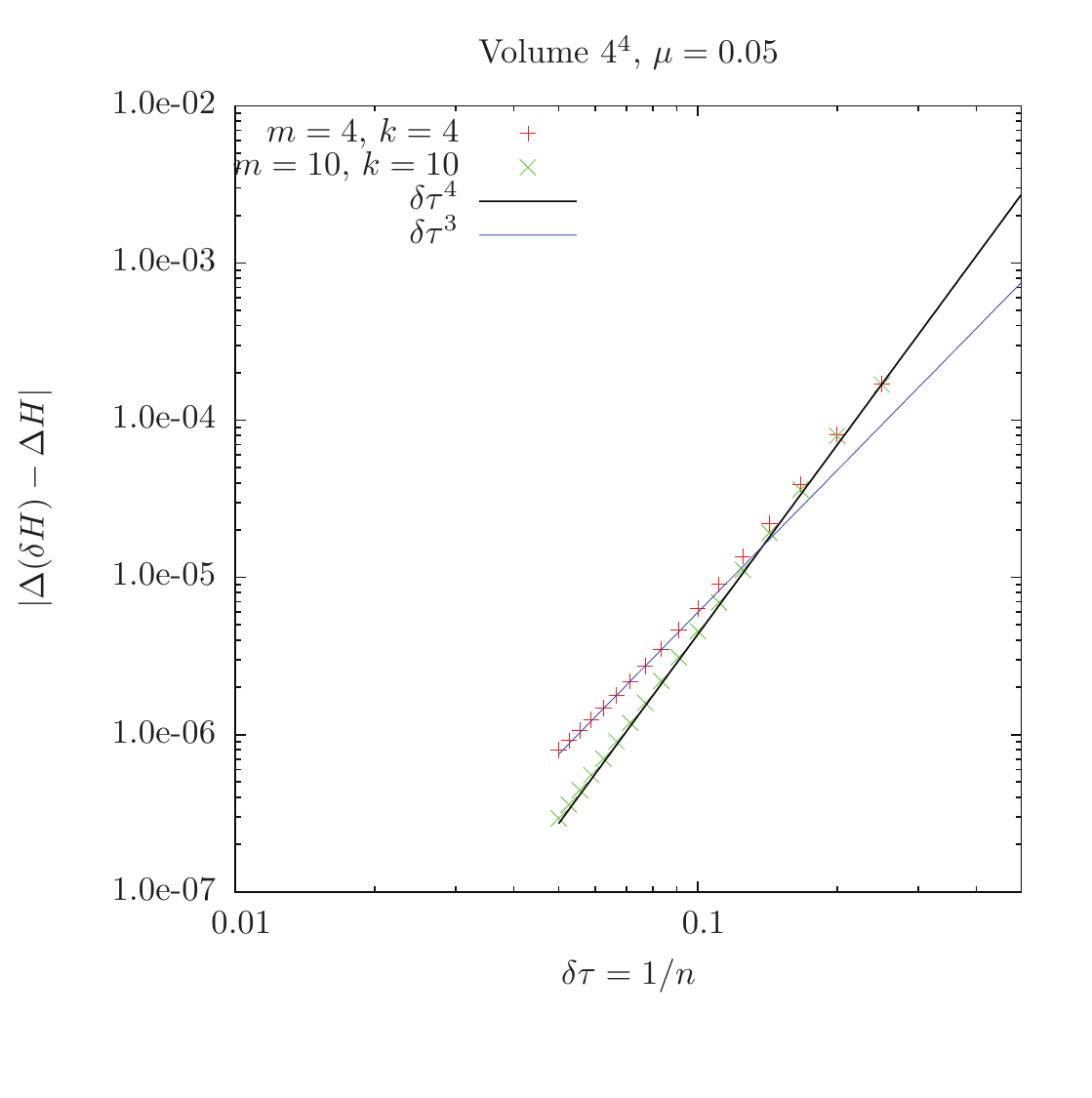}\label{fig:delta_tau3}}
\end{minipage}
\caption{\emph{Scaling behavior with $\delta\tau = 1/n$ for $\Delta H$ and $\Delta(\delta H)$ in small volumes. $\Delta(\delta H)$ is built from the knowledge of the forces at the beginning and the end of the trajectory. Lines are shown to guide the eye.}}\hspace{0.1cm}
\label{fig:benchmarks1}
\end{center}
\end{figure}

In order to test the measurements of the Poisson brackets we check the scaling with $\delta\tau$ of $|\Delta H|\propto\delta\tau^2$, since the Omelyan integrator 
is a \emph{second order integrator}.
Furthermore we test the scaling of $|\Delta(\delta H) + \Delta H|\propto\delta\tau^4$,
indeed, since $\widetilde{H}$ is conserved along the trajectory, the following relation holds
\begin{align}
\Delta \widetilde{H} = 0 = \Delta H + \Delta(\delta H) + \ord(\delta\tau^4) \Longrightarrow \Delta H = - \Delta(\delta H) + \ord(\delta\tau^4).
\label{eq:delta}
\end{align}
The dif\mbox{}ference at the end of the trajectory is given by
\begin{align}
\label{eq:storto}
\Delta(\delta H) & = 
\frac{\delta\tau^2}{72} \left(|\mathcal{F}_{1, f}|^2-|\mathcal{F}_{1, i}|^2  + \frac{|\mathcal{F}_{2, f}|^2 - |\mathcal{F}_{2, i}|^2}{4m^2} + \frac{|\mathcal{F}_{3, f}|^2 - |\mathcal{F}_{3, i}|^2}{16m^2k^2}\right),
\end{align}
where the forces are evaluated at the beginning $i$ and the end $f$ of the trajectory.\\
For the tests we have run one trajectory from a thermalized configuration with the following set-up for the levels in the
integration, $\tau_f = 1$,
\begin{itemize}
\item level 1: Hasenbusch, $n = 4,5,\dots,20$,
\item level 2: HMC, $m = 10$,
\item level 3: Gauge, $k = 10$.
\end{itemize}
The results in small volumes are shown in Fig.~\ref{fig:benchmarks1}. 
In Fig.~\ref{fig:delta_tau3} we see in the small $\delta \tau$ region the emergence of a contribution of order $\delta\tau^3$. 
%%%%%%%%%%%%%%%%%%%%%%%%%%%%%%%%%%%%%%%%%%%%%%%%%%%%%%%%%%%%%%%%%%%%%%%%%%
%%%%%%%%%%%%%%%%%%%%%%%%%%%%%%%%%%%%%%%%%%%%%%%%%%%%%%%%%%%%%%%%%%%%%%%%%%
This comes from the ambiguity in the measurement of $\mathcal{F}_{2}$, that is the time in which we measure it in the sub-level steps.
In other words dif\mbox{}ferent forces are measured at dif\mbox{}ferent times.
In formulae we can write
\begin{align}
\label{eq:inf_force}
|\mathcal{F}_2|^2 = |\mathcal{F}^{m\rightarrow\infty}_2|^2 + \ord \left(\frac{\delta\tau}{m}\right),
\end{align}
where $\delta\tau / m$ gives a correction to $\delta H$ of order $1/m^3$.  
This term induces a contribution of order $\delta\tau^3$ in $|\Delta(\delta H) + \Delta H|$ that can be suppressed by choosing a large number of steps $m$.
The same phenomenon may appear for $\mathcal{F}_3$ but the extra term will be suppressed by a power $mk$ and gives a correction to $\delta H$ of order $1/m^3k^3$,
which is completely negligible for $k=10$.

\begin{figure}[h!t]
\begin{center}
\subfigure[\emph{$\Delta H$ history along one trajectory with $n=10$.}]{\includegraphics[scale=0.75]{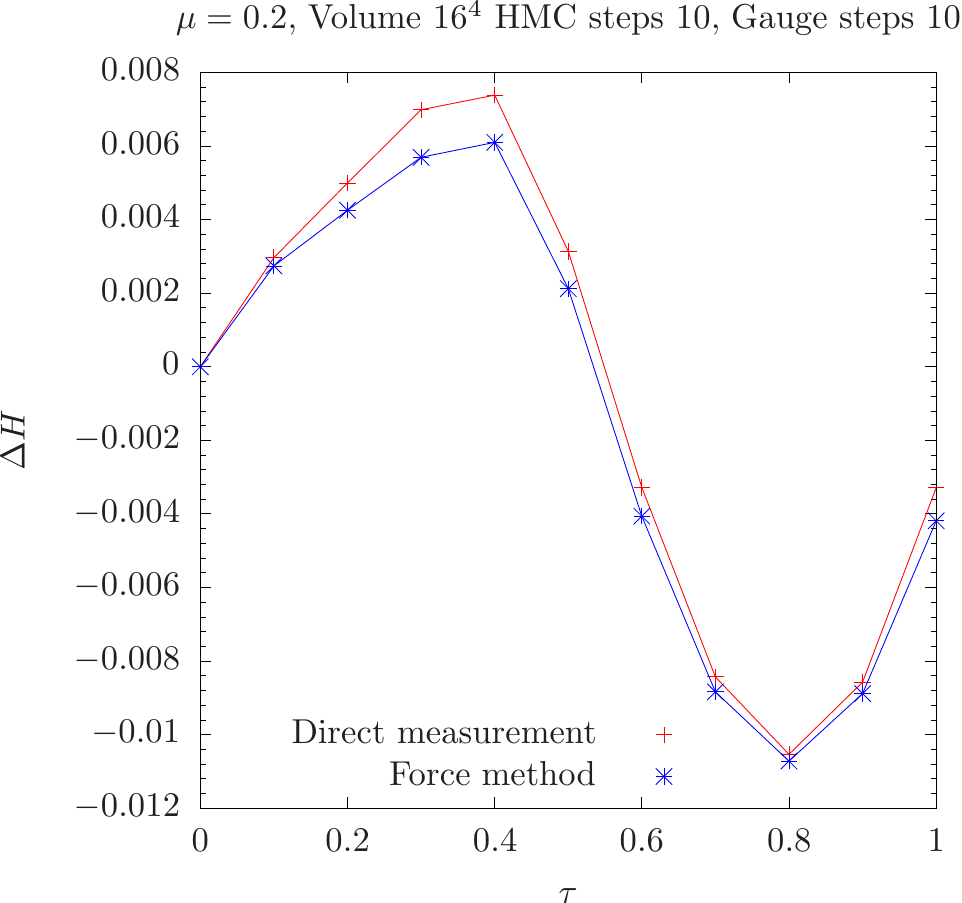}\label{fig:delta_h_10}}
\subfigure[\emph{$\Delta H$ history along one trajectory with $n=20$.}]{\includegraphics[scale=0.75]{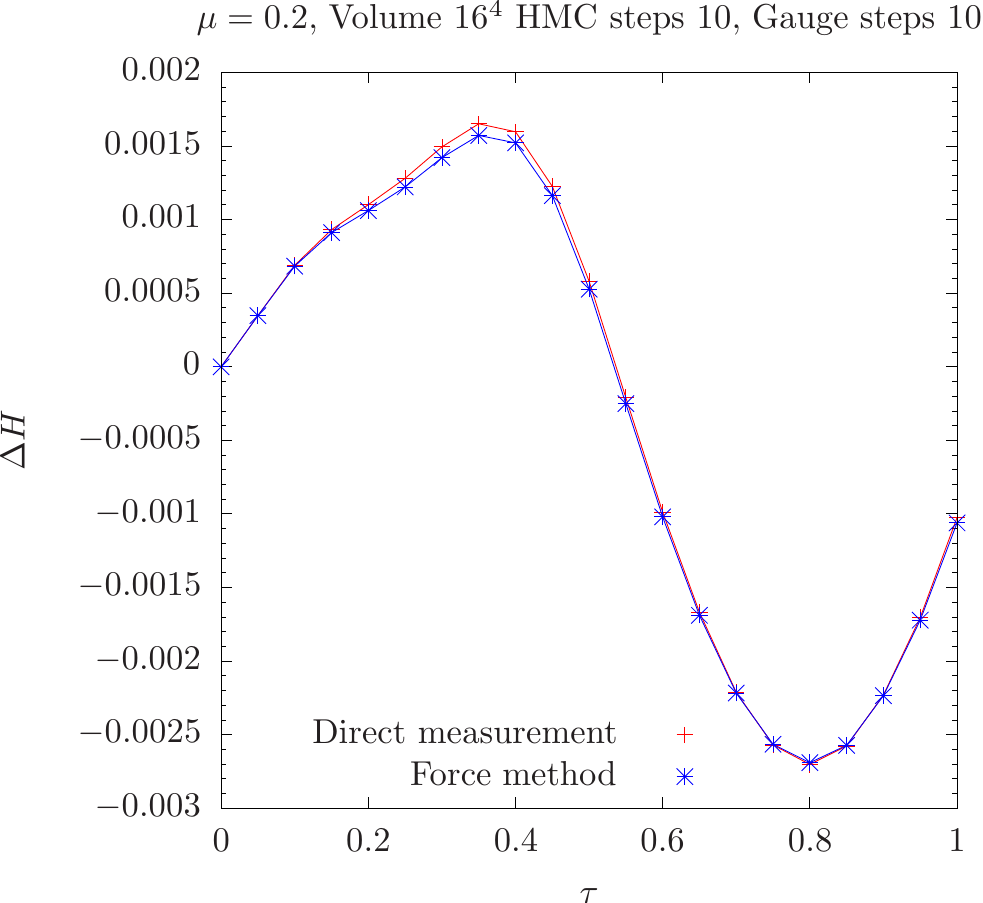}\label{fig:delta_h_20}}
\caption{\emph{Benchmarks for the Poisson brackets measurements.
Direct measurement (red crosses) corresponds to the measure of $\Delta H$ during the trajectory. Force method (blue stars) employs the knowledge of the forces. From the left to the right panel the minimum of $\Delta H$ is scaling as expected.}}
\label{fig:benchmarks2}
\end{center}
\end{figure}

Another useful test is to measure directly $\Delta H$ along the trajectory and compare it with the one built from
the knowledge of the forces, Eq.~\ref{eq:delta}. 
The results for $16^4$ volume are shown in Fig.~\ref{fig:benchmarks2}.
It is worth to notice that the minimum of $\Delta H$ scales as predicted by increasing $n$.
When the minimum (or a
maximum) is attained, then $\Delta H$ cannot decrease (increase) any further due to the existence of the shadow Hamiltonian, 
and it is well understood in terms of the various underlying force contributions.

%%%%%%%%%%%%%%%%%%%%%%%%%%%%%%%%%%%%%%%%%%%%%%%%%%%%%%%%%%%%%%%%%%%%%%%%%%
%%%%%%%%%%%%%%%%%%%%%%%%%%%%%%%%%%%%%%%%%%%%%%%%%%%%%%%%%%%%%%%%%%%%%%%%%%

\section{Cost of a simulation and its minimization}

Although the cost of a simulation is not unique we define it as
\begin{align}
{\rm Cost} = \frac{\# {\rm MVM}}{P_{\rm acc}}.
\end{align}
The number of Matrix-Vector-Multiplications (\# MVM) is a machine independent quantity\footnote{The
gauge part contributes for a maximum of about 5\% of the cost, hence is negligible. We also took into account the
gauge part as a check and it does not af\mbox{}fect the results of this work.}. 
Furthermore we
neglect the autocorrelation since that conceivably has a mild dependence on $\mu$ and therefore should be mostly contribute as an overall factor to the cost.

\subsection{Acceptance \& Matrix-Vector-Multiplication}

We link the acceptance $P_{\rm acc}$ to $\Delta H$ through the Creutz formula, see App.~\ref{app:acc_hmc},
\begin{align}
P_{\rm acc} (\Delta H) = {\rm erfc}\left(\sqrt{{\rm Var}(\Delta H)/8}\right),
\end{align}
and the connection between the variances of $\Delta H$ and $\delta H$ is given by \cite{Clark:2008gh, Clark:2011ir}
\begin{align}
\label{eq:clark}
{\rm Var} (\Delta H) \simeq 2{\rm Var}(\delta H).
\end{align}
In the following our goal is to optimize the choice of parameters $\mu, n, m, k$, (for dif\mbox{}ferent bare masses $m_0$), 
while keeping the choice of the integrator, the solver and the number of Hasenbusch splittings fixed.
We can also fix the bare masses, but a stabilization in terms of $\chi$-square of the fits is seen once we fit also in $m_0$.
But we should mention that it is not strictly necessary.\\
Let us illustrate how to get to the above relation Eq.~\ref{eq:clark}.
We trade $\Delta H$ for 
$\Delta(\delta H)$ through Eq.~\ref{eq:delta}.
By using Eq.~\ref{eq:storto} we can compute the variance (and neglecting all the covariances),
\begin{align}
\nn
\text{Var}\left[\Delta(\delta H)\right] &\equiv \frac{\delta\tau^4}{(72)^2} \bigg[\text{Var}\left(|\mathcal{F}_{1, f}|^2\right) + \text{Var}\left(|\mathcal{F}_{1, i}|^2\right) + \frac{\text{Var}\left(|\mathcal{F}_{2, f}|^2\right) + \text{Var}\left(|\mathcal{F}_{2, i}|^2\right)}{(4m^2)^2}\\
\label{eq:consistent}
&\quad\qquad\qquad + \frac{\text{Var}\left(|\mathcal{F}_{3, f}|^2\right) + \text{Var}\left(|\mathcal{F}_{3, i}|^2\right)}{(16m^2k^2)^2}\,\bigg].
\end{align}
Under the further assumption that the final and initial forces are extracted from the same distribution, $\exp(-H)$\cite{Clark:2010qw},
and they are independent for long enough trajectories, we obtain
\begin{align}
\label{eq:force_2}
\text{Var}\left[\Delta(\delta H)\right] 
&\simeq \frac{2\,\delta\tau^4}{(72)^2} \left[\text{Var}\left(|\mathcal{F}_1|^2\right) 
+ \frac{\text{Var}\left(|\mathcal{F}_2|^2\right)}{(4m^2)^2} 
+ \frac{\text{Var}\left(|\mathcal{F}_3|^2 \right)}{(16m^2k^2)^2}\right]
=  2 {\rm Var}(\delta H).
\end{align}
We have tested Eq.~\ref{eq:clark} by calculating
in two dif\mbox{}ferent ways the acceptance.
We refer to the acceptance calculated with the formula in Eq.~\ref{eq:force_2} (${\rm Var}(\delta H)$) as \quotes{Force method 1}, while \quotes{Force method 2} is used 
for the acceptance calculated with Eq.~\ref{eq:consistent} ($\text{Var}\left[\Delta(\delta H)\right]$).
\quotes{Force method 1} is the one going to be used in the rest of the work. 
There the acceptance rate is calculated from the variance of $\delta H$. 
For each trajectory we calulate the average of $\delta H$ over each of the uppermost level step.
The result is then used to build the variance over the trajectories, where the average forces over the sub-levels are used.
In \quotes{Force method 2} the acceptance is calculated from the variance of $\Delta( \delta H)$. 
To do that we calculate the dif\mbox{}ference of $\delta H$ at the beginning and the end of each trajectory and we extract the variance over the trajectories.
%%%%%%%%%%%%%%%%%%%%%%%%%%%%%%%%%%%%%%%%%%%%%%%%%%%%%%%%%%%%%%%%%%%%%%%%%%
\begin{figure}[h!t]
\begin{center}
\includegraphics[scale=0.8]{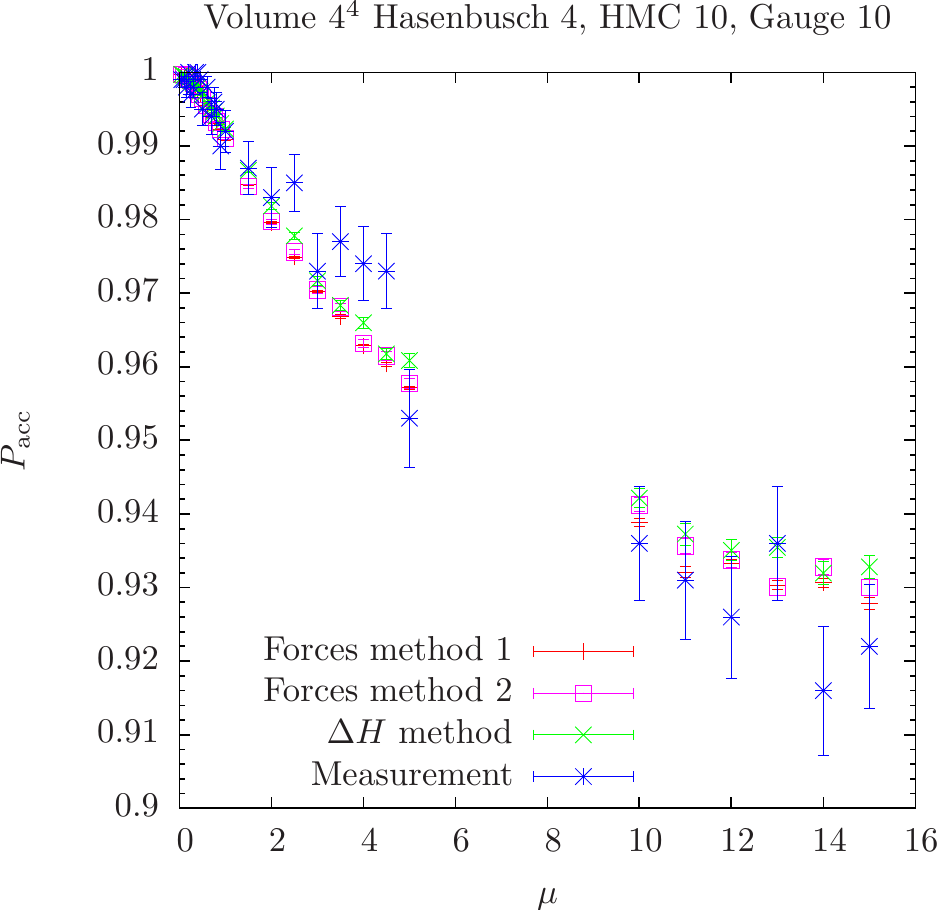}
\caption{\emph{Acceptance as a function of the preconditioning mass. 
Force method 1 and 2 (red segments and violet boxes) are employing the formulae in Eqs.~\ref{eq:force_2}, \ref{eq:consistent}. $\Delta H$ method (green crosses) correspond to the measurement of $H$ (also used for the accept/reject step).
Measurement (blue crosses) correspond to simple counting.}}
\label{fig:compfarison}
\end{center}
\end{figure}
%%%%%%%%%%%%%%%%%%%%%%%%%%%%%%%%%%%%%%%%%%%%%%%%%%%%%%%%%%%%%%%%%%%%%%%%%%
In Fig.~\ref{fig:compfarison} we compare the two methods described above to extract 
the acceptance of a simulation in a small $V=4^4$ volume.
The simulations are carried out starting from a thermalized configuration and running $10^3$ trajectories. 
There the two methods are compared with the direct measurement and the acceptance built directly from $\Delta H$.
Both methods seem to reasonably reproduce the data.

The total average number of MVM is given in terms of the averages at each level by
\begin{align}
\label{eq:mvm}
\# {\rm MVM} = (2n+1) \# {\rm MVM}_1(\mu) +  2n(2m+1) \# {\rm MVM}_2(\mu).
\end{align}

The analysis is done with standard statistical error propagation since each simulation, for dif\mbox{}ferent
choices of $n, m, k, \mu$ and $m_0$, is independent.

\subsection{Fits}

\begin{figure}[t!]
\begin{center}
\includegraphics[scale=0.65]{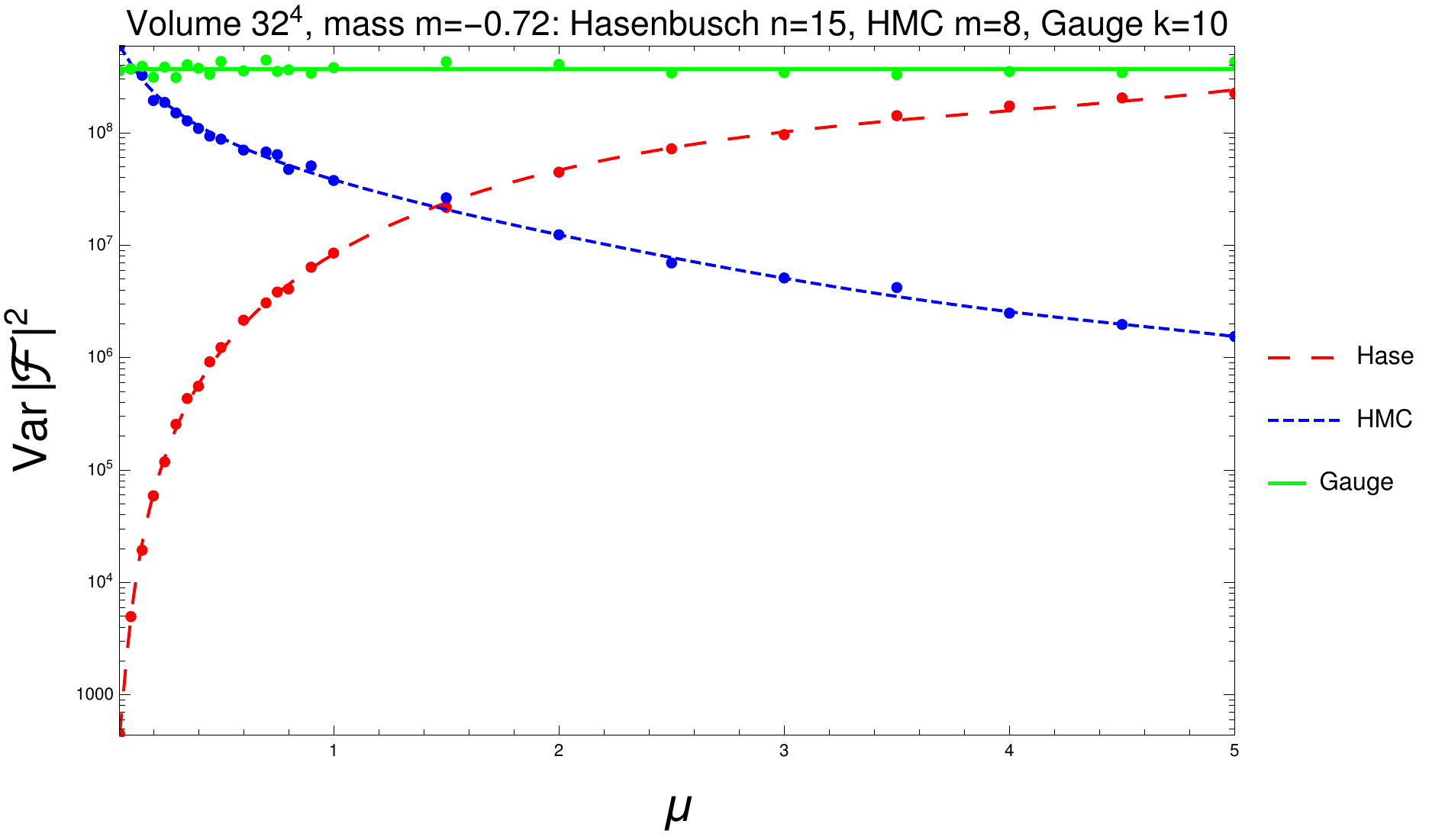}
\caption{\emph{Variances of the forces as a function of the mass preconditioning parameter. 
For very small $\mu$ the HMC variance is dominant, outside this region the Gauge variance is the dominant one. 
In the intermediate $\mu$ region we have an hierarchy of forces that corresponds to the level assignments. For large $\mu$ we have an inverted hierarchy. The figure corresponds to $m_0=-0.72$.}}
\label{fig:var_mu_dep}
\end{center}
\end{figure}

\begin{figure}[h!t]
\begin{center}
\subfigure[\emph{$\#{\rm MVM}_1(\mu)$ data and their fit.}]{\includegraphics[scale=0.35]{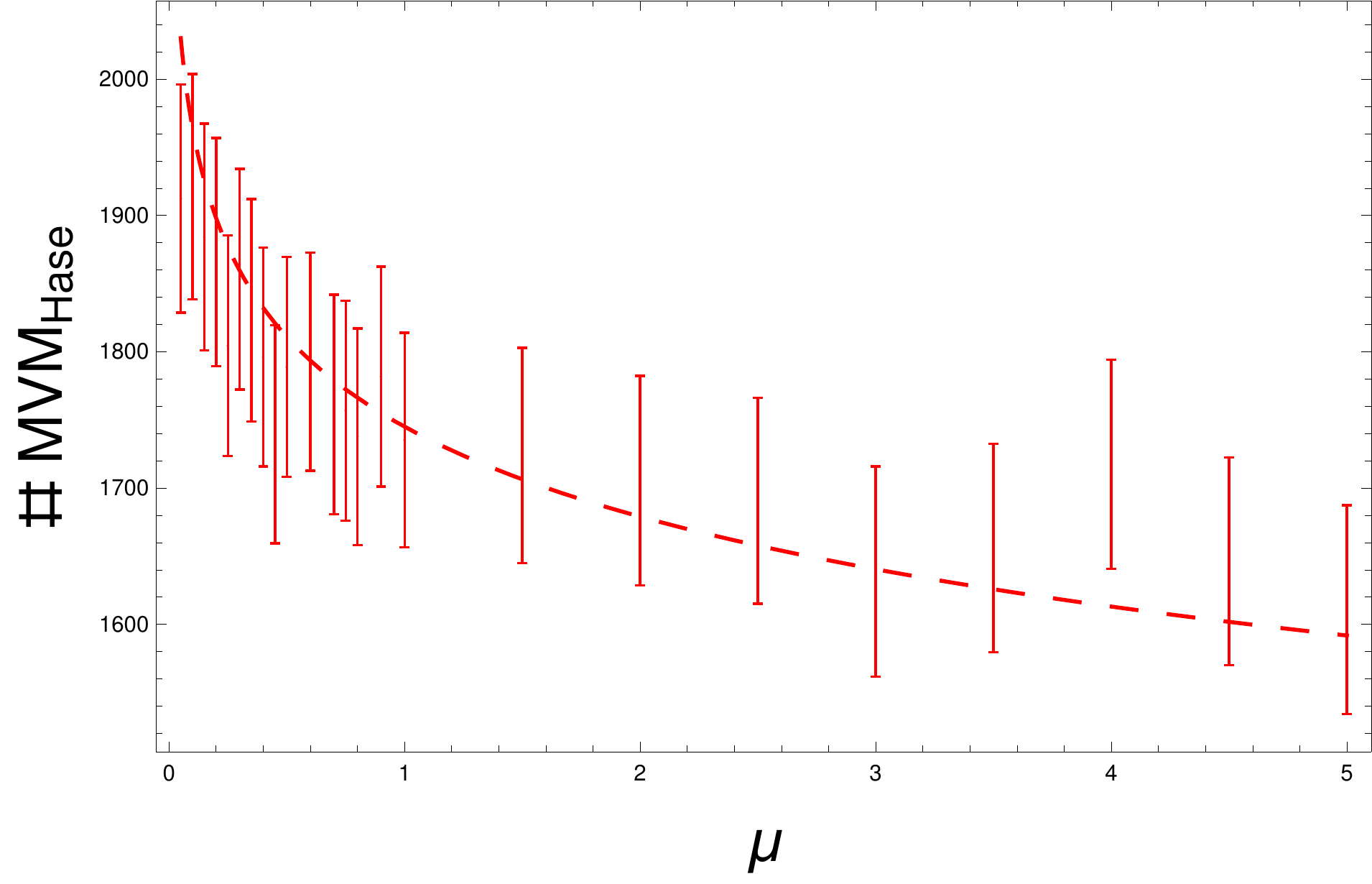}\label{fig:mvm_hase}}
\subfigure[\emph{$\#{\rm MVM}_2(\mu)$ data and their fit.}]{\includegraphics[scale=0.35]{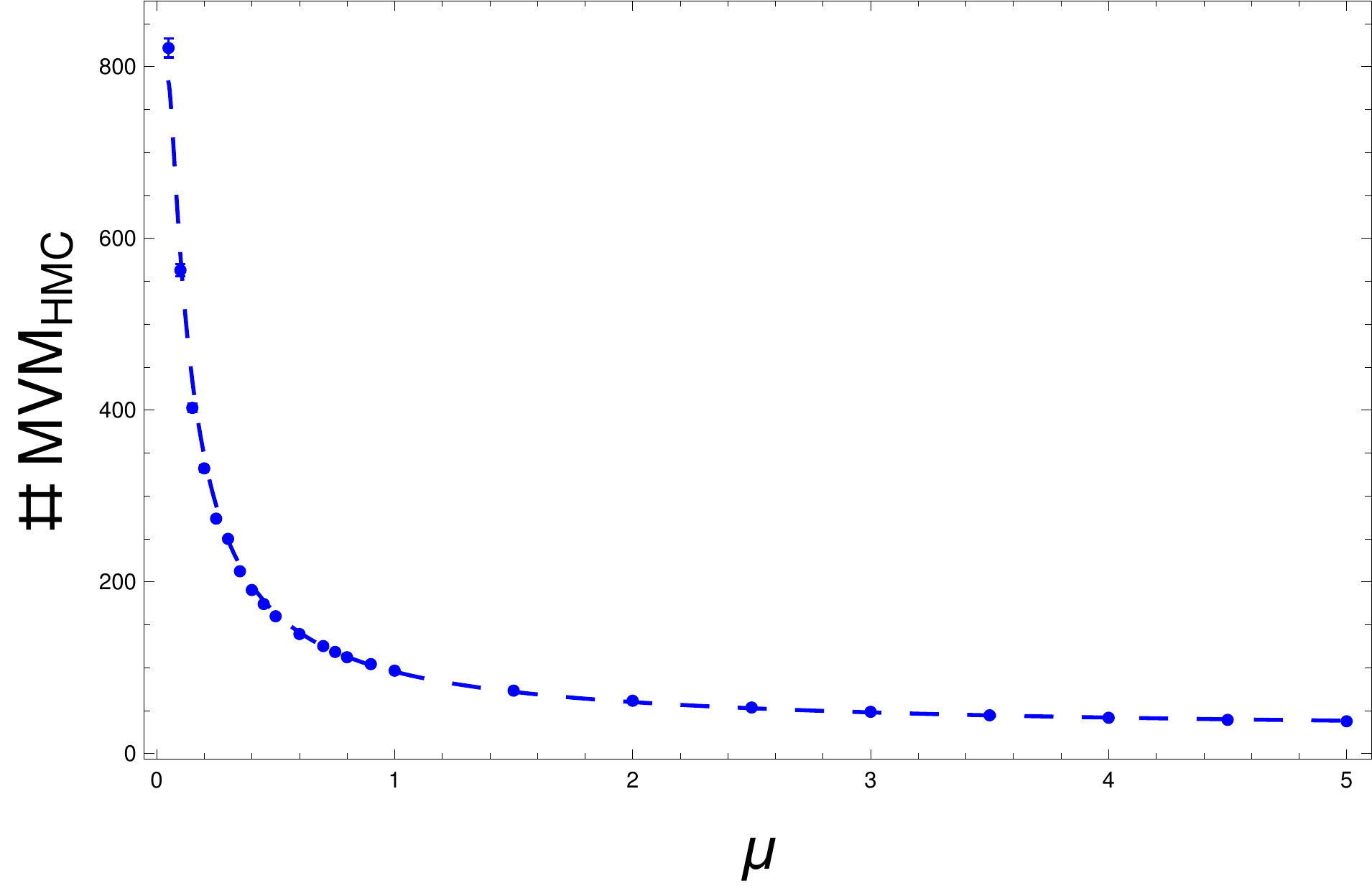}\label{fig:mvm_hmc}}
\caption{\emph{\#\,MVM in Hasenbusch and HMC levels as a function of $\mu$ for $m_0=-0.72$.}}
\label{fig:mvm_mu_dep}
\end{center}
\end{figure}

The idea is to assume that ${\rm Var} (\delta H)$ and \#\,MVM depend explicitly upon $n, m$ and $k$ as in
Eqs.~\ref{eq:force_2}, \ref{eq:mvm} and the dependence on $\mu$ (and $m_0$) of ${\rm Var}(|\mathcal{F}_i|^2)$ 
is the only quantity to be modeled, along with $\# {\rm MVM}_i$.\\
In Figs.~\ref{fig:var_mu_dep}, \ref{fig:mvm_mu_dep} we show the result of the fits in $\mu$ 
and $m_0 - m_{\rm c}$ for large volume $V=34^4$ and $m_0 = -0.72$.
In Fig.~\ref{fig:var_mu_dep} we show the variances for the dif\mbox{}ferent forces and 
their resulting fits. We can identify two dif\mbox{}ferent regions: a \emph{strong} 
dependence region for small $\mu$ and a \emph{weak} dependence region for large $\mu$. In the weak
dependence region we have an \emph{inverted hierarchy} with respect to what was our choice. 
In Fig.~\ref{fig:mvm_mu_dep} we show the number of MVMs per step and sub-step and their fits.\\
The fits were performed in $m_0$ and $\mu$ using the following curves
\begin{align}
\nn
{\rm Var}\left( \mathcal{F}_{\rm Hase} \right) 
= & \frac{a \mu^2}{(m_0-m_{\rm c} )^2}   
+  \left[ b + \frac{c \mu^2}{(m_0-m_{\rm c})^2} + d \mu^4 \right]
\exp\left( -e \mu^2  \right)\\
{\rm Var}\left( \mathcal{F}_{\rm HMC} \right) 
\nn
= & a + b \mu^2 + \frac{c}{\mu}
+ \frac{d\left(m_0 - m_c\right)^2}{\mu}  + e\mu^4 , \\
{\rm Var}\left( \mathcal{F}_{\rm Gauge} \right) 
\nn
= & a + \frac{b}{(m_0 - m_{\rm c})^2} , \\
\#{\rm MVM}_{\rm Hase}
\nn
= & \left( a + \frac{b}{m_0 - m_c} \right) + c \ln \mu , 
\\
\#{\rm MVM}_{\rm HMC}
= & a + \frac{b}{\mu} + \left( c+ \frac{d}{m_0-m_c}\right) \frac{1}{\mu^2}\, .
\end{align}
Only terms that reduced significantly the $\chi^2$ were included in the fitting curves. 

%%%%%%%%%%%%%%%%%%%%%%%%%%%%%%%%%%%%%%%%%%%%%%%%%%%%%%%%%%%%%%%%%%%%%%%%%%
%%%%%%%%%%%%%%%%%%%%%%%%%%%%%%%%%%%%%%%%%%%%%%%%%%%%%%%%%%%%%%%%%%%%%%%%%%
%%%%%%%%%%%%%%%%%%%%%%%%%%%%%%%%%%%%%%%%%%%%%%%%%%%%%%%%%%%%%%%%%%%%%%%%%%
%%%%%%%%%%%%%%%%%%%%%%%%%%%%%%%%%%%%%%%%%%%%%%%%%%%%%%%%%%%%%%%%%%%%%%%%%%

\subsection{Cost function}

\begin{figure}[h!t]
\begin{center}
\subfigure[\emph{${\rm Cost / Cost_{min}}$ with $m=3$ as function of $\mu$ and $n$.}]{\includegraphics[scale=0.5]{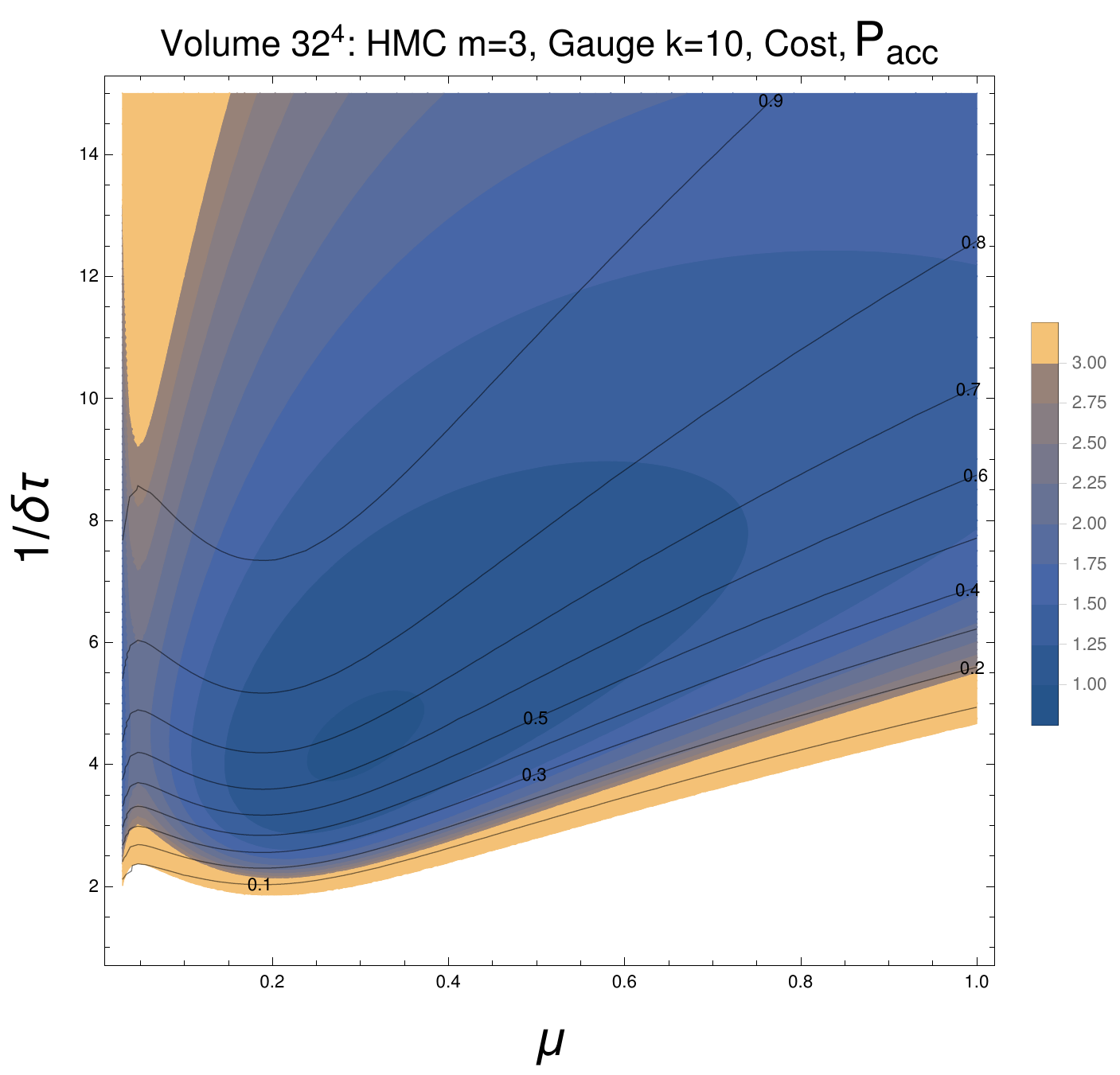}\label{fig:cost_min_hase}}
\hspace{0.5cm}
\subfigure[\emph{${\rm Cost / Cost_{min}}$ with $n=5$ as function of $\mu$ and $n$.}]{\includegraphics[scale=0.5]{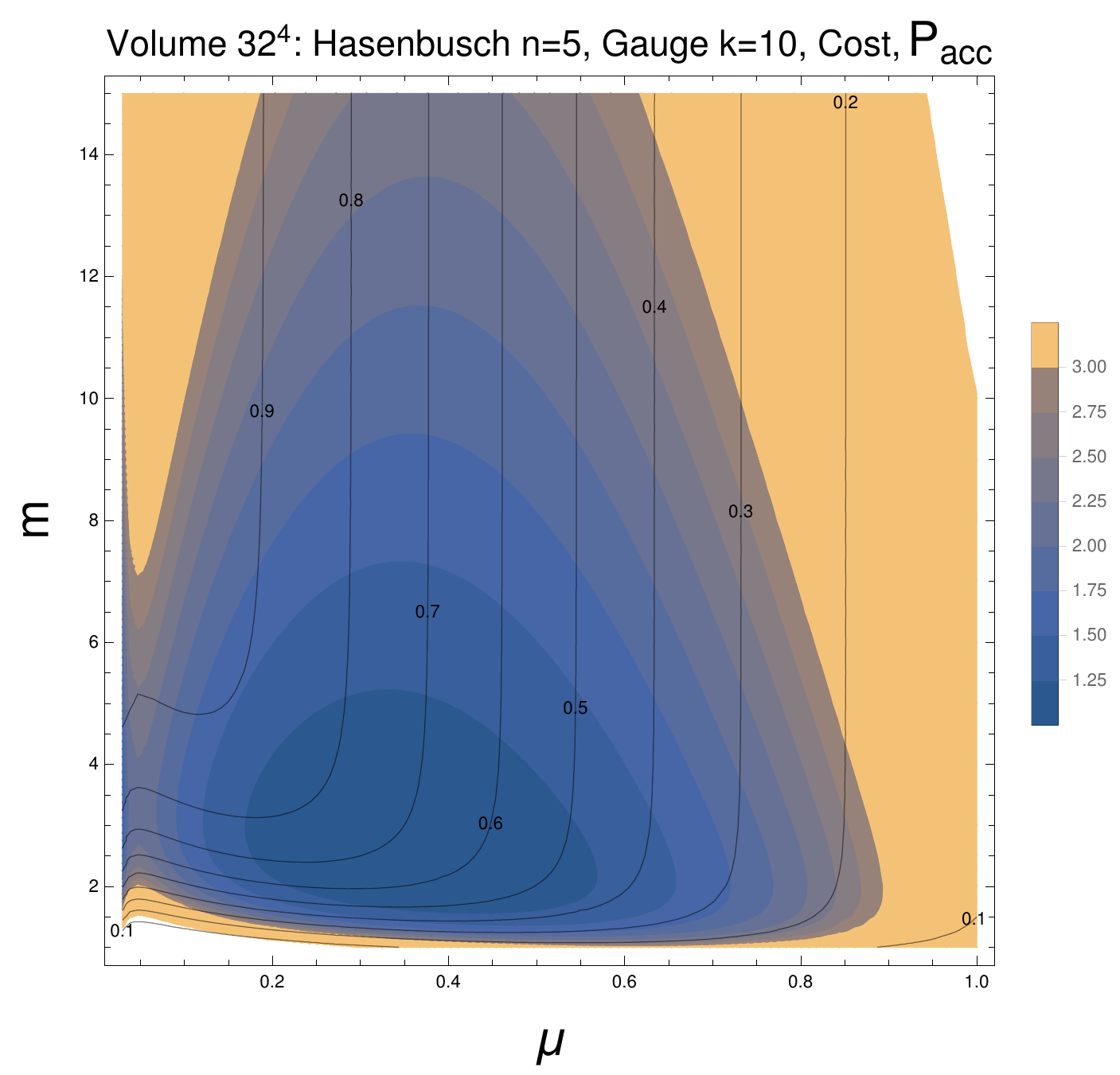}\label{fig:cost_min_hmc}}
\caption{\emph{Ratio ${\rm Cost / Cost_{min}}$ around the minimum $(n,m,\mu)\simeq(5, 3, 0.3)$ is displayed as curve level. Acceptance rate lines are drawn. Bare mass $m_0 = -0.72$.}}
\label{fig:cost_min}
\end{center}
\end{figure}

We can now build the cost as a function of $n, m, k$, $\mu$ and $m_0$. For simplicity we fix $k = 10$, and in
order to find the minimum in the other parameters we require\footnote{This requirement is needed for the Creutz formula
to hold true.} $P_{\rm acc} \gtrsim 70\%$. 
The minimization is carried out at fixed bare mass, in the following we specialize to $m_0 = -0.72$ since the results for other bare masses are qualitatively the same.
With this set-up we
found the minimum, ${\rm Cost_{min}}$, to be at $(n, m, \mu) \approx (5, 3, 0.3)$. In Fig.~\ref{fig:cost_min} we show the cost, normalized to
the minimum, and the acceptance in the plane $(\mu, n)$ and $(\mu, m)$.
In Fig.~\ref{fig:cost_min_hase} we fix $m = 3$ and it can be seen
 that the minimum is close to the boundary $P_{\rm acc}\sim 70\%$. 
 By taking the cost of a simulation to
be $1 < {\rm Cost} / {\rm Cost_{min}} < 1.25$ the window in parameter space is quite broad $0.2\lesssim\mu\lesssim 0.7$ and $3 \lesssim n \lesssim 8$.
Same conclusions can be drawn for the Fig.~\ref{fig:cost_min_hmc}.
%%%%%%%%%%%%%%%%%%%%%%%%%%%%%%%%%%%%%%%%%%%%%%%%%%%%%%%%%%%%%%%%%%%%%%%%%%
%%%%%%%%%%%%%%%%%%%%%%%%%%%%%%%%%%%%%%%%%%%%%%%%%%%%%%%%%%%%%%%%%%%%%%%%%%

\section{Comparison with simulations}

\begin{figure}[h!t]
\begin{center}
\includegraphics[scale=0.7]{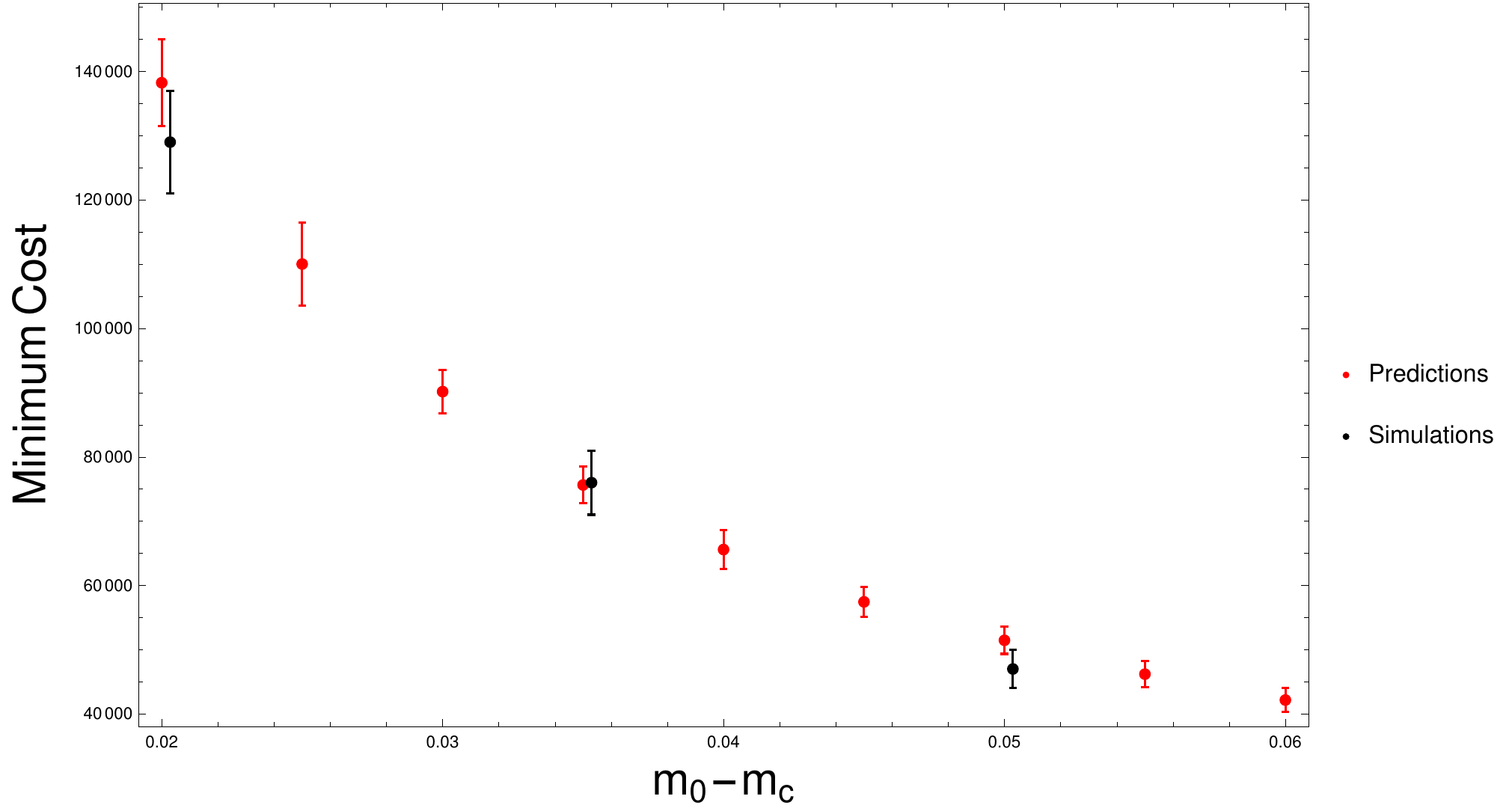}
\caption{\emph{Cost at the minimum, comparison with simulation. Each point correspond to a dif\mbox{}ferent set of parameters $\mu, n, m$. In black comparison with direct simulations is made.}}
\label{fig:minimum_cost_sim}
\end{center}
\end{figure}

We have run simulations around the minima found with the above methodology for each bare mass $m_0$, Tab.~\ref{tab:run_sim_cost}. 
The results for the cost from simulations and predictions are shown in 
Fig.~\ref{fig:minimum_cost_sim} for the dif\mbox{}ferent $m_0 - m_{\rm c}$.
Each prediction point represents the minimum cost with dif\mbox{}ferent 
bare mass $m_0$, mass-preconditioning $\mu$, 
number of steps in the uppermost level $n$ and middle level $m$ ($k$ is fixed to 10).
The result in Fig.~\ref{fig:minimum_cost_sim} represent the central result of this work.
It tests the validity of the assumptions made so far and it nicely predicts 
the value of the cost within factors 3 to 4.

\begin{table}[htb]
\begin{center}
\begin{tabular}{| c | c | c | c | c |} 
 \hline
 $m_0$ & $\mu$ & $n$ & $m$ & $N_{\rm cnf}$ \\ [0.5ex] 
 \hline
 -0.72 & 0.29 & 5 & 3 & 463  \\ 
 \hline
 -0.735 & 0.22 & 5 & 3 & 403  \\ 
 \hline
 -0.75 & 0.18 & 5 & 4 & 167 \\  [1ex]
 \hline
 \end{tabular}
\caption{\emph{Parameters corresponding to the minimum cost of simulations and number of configurations produced.}}
\label{tab:run_sim_cost}
\end{center}
\end{table}

%%%%%%%%%%%%%%%%%%%%%%%%%%%%%%%%%%%%%%%%%%%%%%%%%%%%%%%%%%%%%%%%%%%%%%%%%%
%%%%%%%%%%%%%%%%%%%%%%%%%%%%%%%%%%%%%%%%%%%%%%%%%%%%%%%%%%%%%%%%%%%%%%%%%%

\section{Conclusions}

We presented a strategy to optimize the parameters of the Omelyan integrator with $\alpha = 1/6$,
Hasenbusch mass preconditioning and three time-scales. 
We have studied the scaling behavior of $\Delta H$ in $\delta\tau$ and the goodness 
of the Poisson brackets measurements through the driving forces in a 
HMC simulation. 
We have also generalized the shadow hamiltonian in the presence of multiple-time scales splitting.

We have shown that once the variances of the forces are measured, at 
fixed bare mass, as a function of the mass-preconditioning parameter 
the optimal set-up for the integrator can be found.
Our method relies on the existence of a shadow Hamiltonian.\\
We found that the minimum of the cost is quite broad as a function of 
the mass-preconditioning mass and the number of steps in a level.
Nonetheless we are able to give reliable estimates of the 
cost in agreement with actual simulations. 
An important observation is that we do not need to modify extensively 
preexisting codes, since practically all of them have to measure the forces during 
a trajectory. 

The schematic recipe followed by this work is the following:
\begin{itemize}
\item Start with a \emph{reasonable} choice for the simulation of $(n, m, k)$ at fixed $m_0$.
\item Measure the forces in each level, which we already computed for the evolution, and calculate $|\mathcal{F}_i|^2 = \sum_{x,\mu,a}T_{R,i} \left(F_i^{a\mu}(x)\right)^2$ and its variance ${\rm Var} (|\mathcal{F}_i|^2)$.
\item Measure the number of MVMs in each level.
\item By fitting the dependence in $\mu$ we are able to predict the cost dependence
on $(n, m, k, \mu)$ (and $m_0$) with accuracy within 10\%.
\end{itemize}
Generalizing it to a larger number of Hasenbusch levels on dif\mbox{}ferent quark determinant splitting \cite{Luscher:2005rx, Ce:2016ajy} is rather straightforward, especially as long as covariances can be neglected.
The results are encouraging and we plan to consider dif\mbox{}ferent strongly interacting BSM models.

\afterpage{\blankpage}

\setcounter{equation}{0}

\chapter{QED leading corrections to hadronic observables: the muon anomaly}
\label{chap:qedlc_ho_muon}

The $(g-2)_\mu$ is one of the most precise measurement in particle physics and it serves as a test of the Standard Model (SM).
The anomalous magnetic moment of a lepton $\ell$ mediates helicity flip transitions, which implies that quantum corrections 
are proportional to $\delta a_\ell\propto m_\ell^2 / M^2$, with $M$ being the mass of a heavy particle inside or outside the SM.
This tells us that the anomalous magnetic moment can be used for indirect search of New Physics (NP) beyond the SM.
The sensitivity to new particles grows quadratically with the mass of the lepton of interest. 
The muon is then a good candidate for the search of NP, since the $\tau$ is ruled out because of its short lifetime.
Notice that the above argument can be also used to see that there is an enhanced sensitivity to the hadronic contributions
that are notoriously dif\mbox{}ficult to extract.

The persistent 3-4 $\sigma$ tension between the experimental value and theoretical calculation has generated a lot of
interests in the past years.
In Tab.~\ref{table:a_mu} we give the results for the experimental measure of $a_\mu$ as well as the dif\mbox{}ferent theoretical
contributions in the SM according to Ref.~\cite{Amsler:2008zzb}.
The QED value is calculated with a 5-loop computation, the EW with a 2-loop computation, the HVP at LO and NLO is calculated through
the measurement of the cross section of $e^+ e^-$ into hadrons, $\sigma \left( e^+ e^- \rightarrow {\rm Hadrons}\right)$, and the 
HLbL (light by light scattering) through large $N_c$ arguments.
At this order the contributions to the anomalous magnetic moment of the muon can be separated as
\begin{align*}
a_\mu^{\rm SM} = a_\mu^{\rm QED} + a_\mu^{\rm EW} + a_\mu^{\rm HAD}.
\end{align*}
For a review on the topic see Ref.~\cite{Jegerlehner:2009ry}.
\begin{table}[htb]
\begin{center}
{\renewcommand{\arraystretch}{1.2}%
\begin{tabular}{| c | c | c |} 
 \hline
	Contribution & $a_\mu\times 10^{10}$ & $\delta a_\mu\times 10^{10}$ \\ [1ex] 
 \hline
 $a_\mu^{\rm QED}$ (5-loop) & 11658471.895 & 0.008   \\ 
  $a_\mu^{\rm EW}$ (2-loop) & 15.36 & 0.1   \\ 
  $a_{\mu, {\rm LO}}^{\rm HAD}$ & 692 & 4   \\  
  $a_{\mu, {\rm NLO}}^{\rm HAD}$ & -9.84 & 0.06   \\ 
  $a_{\mu, {\rm NLO}}^{\rm HLbL}$ & 10.5 & 2.6   \\   [1ex]
 \hline
 $a_\mu^{\rm SM}$  & 11659180 & 5   \\ 
 \hline
 $a_\mu^{\rm Exp}$  & 11659209 & 6   \\ 
 \hline
 $\Delta a_\mu$  & 29 &  8  \\ 
  \hline
 \end{tabular}}
\caption{\emph{Dif\mbox{}ferent SM contributions to the theoretical value of $a_\mu$.}}
\label{table:a_mu}
\end{center}
\end{table}
We can clearly see that the dominant contribution to the anomalous magnetic moment is due to QED and at the level we are today in the experiment 
we need to include all the possible contributions from the Standard Model.

Recently $Z'$ models were introduced to explain the violations of lepton flavor universality because of tensions in the measurements of 
$R_K = \mathcal{B} \left( B\rightarrow K \mu^+ \mu^- \right) / \mathcal{B} \left( B\rightarrow K e^+ e^- \right)$ and 
$R_{K^\star}$ compared to the SM expectations.
In Ref.~\cite{DiChiara:2017cjq} the authors argue that a vectorial coupling to the $Z'$ could also alleviate the discrepancy between the $g-2$ measurement
and the SM prediction.
Every extension beyond the SM has to deal with the $(g-2)_\mu$ anomaly to survive, hence its crucial importance.

\begin{figure}[!ht]
\begin{center}
\subfigure[\emph{Light $u$-$d$ contribution to the hadronic $a_\mu$.}]{\includegraphics[scale=0.4,angle=-0]{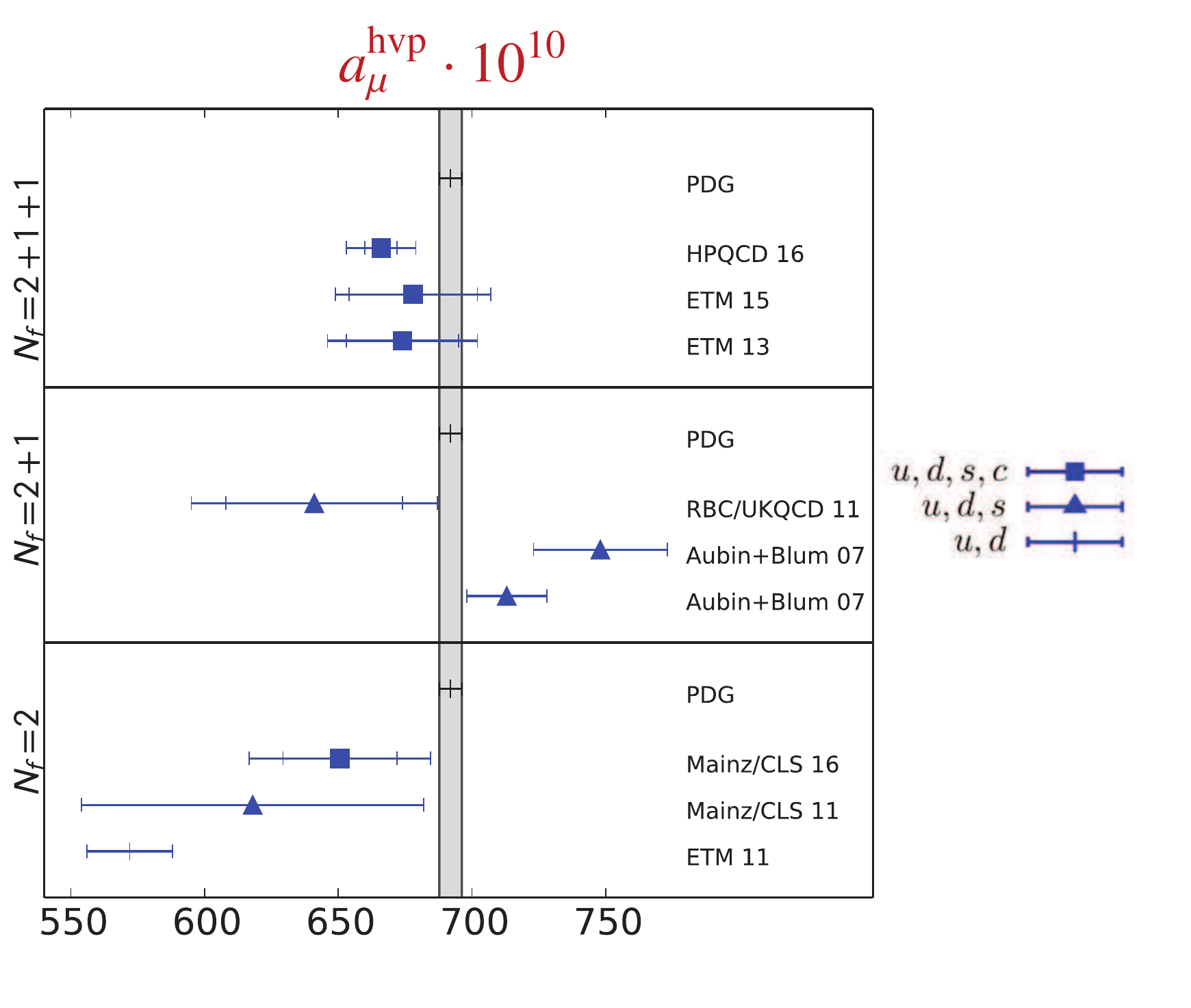}}
\subfigure[\emph{Strange $s$ contribution to the hadronic $a_\mu$.}]{\includegraphics[scale=0.5,angle=-0]{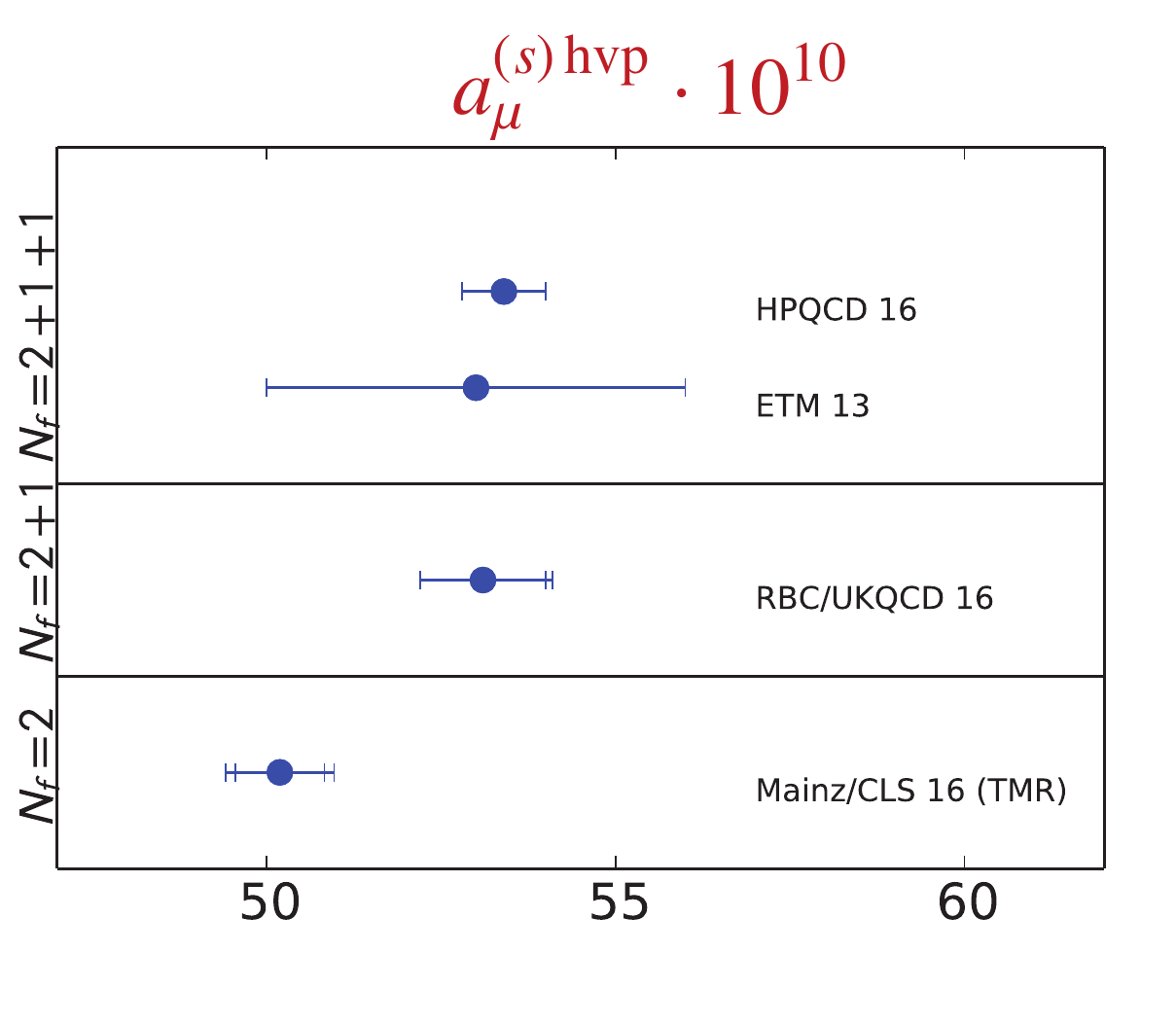}}
\subfigure[\emph{Charm $c$ contribution to the hadronic $a_\mu$.}]{\includegraphics[scale=0.5,angle=-0]{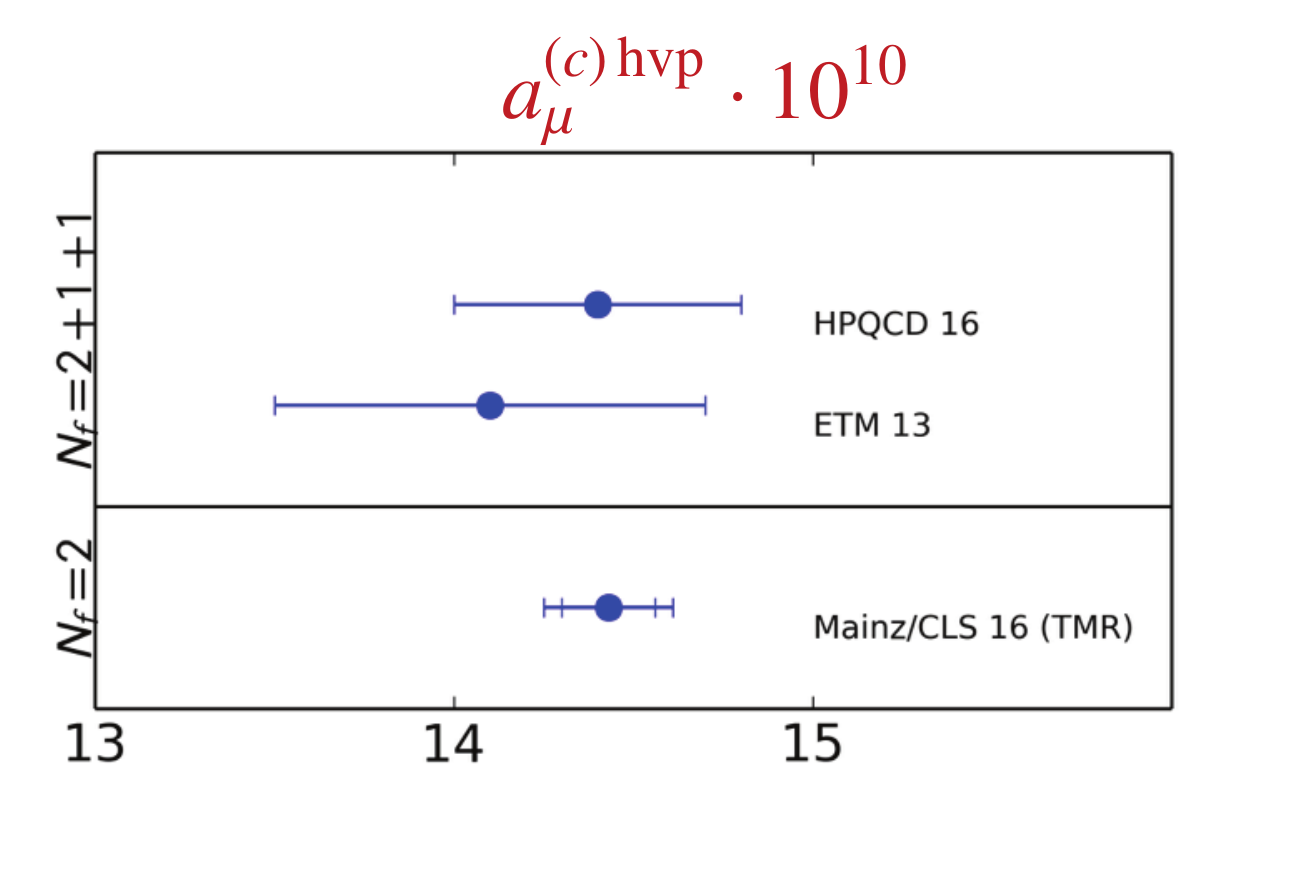}}
\caption{\emph{Summary of the hadronic contributions to $a_\mu$. The light contribution makes the 90\% of the total, while the strange and the charm account for 8\% and 2\% respectively. Plots taken from the plenary talk given by H.~Wittig at 2016 the Lattice Conference.}}
\label{fig:gmuon_summary}
\end{center}
\end{figure}

The lattice regularization can give a non-perturbative answer to the determination of the hadronic vacuum polarization contribution to the muon magnetic anomaly,
that is the one dominating the error and represents the second most important contribution. 
Especially in view of the new planned experiments E989 at FNAL and E34 at J-PARC that will improve the determination of $a_\mu$ by a factor four on the 
experimental side.\\
In the last decade the lattice community has made a huge ef\mbox{}fort to determine the hadronic contribution to $(g-2)_\mu$.
A first attempt was made in \cite{Blum:2002ii} and a complete quenched study with Wilson fermions was made in \cite{Gockeler:2003cw}.\\
A study with two degenerate Wilson fermions in QCD can be found in \cite{DellaMorte:2017dyu}, there a variety of techniques
are employed as Pad\'e fits \cite{Aubin:2012me}, time-momentum representation \cite{Bernecker:2011gh} and time-moments \cite{Chakraborty:2015ugp}.
See Ref.~\cite{Blum:2016xpd} for an analogous use of those techniques  for the calculation of the leading strange quark-connected contribution to
$(g-2)_\mu$.
In Fig.~\ref{fig:gmuon_summary} we give a summary on the status of $a^{\rm HAD}_\mu$ with lattice calculations.
For a review on the subject see Ref.~\cite{Blum:2013qu}.
%%%%%%%%%%%%%%%%%%%%%%%%%%%%%%%%%%%%%%%%%%%%%%%%%%%%%%%%%%%%%%%%%%%%%%%%%%
%%%%%%%%%%%%%%%%%%%%%%%%%%%%%%%%%%%%%%%%%%%%%%%%%%%%%%%%%%%%%%%%%%%%%%%%%%

Recently the inclusion of QED has been considered for the HVP \cite{Boyle:2016lbc}.
Taking into account QED ef\mbox{}fects is important if the lattice estimate has to be competitive with the $e^+ e^-$ cross section method, since the latter 
already include all the contributions from dif\mbox{}ferent sectors of the SM.
Furthermore in \cite{Calame:2015fva} an alternative method to measure the hadronic contribution using experimental data
employing a space-like kinematics is proposed, which allows for a direct comparison with lattice estimates.\\
Here we present an application of the concepts introduced in Chap.~\ref{qedothl}.
We present preliminary results on the Hadronic Vacuum Polarization (HVP) of the photon, required to compute
the leading correction to the anomalous magnetic moment of the muon.

The chapter is organized as follows; in Section~1 we review the muon anomaly in Minkowski space in QFT.
Section~2 is devoted to the details on the chosen lattice set-up to extract electromagnetic corrections to the
vacuum polarization.
There we discuss the electromagnetic shift in the critical mass, and present preliminary results on the 
pseudoscalar meson masses.
We also correct the masses for finite volume and photon mass ef\mbox{}fects.
In Section~3 we present the Vacuum Polarization and its form in lattice regularization with Wilson fermions.
A variety of techniques are presented, as Pad\'e fits and time moments (Section~4).
Section~5 collects the central results of the work and we give preliminary estimates on the 
electromagnetic corrections to the muon anomaly.
Section~6 contains our conclusion.

\section{Introduction}

At the classical level an orbiting particle with electromagnetic charge $e$ and mass $m$ has a dipole magnetic moment given 
by $\underline{\mu}_L = \frac{e}{2m} \underline{L}$,
where $\underline{L} = m \underline{r}\times \underline{v}$ is the orbital angular momentum, and through that interacts with an external magnetic field
in such a way that it is energetically favorable for the particle to align its magnetic dipole with it, i.e.
\begin{align}
H = - \underline{\mu}\cdot \underline{B}.
\end{align}
For a particle $\ell$ with spin $\underline{S}$ and electromagnetic charge $Q $, in units of $e$, the magnetic moment is given by
\begin{align}
\underline{\mu}_\ell = g_\ell \frac{Qe}{2m} \underline{S},
\end{align}
At classical level the Dirac value is exactly $g_\ell = 2$.\\
When considering loop corrections the gyro-magnetic ratio gets quantum contributions. 
We define the deviation from its classical value as the \emph{anomalous magnetic moment}
\begin{align}
a_\ell = \frac{g_\ell - 2}{2}.
\end{align}
The first calculation of the leading contribution to the anomalous magnetic moment was done by \cite{Schwinger:1948iu} and the 
ef\mbox{}fect of the quantum fluctuation due to virtual lepton-photon interaction (vertex) was found to be 
\begin{align*}
a_\ell^{\text{QED}} = \frac{\alpha}{2\pi},\,\text{ where } \ell = e, \mu, \tau.
\end{align*}
We start the discussion in Minkowski space.
At tree-level the lepton-photon vertex is given by following expression
\begin{align*}
&\begin{tikzpicture}[line width=1.5 pt, scale=2]
	\draw[fermion] (0,0.5) -- (0.5,0);
	\draw[fermionbar] (0,-0.5) -- (0.5,0);
	\draw[vector] (0.5,0) -- (1,0);
	% The labels
	\node at (1.2,0){$\gamma$};
	\node at (1.2,-0.2){$k,\mu$};
	\node at (-0.15,0.65){$p$};
	\node at (-0.15,-0.65){$p+k$};
	\node at (1.8,0){$\equiv -i e \gamma_\mu .$};
\end{tikzpicture}
\end{align*}
After quantization the vertex receives corrections and can be written as
\begin{align*}
\begin{tikzpicture}[line width=1.5 pt, scale=2]
	\draw[fermion] (0,0.5) -- (0.4,0.155);
	\draw[fermionbar] (0,-0.5) -- (0.4,-0.155);
	\draw[vector] (0.65,0) -- (1,0);
	% The labels
	\node at (1.2,0){$\gamma$};
	\node at (1.2,-0.2){$k,\mu$};
	\node at (-0.15,0.65){$p$};
	\node at (-0.15,-0.65){$p+k$};
	%\node[circle,fill=black,inner sep=0pt,minimum size=10pt] (a) at (0.5,0) {};
	\draw[pattern=north west lines,draw=black, anchor=base,baseline] (0.63,0) arc (0:360:.15);
	\node at (2.25,0){$\equiv -i e \Gamma^\mu(p, p'=p+k) .$};
\end{tikzpicture}
\end{align*}
The form of $\Gamma^\mu(p, p')$ can be constrained by means of Lorentz invariance and current conservation, i.e.~$k_\mu \Gamma^\mu = 0$, yielding to
\begin{align}
\Gamma_\mu(p, p') =  A \gamma_\mu +B \left( p_\mu + p'_\mu \right),
\end{align}
where $A$ and $B$ are real coef\mbox{}ficients.
With the use of the Gordon identity
\begin{align}
\overline{u}(p') \gamma_\mu u(p) 
& = \overline{u}(p') \frac{1}{2m} \left[  \left(p_\mu + p'_\mu \right) + i \left[\gamma_\mu, \gamma_\nu\right] \left(p^\nu - p'^\nu \right)  \right] u(p) ,
\end{align}
we arrive to 
\begin{align}
\Gamma_\mu(p, p') =  \gamma_\mu F_1(k^2) +  i \left[\gamma_\mu, \gamma_\nu\right]  \frac{k^\nu}{2m} F_2(k^2),
\end{align}
with $F_1$ and $F_2$ form factors.\\
The vertex renormalization in QED gives
\begin{align}
-i e \Gamma_\mu ( p, p'=p) = -i e \gamma_\mu,
\end{align}
from which we infer that $F_1(k^2=0) = 1$, i.e.~the charge of the lepton in units of $e$, to \emph{all orders} in perturbation theory.\\
By considering a non-relativistic electron scattered by a static potential one can show that the gyromagnetic ratio is given by
\begin{align}
g_\ell = 2 + 2 F_2(k^2=0),
\end{align}
and at leading order in QED $F_2(0) = 0$. 
This means that the gyromagnetic ratio becomes
\begin{align}
g_\ell = 2 + \ord(\alpha) \rightarrow a_\ell = \frac{g_\ell - 2 }{2} = \ord(\alpha).
\end{align}

\section{Details on the computation of the vacuum polarization including QED corrections}
\label{sect:dcVPQEDc}

\begin{figure}[htb!]
\begin{center}
\subfigure[]{
\begin{tikzpicture}[line width=1.5 pt, scale=2]
	\draw[fermion] (0.5,0) arc (0:180:.5);
	\draw[fermion] (-0.5,0) arc (-180:0:.5);
	\draw[vector] (0.45,0.2) -- (-0.45, -0.2);
	% The labels
	\node at (-0.5,0) [minimum size=0.1cm,draw] {};
	\node at (0.5,0) [minimum size=0.1cm,draw] {};
\end{tikzpicture}
}
\hspace{0.5cm}
\subfigure[]{
\begin{tikzpicture}[line width=1.5 pt, scale=2]
	\draw[fermion] (0.5,0) arc (0:180:.5);
	\draw[fermion] (-0.5,0) arc (-180:0:.5);
	\draw[vector] (0.45,0.2) -- (-0.45, 0.2);
	% The labels
	\node at (-0.5,0) [minimum size=0.1cm,draw] {};
	\node at (0.5,0) [minimum size=0.1cm,draw] {};
\end{tikzpicture}
}
\hspace{0.5cm}
\subfigure[\emph{Quark-disconnected diagram, flavor-symmetry and Zweig suppressed.}]{
\begin{tikzpicture}[line width=1.5 pt, scale=2]
	\draw[fermion] (0.5,0) arc (0:180:.5);
	\draw[fermion] (-0.5,0) arc (-180:0:.5);
	\draw[vector] (0.5,0)  -- (1,0);
	\draw[fermion] (2,0) arc (0:180:.5);
	\draw[fermion] (1,0) arc (-180:0:.5);
	%The labels
	\node at (-0.5,0) [minimum size=0.1cm,draw] {};
	%\node at (0.5,0) [minimum size=0.1cm,draw] {};
	\node at (2,0) [minimum size=0.1cm,draw] {};
\end{tikzpicture}
}
\hspace{0.5cm}
\subfigure[\emph{Absent diagram due to electroquenched approximation.}]{
\begin{tikzpicture}[line width=1.5 pt, scale=2]
	\draw[fermion] (0.5,0) arc (0:180:.5);
	\draw[fermion] (-0.5,0) arc (-180:0:.5);
	\draw[vector] (-0.2,0.45) -- (0,0.2);
	\draw[fermion] (0.2,0) arc (0:360:.2);
	%The labels
	\node at (-0.5,0) [minimum size=0.1cm,draw] {};
	\node at (0.5,0) [minimum size=0.1cm,draw] {};
	\node at (1,0) {};
		\node at (-1,0) {};
\end{tikzpicture}
}
\caption{\emph{Leading order QED corrections to the HVP in the continuum theory.
Squares corresponds to insertions of conserved current.
All orders of QCD are understood when drawing the diagrams.
Diagrams (a) and (b) are included in our calculation, while (c) and (d) are absent.
}}
\label{fig:qedhvp}
\end{center}
\end{figure}
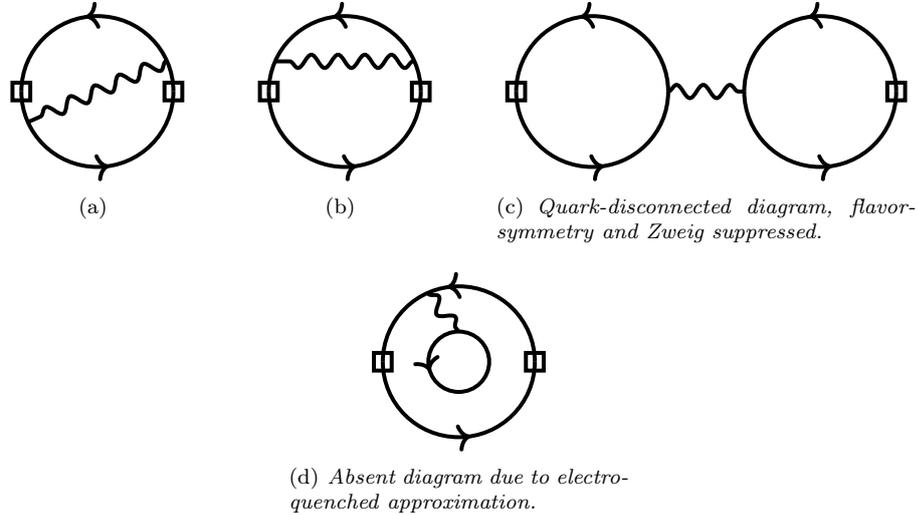

We turn to the actual calculation of the HVP on the lattice.
We work in the electroquenched approximation and use dynamical QCD configurations generated by the CLS initiative 
with two degenerate flavors of non-perturbatively $\ord(a)$ improved Wilson fermions \cite{Fritzsch:2012wq}.
We consider QED$_{\rm L}$ and QED$_{\rm M}$ to deal with the finite-volume zero modes, see Chap.~\ref{qedothl}.
We fix the gauge to the Coulomb one in QED$_{\rm L}$. While in the massive case we fix it to Feynman.\\
In Fig.~\ref{fig:qedhvp} we report the relevant diagrams for the leading order QED corrections to the HVP
(only \quotes{continuum} diagrams are shown).
Notice that those diagrams are at the same order in $\alpha$ as the HLbL, that is not included. 
A computation including QED contribution must be performed to see whether those ef\mbox{}fects increase or decrease the discrepancy 
between the theoretical and the experimental muon anomaly.
In this exploratory study we neglect quark-disconnected diagrams, which are flavor-symmetry and Zweig suppressed, 
and all diagrams involving charged sea quarks (electroquenched approx.).
We would like to compare the HVP with and without electromagnetic ef\mbox{}fects.
The two HVPs will be dif\mbox{}ferent functions of the renormalized parameters, and to have a meaningful
comparison we need to consider them at the same renormalized parameters values.
One way to go could be to rescale the bare values for the change in the
quark mass, the strong coupling (reflected in the change of the lattice spacing) and
the electromagnetic coupling, when considering the electromagnetic corrections to the HVP.
We expect all those changes to be at the percent level. 
For this reason we neglect the change in the lattice spacing and the change in the electromagnetic coupling.
A percent change in the mass for quite light quark masses will cause a large change in the pion masses, 
since we are working with Wilson fermions\footnote{There is a multiplicative plus additive renormalization in the mass.}
 (see Sect.~\ref{sect:crt_mass}).
We take into account this ef\mbox{}fect by matching the charged (or neutral) pion masses in Q(C+E)D 
with the pion mass in QCD as we will see in Sect.~\ref{sect:ps_spect}.\\
In addition we make sure that the volumes and photon masses used in QED$_{\rm M}$ are such that the correct dispersion 
relation is reproduced by the energy levels extracted from the charged pions two-point functions.

It should be noted that the leading electromagnetic order corrections to the HVP can be extracted 
also in the perturbative way as in the spirit of \cite{deDivitiis:2013xla}.

\subsection{Critical mass}
\label{sect:crt_mass}

We start by first calculating the ef\mbox{}fect of the QED inclusion on the critical mass.
This will serve as a cross-check of our implementations of QED$_{\rm L}$ and QED$_{\rm M}$.
We add to preexistent QCD configurations 
the quenched QED configurations by forming 
a $\mathbf{U}(3)$ gauge theory with un-improved fermions.\\
By using lattice perturbation theory one can estimate the QED ef\mbox{}fect on the quark mass. 
We take here the results given in the App.~\ref{app:em_cr_mass_shift} for the QED case.
We calculate the expected shift in the A ensembles with parameters given in Tab.~\ref{table:par}.
\begin{table}[h!]
\begin{center}
 \begin{tabular}{|c |c| c| c| c| c| c| c| c| c|} 
 \hline
  Run & $L/a$ & $\beta$ & $c_{sw}$ & $\kappa$ & $\kappa_c$ & $am_\pi$ & $m_\pi L$ & $a$[fm] & $m_\pi$[MeV] \\ [0.5ex] 
 \hline
  A3 & 32 & 5.20 & 2.01715 & 0.13580 & 0.1360546 & 0.1893(6) & 6.0 & 0.079(3)(2) & 473\\
  A4 & 32 & 5.20 & 2.01715 & 0.13590 & 0.1360546 & 0.1459(6) & 4.7 & 0.079(3)(2) & 364\\
  A5 & 32 & 5.20 & 2.01715 & 0.13594 & 0.1360546 & 0.1265(8) & 4.0 & 0.079(3)(2) & 316\\[1ex]
 \hline
\end{tabular}
\end{center}
\caption{\emph{Gauge configuration parameters and results in QCD, see Ref.~\protect\cite{Capitani:2015sba}.}}
\label{table:par}
\end{table}
In particular by using the approximations given in the App.~\ref{app:em_cr_mass_shift} we arrive to the critical mass calculated 
in the two dif\mbox{}ferent approximations: Eq.~\ref{eq:one-loop-m-c}, where 1-loop QED diagrams are taken only, 
and Eq.~\ref{eq:daisy-m-c}, where the QED tadpole is resummed to all orders. 

\begin{figure}[h!t]
\begin{center}
\includegraphics[scale=0.8]{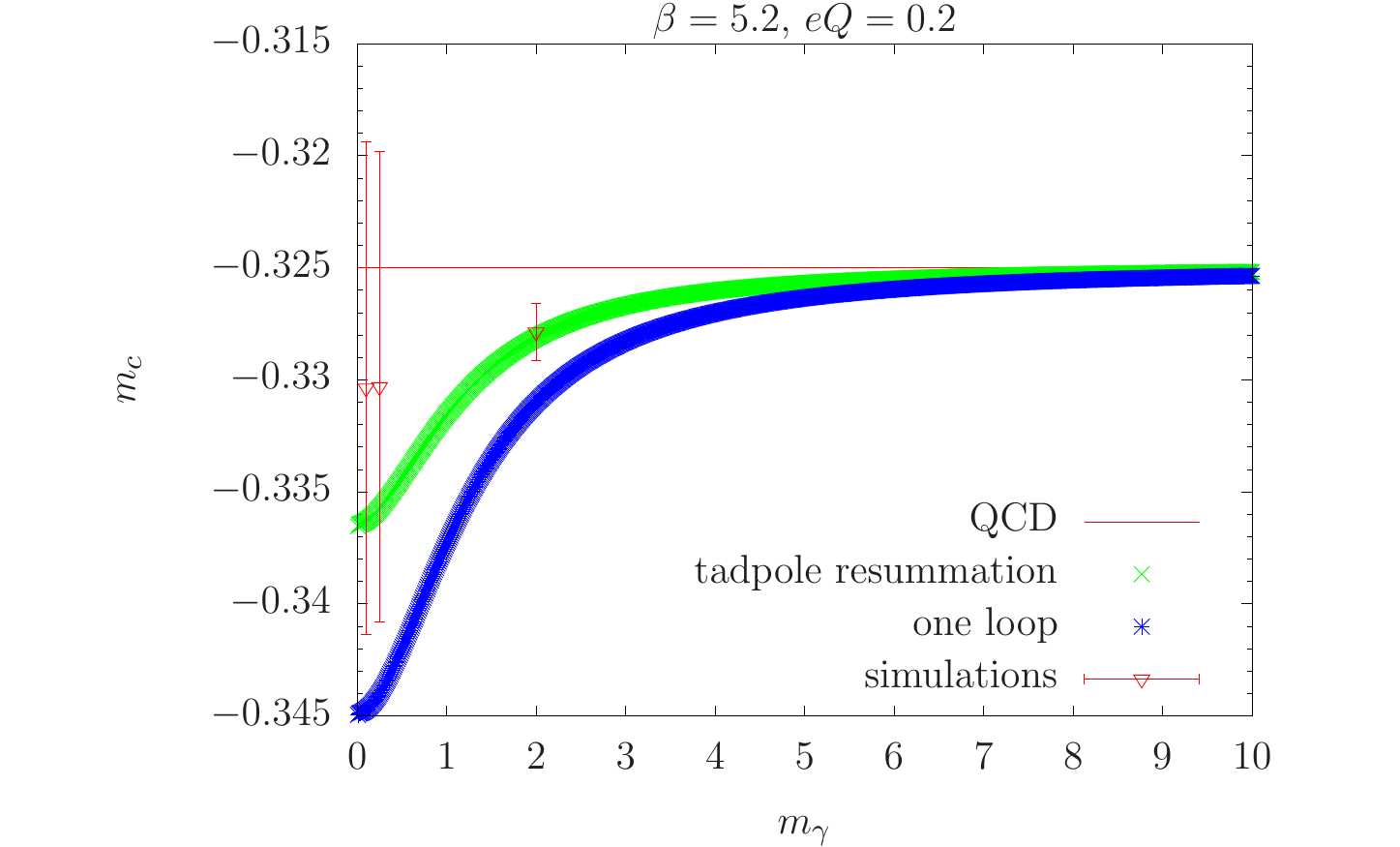}
\caption{\emph{Comparison between the critical mass from lattice perturbation theory in Feynman gauge using the one-loop result Eq.~\ref{eq:one-loop-m-c} (blue stars)
 and the tadpole resummation Eq.~\ref{eq:daisy-m-c} (green crosses). Orange triangles represents values from lattice simulations. The QCD critical mass is shown for comparison (red line).}}
\label{fig:critical_mass}
\end{center}
\end{figure}

In order to do the comparison with the simulations we need to find an operative definition of the critical mass $m_c$ after the inclusion of QED. 
We consider the unphysical theory with two quarks degenerate in mass and charge $\Psi = (u, d)$.
We use the PCAC mass to find the critical mass, since $m_{\rm PCAC} = Z(\beta, e^2) \left( m_0 - m_c(\beta, e^2)\right)$. 
The PCAC quark mass can be defined through the spatially integrated axial Ward identity, as in Eq.~\ref{eq:pcac_bla},
\begin{align}
\label{eq:integrated_pcac}
m_{\rm PCAC} = \frac{\partial_0\langle A^-_0(\underline{p}=\underline{0}, x_0) P^+(0)\rangle}{\langle P^-(\underline{p}=\underline{0}, x_0) P^+(0)\rangle}\;,
\end{align}
where we have the following definitions,
\begin{align}
\nn
P^-(x) &= \Psibar(x) \gamma_5 \frac{1}{2}\left(\tau^1 + i \tau^2\right)\Psi(x) = \overline{u}(x) \gamma_5 d(x), \\
\nn
P^+(x) &= \Psibar(x) \gamma_5 \frac{1}{2}\left(\tau^1 - i \tau^2\right)\Psi(x) = \overline{d}(x) \gamma_5 u(x), \\
A^-_0(x) &= \Psibar(x) \gamma_0\gamma_5 \frac{1}{2}\left(\tau^1 + i \tau^2\right)\Psi(x) = \overline{u}(x) \gamma_0\gamma_5 d(x).
\end{align}
By using Eq.~\ref{eq:integrated_pcac} we extract the PCAC mass for each value of $m_\gamma$. 
At fixed photon mass we then perform an extrapolation, guided by chiral 
perturbation theory, to find the critical mass.\\
In Fig.~\ref{fig:critical_mass} we show a preliminary result for the critical mass in QCD+qQED$_{\rm M}$ simulation, or Q(C+E$_{\rm M}$)D in the following.
The picture emerging is quite clear, once we include QED the quarks become more massive, and the results are compatible with the 
finite-volume lattice perturbation theory predictions.
%%%%%%%%%%%%%%%%%%%%%%%%%%%%%%%%%%%%%%%%%%%%%%%%%%%%%%%%%%%%%%%%%%%%%%%%%%
%%%%%%%%%%%%%%%%%%%%%%%%%%%%%%%%%%%%%%%%%%%%%%%%%%%%%%%%%%%%%%%%%%%%%%%%%%

\subsection{Pseudoscalar spectrum}
\label{sect:ps_spect}

\begin{figure}[!htb]
   \begin{minipage}{0.5\textwidth}
     \hspace{-2.5cm}
     \includegraphics[width=1.5\linewidth]{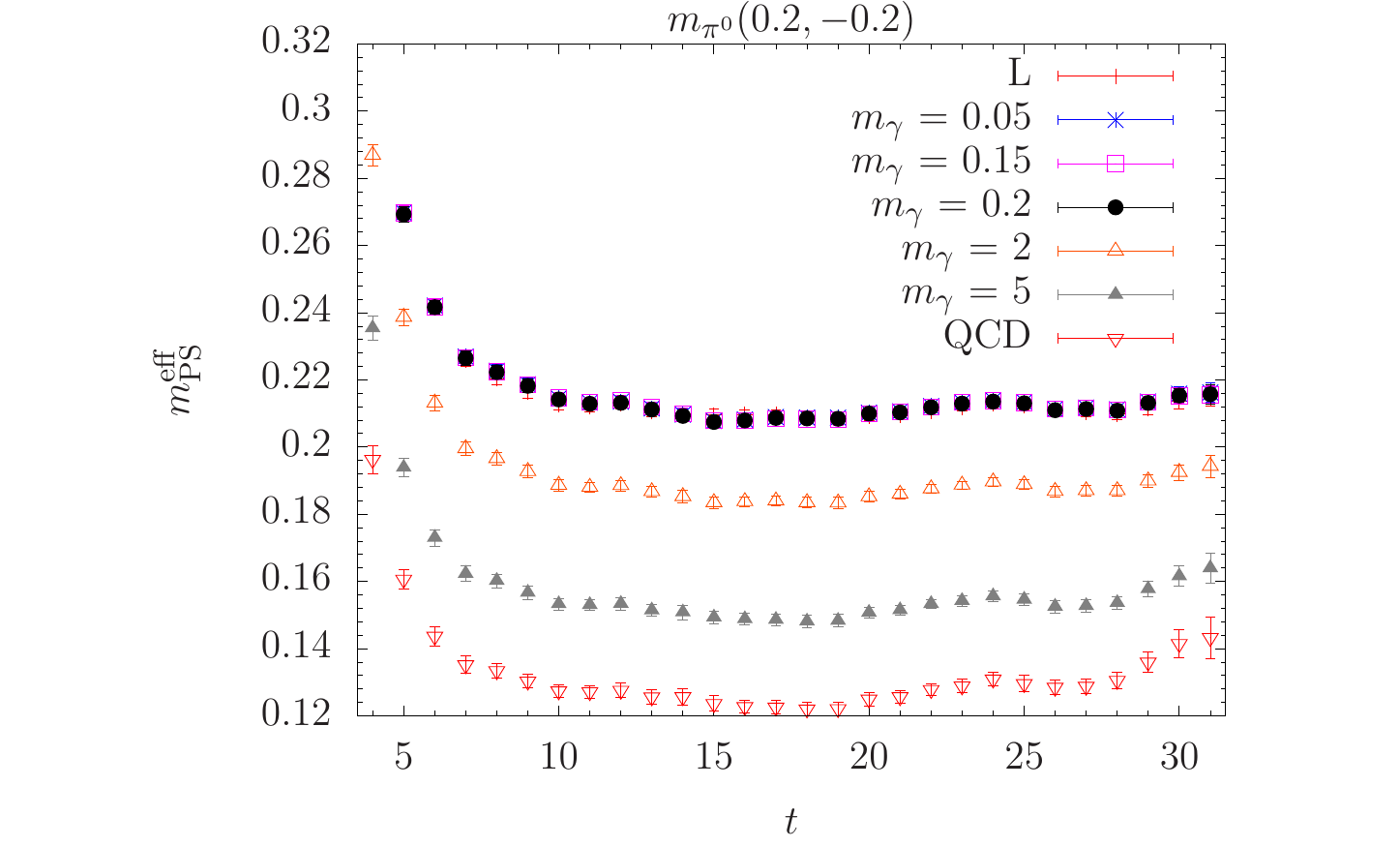}
     \caption{\emph{Ef\mbox{}fective neutral pseudoscalar masses for dif\mbox{}ferent photon masses and comparison with QCD and Q(C+E$_{\it L}$)D.}}\label{fig:mpsdiffgamma}
   \end{minipage}\hfill
   \hspace{0.5cm}
   \begin{minipage}{0.5\textwidth}
     %\centering
     \vspace{0.3cm}
     \hspace{-2.8cm}
     \includegraphics[width=1.5\linewidth]{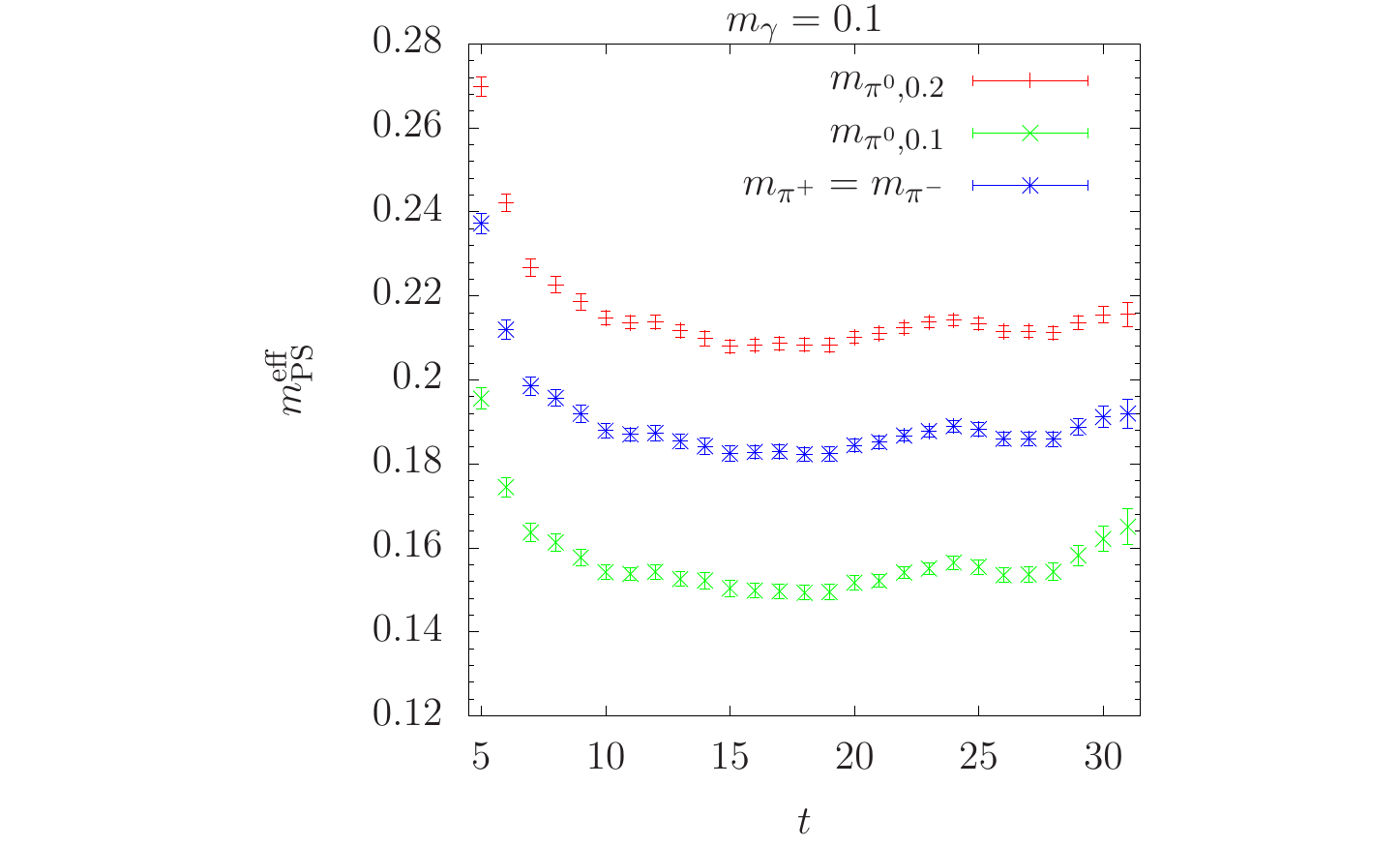}
     \captionsetup{width=0.8\textwidth}
     \caption{\emph{Ef\mbox{}fective pseudoscalar masses for $m_\gamma = 0.1$ and dif\mbox{}ferent charge content.}}\label{fig:mpsdiffcharge}
   \end{minipage}
\end{figure}

\begin{figure}[!htb]
   \begin{minipage}{0.5\textwidth}
     %\centering
     \hspace{-2.cm}
     \includegraphics[width=1.3\linewidth]{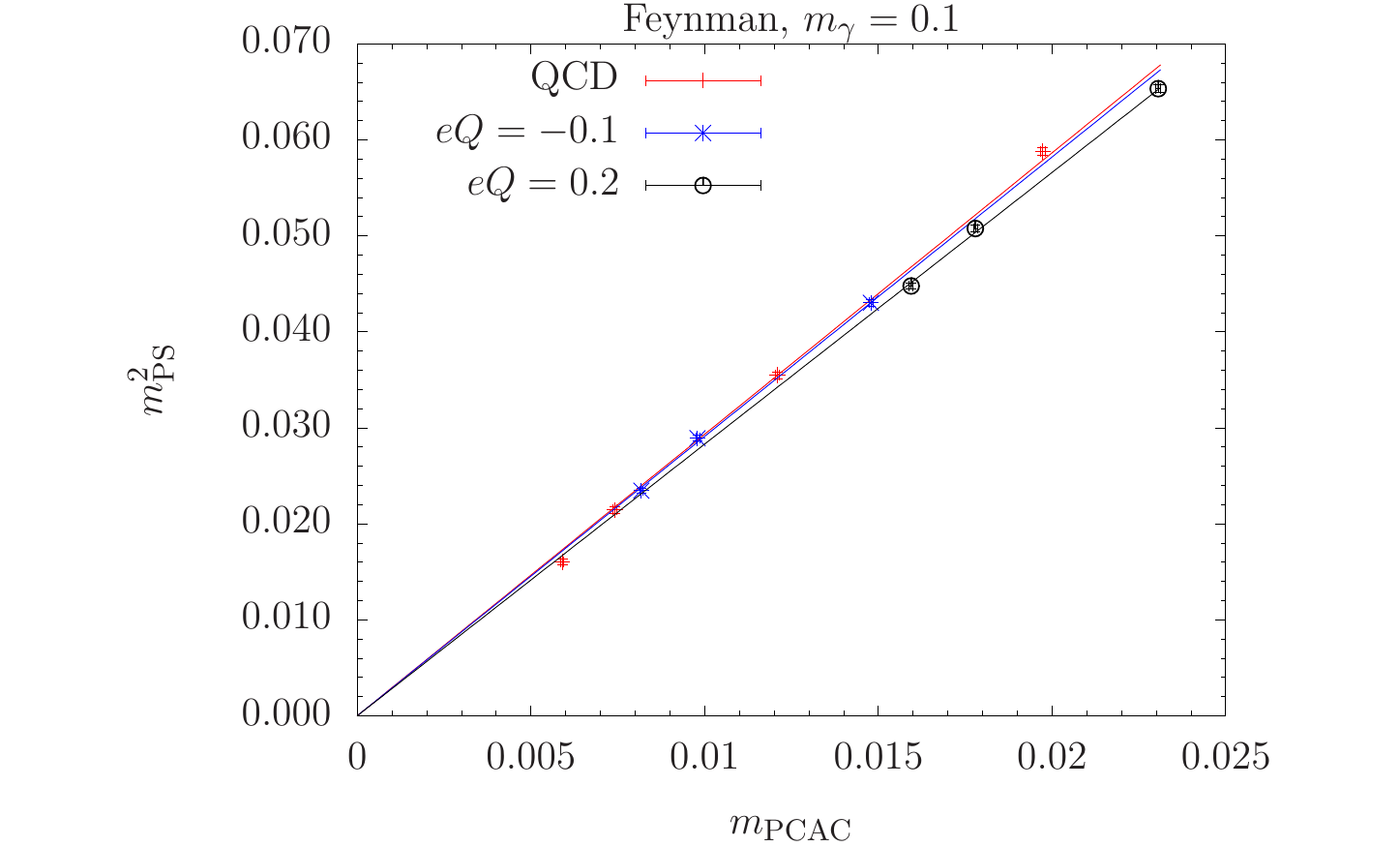}
     \caption{\emph{Neutral pseudoscalar masses versus PCAC mass for dif\mbox{}ferent charges.}}\label{fig:m2ps_pcac}
   \end{minipage}\hfill
   \hspace{0.7cm}
   \begin{minipage}{0.5\textwidth}
     \vspace{0.5cm}
     \hspace{-2.cm}
     \includegraphics[width=1.3\linewidth]{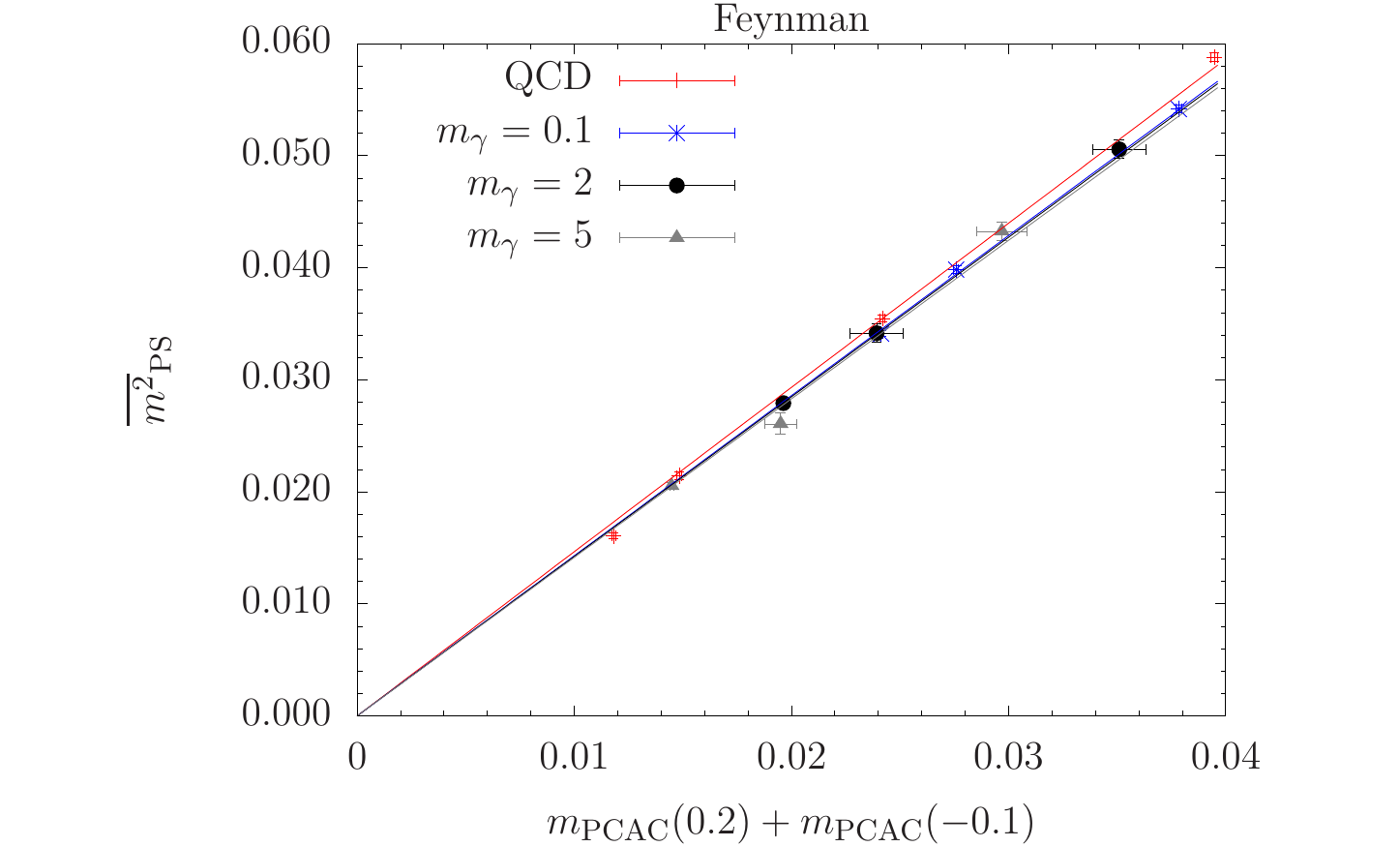}
     \captionsetup{width=0.8\textwidth}
     \caption{\emph{Square average of neutral pseudoscalar masess versus bare PCAC masess for dif\mbox{}ferent photon masses.}}\label{fig:m2ps_ave_pcac}
   \end{minipage}
\end{figure}

Here we give preliminary results on the masses in the pseudoscalar sector. 
The masses are calculated by $\cosh$ ef\mbox{}fective mass and fitted to a constant in the plateau region. 
We used one point source per configuration.
The error is given by using a jackknife procedure. We implemented the QED$_{\rm L}$ and QED$_{\rm M}$ at the physical value of the electric charge $e \simeq 0.3$, $Q_u=2/3$ and $Q_d=-1/3$.
The results are found in Tab.~\ref{table_ps_feyn}.
\begin{table}[h!]
\begin{center}

 \begin{tabular}{|c |c |c |c |c |c |c|} 
 \hline
  Run & $a m_\gamma$ & $m_\gamma L$ & $am_{\pi^0, 0.2}$ & $am_{\pi^0, 0.1}$ & $am_{\pi^-} = am_{\pi^+}$ & $N_{\rm cnf}$\\ [0.5ex] 
 \hline
  A3 & 0 & x & .2549(9) & .2071(9) & .2330(9) & 312 \\
  A3 & 0.1 & 3.2 & .2556(7) & .2074(8) & .2337(8) & 330 \\
  A3 & 0.25 &  8.0 & .2553(7) & .2072(8) & .2331(8) & 330 \\
  \hline
  A4 & 0 & x & .2240(8)  & .1691(9) & .1994(9) & 400 \\
  A4 & 0.1 &  3.2 & .2252(9) & .1699(9) & .2005(9) & 380 \\
  A4 & 0.25 &  8.0 & .2246(8) & .1700(10) & .1998(9) & 380 \\
  \hline
  A5 & 0 & x & .2105(7)  & .1526(9) & .1849(8) & 501 \\
  A5 & 0.1 &  3.2 & .2114(7) & .1528(9) & .1856(8) & 481 \\
  A5 & 0.25 &  8.0 & .2111(7) & .1531(9) & .1852(8) & 481 \\[1ex]
 \hline 
\end{tabular}
\caption{\emph{Pseudoscalar masses in Coulomb gauge QED$_{\it L}$ (denoted with $m_\gamma = 0 $) and QED$_{\it M}$. The resulting pion masses go from 380 MeV to 640 MeV.}}
\label{table_ps_feyn}
\end{center}
\end{table}
In Fig.~\ref{fig:mpsdiffgamma} we show the typical ef\mbox{}fect of the massive QED inclusion and its comparison with QED$_{\rm L}$. 
For such small photon masses the result is quite stable and the comparison is made with the QCD result. 
It is clear from the plot that the pseudoscalar masses become larger, as expected since the quark masses increase. 
In Fig.~\ref{fig:mpsdiffcharge} we show the change with the charge content.
%%%%%%%%%%%%%%%%%%%%%%%%%%%%%%%%%%%%%%%%%%%%%%%%%%%%%%%%%%%%%%%%%%%%%%%%%%
%%%%%%%%%%%%%%%%%%%%%%%%%%%%%%%%%%%%%%%%%%%%%%%%%%%%%%%%%%%%%%%%%%%%%%%%%%
For such chiral masses in QCD we know that the square of the pseudoscalar mass is linear as a function of the bare PCAC mass. We found the same linearity after the inclusion of the massive QED field (for QED case see \cite{Duncan:1996xy}).
In Fig.~\ref{fig:m2ps_pcac} we show the neutral pseudoscalar mass as a function of the PCAC mass for $m_\gamma = 0.1$.
Another example of that is given in Fig.~\ref{fig:m2ps_ave_pcac}, where we plot the average square masses of the $u\overline{u}$ and $d\overline{d}$ states, 
i.e.~$\overline{m^2}_{\rm PS} = \frac{1}{2}\left(m^2_{\rm PS}(eQ=0.2)+m^2_{\rm PS}(eQ=-0.1)\right)$, versus the sum of the two dif\mbox{}ferent PCAC masses, as suggested in \cite{Duncan:1996xy}\footnote{There it is pointed out that the linear average would introduce non analytic terms in the quark masses, while the square averaging respect the chiral symmetry expected from the full theory.}.
This means, upon identifying $eQ_1=0.2$ and $eQ_2=-0.1$, that we expect the following dependence:
\begin{align}
\overline{m^2}_{\rm PS} = A(eQ_1, eQ_2) + m_{\rm PCAC}(eQ_1) B(eQ_1, eQ_2) + m_{\rm PCAC}(eQ_2) B(eQ_2, eQ_1).
\end{align}
%%%%%%%%%%%%%%%%%%%%%%%%%%%%%%%%%%%%%%%%%%%%%%%%%%%%%%%%%%%%%%%%%%%%%%%%%%
%%%%%%%%%%%%%%%%%%%%%%%%%%%%%%%%%%%%%%%%%%%%%%%%%%%%%%%%%%%%%%%%%%%%%%%%%%

\subsection{Finite volume and photon mass ef\mbox{}fects}

We can take into account for finite volume and photon mass ef\mbox{}fects and correct the lattice results.
The relevant formulae are given in Ref.~\cite{Borsanyi:2014jba} for QED$_{\rm L}$, and Ref.~\cite{Endres:2015gda} for QED$_{\rm M}$.
We report them here for completeness and remind the reader that we are using lattice units. \\
Since the preliminary results concern only one volume, in QED$_{\rm L}$, we do not perform the fit but instead 
we give an estimate based on the lattice results
\begin{align}
\delta_V m^{2, {\rm LO}}_{\pi^\pm} = m^2_{\pi^\pm}(L, T \rightarrow\infty) - m^2_{\pi^\pm}(L, T) \simeq  \frac{Q^2 \alpha \kappa\,   m_{\pi^\pm}(L, T)  }{L}\, ,
\end{align}
with $\kappa =  2.837297(1)$. 
Results are given in Tab.~\ref{tab:finite_volume_qel} and we can conclude that finite volume ef\mbox{}fects are completely negligible.
\begin{table}[h!]
\begin{center}
{\renewcommand{\arraystretch}{1.3}%
 \begin{tabular}{|c |c |c |} 
 \hline
  Run & $m^2_{\pi^\pm}$ & $\delta_V m^{2,{\rm LO}}_{\pi^\pm}$ \\ [0.5ex] 
 \hline
  A3 & 0.0537(4) & 0.00015(1) \\
  \hline
  A4 & .0398(4) & 0.00013(1) \\
  \hline
  A5 & .0342(3) & 0.00012(1) \\[1ex]
 \hline
\end{tabular}}
\caption{\emph{QED$_{\rm L}$ leading order finite volume corrections.}}
\label{tab:finite_volume_qel}
\end{center}
\end{table}

In QED$_{\rm M}$ we use the analytic formulae for the extracted masses at each volume
\begin{align}
\nn
\delta_V m^{{\rm LO}}_{\pi^\pm} &= 2\pi Q^2 \alpha\, m_\gamma \left[ \mathcal{I}_1(m_\gamma L) - \frac{1}{(m_\gamma L)^3}\right],\\
\delta_V m^{{\rm NLO}}_{\pi^\pm} &= \pi Q^2 \alpha\, \frac{m^2_\gamma}{m_{\pi^\pm}} \left[ 2\mathcal{I}_{1/2}(m_\gamma L) + \mathcal{I}_{3/2}(m_\gamma L)\right],
\end{align}
where $\mathcal{I}_n(z)$ is defined as
\begin{align}
\mathcal{I}_n(z) = \frac{1}{2^{n+1/2}  \pi^{3/2} \Gamma(n) } 
\sum_{\underline{\nu}\neq\underline{0}} \frac{K_{3/2-n}(z|\underline{\nu}|)}{(z|\underline{\nu}|)^{3/2-n}},
\end{align}
with $\underline{\nu}\in \mathbb{Z}^3$ and $K_n$ modified Bessel function of the second kind.
At leading order we obtain the following corrections
\begin{align}
\nn
\delta_V m^{\rm LO}_{\pi^\pm}(m_\gamma = 0.1) &= -.000097(1) ,\\
\delta_V m^{\rm LO}_{\pi^\pm}(m_\gamma = 0.25) &= -.000023(1) ,
\end{align}
that are negligible at this stage. 
For completeness next-to-leading order results are given in Tab.~\ref{tab:finite_volume_qedm}.
\begin{table}[h!]
\begin{center}
{\renewcommand{\arraystretch}{1.2}%
 \begin{tabular}{|c |c |c |c |} 
 \hline
  Run & $m_\gamma$ & $m_{\pi^\pm}$ & $\delta_V m^{\rm NLO}_{\pi^\pm}$ \\ [0.5ex] 
 \hline
  A3 & 0.1 & .2337(8) & .000022(1)  \\
  A3 & 0.25 & .2331(8) &  3.7(1) $\times 10^{-7}$ \\
  \hline
  A4 & 0.1 & .2005(9) & .000026(1)\\
  A4 & 0.25 & .1998(9) & 4.3(1) $\times 10^{-7}$\\
  \hline
  A5 & 0.1 & .1856(8) &  .000028(1)\\
  A5 & 0.25 & .1852(8) &  4.7(1) $\times 10^{-7}$ \\[1ex]
 \hline
\end{tabular}}
\caption{\emph{QED$_{\rm M}$ next-to-leading order finite volume corrections.}}
\label{tab:finite_volume_qedm}
\end{center}
\end{table}
After we have corrected for the finite volume ef\mbox{}fects we need to remove the photon mass ones.
Here we should consider the fit at next-to-leading order but since we have only two masses we use the formula at leading order, given by
\begin{align}
\delta m^{{\rm LO}}_{\pi^\pm} = m_{\pi^\pm}(m_\gamma\rightarrow\infty) - m_{\pi^\pm}(m_\gamma) =  - \frac{\alpha}{2} Q^2 m_\gamma.
\end{align}
The results are the following
\begin{align}
\nn
\delta m^{\rm LO}_{\pi^\pm}(m_\gamma =0.1) &= .00036(1), \\
\delta m^{\rm LO}_{\pi^\pm}(m_\gamma =0.25) &= .00091(1) ,
\end{align}
we can conclude that photon mass ef\mbox{}fects are also negligible.

\subsection{Massless limit in QED$_{\rm M}$}

\begin{figure}[!htb]
   \begin{minipage}{0.5\textwidth}
     \hspace{-2.5cm}
     \includegraphics[width=1.45\linewidth]{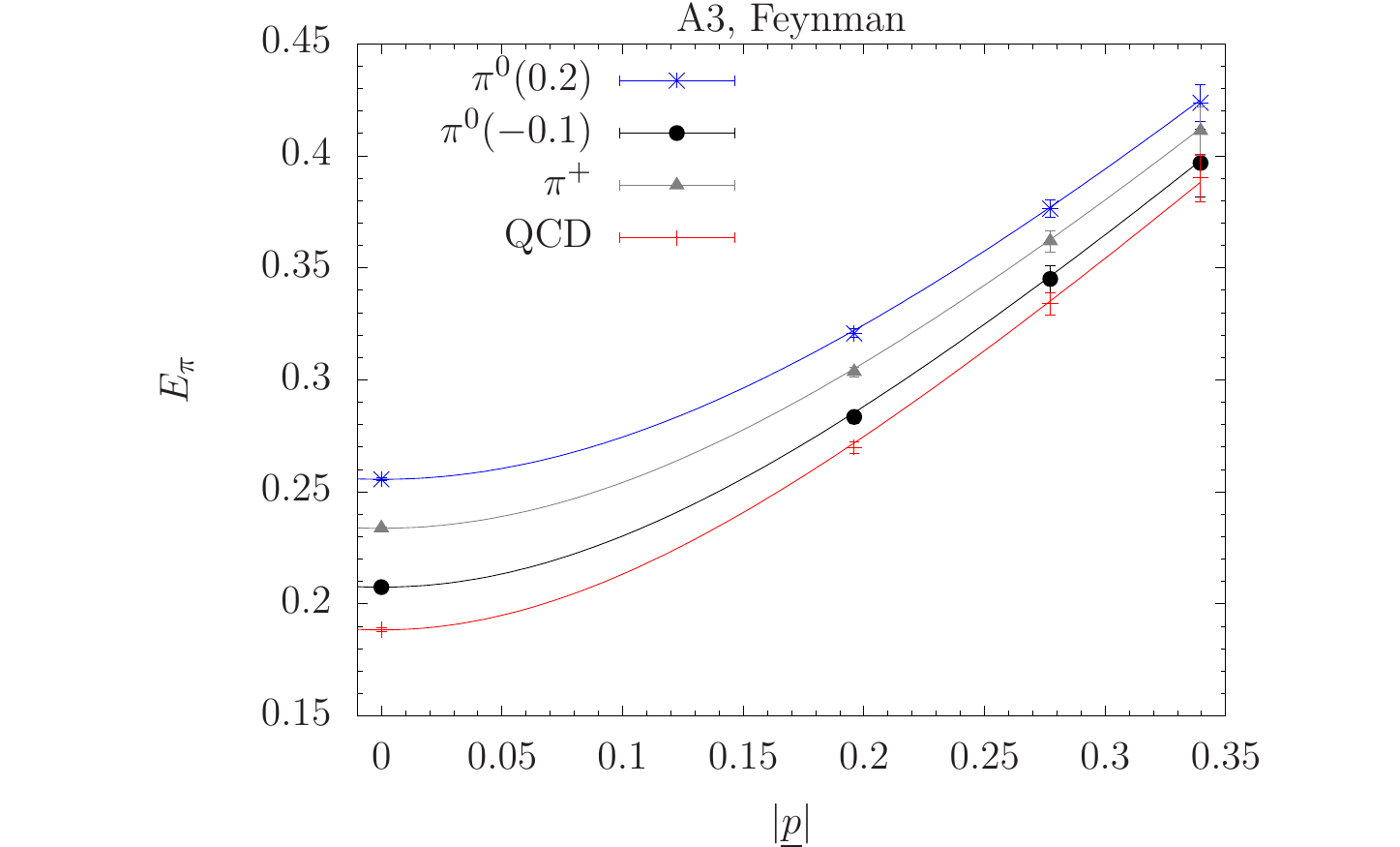}
     \caption{\emph{Dispersion relation for the charged and neutral pions in the A3 ensemble.}}\label{fig:eff_energy}
   \end{minipage}\hfill
   \hspace{0.5cm}
   \begin{minipage}{0.5\textwidth}
     \vspace{0.8cm}
     \hspace{-2.5cm}
     \includegraphics[width=1.45\linewidth]{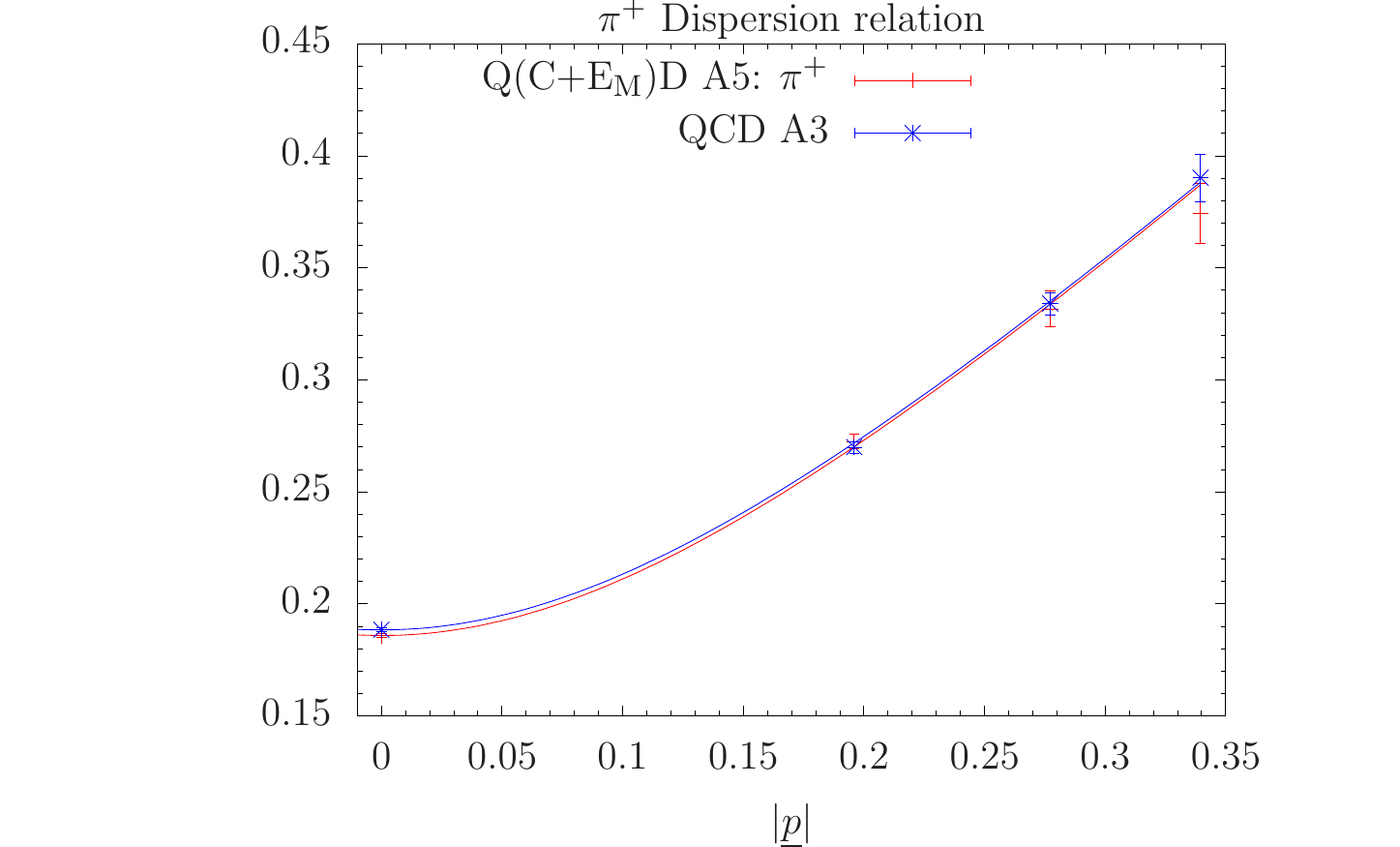}
     \captionsetup{width=0.8\textwidth}
     \caption{\emph{Dispersion relation for charged and neutral pions for $m_\gamma = 0.1$. The solid lines are the expectations from the continuum dispersion relation.}}\label{fig:matching_pions}
   \end{minipage}
\end{figure}

It was pointed out in Ref.~\cite{Patella:2017fgk} that in the limit of small photon mass the ef\mbox{}fective energy has a linear $t$ behavior 
and furthermore is constant as we change the momentum $\underline{p}$. 
Here we look at the behavior of the ef\mbox{}fective energy in the regime of small photon mass.
The result is found by mimicking the calculations given in \cite{Patella:2017fgk} and we report here the result for completeness
\begin{align*}
E_{\rm ef\mbox{}f} (t, \underline{p}) 
\stackrel{m_\gamma\rightarrow 0}{\simeq} \frac{ (Q_u-Q_d)^2 e^2}{m_\gamma^2 V} t 
- \frac{\de}{\de t } \ln\langle \mathcal{O}(t, \underline{0})\overline{\mathcal{O}}(0)
 \delta_{Q^{\rm T},0}\rangle_{\rm TL},
\end{align*}
where $\mathcal{O} = \pi^+$ is the pion interpolating field, the charge is $Q_u-Q_d = 1$, $Q^{\rm T} = Q_u + Q_d$ and 
$\langle \dots \rangle_{\rm TL}$ is the expectation value taken in the QED$_{\rm TL}$ formulation.
In the regime in which the above formula is valid the ef\mbox{}fective mass is independent from the momentum, 
plus a linear term in $t$ must be subtracted by hand before extracting the ef\mbox{}fective energy.
The linear term was recognized and studied in the non-relativistic approach in \cite{Endres:2015gda}
and explicitly subtracted. Notice that in our case $m_\gamma^2 V$ is around three orders of magnitude 
bigger than the one in \cite{Endres:2015gda}, that explains why we do not see the linear 
behavior in $t$, Fig.~\ref{fig:mpsdiffcharge}.\\
In Fig.~\ref{fig:eff_energy} we show the dispersion relation and we observe that it follows the continuum-like curve.\\
In Fig.~\ref{fig:matching_pions} we see that after the inclusion of the massive QED field the 
charged pion ef\mbox{}fective energy in the A5 ensemble matches the one in A3 with QCD only,
this observation will be useful later.
%%%%%%%%%%%%%%%%%%%%%%%%%%%%%%%%%%%%%%%%%%%%%%%%%%%%%%%%%%%%%%%%%%%%%%%%%%
%%%%%%%%%%%%%%%%%%%%%%%%%%%%%%%%%%%%%%%%%%%%%%%%%%%%%%%%%%%%%%%%%%%%%%%%%%

\section{Vacuum polarization}

\begin{figure}[!ht]
\begin{center}
\includegraphics[scale=1,angle=-0]{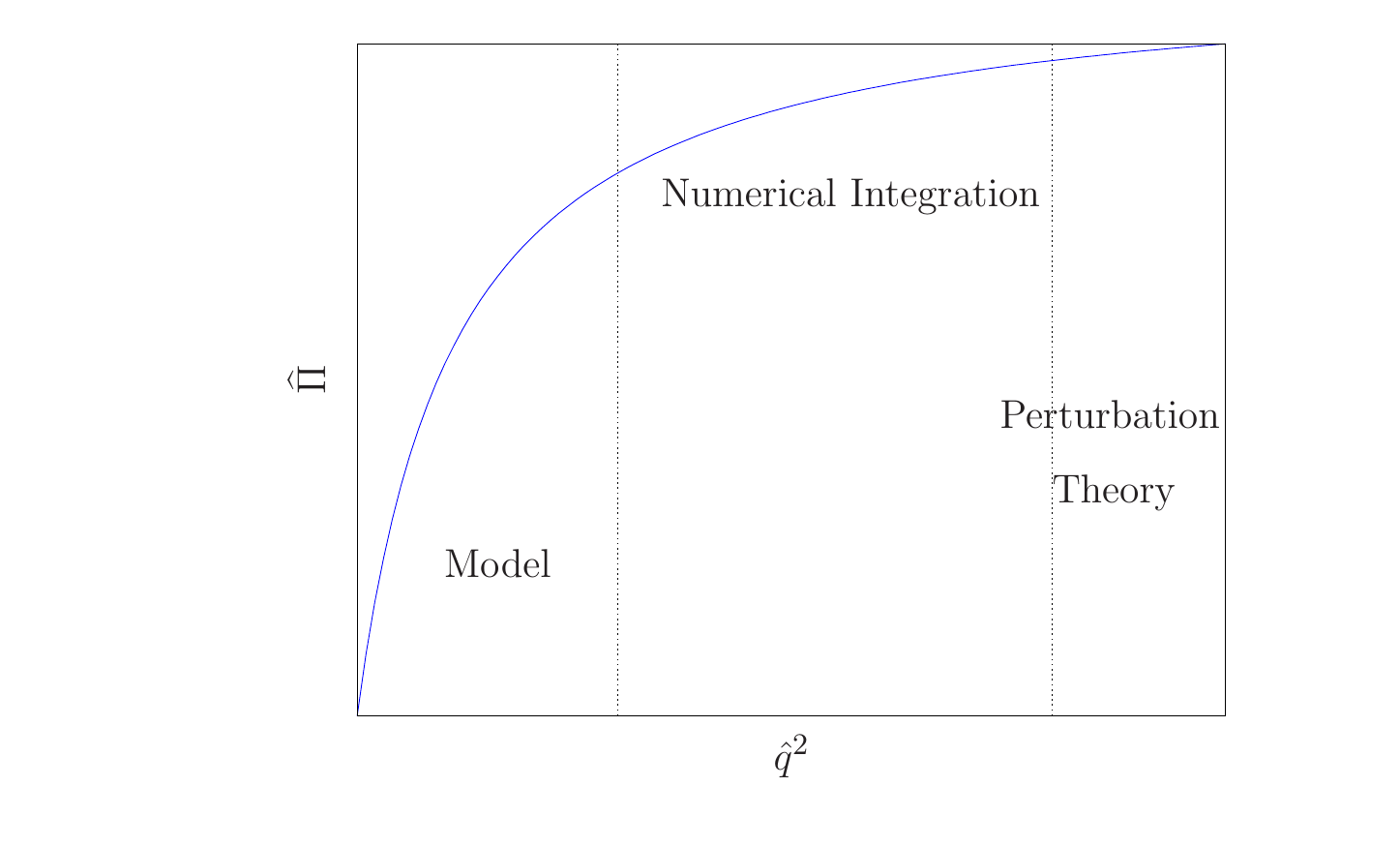}
\caption{\emph{Schematic representation of the subtracted scalar Hadronic Vacuum Polarization. 
In the low $q^2$ region, \quotes{Model}, we use for example Pad\'e fit to estimate the HVP and we integrate the result.
In the mid-$q^2$ region we numerically integrate directly the lattice data.
In the large $q^2$ region we use perturbation theory.}}
\label{fig:schema}
\end{center}
\end{figure}

The quark electromagnetic current is defined as
\begin{align}
\label{eq:em_curr}
J_\mu(x) = \sum_f^{N_f} Q_f J^f_\mu(x)= \sum_f^{N_f} Q_f\, \psibar_f(x) \gamma_\mu \psi_f(x),
\end{align}
where $f$ denotes the flavor index and $Q_f$ the associated charge ($e=1$), 
see App.~\ref{app:opscurr_cont} for a derivation of the one-point-split current in the formal theory.
In the case of $N_f=2$ with $\Psi = (u, d)$, $Q_u = 2/3$ and $Q_d = -1/3$, the electromagnetic current is a mixture of $\I$ and $\sigma^3$ matrices in flavor space
\begin{align}
J_\mu(x) =  \Psibar(x) \gamma_\mu \left[ \frac{Q_u}{2} \left(\I + \sigma^3\right) + \frac{Q_d}{2} \left(\I - \sigma^3\right)  \right]   \Psi(x).
\end{align}
The Euclidean HVP tensor is defined as
\begin{align}
\Pi_{\mu\nu}^{ff'} (q) = \int \de^4x\, \e^{iq\cdot x}\langle J^f_\mu(x) J^{f'}_\nu(0)\rangle,
\end{align}
and can be decomposed thanks to Lorentz invariance and current conservation as follows
\begin{align}
\label{eq:cont_wti}
\Pi^{ff'}_{\mu\nu} (q) = (\delta_{\mu\nu}q^2 - q_\mu q_\nu) \Pi^{ff'}(q^2),
\end{align}
leading to the scalar VP $\Pi^{ff'}(q^2)$, where the flavors $f$ and $f'$ need not to be the same.
The total vacuum polarization is then a sum over the possible flavors
\begin{align}
\Pi(q^2) = \sum_{f,f'} Q_f Q_{f'} \Pi^{ff'}(q^2).
\end{align}
The leading order contribution to the anomalous magnetic moment of the muon ($a_\mu^\text{HLO}$) 
for space-like momenta is related to the scalar VP as follows
\begin{align}
\label{eq:muon_anom}
a^\text{HAD}_{\mu, {\rm LO}} = \left(\frac{\alpha}{\pi}\right)^2\int_0^\infty \de q^2 f(q^2) \hat{\Pi}(q^2),
\end{align}
where $\hat{\Pi}$ is the renormalized scalar VP and
we have defined the following quantities
\begin{align}
\nn
\widehat{\Pi}(q^2) &\equiv 4\pi^2\left[\Pi(q^2) - \Pi(0)\right],\\
\nn
Z &\equiv \frac{\sqrt{q^4+4m_\mu^2 q^2} - q^2}{2m_\mu^2 q^2},\\
f(q^2) &\equiv \frac{m_\mu^2 q^2 Z^3(1-q^2 Z)}{1+ m_\mu^2 q^2 Z^2}\, .
\end{align}
The integrand function $f(q^2)\, \widehat{\Pi}(q^2)$ is peaked around values of $q^2 \sim \left(\sqrt{5} - 2\right)  m_\mu^2$, 
but the lowest momentum on the lattice is quite far from this value for typical lattice sizes\footnote{That is the reason why TBCs 
are used in this kind of calculations, see Chap.~\ref{chap:rtbc}}.
We divide the $\widehat{\Pi}(q^2)$ in three regions to reduce its model dependence \cite{Golterman:2014ksa}.
For small $q^2$ we should rely on a parametrization of $\widehat{\Pi}(q^2)$, which is 
then convoluted with $f(q^2)$ to find $a_\mu$. The extraction of $\Pi(0)$ is quite dif\mbox{}ficult since the statistical accuracy deteriorates as $q\rightarrow0$.
In the mid-$q^2$ region one integrates directly the lattice data multiplied by $f(q^2)$.
The muon anomaly $a_\mu$ in the large $q^2$ region is computed using perturbation theory (a 3-loop computation is 
available in Ref.~\cite{Chetyrkin:1996cf}). The situation is schematically shown in Fig.~\ref{fig:schema}.
%%%%%%%%%%%%%%%%%%%%%%%%%%%%%%%%%%%%%%%%%%%%%%%%%%%%%%%%%%%%%%%%%%%%%%%%%%
%%%%%%%%%%%%%%%%%%%%%%%%%%%%%%%%%%%%%%%%%%%%%%%%%%%%%%%%%%%%%%%%%%%%%%%%%%

\subsection{Lattice regularization}

We build the electromagnetic current in Q(C+E)D as the Noether current of the infinitesimal $\mathbf{U}(1)$ vector transformation
\begin{align}
\nn
\delta_V \psi_f(x) &= i \alpha(x) Q_f \psi_f(x),\\
\delta_V \overline{\psi}_f(x) &= -i \alpha(x) \overline{\psi}_f(x) Q_f.
\end{align} 
On the lattice we implement the one-point-split current, which is the conserved current of the $\mathbf{U}_\text{V}(1)$\footnote{For $N_f = 2$ 
Eq.~\ref{eq:sub_qcd_qed} modifies as $\mathbf{H}_\text{Q(C+E)D}(N_f=2) = \mathbf{U}'(1)_{\rm V} \otimes \mathbf{U}(1)_{\rm V}$, where $\mathbf{U}'(1)_{\rm V}$
is the subgroup of $\mathbf{SU}(2)_{\rm V}$ generated by $\sigma^3$.}
, and given as
\begin{align}
\label{eq:opsplit}
V_\mu^f(x) = \frac{1}{2} \left[\psibar_f(x)U_\mu(x)(\gamma_\mu-r)\psi_f(x+\hat{\mu}) + \psibar_f(x+\hat{\mu})U^\dagger_\mu(x)(\gamma_\mu+r)\psi_f(x) \right],
\end{align}
in the case of Wilson fermions\footnote{For both the improved and un-improved theory.} with $r$ parameter of the Wilson term, see App.~\ref{app:opscurr_latt} for the derivation of the current.
The VP tensor is found to be
\begin{align}
\Pi_{\mu\nu} (x) = \bigg\langle \left(\sum_f^{N_f} Q_f V_\mu^f(x)\right)\left(\sum_{f'}^{N_f} Q_{f'} V_\mu^{f'}(0)\right) \bigg\rangle
\end{align}
and by Fourier transforming the above expression we get 
\begin{align}
\label{eq:off_diag}
\Pi_{\mu\nu} (\hat{q}) = \sum_{n\in\Lambda} \e^{iq\cdot (x+\hat{\mu}/2+\hat{\nu}/2)} \Pi_{\mu\nu} (n),
\end{align}
where we recall $q_\mu \in\widetilde{\Lambda}$ and $\hat{q}_\mu = 2\sin(q_\mu/2)$.\\
When considering the diagonal term, $\mu=\nu$, we need to take
into account contact terms that modify the expression 
we have given in the formal theory\footnote{Contact terms arise from the overlapping of composite operators.}. 
The contact term is given by
\begin{align}
\nn
J_\mu^{f,\text{ct}} (x) &= \frac{1}{2} \left[\psibar_f(x)U_\mu(x)(r-\gamma_\mu )\psi_f(x+\hat{\mu}) + \psibar_f(x+\hat{\mu})U^\dagger_\mu(x)(\gamma_\mu+r)\psi_f(x) \right]\, ,\\
\label{eq:diag}
\Pi_{\mu\nu}^{f,\text{ct}} (\hat{q}) & = -\delta_{\mu\nu} \langle J_\mu^{f,\text{ct}}(0)\rangle .
\end{align}
The HVP is now given by Eqs.~\ref{eq:diag}, \ref{eq:off_diag}, namely
\begin{align}
\Pi_{\mu\nu}^{ff'} (\hat{q}) =\left[ \sum_{x\in\Lambda}  \e^{iq\cdot (x+\hat{\mu}/2+\hat{\nu}/2)} \langle V_\mu^f(x) V_\nu^{f'}(0) \rangle\right] -\delta_{\mu\nu} \delta_{ff'} \langle J_\mu^{f,\text{ct}}(0)\rangle,
\end{align}
which fulfills the WTI
\begin{align}
\hat{q}_\mu \Pi_{\mu\nu}^{ff'} = 0 = \hat{q}_\nu \Pi_{\mu\nu}^{ff'}.
\end{align}
For the explicit form of the HVP after Wick contractions see App.~\ref{app:HVP}. \\
We relate the HVP to the scalar VP by using the continuum relation
\begin{align}
\label{eq:scalar_vp}
\Pi_{\mu\nu}^{ff'} (\hat{q}) = \left(\delta_{\mu\nu} \hat{q}^2 - \hat{q}_\mu\hat{q}_\nu \right) \Pi^{ff'}(\hat{q}^2) + \ord\left(\hat{q}^4\right).
\end{align}
The term $\ord\left(\hat{q}^4\right)$ denotes all the higher order terms invariant under hypercubic transformations.
Those vanish in the continuum limit and in an anisotropic finite volume, e.g.~$L\neq T$, the hypercubic invariance is further broken.\\
The scalar VP has order $a$ ef\mbox{}fects since the conserved vector current is not improved.
One way to improve the current would be to consider the following modification
\begin{align}
V_\mu^\text{impr}(\hat{q}) = V_\mu(\hat{q}) + c \sin(q_\nu/2) \sum_n \e^{iq\cdot x} T_{\mu\nu}(x),
\end{align}
where $T_{\mu\nu}(x) = \psibar(x) \sigma_{\mu\nu} \psi(x)$ and
$c$ is a constant to be fitted to remove the order $a$ ef\mbox{}fects for on-shell improvement of the matrix elements of the current $V$. 
See Ref.~\cite{Gockeler:2003cw} for Wilson fermions in the qQCD case.

\subsection{Scalar vacuum polarization}

In order to extract the scalar VP we have to analyze the relation Eq.~\ref{eq:scalar_vp}.
\paragraph{Of\mbox{}f diagonal components}
For $\mu \neq \nu$ the equation becomes
\begin{align}
\Pi_{\mu\nu}^{ff'} (\hat{q}) = - \hat{q}_\mu\hat{q}_\nu  \Pi^{ff'}(\hat{q}^2),
\end{align}
and by summing over all the possible $N$ permutations of the indexes with $\mu\neq\nu$ and the same $\hat{q}^2$ we obtain
\begin{align}
\Pi^{ff'}(\hat{q}^2) = -\frac{1}{\hat{q}_\mu\hat{q}_\nu} \Pi_{\mu\nu}^{ff'} (\hat{q})
 \Longrightarrow 
 \sum_{\mu\neq\nu} \Pi^{ff'}(\hat{q}^2) = N\, \Pi^{ff'}(\hat{q}^2) = -\sum_{\mu\neq\nu} \frac{1}{\hat{q}_\mu\hat{q}_\nu} \Pi_{\mu\nu}^{ff'} (\hat{q}),
\end{align}
from which we extract the scalar VP,
\begin{align}
\Pi^{ff'}(\hat{q}^2) = -\frac{1}{N} \sum_{\mu\neq\nu} \frac{\Pi_{\mu\nu}^{ff'} (\hat{q})}{\hat{q}_\mu\hat{q}_\nu }.
\end{align}
\paragraph{Diagonal components}
For the diagonal component $\mu=\nu$ the Eq.~\ref{eq:scalar_vp} becomes
\begin{align}
\Pi_{\mu\mu}^{ff'} (\hat{q}) = \left( \hat{q}^2 - \hat{q}^2_\mu\right)  \Pi^{ff'}(\hat{q}^2),
\end{align}
and again by summing over $\mu$ we find the equation for the extraction of the scalar vacuum polarization
\begin{align}
\sum_\mu \Pi_{\mu\mu}^{ff'} (\hat{q}) = \sum_\mu \left( \hat{q}^2 - \hat{q}^2_\mu\right)  \Pi^{ff'}(\hat{q}^2) =  \big( 4\hat{q}^2 - \underbrace{\sum_\mu\hat{q}^2_\mu}_{q^2} \big)  \Pi^{ff'}(\hat{q}^2) = 3\hat{q}^2 \Pi^{ff'}(\hat{q}^2),
\end{align}
that is given by
\begin{align}
\Pi^{ff'}(\hat{q}^2) = \frac{1}{3\hat{q}^2} \sum_\mu \Pi^{ff'}_{\mu\mu}(\hat{q}).
\end{align}

\subsection*{Tensor zero mode subtraction}

In this work we use a modified version of the HVP tensor, 
in particular we subtract the zero mode contribution (ZMS) from it
\begin{align}
\Pi_{\mu\nu}^\text{ZMS} (\hat{q}) = \sum_{n\in\Lambda} \e^{iq\cdot (n+\hat{\mu}/2+\hat{\nu}/2)} \Pi_{\mu\nu} (n) - \sum_{n\in\Lambda} \Pi_{\mu\nu} (n) = \Pi_{\mu\nu} (\hat{q}) - \Pi_{\mu\nu} (\hat{q}=0).
\end{align}
It has been shown in \cite{Bernecker:2011gh} that in the infinite volume limit, thanks to Lorentz symmetry, the zero mode contribution vanishes whereas in the finite volume it is not constrained and can dif\mbox{}fer from zero. 
By using the ZMS version improvements in the signal-to-noise in the low  $\hat{q}^2$ region can be seen, see for example Ref.~\cite{Blum:2016xpd}.

\subsection{Pad\'e fits}

Since the integrand in Eq.~\ref{eq:muon_anom} is strongly peaked around the muon mass we need a good description of the scalar VP in low $\hat{q}^2$ region. Furthermore what enters in the integral for the muon anomaly is the subtracted VP, meaning that we need to extrapolate the value $\Pi(\hat{q}^2 = 0)$.
To do so we need to parametrize the VP as a function of $\hat{q}^2$.\\
It is known that the Taylor expansion of the VP does not converge and yields to very poor fits in terms of $\chi^2$.
The Pad\'e expansion ef\mbox{}fectively increases the radius of convergence of the sum. 
We can recast the Pad\'e fits in the following form
\begin{align*}
R_{mn}(\hat{q}^2) = \Pi(0) + \hat{q}^2\left(\delta_{mn} c + \sum_{i=0}^{m-1} \frac{a_i}{b_i+\hat{q}^2}\right),
\end{align*}
where $n = m, m+1$ and $\Pi(0)$, $a_i$'s and $b_i$'s are the parameters to be fitted.

\section{Time moments}

Another way to extract the scalar VP at zero momentum involves time moments.
Those give derivatives of the scalar VP with respect to $\hat{q}^2$ at zero \cite{Chakraborty:2015ugp}. 
We start from Eq.~\ref{eq:scalar_vp} and we consider only the diagonal spatial components $\mu = \nu = j$
\begin{align}
\Pi^{ff'}_{jj}(\hat{q}) = \left( \hat{q}^2 -\hat{q}_j\right) \Pi^{ff'}(\hat{q}^2).
\end{align}
If we restrict to momenta of the kind $\hat{q} = (\hat{q}_0, \underline{0})$ and we sum over $j$ we obtain
\begin{align*}
\sum_j \Pi^{ff'}_{jj}(\hat{q}_0) = 3\hat{q}_0^2 \Pi^{ff'}(\hat{q}_0^2).
\end{align*}
The scalar VP is then given by
\begin{align}
\label{eq:time_scalar_vp}
\hat{q}_0^2 \Pi^{ff'}(\hat{q}_0^2) = 
\frac{1}{3} \sum_j \Pi_{jj}^{ff'} (\hat{q}_0) =  
\frac{1}{3}\sum_j \left\{ \left[ \sum_{x_0,\underline{x}\in\Lambda}  \e^{iq_0 x_0} \langle V_j^f(n) V_j^{f'}(0) \rangle\right] - \delta_{ff'} \langle J_j^{f,\text{ct}}(0)\rangle  \right\}
\end{align}

\subsection{Infinite volume}

In order to obtain the time moments relations we consider derivatives with respect to $q_0$ evaluated at zero\footnote{The contact term does not contribute since it is a constant.}, 
\begin{align}
\nn
\frac{\partial^{2k}}{\partial q_0^{2k}} \left( q_0^2\Pi^{ff'}(q_0^2) \right)\bigg|_{q_0 = 0} & = 
 \frac{1}{3} \sum_{j, x_0,\underline{x}}  \langle V_j^f(x) V_j^{f'}(0) \rangle  \left( \frac{\partial^{2k}}{\partial q_0^{2k}} \e^{iq_0 x_0}\right)\bigg|_{q_0 = 0}\\
 \nn
 &= \frac{1}{3} \sum_{j, x_0,\underline{x}}  \langle V_j^f(x) V_j^{f'}(0) \rangle  \, (i)^{2k} x_0^{2k} \e^{iq_0 x_0}\bigg|_{q_0 = 0}\\
 &= \frac{1}{3} \sum_{j, x_0,\underline{x}} (-)^{k} x_0^{2k}\, \langle V_j^f(x) V_j^{f'}(0) \rangle \equiv G_{2k}^{ff'}. 
\end{align}
We have an infinite tower of relations from which we can build the Taylor expansion of the scalar VP
\begin{align}
\Pi^{ff'}( q_0^2 ) = 
\sum_{k=1}^\infty G^{ff'}_{2k} \frac{q^{2k-2}}{(2k)!} = 
 \frac{1}{3} \sum_{j, x_0,\underline{x}} \, \langle V_j^f(x) V_j^{f'}(0)\rangle   \sum_{k=0}^\infty (-)^{k} x_0^{2k} \frac{q^{2k-2}}{(2k)!}\, .
\end{align}
Notice that for $k=1$ in the above formula we obtain
\begin{align}
\Pi^{ff'}( q_0^2=0 ) = 
\frac{1}{2} \frac{\partial^{2}}{\partial q_0^{2}} \left( q_0^2\Pi^{ff'}(q_0^2) \right)\bigg|_{q_0 = 0} =
-\frac{1}{6} \sum_{j} \sum_{x_0,\underline{x}} \, x_0^2 \langle V_j^f(x) V_j^{f'}(0)\rangle\, ,
\end{align}
for any parametrization of $\Pi(q_0^2)$, used to estimate $\Pi(0)$ for the subtracted scalar VP.

\subsection{Finite volume}

On the lattice we are forced to work with finite volumes, 
meaning that momenta are quantized and derivatives have to be replaced by finite dif\mbox{}ferences.\\
We use the symmetric derivative definition because the error is proportional to $h^2$, with $h$ being the separation between adjacent momenta
\begin{align}
\nonumber
\frac{\Delta f(q_0)}{\Delta q_0} & 
= \frac{1}{2}\left(\frac{T}{2\pi}\right) \left[ f\left(q_0 + \frac{2\pi}{T}\right) - f\left(q_0 - \frac{2\pi}{T}\right) \right], \\
\nonumber
\frac{\Delta^2 f(q_0)}{\Delta q_0^2} & 
= \frac{1}{2^2}\left(\frac{T}{2\pi}\right)^2 \left[ f\left(q_0 + \frac{4\pi}{T}\right) - 2f(q_0) + f\left(q_0 - \frac{4\pi}{T}\right) \right], \\
\nonumber
\vdots\\
\label{eq:finite_diff}
\frac{\Delta^{2n} f(q_0)}{\Delta q_0^{2n}} & 
= \frac{1}{2^{2n}}\left(\frac{T}{2\pi}\right)^{2n} \sum_{k=0}^n (-)^k \binom{n}{k} f\left[ q_0 + (n-2k) \frac{2\pi}{T} \right].
\end{align} 
By applying the above formulae to the expression of the scalar VP in Eq.~\ref{eq:time_scalar_vp} 
we have to evaluate the finite dif\mbox{}ference of the exponential
\begin{align}
\nn
\frac{\Delta \e^{iq_0 x_0}}{\Delta q_0}\bigg|_{q_0=0} & 
= \frac{1}{2}\left(\frac{T}{2\pi}\right) \e^{iq_0 x_0} \left[ \e^{i2\pi x_0/T} - \e^{-i2\pi x_0/T} \right]\bigg|_{q_0=0} \\
\nn
&=\frac{T}{2\pi} \e^{iq_0 x_0} \, i \sin\left(\frac{2\pi}{T}x_0\right)\bigg|_{q_0=0} =\frac{T}{2\pi} i \sin\left(\frac{2\pi}{T}x_0\right), \\
\nn
\frac{\Delta^2 \e^{iq_0 x_0}}{\Delta q_0^2}\bigg|_{q_0=0} & 
= \left(\frac{T}{2\pi}\right)^2 i^2 \sin^2\left(\frac{2\pi}{T} x_0\right) , \\
\nn
\vdots\\
\frac{\Delta^{2n} \e^{iq_0 x_0}}{\Delta q_0^{2n}}\bigg|_{q_0=0} & 
= \left(\frac{T}{2\pi}\right)^{2n} (-)^n \sin^{2n}\left(\frac{2\pi}{T}x_0\right).
\end{align} 
By combining all the results the master formula for the time moments is given by
\begin{align}
\label{eq:time_moments_explicit}
\frac{\Delta^{2k} }{\Delta q_0^{2k}} \hat{q}_0^2 \Pi^{ff'}(\hat{q}_0^2)\bigg|_{q_0=0}
= \frac{1}{3} \sum_{j, x_0, \underline{x}} \left(\frac{T}{2\pi}\right)^{2k} (-)^k \sin^{2k}\left(\frac{2\pi}{T}x_0\right) \langle V_j^f(x) V_j^{f'}(0)\rangle.
\end{align}
Now we can give a specific Pad\'e approximation of $\Pi^{ff'}(\hat{q}_0^2)$, put it in the l.h.s.~of the above formula and determine the 
parameters by equating the results with the r.h.s.~of Eq.~\ref{eq:time_moments_explicit}. 
Given a representation of the scalar VP with $N$ parameters we need the same number of equations 
(given by the master formula).
Then we solve the highly non-linear system in order to extract the 
relevant parameters. 
By increasing the number of parameters we are forced to increase the number of derivatives to take into account 
and therefore the number of momenta to consider, see Eq.~\ref{eq:finite_diff}.
This af\mbox{}fects the estimate of an extremely local, in Fourier space, quantity as $\Pi^{ff'}(\hat{q}^2=0)$.

\section{Subtracted scalar HVP}

\begin{figure}[!ht]
\begin{center}
\subfigure[\emph{Subtracted HVP in QCD.}]{\includegraphics[scale=1,angle=-0]{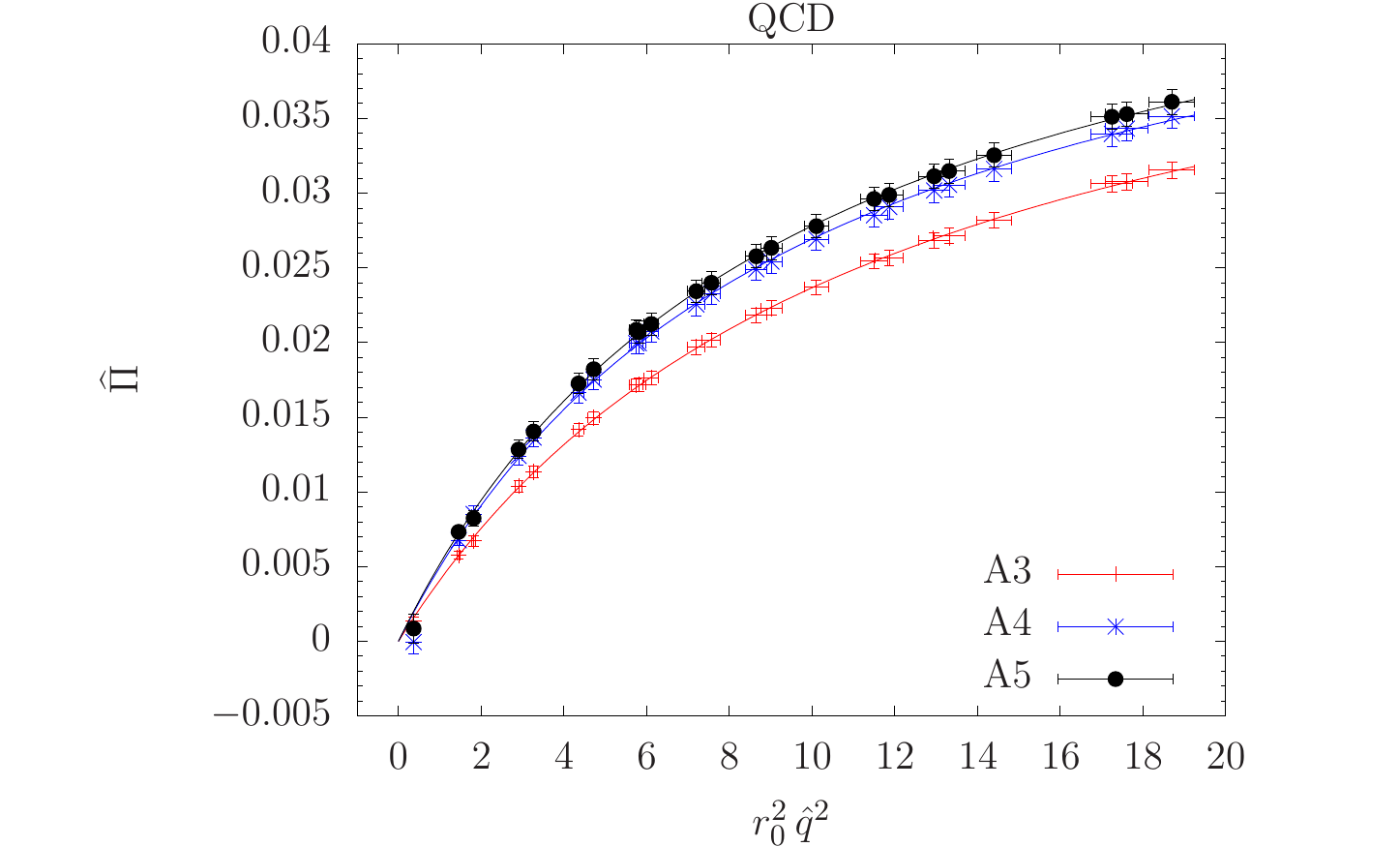}\label{fig:hvp_qcd_qed_a}}
\subfigure[\emph{Unsubtracted HVP in QCD + qQ(C+E)D in A5 ensemble.}]{\includegraphics[scale=1,angle=-0]{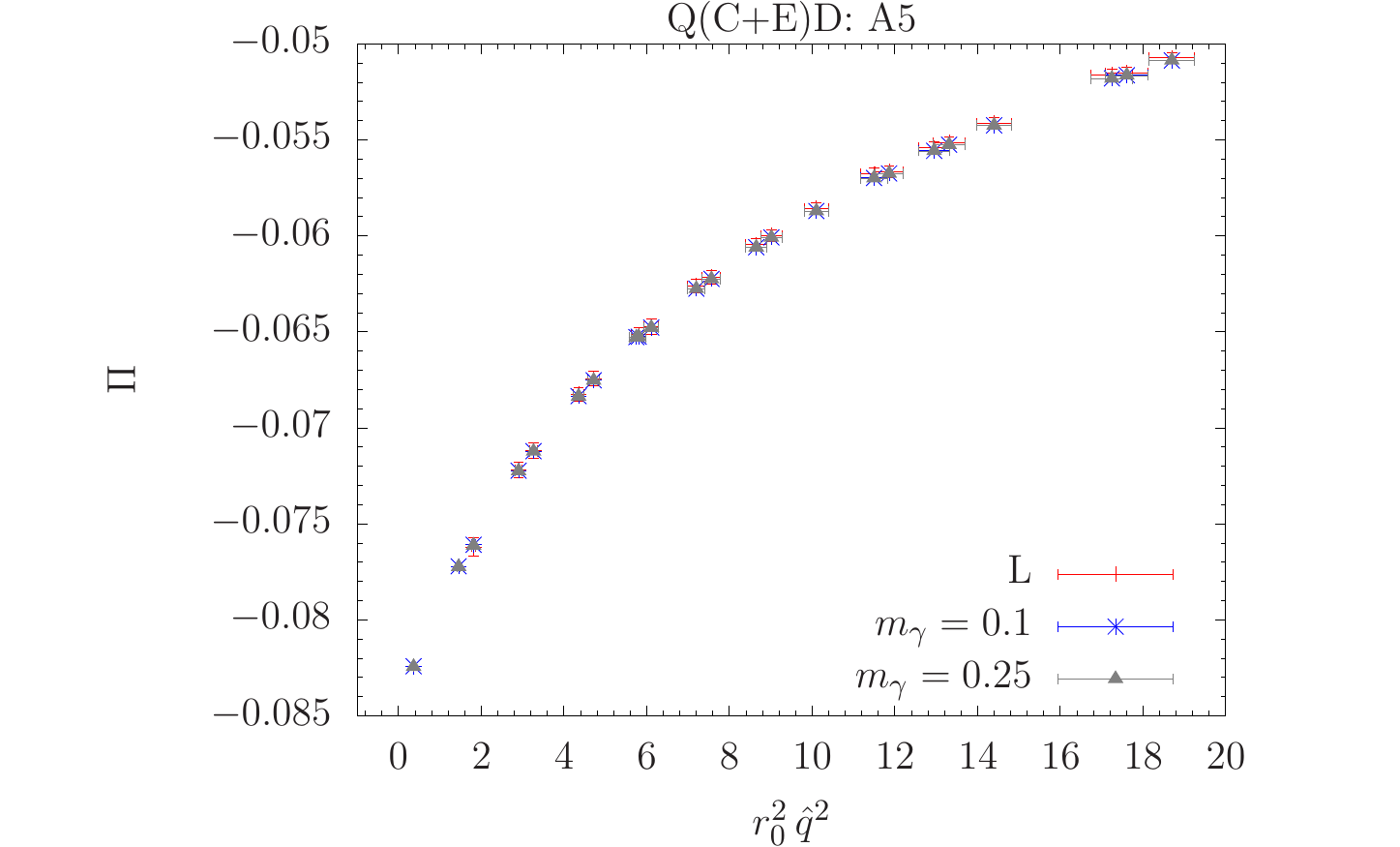}\label{fig:hvp_qcd_qed_b}}
\caption{\emph{HVP as a function of $r_0 \hat{q}^2$ with and without the inclusion of QED.}}
\label{fig:hvp_qcd_qed}
\end{center}
\end{figure}

In this section we collect the main results for the subtracted scalar HVP.
In Fig.~\ref{fig:hvp_qcd_qed_a} we show the results in the case of QCD, while in Fig.~\ref{fig:hvp_qcd_qed_b}
the unsubtracted HVP with inclusion of massive and massless QED.
Notice that $\widehat{\Pi}$ versus $r_0^2\,\hat{q}^2$ is a renormalized plot.
The value of the Sommer parameter $r_0 / a$ is calculated starting from gluonic observables that do not get 
electromagnetic contributions in qQED.
We therefore use the values of $r_0 /a$ computed in QCD, those can be found in Ref.~\cite{Fritzsch:2012wq}.

%%%%%%%%%%%%%%%%%%%%%%%%%%%%%%%%%%%%%%%%%%%%%%%%%%%%%%%%%%%%%%%%%%%%%%%%%%
\begin{figure}[!ht]
\begin{center}
\subfigure[\emph{Comparison of the scalar VP with and without QED.}]{\includegraphics[scale=0.7,angle=-0]{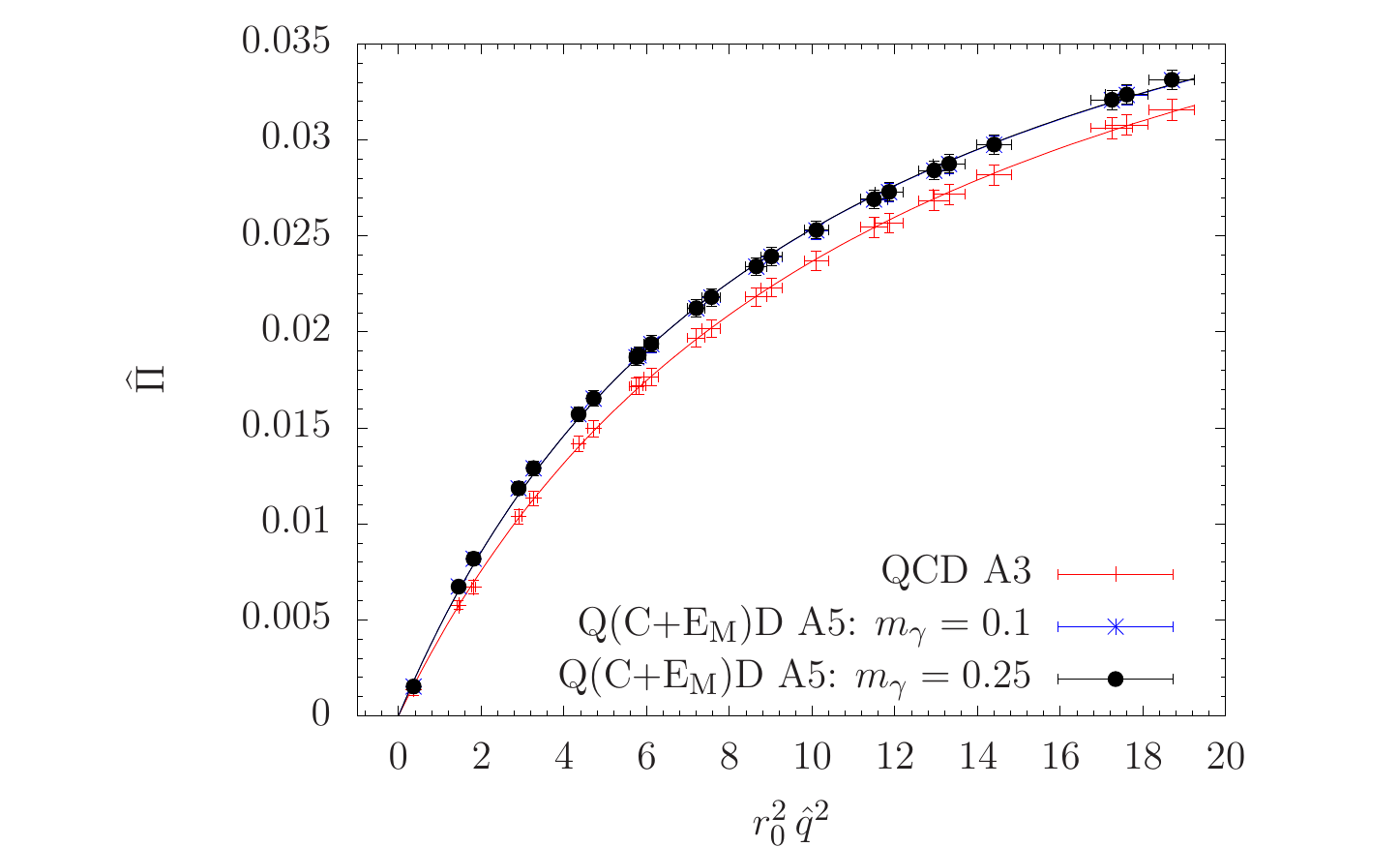}\label{fig:hvp_qecd_A5}}
\subfigure[\emph{Dif\mbox{}ference of the HVP with and without QED.}]{\includegraphics[scale=0.7,angle=-0]{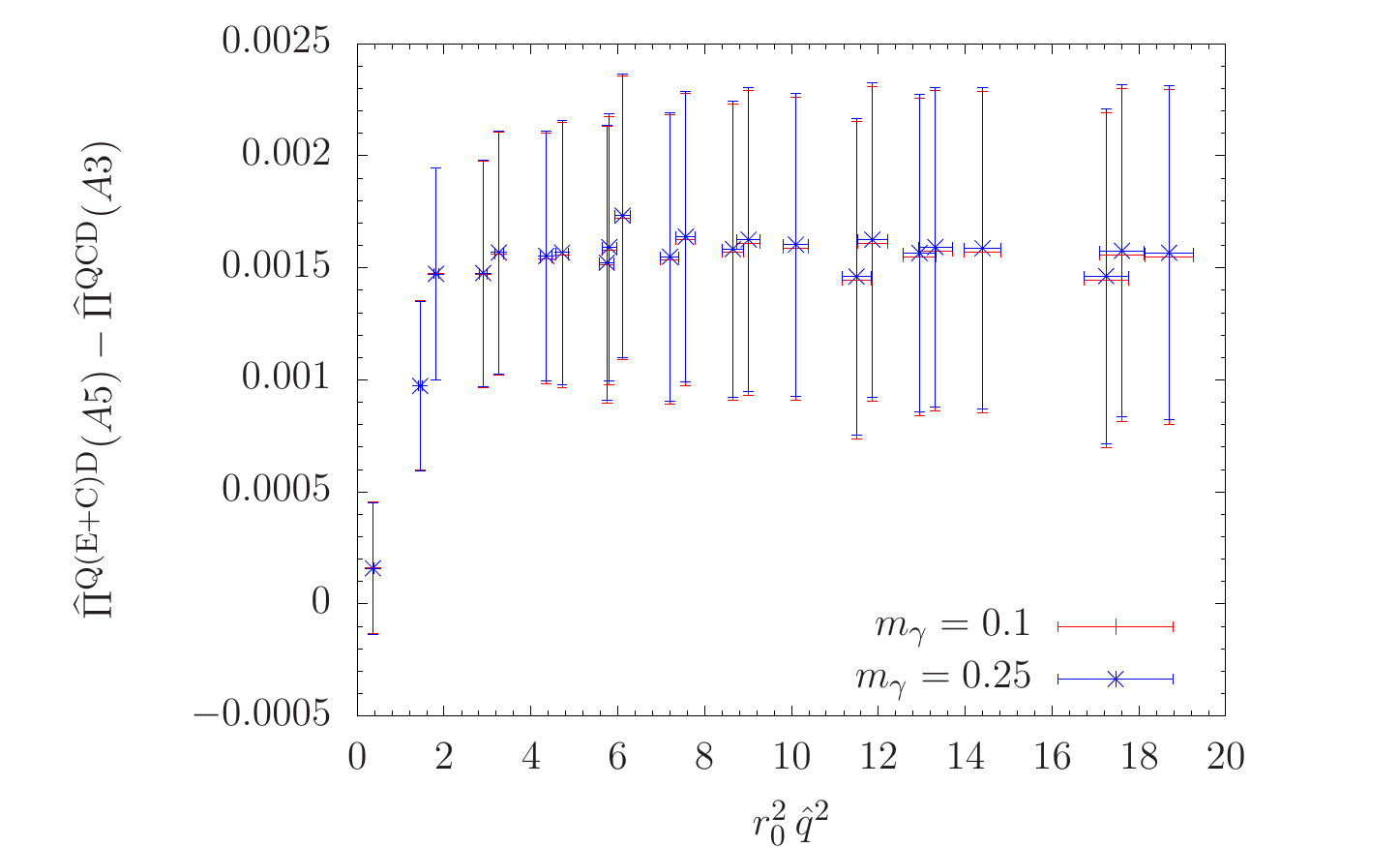}\label{fig:conf_conf_QECD_A5}}
\subfigure[\emph{Relative dif\mbox{}ference of the QCD abd Q(C+E)D HVP.}]{\includegraphics[scale=0.7,angle=-0]{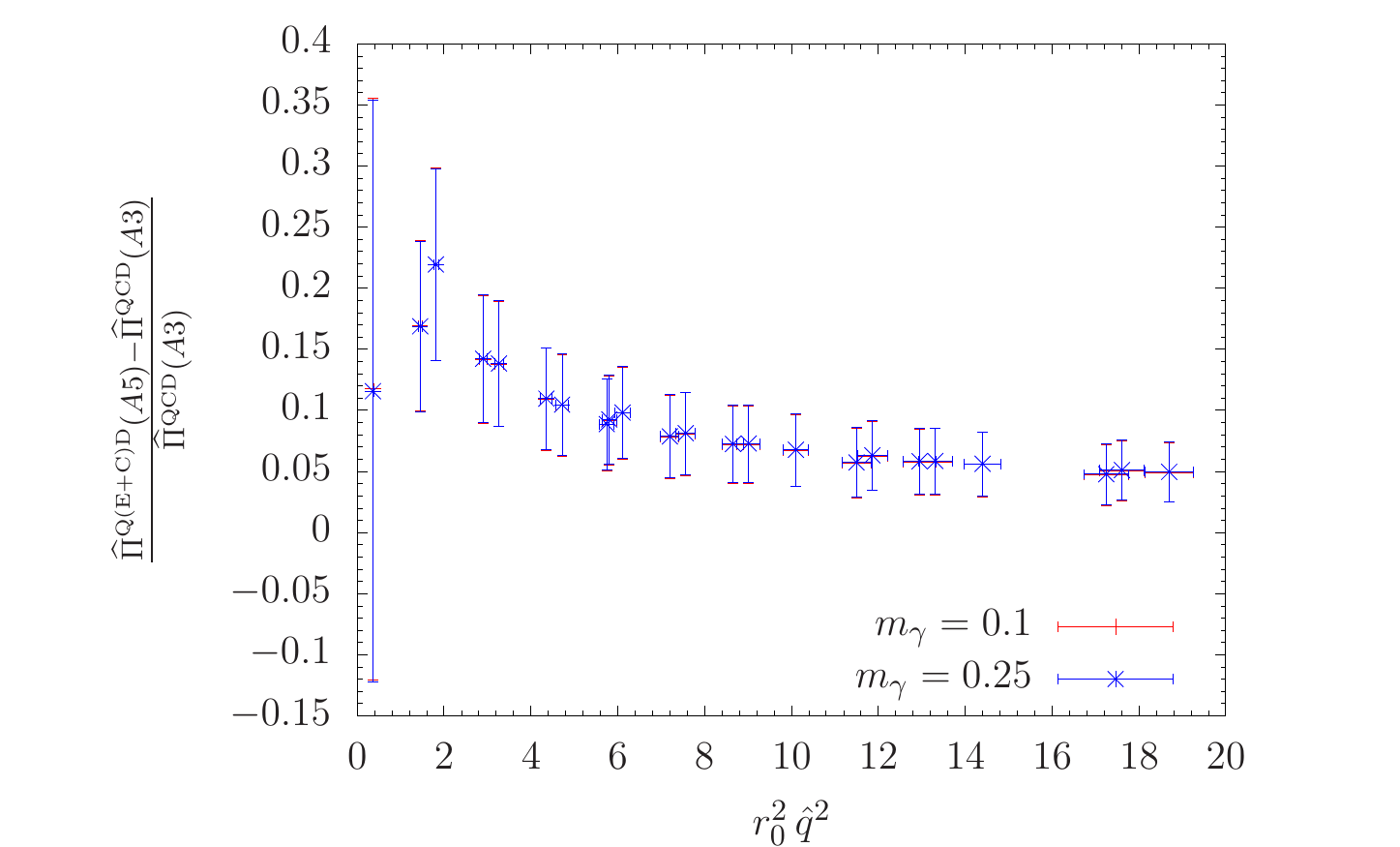}\label{fig:conf_QECD_A5}}
\caption{\emph{Comparisons of the HVP on the matched ensembles.}}
\label{fig:hvp_qecd_sub}
\end{center}
\end{figure}

%%%%%%%%%%%%%%%%%%%%%%%%%%%%%%%%%%%%%%%%%%%%%%%%%%%%%%%%%%%%%%%%%%%%%%%%%%
The shape of the scalar VP is qualitatively the same with or without QED.
In Fig.~\ref{fig:hvp_qecd_A5} we show the ef\mbox{}fect of QED inclusion and the comparison is made thanks to the matching condition presented in Sect.~\ref{sect:dcVPQEDc}.
The usefulness of the matching condition is twofold; on one side it allows for a comparison
between dif\mbox{}ferent ensembles, on the other side we know that the HVP is strongly dependent on the pion masses \cite{DellaMorte:2011aa}, hence we need to match the pion masses if we
want to extract the EM ef\mbox{}fects.\\
It can be seen in Fig.~\ref{fig:conf_QECD_A5} how the QED ef\mbox{}fect is consistent with a constant within errors.
More precisely the ef\mbox{}fect is about 7\% and compatible with zero within two/three combined sigmas.
%%%%%%%%%%%%%%%%%%%%%%%%%%%%%%%%%%%%%%%%%%%%%%%%%%%%%%%%%%%%%%%%%%%%%%%%%%
%%%%%%%%%%%%%%%%%%%%%%%%%%%%%%%%%%%%%%%%%%%%%%%%%%%%%%%%%%%%%%%%%%%%%%%%%%
That suggests to look directly at the dif\mbox{}ference of the subtracted scalar HVPs, Fig.~\ref{fig:conf_conf_QECD_A5}.
A na\"ive way to proceed is to model the HVPs once in QCD and once in Q(C+E)D and calculate $a_\mu$ in the two theories.
The dif\mbox{}ference encodes the QED corrections once the results have been extrapolated to the same pion masses.
This introduces large systematics due to the two fits and washes out the QED ef\mbox{}fects.
Instead we follow a dif\mbox{}ferent approach where we first match the pion masses and then we model the dif\mbox{}ference for the HVPs directly.
Notice this represents a one less-fit procedure dependent approach.
After a good description of the dif\mbox{}ference of the HVPs is found we convolute that with the $f(\hat{q}^2)$ and get the result for $a_{\mu}^{\rm HAD, QED}$.

For a preliminary estimate of the electromagnetic ef\mbox{}fects we model the dif\mbox{}ference with Pad\'e fits $R_{10}$ and $R_{11}$ and a mixed fit, namely
linear rising in $\hat{q}^2$ and then constant.
By integrating the dif\mbox{}ferent fits numerically we find
\begin{align}
\nn
a_{\mu}^{\rm HAD, QED} (R_{10}) &= (60 \pm 28) \times 10^{-10}\, ,\\
\nn
a_{\mu}^{\rm HAD, QED} (R_{11}) &= (50 \pm 24) \times 10^{-10}\, ,\\
a_{\mu}^{\rm HAD, QED} ({\rm mixed}) &= (35 \pm 14) \times 10^{-10}\, .
\end{align}
We see that at this level the systematics, as expected, still dominate the error.
One crucial thing to notice is that the electromagnetic ef\mbox{}fects go in the direction of reduce the tension between the theoretical and experimental value.
We checked that the conclusions do not change when considering QED$_{\rm L}$ as we can see from Fig.~\ref{fig:hvp_qcd_qed_b}.
It appears that all the issues related to the QED$_{\rm L}$ formulation \cite{Patella:2017fgk} are under control or do not have ef\mbox{}fect on the quantities we have analyzed so far.

\section{Conclusions}

We presented preliminary results for the electromagnetic ef\mbox{}fects on several observables.
We discussed the change in the critical mass due to the inclusion of electromagnetic interactions and compared 
with perturbation theory calculations.
We found the linearity of the square pseudoscalar mass in the PCAC mass.
We gave results for the light pseudoscalar spectrum and we found completely negligible finite volume and photon mass ef\mbox{}fect.
We made sure that in QED$_{\rm M}$ we extracted the right energy values, and that the massive approach does not reduce to a \quotes{very complicated way}
to use QED$_{\rm TL}$.
For the HVP we presented a strategy to give an estimate on the electromagnetic corrections, which schematically consists in:
\begin{itemize}
\item matching the (charged) pion masses in dif\mbox{}ferent ensembles, with and without QED,
\item looking directly at the dif\mbox{}ference of the subtracted HVPs in the Q(C+E)D and QCD matched ensembles.
\end{itemize}
In this way we were able to see a clear signal for the HVP that at the end reflected in an estimate for the muon anomaly.
Perhaps the most important result is that the electromagnetic corrections in the muon anomaly clearly ameliorate the tension between theory and experiment.

The results are very encouraging and we plan to consider the inclusion of twisted boundary conditions, in order to stabilize the fits in $\hat{q}^2$, as well as
analyze dif\mbox{}ferent ensembles available in the CLS initiative, in order to study the cutof\mbox{}f dependence, but mostly the pion mass dependence
of the EM ef\mbox{}fects.

\afterpage{\blankpage}

\setcounter{equation}{0}

\chapter{Conclusions and Outlooks}
\label{chap:concl_outl}

The goal of the dif\mbox{}ferent projects presented in this thesis was to accurately assess some residual systematics ef\mbox{}fects in high-precision non-perturbative applications of
Lattice Gauge Theories.

{\noindent We} studied the viability of reweighting techniques for twisted boundary conditions in the spatial directions for fermionic fields.
There we suggested to compensate for the breaking of unitarity due to the choice of dif\mbox{}ferent boundary conditions in the valence and
sea sector.
This project started in order to estimate the above mentioned violations in the computation of
renormalization factors for the minimal walking Technicolor theory, using the RI-MOM scheme and partially twisted boundary conditions.
We performed a complete study of the reweighting approach from tree-level to actual Monte Carlo simulations.
We studied gluonic and fermionic observables in large and small volumes and
we saw that the ef\mbox{}fect of the reweighting is at the sub-percent level.
In small volumes we found a quite pronounced sensitivity of the critical mass for un-improved Wilson fermions to changes in the twisting angle.
That is clearly a cutof\mbox{}f and finite volume ef\mbox{}fect that could be rather large for large twisting angles.\\

{\noindent In} the second part of the thesis we presented an optimization of the HMC algorithm.
No optimization study exists for BSM theories on the lattice.
The lattice BSM field is relatively young and the time for more precise result has come.
We proposed a new strategy to find optimal parameters in the case of the Omelyan integrator
and Hasenbusch mass-preconditioning.
The entire method is based on the existence of an exactly conserved quantity, the Shadow Hamiltonian.
Its implementation is straightforward and we only needed to calculate the driving forces used during the Molecular Dynamics step.
We implemented our strategy in the case of multi-time scales and mass-preconditioning, and we found the general
form of the Shadow Hamiltonian in that case.
Once the forces during the trajectory were measured we were able to give predictions on the cost and acceptance
of a simulation within 10\% accuracy.
Before that a rule of thumb was used to find optimal parameters.
The method can be also implemented on-line during the generation of configurations.
That is quite an appealing possibility, were the algorithm correct itself changing the optimized parameters for the trajectories during thermalization.\\
Since the twisted boundary conditions represent an IR regularization as a mass term, 
during the time of the optimization project I also developed a twisted-boundary-preconditioning.
Unfortunately, as we saw in the reweighting study, the pion mass becomes smaller by increasing the 
twisting angle.
This means that there is an even worse conditioning number for the Dirac-Wilson operator, hence no speed-up
was achieved.
Furthermore larger forces were found, compared to the mass-preconditioning study, and the algorithm ran 
earlier into instabilities.\\

{\noindent In} the last part of the thesis we presented a new way to isolate electromagnetic ef\mbox{}fects for
the hadronic contribution to $(g-2)_\mu$.
We added quenched QED configurations to preexistent QCD ones.
We considered two dif\mbox{}ferent regularizations of the finite volume zero mode.
We were able to see a clear ef\mbox{}fect, even for physical quark charges and 
electromagnetic coupling.
The crucial question to be answered in the future is whether the ef\mbox{}fect we have seen is going to be larger
for smaller pion masses or not.
Within the limitations of the computation (single lattice spacing, pion mass around 400 MeV), 
we saw an ef\mbox{}fect of the same size as the discrepancy (in $a_\mu$) between theory and experiments.
That is intriguing and it is clearly a motivation for continuing and improving this study.
In order to reduce the systematics, the first thing to do is to analyze more volumes, pion masses and dif\mbox{}ferent lattice spacings, 
already available from the CLS initiative.
Other improvements can be achieved by considering the addition of a clover term for the electromagnetic part of the action,
as well as improving the vector current, in order to have better control on the continuum extrapolation.
Another crucial ef\mbox{}fect to include in the hadronic muon anomaly in the future is the quark mass splitting. 
This ef\mbox{}fect can go in the opposite direction of the EM ef\mbox{}fects. 
An example of that is visible in the proton-neutron mass splitting.
The reweighting technique can be employed to include those ef\mbox{}fects 
and the experience acquired in the reweighting of twisted boundary conditions can be useful.\\
During the work on the muon anomaly we started to explore the possibility 
of using dif\mbox{}ferent gauge conditions in order to eliminate the zero mode problem.
So far in the literature no one has considered the use of temporal gauge.
On a lattice with periodic boundary conditions it is not possible to fix
the temporal gauge in all the sites.
We need to fix two dif\mbox{}ferent gauges, for example temporal in all the sites excluding one and Coulomb gauge on the remaining.
Furthermore we would like to have a massive approach since it gives exponentially suppressed finite volume ef\mbox{}fects.
That possibility is even less clear since no simil-Stueckelberg mechanism is developed in this case.
The temporal gauge may be particularly useful there as it would presumably
remove time-dependent corrections to the ef\mbox{}fective masses, which are due to the zero mode in $A_0$.
This is definitely a new possibility worth to explore.

\afterpage{\blankpage}

\appendix
\setcounter{chapter}{0}
\renewcommand{\chaptername}{Appendix}
\renewcommand{\theequation}{\Alph{chapter}.\arabic{equation}}
\addcontentsline{toc}{chapter}{\numberline{}Appendixes}
\setcounter{equation}{0}

\newpage
\chapter{Ward-Takahashi Identities}
\label{app:wti}

A multilocal operator is given by
\begin{equation}
\mathcal{O} (x_1,x_2, \dots, x_n)\equiv \mathcal{O}_1(x_1) \mathcal{O}_2(x_2) \dots \mathcal{O}_n(x_n) = \prod_{k=1}^n \mathcal{O}_k(x_k),
\end{equation}
and its expectation value is formally given by the functional integral
\begin{align}
\label{formalism}
\big\langle\mathcal{O}(x_1,x_2, \dots, x_n)\big\rangle& = 
\frac{1}{\mathcal{Z}}\int\mathcal{D}\left[\text{A}_\mu,\psi,\overline{\psi}\right]\e^{-\text{S}_\text{YM}(\text{A}_\mu)-\text{S}_\text{F}(A, \psi,\overline{\psi})}\mathcal{O}(x_1, x_2, \dots, x_n)\, ,
\end{align}
where $\text{S}_\text{YM}$ is the pure gauge Yang-Mills action and $\text{S}_\text{F}$ stands for the fermionic action coupled to gauge fields, see eq.~\ref{QCD_continuum_lagrangian} for the corresponding Lagrangians in QCD-like theories.\\
As we know the functional integral is ill-defined and we need a regularization. 

\section{WTIs proof}
\label{app:WTI_proof}
The Ward-Takahashi identity (WTI) are found by requiring the invariance of the expectation value under symmetry transformations of fermionic variables. 
For the sake of simplicity we will omit in the following the pure gauge action in the functional integral and in the partition function.\\

In order to derive the WTI we consider a generic case where $\mathbf{G}$ is the \emph{global} symmetry group under investigation. The spinors transform linearly with respect the representation $\mathcal{G}(g)$ of the group, with $g\in\mathbf{G}$, i.e.
\begin{equation}
\psi'_\alpha=\mathcal{G}_{\alpha\beta}(g)\psi_\beta.
\end{equation}
We consider the infinitesimal transformation as
\begin{equation}
\delta\psi_\alpha=i\omega^at_{\alpha\beta}^a\psi_\beta\, ,
\end{equation}
with $t^a$ hermitian and antisymmetric generators of the algebra corresponding to the symmetry group.

By hypothesis the action is invariant under the symmetry transformation and we suppose that also the functional measure is\footnote{We exclude anomalous symmetries.}
\begin{align}
\nn
&\text{S}(\overline{\psi}\phantom{}',\psi')=\text{S}(\overline{\psi},\psi)\, ,\\
&\mathcal{D}[\psibar\phantom{}',\psi']=\mathcal{D}[\psibar,\psi]\, .
\end{align}
The $k$-th operator $\mathcal{O}_k(x_k)$, in the multilocal, transforms as follows
\begin{equation}
\delta\left(\mathcal{O}_k(x_k)\right)_\alpha=i\omega^a\left(\mathcal{R}_k^a\right)_{\alpha\beta}\left(\mathcal{O}_k(x_k)\right)_\beta
\end{equation}
where $\mathcal{R}^a_k$ is the representation under which the operator transforms, that can be dif\mbox{}ferent from the one of the fundamental fields.\\
From these definitions it follows that the multilocal operator transforms as
\begin{equation}
\delta\left(\mathcal{O}(x_1, x_2, \dots, x_n)\right)_\alpha=\sum_{k=1}^n\mathcal{O}_1(x_1)\dots\delta\left(\mathcal{O}_k(x_k)\right)_\alpha\dots\mathcal{O}_n(x_n).
\end{equation}

We make the transformation \underline{local}, i.e.~$\omega^a\equiv\omega^a(x)$ smooth functions, and in the following we omit the matrix indexes $\alpha, \beta$. The action will now change as follows 
\begin{equation}
\text{S}_\text{F}(\psi', \overline{\psi}')=\text{S}_\text{F}(\psi,\overline{\psi})-i\int\de^4x\left(\partial_\mu\omega^a(x)\right) J^a_\mu(x)\stackrel{\text{by p.}}{=}\text{S}_\text{F},(\psi,\overline{\psi})+i\int\de^4x\left(\partial_\mu J^a_\mu(x)\right)\omega^a(x)\, ,
\end{equation}
where we assumed that $\omega$ is a smooth function that vanishes outside some bounded region $R$, and $J_\mu^a$ the current associated to the transformation. 
The operator will change as
\begin{align}
\mathcal{O}'(x_1, x_2, \dots, x_n)
=
\mathcal{O}(x_1, x_2, \dots, x_n)
+i\sum_{k=1}^n\omega^a(x_k)\mathcal{O}_1(x_1)\mathcal{O}_2(x_2)\dots
 \mathcal{R}_k^a \mathcal{O}_k(x_k) 
\dots\mathcal{O}_n(x_n).
\end{align}
The expectation value cannot change under the symmetry transformation, hence
\begin{align}
\big\langle\mathcal{O}'(x_1, \dots, x_n)\big\rangle&=\big\langle\mathcal{O}(x_1, \dots, x_n)\big\rangle.
\end{align}
The transformed expectation value has the form
\begin{align}
\nn
\big\langle\mathcal{O}'(x_1,x_2,\dots,x_n)\big\rangle&=\frac{1}{\mathcal{Z'}}\int\mathcal{D}[\psibar\phantom{}',\psi']\e^{-\text{S}_\text{F}(\psi',\overline{\psi}')}\mathcal{O}'(x_1,x_2,\dots,x_n)\\
\nn
&\simeq\frac{1}{\mathcal{Z}}\int\mathcal{D}[\psibar,\psi]\exp\left[-\text{S}_\text{F}(\psi,\overline{\psi})-i\int\de^4x\left(\partial_\mu\text{J}^a_\mu(x)\right)\omega^a(x)\right]\\
&\phantom{==}\times\left[\mathcal{O}(x_1,x_2,\dots,x_n)+i\sum_{k=1}^n\omega^a(x_k)\mathcal{O}_1(x_1)\mathcal{O}_2(x_2)\dots 
\mathcal{R}_k^a\mathcal{O}_k(x_k) 
\dots\mathcal{O}_n(x_n)\right],
\end{align}
and we functionally expand it with respect to the parameter $\delta\omega^b(y)$
\begin{align}
\nn
\big\langle\mathcal{O}'\big\rangle&\simeq\frac{1}{\mathcal{Z}}\int\mathcal{D}[\psibar,\psi]\e^{-\text{S}_\text{F}(\psi,\overline{\psi})}\left(1-i\int\de^4x\left(\partial_\mu J^a_\mu(x)\right)\frac{\delta\omega^a(x)}{\delta\omega^b(y)}\delta\omega^b(y)\right)\\
\nn
&\phantom{=}\times\left[\mathcal{O}+i\sum_{k=1}^n\frac{\delta\omega^a(x_k)}{\delta\omega^b(y)}\mathcal{O}_1(x_1)\mathcal{O}_2(x_2)\dots \mathcal{R}_k^a\mathcal{O}_k(x_k) \dots\mathcal{O}_n(x_n)\delta\omega^b(y)\right]\\
&\simeq\big\langle\mathcal{O}\big\rangle-i\delta\omega^b(y)\bigg\{\big\langle\partial_\mu^y J_\mu^b(y)\mathcal{O}\big\rangle-\sum_k\delta^4(y-x_k)\big\langle\mathcal{O}_1\mathcal{O}_2\dots \mathcal{R}_k^b\mathcal{O}_k \dots\mathcal{O}_n\big\rangle\bigg\}.
\end{align}
By taking into account that the expectation values on l.h.s.~and r.h.s.~are identical and that the parameter $\delta\omega^b(y)$ can be arbitrary varied we obtain that the quantity in curly brackets must vanish. After renaming variables and indexes we get
\begin{align}
\big\langle\partial_\mu^x J_\mu^a(x)\mathcal{O}(x_1,x_2,\dots,x_n)\big\rangle=\sum_{k=1}^n\delta^4(x-x_k)\big\langle\mathcal{O}_1(x_1)\mathcal{O}_2(x_2)\dots \mathcal{R}_k^a\mathcal{O}_k(x_k) \dots\mathcal{O}_n(x_n)\big\rangle,
\end{align}
which we can write in a more compact form in the simple case of one local operator
\begin{equation}
\label{WTI}
\langle\delta S_x\mathcal{O}_y\rangle=\delta_{xy}\langle\delta\mathcal{O}_y\rangle.
\end{equation}
The WTI in its final form is 
\begin{align}
\label{path_integral_formalism}
\partial_\mu^x\big\langle J_\mu^a(x)\mathcal{O}(x_1,x_2,\dots,x_n)\big\rangle=\sum_{k=1}^n\delta^4(x-x_k)\big\langle\mathcal{O}_1(x_1)\mathcal{O}_2(x_2)\dots \mathcal{R}_k^a\mathcal{O}_k(x_k) \dots\mathcal{O}_n(x_n)\big\rangle.
\end{align}

\section{Singlet One-point-split current}

Here we derivate the WTI in the case of a vector transformation $\mathbf{U}_{\rm V}(1)$, in the formal theory and on the lattice with un-improved Wilson fermions in the Euclidean space.

\subsection{Formal derivation}
\label{app:opscurr_cont}

By considering $\mathcal{O}_y = \I$, and recalling that $\delta \text{S}_x / \delta \alpha= \partial_\mu J_\mu(x)$, $\alpha$ being the parameter of the symmetry transformation, we have
\begin{align}
\partial_\mu \langle J_\mu(x) \rangle = 0.
\end{align}
The vector transformation $\mathbf{U}(1)_{\rm V}$ is given by
\begin{align}
\nn
\delta_V \psi_f(x) &= i\alpha(x) Q_f \psi_f(x),\\
\delta_V \psibar_f(x) &= -i\alpha(x) \psibar_f(x) Q_f,
\end{align}
and the fermionic action is
\begin{align}
\text{S} = \int\de^4 x\, \psibar_f(x) \left( \slashed{D} + m_f\right) \psi_f(x).
\end{align}
The variation of the action reads
\begin{align}
\nn
\delta_V \text{S} & = i \int \de^4 x\, \psibar_f(x) Q_f \left[\alpha(x) \left(\slashed{D}+m_f\right) -  \left(\slashed{D}+m_f\right) \alpha(x) \right] \psi_f(x)\\
\nn
& = -i \int \de^4 x\, \psibar_f(x) Q_f \gamma_\mu\psi_f(x) \left(\partial_\mu \alpha(x)\right)\\
& \stackrel{\text{by p.}}{=} i \int \de^4 x\, \alpha(x) \partial_\mu\left(\psibar_f(x) Q_f \gamma_\mu\psi_f(x)\right)  \equiv 0,
\end{align}
that gives $\partial_\mu \langle V^f_\mu(x) \rangle = 0$ with the following definition
\begin{align}
V_\mu^f(x) = \psibar_f(x) Q_f \gamma_\mu\psi_f(x),
\end{align}
which is the electromagnetic current used in Eq.~\ref{eq:em_curr} with the identification $V_\mu^f(x) = Q_f J^f_\mu(x)$.

\subsection{Lattice derivation}
\label{app:opscurr_latt}

The Dirac-Wilson action on the lattice is, with $r=1$, is given in Eq.~\ref{eq:dw_act} and its variation under infinitesimal vector transformations on the lattice is found to be
\begin{align}
\nn
\delta_V \text{S}_{\rm F} &= \sum_{x,\mu}\frac{i}{2} Q_f \bigg\{ \psibar_f(x) \left[-\alpha(x)\right] (\gamma_\mu-1) U_\mu(x) \psi_f(x+a\hat{\mu}) + \psibar_f(x) (\gamma_\mu-1) U_\mu(x) \alpha(x+a\hat{\mu}) \psi_f(x+a\hat{\mu}) \\
\nn
& - \psibar_f(x) \left[-\alpha(x)\right] (\gamma_\mu+1) U^\dagger_\mu(x-a\hat{\mu}) \psi_f(x-a\hat{\mu}) - \psibar_f(x) (\gamma_\mu+1) U^\dagger_\mu(x-a\hat{\mu}) \alpha(x-a\hat{\mu}) \psi_f(x-a\hat{\mu})      \bigg\}\\
\nn
& = \sum_{\mu}\frac{i}{2} Q_f \bigg\{ -\sum_x\alpha(x) \psibar_f(x)  (\gamma_\mu-1) U_\mu(x) \psi_f(x+a\hat{\mu})\\
\nn
&\phantom{= \sum_{\mu}\frac{i}{2} Q_f \bigg\{} + \sum_{x' = x+a\hat{\mu}} \alpha(x') \psibar_f(x'-a\hat{\mu}) (\gamma_\mu-1) U_\mu(x'-a\hat{\mu})  \psi_f(x') \\
\nn
&\phantom{= \sum_{\mu}\frac{i}{2} Q_f \bigg\{} + \sum_{x}\alpha(x) \psibar_f(x) (\gamma_\mu+1) U^\dagger_\mu(x-a\hat{\mu}) \psi_f(x-a\hat{\mu}) \\
\nn
&\phantom{= \sum_{\mu}\frac{i}{2} Q_f \bigg\{} - \sum_{x' = x-a\hat{\mu}} \alpha(x') \psibar_f(x'+a\hat{\mu}) (\gamma_\mu+1) U^\dagger_\mu(x')  \psi_f(x')      \bigg\}\\
\nn
& = \sum_{x,\mu}\frac{i}{2} Q_f \alpha(x) \bigg\{ - \psibar_f(x) (\gamma_\mu-1) U_\mu(x) \psi_f(x+a\hat{\mu}) + \psibar_f(x-a\hat{\mu}) (\gamma_\mu-1) U_\mu(x-a\hat{\mu}) \psi_f(x) \\
& \phantom{ = \sum_{x,\mu}\frac{i}{2} Q_f \alpha(x) \bigg\{ }    - \psibar_f(x) (\gamma_\mu+1) U^\dagger_\mu(x-a\hat{\mu}) \psi_f(x-a\hat{\mu}) - \psibar_f(x+a\hat{\mu}) (\gamma_\mu+1) U^\dagger_\mu(x) \psi_f(x)      \bigg\},
\end{align}
which can be rewritten as a total divergence
\begin{align}
\nn
\delta_V \text{S}_{\rm F} & = - \sum_{x,\mu}\frac{i}{2} Q_f \alpha(x) \partial_\mu^- \big[ \psibar_f(x) (\gamma_\mu-1) U_\mu(x) \psi_f(x+a\hat{\mu}) + \psibar_f(x+a\hat{\mu}) (\gamma_\mu+1) U^\dagger_\mu(x) \psi_f(x) \big]\\
& = -i \sum_{x,\mu} Q_f \alpha(x) \partial_\mu^- V_\mu^f(x)\, ,
\end{align}
where $V_\mu^f$ matches the definition of the singlet one-point-split current in Eq.~\ref{eq:opsplit} with $r=1$

\section{Bare PCAC relation on the lattice}
\label{app:lattice_PCAC}

We give here a derivation of the bare PCAC relation on the lattice. To this end
we recall the interaction part of the Lagrangian in the case of $\Psi$ flavor multiplet
\begin{equation}
\mathcal{L}^\text{I}_{\rm F}=\frac{1}{2a}\sum_\mu\left[\overline{\Psi}(x)U_\mu(x)(\gamma_\mu-r)\Psi(x+a\hat{\mu})-\overline{\Psi}(x+a\hat{\mu})U^\dagger_\mu(x)(\gamma_\mu+r)\Psi(x)\right].
\end{equation}
The variation of the Lagrangian under Eq.~\ref{SU_2_A_transformation} axial transformations is found to be
\begin{align}
\nn
\delta\mathcal{L}^\text{I}_{\rm F}=&\frac{1}{2a}\sum_\mu\bigg[\overline{\Psi}(x)\left(i\omega_f(x)\frac{\sigma_f}{2}\gamma_5\right)U_\mu(x)(\gamma_\mu-r)\Psi(x+a\hat{\mu})\\
\nn
&\phantom{\frac{1}{2a}\sum_\mu\bigg[}+\overline{\Psi}(x)U_\mu(x)(\gamma_\mu-r)\left(i\omega_f(x+a\hat{\mu})\frac{\sigma_f}{2}\gamma_5\right)\Psi(x+a\hat{\mu})\\
\nn
&\phantom{\frac{1}{2a}\sum_\mu\bigg[}-\overline{\Psi}(x+a\hat{\mu})\left(i\omega_f(x+a\hat{\mu})\frac{\sigma_f}{2}\gamma_5\right)U^\dagger_\mu(x)(\gamma_\mu+r)\Psi(x)\\
&\phantom{\frac{1}{2a}\sum_\mu\bigg[}-\overline{\Psi}(x+a\hat{\mu})U^\dagger_\mu(x)(\gamma_\mu+r)\left(i\omega_f(x)\frac{\sigma_f}{2}\gamma_5\right)\Psi(x)\bigg].
\end{align}
By renaming the variables of the second and third term we obtain
\begin{align}
\nn
\delta\mathcal{L}^\text{I}_L=&\frac{1}{2a}\sum_\mu\bigg[\overline{\Psi}(x)\left(i\omega_f(x)\frac{\sigma_f}{2}\gamma_5\right)U_\mu(x)(\gamma_\mu-r)\Psi(x+a\hat{\mu})\\
\nn
&\phantom{\frac{1}{2a}\sum_\mu\bigg[}+\overline{\Psi}(x-a\hat{\mu})U_\mu(x-a\hat{\mu})(\gamma_\mu-r)\left(i\omega_f(x)\frac{\sigma_f}{2}\gamma_5\right)\Psi(x)\\
\nn
&\phantom{\frac{1}{2a}\sum_\mu\bigg[}-\overline{\Psi}(x)\left(i\omega_f(x)\frac{\sigma_f}{2}\gamma_5\right)U^\dagger_\mu(x-a\hat{\mu})(\gamma_\mu+r)\Psi(x-a\hat{\mu})\\
&\phantom{\frac{1}{2a}\sum_\mu\bigg[}-\overline{\Psi}(x+a\hat{\mu})U^\dagger_\mu(x)(\gamma_\mu+r)\left(i\omega_f(x)\frac{\sigma_f}{2}\gamma_5\right)\Psi(x)\bigg].
\end{align}
Now we use the anticommutator relation $\left\{\gamma_\mu,\gamma_5\right\}=0$ and we find
\begin{align}
\nn
\delta\mathcal{L}^\text{I}_{\rm F}=&\frac{i\omega_f(x)}{2a}\sum_\mu\bigg[\overline{\Psi}(x)U_\mu(x)(-\gamma_\mu-r)\gamma_5\frac{\sigma_f}{2}\Psi(x+a\hat{\mu})+\overline{\Psi}(x-a\hat{\mu})U_\mu(x-a\hat{\mu})(\gamma_\mu-r)\gamma_5\frac{\sigma_f}{2}\Psi(x)\\
\nn
&\phantom{\frac{i\omega_f(x)}{2a}}-\overline{\Psi}(x+a\hat{\mu})U^\dagger_\mu(x)(\gamma_\mu+r)\gamma_5\frac{\sigma_f}{2}\Psi(x)-\overline{\Psi}(x)U^\dagger_\mu(x-a\hat{\mu})(-\gamma_\mu+r)\gamma_5\frac{\sigma_f}{2}\Psi(x-a\hat{\mu})\bigg]\\
\equiv & i\omega_f(x)\left[\partial_\mu^- A_\mu^f - \chi_{\rm A}^f\right],
\end{align}
where we have defined the variation of the Wilson term as $\chi_{\rm A}^f$, which cannot be cast in form of a total divergence, and the axial current in the following way
\begin{equation}
A_\mu^f=-\frac{1}{2}\left[\Psibar(x+a\hat{\mu})U^\dagger_\mu(x)\gamma_\mu\gamma_5\frac{\sigma_f}{2}\Psi(x)+\Psibar(x)U_\mu(x)\gamma_\mu\gamma_5\frac{\sigma_f}{2}\Psi(x+a\hat{\mu})\right].
\end{equation}
The variation is found to be
\begin{equation}
%\label{WTI_lattice_axial}
\frac{1}{i}\frac{\delta\mathcal{L}_{\rm F}}{\delta\omega_f(x)}\bigg|_{\omega_f(x)=0}=\partial_\mu^- A_\mu^f(x) - \chi_{\rm A}^f-\overline{\Psi}(x)\gamma_5\bigg\{\frac{\sigma_f}{2}, M_0\bigg\}\Psi(x).
\end{equation}
We arrived at the PCAC relation on the lattice between bare quantities, $M_0 = m_0\I$,
\begin{align}
\label{eq:PCAC_lattice}
\partial_\mu^-\langle A_\mu^f(x)\mathcal{O}(y)\rangle = 2m_0\langle P^f(x)\mathcal{O}(y)\rangle + \langle\chi_{\rm A}^f(x)\mathcal{O}(y)\rangle .
\end{align}

\appendix
\setcounter{chapter}{1}
\renewcommand{\chaptername}{Appendix}
\renewcommand{\theequation}{\Alph{chapter}.\arabic{section}.\arabic{equation}}
\setcounter{equation}{0}

\chapter{Reweighting factors and reweighted observables}

\section{Tree-level reweighting factors}
\label{app:tree-level_rew_fact}

In the following we give the Dirac-Wilson spectrum in the case of PBCs and TBCs and the analytic evaluation of the 
tree-level reweighting factors. The lattice spacing is fixed to one.

\subsection{Dirac-Wilson operator and eigenvalues}

\begin{figure}[h!t]
\begin{center}
\includegraphics[scale=0.8]{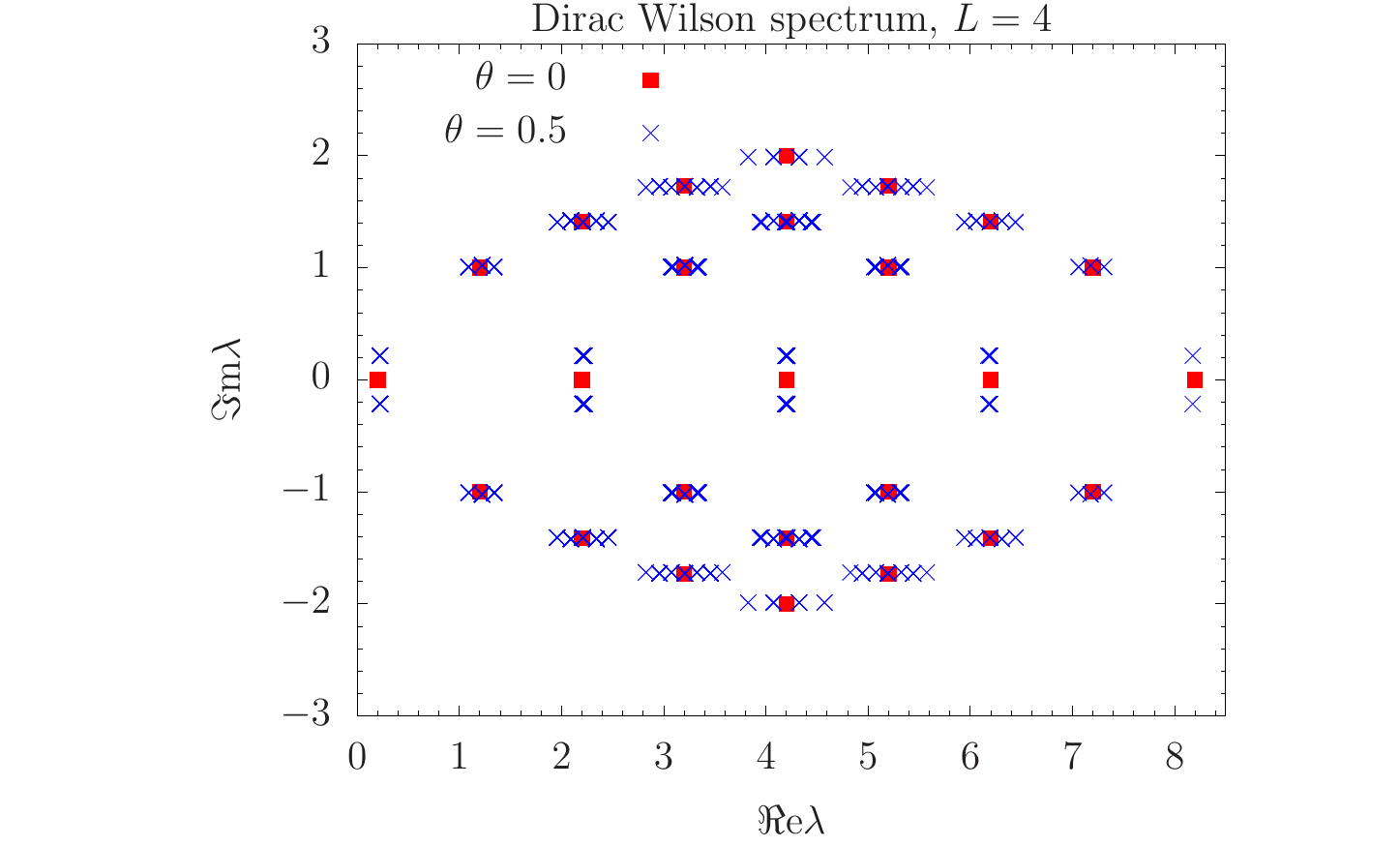}
\caption{\emph{Dirac-Wilson spectrum for dif\mbox{}ferent values of $\theta$ in $V=4^4$ volume and mass $m_0=0.2$.}}
\label{fig:dw_spect}
\end{center}
\end{figure}

We replace the standard links using Eq.~\ref{eq:modified_links}. In the free case ($U=\I$), the forward and backward derivatives become
\begin{align}
\nn
\nabla^+_\mu(\theta)\psi(x)&=\left[\mathcal{U}_\mu(x)\psi(x+\hat{\mu})-\psi(x)\right]=\left[\e^{i\theta_\mu/L_\mu}\psi(x+\hat{\mu})-\psi(x)\right],\\
\nabla^-_\mu(\theta)\psi(x)&=\left[\psi(x)-\mathcal{U}^\dagger_\mu(x-\hat{\mu})\psi(x-\hat{\mu})\right]=\left[\psi(x)-\e^{-i\theta_\mu/L_\mu}\psi(x-\hat{\mu})\right].
\end{align}
The Dirac-Wilson action in momentum space reads
\begin{align}
\nn
\text{S}_W(\theta)&
=\sum_{k\in\tilde{\Lambda}}
\widetilde{\overline{\psi}}(k)
\Bigg\{\Bigg[\left(m+4\right)-
\sum_{\mu=0}^3\cos\left(k_\mu+\theta_\mu/L_\mu\right)\Bigg]\I_{4\times 4}\\
&\phantom{=\sum_{k\in\tilde{\Lambda}}\widetilde{\overline{\psi}}(k)\Bigg[\Bigg[}
+i\Bigg[
\sum_{\mu=0}^3\gamma_\mu\sin\left(k_\mu+\theta_\mu/L\right)\Bigg]\Bigg\}\widetilde{\psi}(k).
\end{align}
We recall that the $k$'s are now in the first Brillouin zone.\\
The Dirac-Wilson eigenvalues, at fixed momentum for each color degrees of freedom, are
\begin{align}
\nn
\lambda_{1,2}&=\left(m+4\right)-\sum_{\mu=0}^3\cos\left(k_\mu+\theta_\mu/L_\mu\right) + i\left(\sum_{\mu=0}^3\sin^2 \left(k_\mu+\theta_\mu/L_\mu\right)\right)^{1/2},\\
\label{eq:eigenvalues_dw}
\lambda_{3,4}&=\left(m+4\right)-\sum_{\mu=0}^3\cos\left(k_\mu+\theta_\mu/L_\mu\right) - i\left(\sum_{\mu=0}^3\sin^2 \left(k_\mu+\theta_\mu/L_\mu\right)\right)^{1/2}.
\end{align}
In Fig.~\ref{fig:dw_spect} we show the ef\mbox{}fect of $\theta_\mu = (0, \underline{\theta})$ on the Dirac-Wilson spectrum.

%%%%%%%%%%%%%%%%%%%%%%%%%%%%%%%%%%%%%%%%%%%%%%%%%%%%%%%%%%%%%%%%%%%%%%%%%%
%%%%%%%%%%%%%%%%%%%%%%%%%%%%%%%%%%%%%%%%%%%%%%%%%%%%%%%%%%%%%%%%%%%%%%%%%%

\subsection{Reweighting factors: tree-level analytic evaluation}

From the knowledge of the eigenvalues Eq.~\ref{eq:eigenvalues_dw} we write down the tree-level reweighting factor Eq.~\ref{eq2:rew_fact}
\begin{align}
\label{eq:rew}
W_\theta=&\prod_{k\in\widetilde{\Lambda}}\left\{\frac{\left[\left(m+4\right)-\sum_{\mu}\cos\left(k_\mu+\theta_\mu/L_\mu\right)\right]^2+\sum_{\mu}\sin^2\left(k_\mu+\theta_\mu/L_\mu\right)}
{\left[\left(m+4\right)-\sum_{\mu}\cos\left(k_\mu\right)\right]^2+\sum_{\mu}\sin^2\left(k_\mu\right)}\right\}^{2N_c},
\end{align}
recalling that $\theta_0=0$.

\paragraph{Small $\theta$ expansion}

We perform a perturbative expansion around $\theta_\mu = 0$. The exponential of the action becomes
\begin{align}
\exp\left(-\text{S}_\theta\right) \simeq \exp\left(-\text{S} + \theta_\mu J_\mu\right) \simeq \exp\left(-\text{S}_\theta\right) \left(1+\theta_\mu J_\mu \right),
\end{align}
with a suitable current $J_\mu$.
The vacuum expectation value of an operator $\mathcal{O}$ becomes $\langle\mathcal{O}\rangle_\theta \simeq \langle\mathcal{O}\rangle + \theta_\mu \langle\mathcal{O} J_\mu\rangle$.
We know that $\langle\mathcal{O} J_\mu\rangle \equiv 0$ vanishes because of Lorentz invariance, so we conclude that any expectation value must start 
with a correction\footnote{Similar argument is made for the expansion of the observable in $\theta$, if it depends on the twisting angle.} of order $\theta^2$
\begin{equation}
\langle\mathcal{O}\rangle_\theta \simeq \langle\mathcal{O}\rangle + \ord\left(\theta^2\right).
\end{equation}

\paragraph{Large $m$ expansion}

We analyze the behavior of the tree-level reweighting factor in Eq.~\ref{eq:rew}
 for $m\rightarrow\infty$.\\
We start by looking at the numerator
\begin{align}
\nonumber
&\left[\left(m+4\right)-\sum_{\mu=0}^3\cos\left(k_\mu+\theta_\mu/L_\mu\right)\right]^2+\sum_{\mu=0}^3\sin^2\left(k_\mu+\theta_\mu/L_\mu\right)\\
\label{eq_app:d_star}
&\simeq\left(m+4\right)^2-2m\sum_{\mu=1}^4\cos\left(k_\mu+\theta_\mu/L_\mu\right)\simeq m^2+2m\left(4-\sum_{\mu=0}^3\cos\left(k_\mu+\theta_\mu/L_\mu\right)\right)\, .
\end{align}
We build the ratio and we obtain 
\begin{align}
\nn
&\frac{m^2+2m\left(4-\sum_{\mu=0}^3\cos\left(k_\mu+\theta_\mu/L_\mu\right)\right)}
{m^2+2m\left(4-\sum_{\mu=0}^3\cos k_\mu \right)}\simeq
\frac{m^2+2m\left(4-\sum_{\mu=0}^3\cos\left(k_\mu+\theta_\mu/L_\mu\right)\right)}
{m^2\left[1+\frac{2}{m}\left(4-\sum_{\mu=0}^3\cos k_\mu \right)\right]}\\
\nn
&\simeq
\left[1+\frac{2}{m}\left(4-\sum_{\mu=0}^3\cos\left(k_\mu+\theta_\mu/L_\mu\right)\right)\right]
\left[1-\frac{2}{m}\left(4-\sum_{\mu=0}^3\cos k_\mu \right)\right]\\
&\simeq
1+\frac{2}{m}\sum_{\mu=0}^3\left[\cos k_\mu -\cos\left(k_\mu+\theta_\mu/L_\mu\right)\right].
\end{align}
The reweighting factor become
\begin{align}
\nonumber
W_\theta&\simeq\prod_{k\in\widetilde{\Lambda}}\left\{1+\frac{2}{m}\sum_{\mu=0}^3\left[\cos k_\mu -\cos \left(k_\mu+\theta_\mu/L_\mu\right)\right]\right\}^{2N_c}\\
\nonumber
&\simeq\prod_{k\in\widetilde{\Lambda}}\left\{1+\frac{4N_c}{m}\sum_{\mu=0}^3\left[\cos k_\mu-\cos\left(k_\mu+\theta_\mu/L_\mu\right)\right]\right\}\\
\nonumber
&\simeq 1+\frac{4N_c}{m}\sum_{k\in\widetilde{\Lambda}}\left\{\sum_{\mu=0}^3\left[\cos k_\mu -\cos\left(k_\mu+\theta_\mu/L_\mu\right)\right]\right\}\\
&\simeq 1+\frac{4N_c}{m}\sum_{k\in\widetilde{\Lambda}}\left\{\sum_{j=1}^3\left[\cos k_j -\cos\left(k_j+\theta/L_j\right)\right]\right\}.
\end{align}
Here the integral over the interval $\left(-\pi, \pi\right)$ of the function in the curly brackets is vanishing, which implies that the correction in $1/m$ starts to the second order,
\begin{equation}
\label{eq_app:m_inf}
W_\theta = 1+\ord\left(m^{-2}\right).
\end{equation}

\section{Observables with TBCs}
\label{app:obs_tbcs}

Here we give the relevant formulae for the observables, i.e. the plaquette and the pion dispersion relation, studied in the reweighting project in Chap.~\ref{chap:rtbc}.

\subsection{Plaquette}

The most simple observable that we can think to reweight is the \emph{plaquette} because it has no dependence on the field $B = \theta / L$.\\
The plaquette on the lattice is defined as the shortest loop path and is the product of links
\begin{equation}
P[U]=\tr\bigg\{\prod_{\left(x, \mu\right)\in\text{loop}}U_\mu\bigg\}.
\end{equation}
In the case of constant $\mathbf{U}(1)$ interaction (Eq.~\ref{eq:modified_links}) the plaquette becomes
\begin{equation}
\label{eqapp:plaquette}
P[\mathcal{U}] = \tr\bigg\{\prod_{\left(x, \mu\right)\in\text{loop}}\mathcal{U}_\mu\bigg\} = \tr\bigg\{\prod_{\left(x, \mu\right)\in\text{loop}}\e^{i B_\mu}U_\mu\bigg\} = \tr\bigg\{\prod_{\left(x, \mu\right)\in\text{loop}}U_\mu\bigg\}=P[U],
\end{equation}
because we always find two dif\mbox{}ferent modified links pointing in opposition direction and therefore the constant $B$-term disappear.\\
The reweighted plaquette expectation value is then
\begin{align}
\langle P[\mathcal{U}]\rangle_\theta = \langle P[U]\rangle_\theta = \frac{\langle P[U]W_\theta\rangle_0}{\langle W_\theta\rangle_0}.
\label{eq:rew_plaq}
\end{align}

\subsection{Valence twisting and pion dispersion relation}

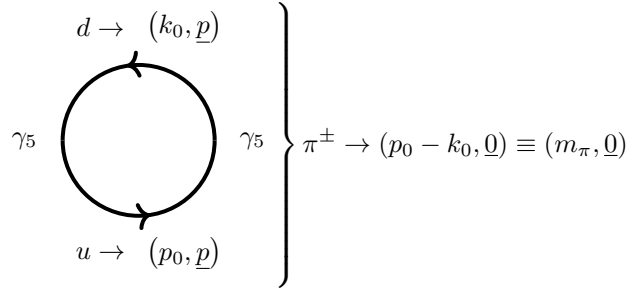
\begin{figure}[htb!]
\begin{equation*}
\begin{rcases}
\begin{tikzpicture}[line width=1.5 pt, scale=2]
	\draw[fermion] (0.5,0) arc (0:180:.5);
	\draw[fermion] (-0.5,0) arc (-180:0:.5);
	% The labels
	\node at (0.75,0){$\gamma_5$};
	\node at (-0.75,0){$\gamma_5$};
	\node at (-0.25,0.75){$d\rightarrow$};
	\node at (-0.25,-0.75){$u\rightarrow$};
	\node at (0.3,0.75){$\left(k_0, \underline{p}\right)$};
	\node at (0.3,-0.75){$\left(p_0, \underline{p}\right)$};
\end{tikzpicture}
\end{rcases}
\pi^\pm\rightarrow\left(p_0-k_0, \underline{0}\right)\equiv\left(m_\pi, \underline{0}\right)
\end{equation*}
\caption{\emph{Diagrammatic expression of the Eq.~\ref{eq:trace_wick}. We show on the diagram the momentum flow.}}
\label{fig:pion_untwist}
\end{figure}

We are interested in expectation values of meson interpolators when two flavors $f_1, f_2$ are involved. The most general form for such interpolators is
\begin{equation*}
\mathcal{O}_\text{M}(x)=\overline{\psi}^{(f_1)}(x)\Gamma\psi^{(f_2)}(x), \quad \overline{\mathcal{O}}_\text{M}(y)=\overline{\psi}^{(f_2)}(y)\Gamma\psi^{(f_1)}(y),
\end{equation*}
where $\Gamma$ is an element of the Clif\mbox{}ford algebra, i.e.~combination of the $\gamma$ matrices. The operator $\overline{\mathcal{O}}$ creates a meson, with the right quantum numbers, from the vacuum and $\mathcal{O}$ annihilates that state.\\
We need to compute the Grassmann integrals in order to calculate the fermionic expectation value $\langle\dots\rangle_\text{F}$ . We choose the two flavors to be $f_1=u$, $f_2=d$, in order to have an \emph{iso-triplet} operator\footnote{To avoid the contribution of disconnected diagrams that suf\mbox{}fer from the signal-to-noise problem.}.\\
The fermionic expectation values can be computed by employing Wick contractions as follows
\begin{align}
\nonumber
\langle\mathcal{O}(x)\overline{\mathcal{O}}(y)\rangle_\text{F} &= \big\langle \big(\overline{d}\Gamma u\big)(x)\left(\overline{u}\Gamma d\right)(y)\big\rangle_\text{F}\\
\nonumber
&=\Gamma_{\alpha_1\beta_1}\Gamma_{\alpha_2\beta_2}\big\langle \big(\overline{d}_{\alpha_1,c_1} u_{\beta_1,c_1}\big)(x)\big(\overline{u}_{\alpha_2,c_2} d_{\beta_2,c_2}\big)(y)\big\rangle_\text{F}\\
\nonumber
&=-\Gamma_{\alpha_1\beta_1}\Gamma_{\alpha_2\beta_2}\big\langle u_{\beta_1,c_1}(x) \overline{u}_{\alpha_2,c_2}(y)\big\rangle_u\, \big\langle d_{\beta_2,c_2}(y)\overline{d}_{\alpha_1,c_1}(x)\big\rangle_d\\
\nonumber
&=-\Gamma_{\alpha_1\beta_1}\Gamma_{\alpha_2\beta_2}D^{-1}_u(y,x)_{\beta_1,\alpha_2;c_1,c_2} D^{-1}_{d}(y,x)_{\beta_2,\alpha_1;c_2,c_1}\\
\label{eq:trace_wick}
&= -\tr\left[\Gamma D^{-1}_u(x,y)\Gamma D^{-1}_{d}(y,x)\right],
\end{align}
where in the second line we explicitly showed Dirac indexes with Greek letters and color indexes with Latin ones; in third line we used the factorization property of the fermionic expectation value with respect to flavor\footnote{The action 
and the partition function are separable in the flavor space, 
i.e.~$\text{S}_W = \sum_f\overline{\psi}_f D_W \psi_f = \text{S}_u + \text{S}_d$, $\mathcal{Z}_{\text{F}} = \mathcal{Z}_u \mathcal{Z}_d$. 
That implies that the fermionic expectation value can be written as 
$\langle\dots\rangle_{\text{F}} = \mathcal{Z}_{\text{F}}^{-1}\int\De\left[u, \overline{u}\right] \De\big[d, \overline{d}\big]\e^{-\text{S}_{\text{F}}}\left(\dots\right) = \left[\mathcal{Z}_u^{-1} \int\De\left[u, \overline{u}\right] \e^{-S_u}\left(\dots\right)_u\right] \left[\mathcal{Z}_d^{-1} \int\De\big[d, \overline{d}\big] \e^{-S_d}\left(\dots\right)_d\right] = \langle\dots\rangle_u \langle\dots\rangle_d$.}  $\langle\dots\rangle_\text{F} =\langle\dots\rangle_\text{u} \langle\dots\rangle_\text{d}$
and we interchanged the fermion spinors according to their Grassmann nature. 
In Fig.~\ref{fig:pion_untwist} we draw in a diagrammatic way the result of the above Wick contractions in the case of $\mathcal{O}=\overline{d}\gamma_5 u$ at zero momentum, which correspond to a charged pion $\pi^\pm$.
In order to extract the \emph{dispersion relation} from the correlator we project the interpolators to a definite momentum, say $p$.
The correlator has the following behaviour for large time separations ($x_0\gg 1$)
\begin{align}
\nn
\langle \widetilde{\mathcal{O}}(x_0, \underline{p})\overline{\mathcal{O}}(0)\rangle &= \frac{1}{\sqrt{|V|}}\sum_{\underline{x}\in V_3}\e^{-i \underline{x}\cdot\underline{p}}\langle\mathcal{O}(x)\overline{\mathcal{O}}(0)\rangle\\
\label{eq:latt_zero_mom_pion_corr}
&=\sum_k\langle 0|O|k\rangle\langle k|O^\dagger|0\rangle\e^{-x_0 E_k} \underset{x_0\gg 1}{=} A\e^{-x_0 E(\underline{p})},
\end{align}
where $E(\underline{0})=m_\pi$.\\
The free dispersion relation on the lattice is obtained from the free boson lattice propagator in momentum space (inverse of the Klein-Gordon operator on the lattice), and restoring the lattice spacing $a$ we get
\begin{equation}
\cosh (aE) = \cosh (am_\pi) + \sum_{k=1}^3\left[1-\cos (ap_k)\right]
\underset{ap\rightarrow 0}{\longrightarrow}E^2=m_\pi^2+|\underline{p}|^2.
\end{equation}

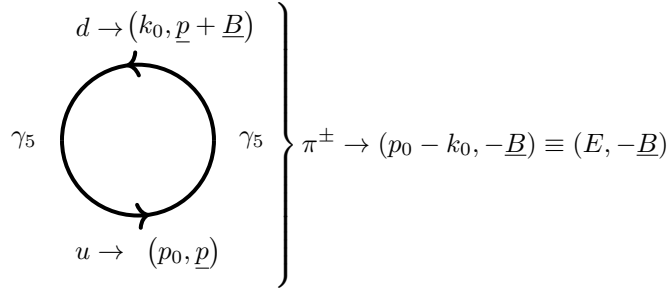
\begin{figure}[htb!]
\begin{equation*}
\begin{rcases}
\begin{tikzpicture}[line width=1.5 pt, scale=2]
	\draw[fermion] (0.5,0) arc (0:180:.5);
	\draw[fermion] (-0.5,0) arc (-180:0:.5);
	% The labels
	\node at (0.75,0){$\gamma_5$};
	\node at (-0.75,0){$\gamma_5$};
	\node at (-0.25,0.75){$d\rightarrow$};
	\node at (-0.25,-0.75){$u\rightarrow$};
	\node at (0.35,0.75){$\left(k_0, \underline{p}+\underline{B}\right)$};
	\node at (0.3,-0.75){$\left(p_0, \underline{p}\right)$};
\end{tikzpicture}
\end{rcases}
\pi^\pm\rightarrow\left(p_0-k_0, -\underline{B}\right)\equiv\left(E, -\underline{B}\right)
\end{equation*}
\caption{\emph{Diagrammatic expression of the Eq.~\ref{eq:trace_wick} in the presence of the constant $\mathbf{U}(1)$ interaction. The state correspond to a charged pion with momentum \underline{B}.}}
\label{fig:pion_twist}
\end{figure}

In the presence of a constant $\mathbf{U}(1)$ interaction in the spatial direction we perform the substitution $\underline{p}\rightarrow \underline{p}+\underline{B}$.
That corresponds to 
a twisting in the valence only.
In Fig.~\ref{fig:pion_twist} we show the change on the momentum flow in the diagram in the zero momentum pion case.
The dispersion relation in this case is
\begin{equation}
\cosh (aE) = \cosh (am_\pi) + \sum_{k=1}^3\left[1-\cos(ap_k+aB_k)\right]
\underset{ap\rightarrow 0}{\longrightarrow}E^2=m_\pi^2+|\underline{p}+\underline{B}|^2.
\label{eq:rew_disp_rel}
\end{equation}
It is clear that the charged pion momentum on the lattice becomes
\begin{equation}
\underline{p}_\pi=\underline{p}+\underline{B}=\frac{2\pi\underline{n}+\underline{\theta}}{aL}.
\end{equation}
\appendix
\setcounter{chapter}{2}
\renewcommand{\chaptername}{Appendix}
\renewcommand{\theequation}{\Alph{chapter}.\arabic{section}.\arabic{equation}}
\setcounter{equation}{0}

\chapter{HMC simulations}

\section{Acceptance in HMC simulations}
\label{app:acc_hmc}

We follow Ref.~\cite{Gupta:1990ka} to work out the acceptance probability.\\
The acceptance is found from the Creutz equality, which comes from the area preserving requirement,
\begin{align}
\langle \e^{-\Delta H} \rangle = 1,
\end{align}
where $\Delta H = H_f - H_i$, and by expanding it into cumulants we get
\begin{align}
\label{eq:cumulants_exp}
\langle \Delta H \rangle = \frac{1}{2} {\rm Var}(\Delta H) + {\rm higher\,\, cumulants}.
\end{align}
When the volume is large enough, i.e.~compared with the relevant correlation length, the cumulants will grow linearly while their dependence on  the step size is dif\mbox{}ferent in $\delta\tau$. In order to have a finite $\langle \Delta H \rangle$ then $\delta\tau$ should be varied to keep the variance fixed, ${\rm Var}(\Delta H)={\rm const}$.\\
The distribution for the acceptance will be a convolution of a Gaussian with mean and width related by $\langle \Delta H \rangle = {\rm Var}(\Delta H)/2 \equiv \sigma^2_0/2$ and an Metropolis test
\begin{align}
\nn
P_{\rm acc}(\Delta H) &= \frac{1}{\sqrt{2\pi}\sigma_0}\int_{-\infty}^\infty {\rm min}(1, \e^{-x}) \exp\left[ -\frac{(x-\langle \Delta H\rangle )^2}{2\sigma_0^2} \right]\de x\\
&= \frac{1}{\sqrt{2\pi}\sigma_0} \left\{ \int_{-\infty}^0 \exp\left[ -\frac{(x-\langle \Delta H\rangle )^2}{2\sigma_0^2} \right]\de x + \int_{0}^\infty \exp\left[ -\frac{(x-\langle \Delta H\rangle )^2}{2\sigma_0^2} - x\right]\de x  \right\}.
\end{align}
By changing the coordinates in the first integral to $z = \frac{x-\langle \Delta H\rangle}{\sqrt{2}\sigma_0} $ we get
\begin{align}
\nn
{\rm first\,\, integral} &= 2\langle\Delta H\rangle^{1/2} \int_{-\infty}^{- \frac{\sqrt{\langle\Delta H\rangle}}{2}} \e^{-z^2}\de z = 2\langle\Delta H\rangle^{1/2} \int_{ \frac{\sqrt{\langle\Delta H\rangle}}{2}}^{\infty} \e^{-z^2}\de z\\
& =  \sqrt{\pi}\langle\Delta H\rangle^{1/2} {\rm erfc}\left( \frac{\sqrt{\langle\Delta H\rangle} }{2}  \right),
\end{align}
where we used the symmetry of the Gaussian integral in the first step.\\
For the second integral we first recognize the square and use the relation between the variance and the mean
\begin{align}
\frac{(x-\langle \Delta H\rangle )^2}{2\sigma_0^2} + x = \frac{(x+\langle \Delta H\rangle)^2}{4\langle \Delta H\rangle},
\end{align}
and then change the variable to $z = \frac{x+\langle \Delta H\rangle}{2 \sqrt{\langle \Delta H\rangle}} $, obtaining
\begin{align}
{\rm second\,\, integral} = 2\langle\Delta H\rangle^{1/2} \int_{ \frac{\sqrt{\langle\Delta H\rangle}}{2}}^{\infty} \e^{-z^2}\de z =  \sqrt{\pi}\langle\Delta H\rangle^{1/2} {\rm erfc}\left( \frac{\sqrt{\langle\Delta H\rangle} }{2}  \right).
\end{align}
By putting all together, with the right pre-factors, we get the formula
\begin{align}
P_{\rm acc}(\Delta H) = {\rm erfc}\left( \frac{\sqrt{\langle\Delta H\rangle} }{2}  \right).
\end{align}
We use \ref{eq:cumulants_exp} to trade $\langle \Delta H \rangle$ with ${\rm Var}(\Delta H)$ because the latter is a better behaved quantity\footnote{In the mean $\langle \Delta H \rangle$ we can have negative and positive contributions and that may lead to large uncertainties.}
\begin{align}
\label{app:acc_rate}
P_{\rm acc}(\text{Var} (\Delta H)) = {\rm erfc}\left( \sqrt{\frac{\text{Var}\left(\Delta H\right) }{8}}  \right).
\end{align}

\section{Multiple time-scale Omelyan shadow Hamiltonian}
\label{app:mts_osh}

Following Ref.~\cite{Clark:2008gh} we give the expression for the shadow 
Hamiltonian in the case of multi-time scale.\\
Let the Hamiltonian be 
\begin{align}
H = {\rm T} + \sum_{i} {\rm S}_i\, ,
\end{align}
with $i$ index running to as many multi-time scales are present.\\
Let be $\exp Z = \exp A \exp B$, then the Baker-Campbell-Hausdorf\mbox{}f formula is given by
\begin{align}
\nn
Z = &(A + B) 
+ [A, B] 
+\frac{1}{12} \left( [A, [A, B]] 
+ [B, [B, A]] \right) \\
& - \frac{1}{24} [B, [A, [A, B]]] 
- \frac{1}{720} \left( [B, [B, [B, [B, A]]]]
+  [A, [A, [A, [A, B]]]]\right) + \dots
\end{align}
From the above formula we can evaluate the following expression
\begin{align}
\label{eqapp:bchbch}
\ln\left(\exp(X) \exp(Y) \exp(X)\right) 
= (2 X + Y) 
+ \frac{1}{6} \left( [[X, Y], X] + [[X, Y], Y] \right) + \dots
\end{align}
We use the work done in Ref.~\cite{Reinsch:2000} in MATHEMATICA to calculate BCH and consequently the shadow Hamiltonian.

\subsection*{One level Leap-frog integrator}

The Leap-frog integrator is given by the following evolution operator
\begin{align}
\left[ \exp\left( \frac{\delta \tau}{2} \hat{S} \right) 
\exp\left( \delta\tau \hat{T} \right) 
\exp\left( \frac{\delta \tau}{2} \hat{S} \right)\right]^n,
\end{align}
with $\delta\tau = 1/n$, and by employing the BCH formula in Eq.~\ref{eqapp:bchbch} we get
\begin{align}
\left(\exp \left[ \hat{H}\delta\tau 
+ \left( \frac{1}{12}   [ [\hat{S}, \hat{T}], \hat{T}]
+  \frac{1}{24}  [[\hat{S}, \hat{T}], \hat{S}]  \right) \delta\tau^3 \right]\right)^n.
\end{align}
By applying the BCH formula again we get
\begin{align}
\widetilde{H} 
= H 
+ \left( \frac{1}{12} [[\hat{S}, \hat{T}], \hat{T}]
+ \frac{1}{24} [[\hat{S}, \hat{T}], \hat{S}] \right) \delta\tau^2.
\end{align}

\subsection*{One-level Omelyan integrator}

The one-level Omelyan integrator is given by
\begin{align}
\left[
\exp\left(\alpha \delta\tau \hat{S}_1\right)
\exp\left(\frac{\delta\tau}{2} \hat{T}\right)
\exp\left(\delta\tau (1-2\alpha) \hat{S}_1\right) 
\exp\left(\frac{\delta\tau}{2} \hat{T}\right)
\exp\left(\alpha \delta\tau \hat{S}_1\right)
\right]^n.
\end{align}
The shadow Hamiltonian is found to be
\begin{align}
\widetilde{H} 
= H 
+  \left[  \frac{6\alpha^2 - 6 \alpha + 1 }{12} [[\hat{T}, \hat{S}_1], \hat{S}_1]
+  \frac{6\alpha - 1 }{24} [[\hat{S}_1, \hat{T}], \hat{T}] \right] \delta\tau^2.
\end{align}

\subsection*{Two-level Omelyan integrator}

The two-level Omelyan integrator is obtained by operating the 
following formal substitution to the one-level Omelyan integrator
\begin{align}
\exp\left(\frac{\delta\tau}{2} \hat{T}\right)
\longrightarrow
\left[
\exp\left(\alpha \frac{\delta\tau}{2m} \hat{S}_2\right)
\exp\left( \frac{\delta\tau}{4m} \hat{T}\right)
\exp\left( \frac{\delta\tau}{2m} (1-2\alpha) \hat{S}_2\right) 
\exp\left(\frac{\delta\tau}{4m} \hat{T}\right)
\exp\left(\alpha \frac{\delta\tau}{2m} \hat{S}_2\right)
\right]^m\, .
\end{align}
The shadow Hamiltonian is found to be
\begin{align}
\nn
\widetilde{H} 
= &H 
-  \delta\tau^2\frac{6\alpha^2 - 6 \alpha + 1 }{12}
\left[ [[\hat{S}_1, \hat{T}], \hat{S}_1]
+ \frac{1}{2 m^2} [[\hat{S}_2, \hat{T}], \hat{S}_2] \right]\\
&+  \delta\tau^2 \frac{6\alpha - 1 }{24} 
\left[  [[\hat{S}_1, \hat{T}], \hat{T}] +
 +\frac{1}{2m^2}[[\hat{S}_2, \hat{T}], \hat{T}]
 +\frac{1}{2m^2}[[\hat{S}_2, \hat{T}], \hat{S}_1]
 \right] .
\end{align}

\subsection*{Multi-level Omelyan integrator}

By playing the same game we can recognize a structure in the shadows 
Hamiltonian, which will be given by
\begin{align}
\nn
\widetilde{H} =
 H 
+ \tau^2 \sum_{i=1} \Bigg[ &\frac{6\alpha^2 - 6 \alpha + 1 }{3} 
  \left( \frac{ [\hat{S}_i, [\hat{S}_i, \hat{T}]] }{\prod_{j=1}^{i} (2 n_j)^2}\right)\\
&+ \frac{6\alpha - 1 }{6} 
 \left( \frac{ [\hat{T}, [\hat{T}, \hat{S}_i]] 
 - \sum_{j<i} [\hat{S}_j, \{\hat{S}_i, \hat{T}]] }
 {\prod_{j=1}^{i} (2 n_j)^2}\right)
 \Bigg]\, ,
\end{align}
where $\tau = \delta\tau \times n_0$ and $n_0, n_1, \dots n_i$ are the $i$ number of integration steps per level.

\section{Forces in HMC}
\label{app:mts_osh}

The forces in HMC are derived as in Ref.~\cite{DelDebbio:2008zf}.
In the following we omit the sum over space-time, color and spin.
The Hamiltonian in the HMC evolution is given by $H = {\rm T} + {\rm S}_{\rm G} + {\rm S}_{\rm F}$,
\begin{itemize}
\item Gaussian weight for the momenta $\pi$, 
\begin{align}
{\rm T} = \frac{\pi^2}{2},
\end{align}
where $\pi\in\mathfrak{su}(N_c)$, $\pi_\mu(x) = i \pi^a_\mu(x) T^a_f$, where $T^a_f$ are the hermitian generators in the fundamental representation.
\item Wilson plaquette gauge action 
\begin{align}
{\rm S}_{\rm G} = \beta \sum_{\mu < \nu} \left( 1- \frac{1}{N_c}\Ree\tr P_{\mu\nu}  \right),
\end{align}
where $\beta = 2 N_c/g^2$, and $P_{\mu\nu}$ the plaquette in the $(\mu, \nu)$ plane
\begin{align}
P_{\mu\nu} = U_{\mu}(x) U_\nu (x+\hat{\mu}) U_\mu^\dagger (x+\nu) U_\nu^\dagger (x).
\end{align}
\item Fermionic action
\begin{align}
{\rm S}_{\rm F} = \phi^\dagger \left(Q^2\right)^{-1} \phi,
\end{align}
where $\phi$ is a pseudofermion, $Q= \gamma_5 D_m$ is the $\gamma_5$ Dirac-Wilson operator, $D_W$ that is given by the following equation
\begin{align}
D_W\psi(x) &= 8 \psi(x) - \frac{1}{2} \sum_\mu  \left[  (1-\gamma_\mu) U^{R}_\mu(x) \psi(x+\hat{\mu}) + (1+\gamma_\mu) U^{R, \dagger}_\mu (x-\hat{\mu}) \psi(x-\hat{\mu})  \right]  \\
D_m\psi(x) &= (D_W + m_0)\psi(x).
\end{align}
The $U^R_\mu$ are the link variables in the representation $R$.
\end{itemize}
A generic element $x$ in the algebra $\mathfrak{su}(N_c)$ is written as $x = i x^a T^a_f$,
and its norm is given by $||x||^2 = \tr(x^\dagger x) = T_{f} \sum_{i,j} |x_{ij}|^2$. 
$T_f$ is the normalization of the generators in the fundamental representation, for other representation we write $T_R$.

\subsection*{Equation of motion}

The equation of motion for the links is found by taking the derivative 
with respect to $\tau$ (the Markov time) of $\I = U_\mu(\tau) U_\mu^\dagger(\tau)$
\begin{align}
\dot{U}_\mu(\tau) U_\mu^\dagger(\tau) + U_\mu(\tau) \dot{U}_\mu^\dagger(\tau) = 0.
\end{align}
The equation is fulfilled by taking
\begin{align}
\dot{U}_\mu(x) = \pi_\mu (x) U_\mu(x).
\end{align}
The equation of motion for the momenta can be obtained by requiring that the Hamiltonian $H$ is 
a conserved quantity
\begin{align}
\dot{H} = 0 = \dot{{\rm T}} + \dot{{\rm S}}_{\rm G} + \dot{{\rm S}}_{\rm F}.
\end{align}
For the first two derivatives we have
\begin{align}
\dot{{\rm T}} &= T_{f} \sum_{x,\mu,a} \pi^a_\mu(x) \dot{\pi}^a_\mu(x)\, ,\\
\nn
\dot{{\rm S}}_{\rm G} &= -\frac{\beta}{N_c} \sum_{x, \mu}  \Ree \tr \left[   \dot{U}_\mu(x)  V^\dagger_{\mu}(x)    \right]	\\
\nn
					&= -\frac{\beta}{N_c} \sum_{x, \mu}  \Ree \tr \left[   \pi_\mu(x) U_\mu(x)  V^\dagger_{\mu}(x)    \right]\\
					&= -\frac{\beta}{N_c} \sum_{x, \mu, a}  \pi^a_\mu(x) \Ree \tr \left[   iT^a_{f} U_\mu(x)  V^\dagger_{\mu}(x)    \right]\, ,
\end{align}
where $V_\mu(x)$ is sum of the staples around the link $U_\mu(x)$.\\
For the fermionic force we write
\begin{align}
\dot{{\rm S}}_{\rm F} &= -\phi^\dagger \left(Q^2\right)^{-1} \dot{(Q^2)}  \left(Q^2\right)^{-1} \phi\, ,
\end{align}
and we define $\eta = \left(Q^2\right)^{-1}\phi$ and $\xi = Q \eta$.\\
By using the hermiticity of $Q^2$ we rewrite the derivative as 
\begin{align}
\label{eq:standard_hmc_force}
\dot{{\rm S}}_{\rm F} &= -2 \xi^\dagger \dot{Q}\, \eta.
\end{align}
Inserting the explicit form of $Q = \gamma_5 D_W$ we get
\begin{align}
\nn
\dot{{\rm S}}_{\rm F} = \Ree \sum_{x, \mu} &\xi^\dagger(x) \dot{U}^R_\mu(x) \gamma_5 (1- \gamma_\mu) \eta(x+\hat{\mu}) \\
&+  \eta^\dagger(x) \dot{U}^{R,\dagger}_\mu(x) \gamma_5 (1- \gamma_\mu) \xi(x+\hat{\mu}) .
\end{align}
By collecting the results we get the equation of motion for the momenta
\begin{align}
\dot{\pi}^a_\mu(x) = & \,\dot{\pi}^{a, {\rm G}}_\mu(x)  + \dot{\pi}^{a, {\rm F}}_\mu(x),\\
\dot{\pi}^{a, {\rm G}}_\mu(x) = & \frac{\beta}{N_c}\frac{1}{T_f} \Ree \tr \left[ i T^a_f U_\mu(x) V^\dagger_\mu(x) \right]
\equiv F^{a\mu}_{\rm G} (x) , \\
\dot{\pi}^{a, {\rm F}}_\mu(x) = & -\frac{1}{T_f} \Ree\tr\left\{  iT^a_R U^R_\mu(x) \gamma_5 (1-\gamma_\mu) \left[ \eta(x+\hat{\mu})\otimes\xi^\dagger(x)
+ \xi(x+\hat{\mu})\otimes\eta^\dagger(x)  \right]    \right\} 
\equiv F^{a\mu}_{\rm F} (x).
\end{align}
Where we denoted as $F$ the associated driving forces to the dif\mbox{}ferent part of the action.

\subsection{Mass preconditioning forces}

The mass preconditioning is given in Eq.~\ref{eq:mass_prec}. The $Q_{\rm HMC}$ part follows exactly as before but with a shift in the mass given by $m_0 + \mu$.
We derive here the force associated to the Hasenbusch term.
We write the associated action as, with $\phi_2$ dif\mbox{}ferent set of pseudofermions, 
\begin{align}
{\rm S}_{\rm Hase} = \phi_2^\dagger\, Q^{-1} \left(D^\dagger_m + \mu\right)  \left(D_m + \mu\right) Q^{-1} \phi_2.
\end{align}
The operator can be rewritten as follows
\begin{align}
\nn
Q^{-1} \left(D^\dagger_m + \mu\right)  \left(D_m + \mu\right) Q^{-1} 
\nn
&= D_m^{-1}\gamma_5 \left(\gamma_5 D_m\gamma_5 + \mu\right)  \left(D_m + \mu\right) D_m^{-1} \gamma_5\\
\nn
&=  \left(\gamma_5 + \mu D_m^{-1} \gamma_5 \right)  \left(\gamma_5 + \mu D_m^{-1}\gamma_5\right)\\
&=  \left(\gamma_5 + \mu Q^{-1} \right)  \left(\gamma_5 + \mu Q^{-1} \right),
\end{align}
and the action becomes
\begin{align}
{\rm S}_{\rm Hase} = \phi_2^\dagger \left(\gamma_5 + \mu Q^{-1} \right)  \left(\gamma_5 + \mu Q^{-1} \right) \phi_2.
\end{align}
Now we take the derivative with respect to the Markov time
\begin{align}
\nn
\dot {{\rm S}}_{\rm Hase} &= \phi_2^\dagger \left[ \mu\dot{\left(Q^{-1} \right)}  \left(\gamma_5 + \mu Q^{-1} \right)
+  \mu\left(\gamma_5 + \mu Q^{-1} \right)  \dot{\left( Q^{-1} \right)}\right] \phi_2\\
&= -\mu\phi_2^\dagger \left[ Q^{-1} \dot{Q} Q^{-1}  \left(\gamma_5 + \mu Q^{-1} \right)
+  \left(\gamma_5 + \mu Q^{-1} \right)  Q^{-1} \dot{Q}   Q^{-1}   \right] \phi_2.
\end{align}
Note that the above equation has the standard form of a fermionic force in the HMC algorithm provided that
\begin{align}
\nn
X &\equiv Q^{-1} \phi_2,\\
Y & \equiv Q^{-1} \left( \gamma_5 + \mu Q^{-1}\right) \phi_2,
\end{align}
in particular, see Eq.~\ref{eq:standard_hmc_force},
\begin{align}
\dot {{\rm S}}_{\rm Hase} &= -2 \mu X^\dagger \dot{Q} Y.
\end{align}
The forces given by the dif\mbox{}ferent pieces of the action are
\begin{align}
F^{a\mu}_{\rm G} (x) = & \frac{\beta}{N_c}\frac{1}{T_f} \Ree \tr \left[ i T^a_f U_\mu(x) V^\dagger_\mu(x) \right], \\
F^{a\mu}_{\rm HMC} (x) = & -\frac{1}{T_f} \Ree\tr\left\{  iT^a_R U^R_\mu(x) \gamma_5 (1-\gamma_\mu) \left[ \eta(x+\hat{\mu})\otimes\xi^\dagger(x)
+ \xi(x+\hat{\mu})\otimes\eta^\dagger(x)  \right]    \right\}, \\
F^{a\mu}_{\rm Hase} (x) = & -\frac{\mu}{T_f} \Ree\tr\left\{  iT^a_R U^R_\mu(x) \gamma_5 (1-\gamma_\mu) \left[ Y(x+\hat{\mu})\otimes X^\dagger(x)
+ X(x+\hat{\mu})\otimes Y^\dagger(x)  \right]    \right\} ,
\end{align}
where 
\begin{align}
\xi &\equiv Q^{-1} \phi_1,\\
\eta &\equiv \left(Q^2\right)^{-1}\phi_1,\\
X &\equiv Q^{-1} \phi_2,\\
Y & \equiv Q^{-1} \left( \gamma_5 + \mu Q^{-1}\right) \phi_2,
\end{align}
and we recall that $Q$ in this case is the mass shifted $\gamma_5$ Dirac-Wilson operator
\begin{align}
Q= \gamma_5 (D_m + \mu) = \gamma_5 (D_W + m_0 + \mu).
\end{align}

\section{Number of configurations and parameters in the optimization study}
\label{app:ncp_op}

Here we collect the number of configurations analyzed in the force study in Chap.~\ref{chap:oHMCp}.
Simulations correspond to a $\mathbf{SU}(2)$ gauge group with 
a doublet of un-improved Wilson fermions in the fundamental representation
and Wilson plaquette gauge action in a $V=32^4$ volume.
The value $m_c$ 
of the bare mass parameter yielding massless fermions
is estimated to be $-0.77(2)$ at $\beta = 2.2$ \cite{Lewis:2011zb, Hietanen:2014xca, Arthur:2016dir}.
The integrator choice is the Omelyan with $\alpha = 1/6$. 
The inversions of the $\gamma_5$-hermitian Dirac-Wilson operator
are performed using a version of the Quasi-Minimal Residual (QMR).
The algorithm is not absolutely convergent and in the case
of no convergence our implementation switch to the BiCGstab algorithm.
For the evolution of the gauge field we use the Minimal Residual
Extrapolation (MRE) algorithm to build an initial guess for
the inverter based on the past solutions \cite{Brower:1995vx}.
In particular 5 past solutions where stored throughout the HMC simulation.
The force precision was set to $10^{-14}$ and the $\Delta H$ measurement
precision 
for the accept-reject step to $10^{-14}$, which is enough to ensure 
reversibility of the algorithm.
The level parameters are $n$ for the Hasenbusch, $m$ HMC and $k$ Gauge as described
in Sect.~\ref{sec:multiscale}.

\subsection*{$m_0 = -0.72$}
\label{app:m072}

\begin{align*}
n = 15, \quad m = 8, \quad k=10.
\end{align*}

\parbox{0.45\linewidth}{
\centering
\begin{tabular}{|c| c |} 
 \hline
 $\mu$ & $N_{\rm cnf}$ \\ [0.5ex] 
 \hline\hline
 0.05 & 109  \\ 
 \hline
 0.1 & 98  \\ 
 \hline
 0.15 & 128  \\ 
 \hline
 0.2 & 153  \\ 
 \hline
 0.25 & 128  \\ 
 \hline
 0.3 & 136  \\ 
 \hline
 0.35 & 144  \\ 
 \hline
 0.4 & 155  \\ 
 \hline
 0.45 & 112  \\ 
 \hline
 0.5 & 115  \\ [1ex]
 \hline
 \end{tabular}
%\caption{Foo}
}
\hfill
\parbox{0.45\linewidth}{
\centering
\begin{tabular}{|c| c |} 
 \hline
 $\mu$ & $N_{\rm cnf}$ \\ [0.5ex] 
 \hline\hline
 0.6 & 124  \\ 
 \hline
 0.7 & 130  \\ 
 \hline
 0.75 & 133  \\ 
 \hline
 0.8 & 69  \\ 
 \hline
 0.9 & 137  \\ 
 \hline 
 1 & 1193  \\ 
 \hline
 1.5 & 80  \\ 
 \hline
 2 & 1487  \\ 
 \hline
 2.5 & 88  \\ 
 \hline
 3 &  92 \\[1ex] 
 \hline
\end{tabular}
%\caption{Bar}
}

\subsection*{$m_0 = -0.735$}
\label{app:m0735}

\begin{align*}
n = 5, \quad m = 4, \quad k=10.
\end{align*}

\parbox{0.45\linewidth}{
\centering
\begin{tabular}{|c| c |} 
 \hline
 $\mu$ & $N_{\rm cnf}$ \\ [0.5ex] 
 \hline\hline
 0.05 & 133  \\ 
 \hline
 0.1 & 189  \\ 
 \hline
 0.15 & 202  \\ [1ex]
 \hline
 \end{tabular}
%\caption{Foo}
}
\hfill
\parbox{0.45\linewidth}{
\centering
\begin{tabular}{|c| c |} 
 \hline
 $\mu$ & $N_{\rm cnf}$ \\ [0.5ex] 
 \hline\hline
 0.2 & 202  \\ 
 \hline
 0.25 & 202  \\ 
 \hline
 0.3 & 202  \\ 
 \hline
 0.35 & 202 \\ [1ex] 
 \hline
\end{tabular}
}

\subsection*{$m_0 = -0.75$}
\label{app:m075}

\begin{align*}
n = 15, \quad m = 8, \quad k=10.
\end{align*}

\parbox{0.45\linewidth}{
\centering
\begin{tabular}{|c| c |} 
 \hline
 $\mu$ & $N_{\rm cnf}$ \\ [0.5ex] 
 \hline\hline
 0.05 & 57  \\ 
 \hline
 0.1 & 79  \\
 \hline
 0.15 & 119  \\ 
 \hline
 0.2 & 147  \\ 
 \hline
 0.25 & 152  \\ [1ex]
 \hline
 \end{tabular}
}
\hfill
\parbox{0.45\linewidth}{
\centering
\begin{tabular}{|c| c |} 
 \hline
 $\mu$ & $N_{\rm cnf}$ \\ [0.5ex] 
 \hline\hline
 0.3 & 178  \\ 
 \hline
 0.35 & 178  \\
 \hline
 0.4 & 196  \\ 
 \hline
 0.45 & 186  \\ 
 \hline
 0.5 & 201  \\ 
 \hline
 0.6 & 193  \\ [1ex]
 \hline
 \end{tabular}
}

\appendix
\setcounter{chapter}{3}
\renewcommand{\chaptername}{Appendix}
\renewcommand{\theequation}{\Alph{chapter}.\arabic{section}.\arabic{equation}}
\setcounter{equation}{0}

\chapter{Quantum electrodynamics on the lattice: supplementary material}

\section{Fourier transform}
\label{app:AFG}

We set the notation for the Fourier transform on the lattice. 
The reciprocal lattice $\widetilde{\Lambda}$, with periodic boundary conditions for a functions $f(n+\hat{\mu}L_\mu) = f(n)\;\forall\mu$, is given by
\begin{align*}
\widetilde{\Lambda} = \left\{ p=(p_0, p_1, p_2, p_3): p_\mu = \frac{2\pi}{aL_\mu} k_\mu, \text{ with } k_\mu\in\left(-\frac{L_\mu}{2}, \frac{L_\mu}{2}\right]\subset\mathbb{Z} \right\}.
\end{align*}
The delta functions are defined as
\begin{align}
\nonumber
\delta(x-x') &= \frac{1}{V}\sum_{p\in\widetilde{\Lambda}} \exp\left(ip\cdot (x-x')\right),\\
\label{app:delta_funct_mom}
\delta(p-p') &= \frac{1}{V}\sum_{n\in\Lambda} \exp\left(i(p-p')\cdot x\right).
\end{align}
The Fourier transform and the inverse are defined as
\begin{align}
\nn
\widetilde{f}(p) &= \frac{1}{\sqrt{V}}\sum_{n\in\Lambda} \exp\left(-ip\cdot x\right) f(x),\\
f(x) &= \frac{1}{\sqrt{V}}\sum_{p\in\widetilde{\Lambda}} \exp\left(ip\cdot x\right)\widetilde{f}(p).
\end{align}

\subsection{Reality condition}

Let us consider the Fourier transform of a real quantity such as the electromagnetic field, $A_\mu(x+a\hat{\mu}/2)$.
This will add some conditions on the Fourier coef\mbox{}ficients $\widetilde{A}_\mu(p)$ \cite{Blum:2007cy}.\\
The Fourier transform of the gauge field is
\begin{align}
\label{app:fourier_transf}
A_\mu(x+a\hat{\mu}/2) = \frac{1}{\sqrt{V}} \sum_{p\in\widetilde{\Lambda}} \e^{ip\cdot (x+a\hat{\mu}/2)} \widetilde{A}_\mu(p),
\end{align}
and the reality condition means that the imaginary part of the Fourier transform has to vanish, i.e.
\begin{align}
\label{eq:star}
\frac{1}{\sqrt{V}} \sum_{p\in\widetilde{\Lambda}} \left\{ \sin\left(p\cdot x + ap_\mu/2\right) \Ree \widetilde{A}_\mu(p) + \cos\left(p\cdot x + ap_\mu/2\right) \Imm \widetilde{A}_\mu(p)\right\} = 0.
\end{align}
We introduce the reflection operator in the first Brillouin zone
\begin{align}
\label{app:reflection_op}
R(p)_\mu = \begin{cases}
-p_\mu &\text{ if } p_\mu\neq\pi\\
\pi &\text{ if } p_\mu = \pi
\end{cases}.
\end{align}
We can now write the reality condition into three dif\mbox{}ferent ones:
\begin{enumerate}
\item $p_\mu = 0$, the zero mode is real, $\Imm\widetilde{A}_\mu(0) = 0$.
\item $p_\mu \neq 0,\pi$, there is a pairwise cancellation between one mode and the reflected-conjugate one, i.e.
\begin{align*}
\widetilde{A}_\mu(p) = \widetilde{A}^*_\mu\left(R(p)\right) \longleftrightarrow
\begin{cases}
\Ree\widetilde{A}_\mu(p)&= \Ree\widetilde{A}_\mu(R(p))\\
\Imm\widetilde{A}_\mu(p)&= -\Imm\widetilde{A}_\mu(R(p))
\end{cases}.
\end{align*}
\item $p_\mu = \pi$, $p_\nu = 0, \pi$ for $\nu \neq \mu$, there is no partner for these kind of modes, which means no pairwise cancellation and $\Imm\widetilde{A}_\mu(p) = 0$.
While the real part is left unfixed since the sine in Eq.~\ref{eq:star} is vanishing.
\end{enumerate}
We can incorporate these conditions in one, which reads
\begin{align}
\label{app:reality_condition}
\e^{iR(p)_\mu/2} \widetilde{A}_\mu\left(R(p)\right) = \left( \e^{ip_\mu/2} \widetilde{A}_\mu(p) \right)^*.
\end{align}

\section{Generation in Feynman gauge}
\label{gen_feyn_gauge}

We recall that the action in Feynman gauge in coordinate space is
\begin{align}
\text{S} = -\frac{1}{2}\sum_{n\in\Lambda}\sum_{\mu, \nu} A_\nu(x+a\hat{\nu}/2) \partial_\mu^-\partial_\mu^+ A_\nu(x+a\hat{\nu}/2).
\end{align}
To find the action in the momentum space we Fourier transform by inserting Eq.~\ref{app:fourier_transf}
\begin{align}
\text{S} = -\frac{1}{2V}\sum_{n\in\Lambda}\sum_{\mu, \nu}\sum_{p, p'\in\widetilde{\Lambda}} \e^{ip\cdot (x+a\hat{\nu}/2)}  \partial_\mu^-\partial_\mu^+ \e^{ip'\cdot (x+a\hat{\nu}/2)} \widetilde{A}(p)\widetilde{A}(p').
\end{align}
The Laplacian of the exponential is found to be
\begin{align}
\nn
\partial_\mu^+ \e^{ip'\cdot x} &= \frac{\e^{ip'\cdot x}}{a}\left( \e^{ip'_\mu}-1 \right),\\
\nn
\partial_\mu^-\partial_\mu^+ \e^{ip'\cdot x} &= \frac{\e^{ip'\cdot x}}{a^2}\left( 1-\e^{-ip'_\mu} \right)\left( \e^{ip'_\mu}-1 \right) = \left[2i\sin(p'_\mu/2)\right]^2 \equiv -\widehat{p}_\mu^{'2},\\
\sum_\mu \partial_\mu^-\partial_\mu^+ \e^{ip'\cdot x} &= -\sum_\mu \widehat{p}_\mu^{'2} \equiv -\widehat{p}^{'2}\, ,
\end{align}
and the action becomes
\begin{align}
\text{S} = \frac{1}{2V}\sum_{n\in\Lambda}\sum_{\nu}\sum_{p, p'\in\widetilde{\Lambda}} \e^{i(p+p')\cdot (x+a\hat{\nu}/2)}  \widehat{p}' \widetilde{A}(p)\widetilde{A}(p').
\end{align}
We recognize the the delta function, Eq.~\ref{app:delta_funct_mom}, hence we are left with
\begin{align}
\text{S} = \frac{1}{2V}\sum_{\nu}\sum_{p, p'\in\widetilde{\Lambda}} V\delta(p+p')\e^{iap_\nu/2} \e^{-iap'_\nu/2}\, \widehat{p}^{'2}\, \widetilde{A}(p)\widetilde{A}(p').
\end{align}
The delta function in the first Brillouin zone is equivalent to the reflection operation given in Eq.~\ref{app:reflection_op}
\begin{align}
\text{S} = \frac{1}{2}\sum_{p\in\widetilde{\Lambda}} \widehat{p}^{2} \sum_{\nu} \e^{iap_\nu/2} \, \widetilde{A}(p) \, \e^{-iaR(p)_\nu/2} \, \widetilde{A}\left(R(p)\right),
\end{align}
and by using the reality condition, in Eq.~\ref{app:reality_condition} we obtain the result
\begin{align}
\text{S} = \sum_{p\in\widetilde{\Lambda}} \frac{\widehat{p}^{2}}{2} \sum_{\nu} \big|\widetilde{A}(p)\big|^2 .
\end{align}

\section{Extended Coulomb gauge}
\label{app:ecg_gen}

The Coulomb gauge is defined by the constraint (see Faddeev-Popov approach in Sect.~\ref{sect:faddeev-popov}),
\begin{align}
G(A) = \sum_j\partial_j A_j,
\end{align}
and we shall take $\alpha=1$. 
On the lattice the condition is translated
\begin{align}
\sum_j \partial_j^- A_j(x+a\hat{j}/2) = 0\, , \qquad \widetilde{A}_0(p_0\neq 0, \underline{p}=\underline{0}) = 0\, ,
\end{align}
with $j$ spatial index. 
The second condition is a consequence of the non-completeness of the Coulomb gauge in finite volume, hence the name \quotes{extended}, see Sect.~\ref{sec:drstrangelove} for details.

\subsection{Extended Coulomb via Feynman gauge}

We can generate first the Fourier component of the field in Feynman gauge and then switch to Coulomb by employing a gauge transformation
\begin{align}
\label{eq:gauge_transf}
A_\mu (x+a\hat{\mu}/2) = A_\mu^{\rm C} (x+a\hat{\mu}/2) + \partial_\mu^+ \alpha(x),
\end{align}
where $A_\mu$ is the field generated in Feynman gauge, $A_\mu^{\rm C}$ is the field we would like to satisfy the Coulomb gauge condition and $\alpha$ is the gauge transformation that does the job, see Ref.~\cite{Borsanyi:2014jba}.
By imposing the Coulomb gauge condition one can relate $\alpha$ and $A_\mu$,
\begin{align}
\sum_j \partial_j^- A_j (x+a\hat{j}/2) = \cancel{ \sum_j \partial_j^- A_j^{\rm C} (x+a\hat{j}/2) } + \sum_j \partial_j^-\partial_j^+ \alpha(x).
\end{align}
We Fourier transform the above relation
\begin{align}
\nn
\text{l.h.s.}:\,\, &\sum_j \partial_j^-  A_j (x+a\hat{j}/2) = \sum_{p\in\widetilde{\Lambda}} \e^{ip\cdot x} \sum_j \frac{1-\e^{-iap_j}}{a} \e^{iap_j/2} \widetilde{A}_j(p) \\
&\phantom{\sum_j \partial_j^-  A_j (n+\hat{j}/2) }
= \sum_{p\in\widetilde{\Lambda}} \e^{ip\cdot x} \sum_j 2i\sin(ap_j/2)\, \widetilde{A}_j(p),\\
\text{r.h.s.}:\,\, &\sum_j \partial_j^-\partial_j^+ \alpha(x) =  \sum_{p\in\widetilde{\Lambda}} \e^{ip\cdot x}\, \widetilde{\alpha}(p) \sum_j \frac{\left(1-\e^{-iap_j}\right) \left(\e^{iap_j}-1\right)}{a^2}  = -\sum_{p\in\widetilde{\Lambda}} \e^{ip\cdot x}\, |\underline{\hat{p}}|^2\, \widetilde{\alpha}(p).
\end{align}
In momentum space the condition reads
\begin{align}
\label{eq:alpha}
\widetilde{\alpha}(p) = - \frac{i}{|\underline{\hat{p}}|^2} \sum_j \hat{p}_j\, \widetilde{A}_j(p).
\end{align}
The gauge transformation in Eq.~\ref{eq:gauge_transf} in momentum space gives a relation between all the Fourier coef\mbox{}ficients of the fields generated in dif\mbox{}ferent gauges, in particular
\begin{align}
\widetilde{A}_\mu^{\text{C}} (p) = \widetilde{A}_\mu(p) - i \hat{p}_\mu\, \widetilde{\alpha}(p),
\end{align}
and by plugging in the result found for $\alpha$ in Eq.~\ref{eq:alpha} we have
\begin{align}
\nn
\widetilde{A}_\mu^{\text{C}} (p) &\equiv P^{\text{C}}_{\mu\nu} \widetilde{A}_\nu(p) = \widetilde{A}_\mu(p) - \frac{\hat{p}_\mu}{|\underline{\hat{p}}|^2} \sum_j \hat{p}_j\, \widetilde{A}_j(p) \\
&= \widetilde{A}_\mu(p) - \frac{\hat{p}_\mu\,\left(0, \underline{\hat{p}}\right)_\nu}{|\underline{\hat{p}}|^2} \widetilde{A}_\nu(p) = \left( \delta_{\mu\nu} - \frac{\hat{p}_\mu\,\left(0, \underline{\hat{p}}\right)_\nu}{|\underline{\hat{p}}|^2} \right) \widetilde{A}_\nu(p).
\end{align}
Now the question is: what to do with the $|\underline{\hat{p}}|^2=0$ modes?\\
Those are left unchanged with respect to the Feynman gauge because whatever value will not af\mbox{}fect the Coulomb gauge. To match the extended Coulomb gauge we impose the other constraint on the following components
\begin{align}
\widetilde{A}_0(p_0\neq 0, \underline{p}=\underline{0}) = 0.
\end{align}

\subsection{Generation in the extended Coulomb gauge}

The generation in the extended Coulomb gauge is carried out by following the appendix in Ref.~\cite{Blum:2007cy}.\\
A component $\widetilde{A}_j(p)$ with $\underline{p}\neq\underline{0}$ and $p_j\neq0$ is found from the other two spatial components through the Coulomb condition in momentum space (also called \emph{transversality condition})
\begin{align}
\widetilde{A}_j(p) = -\frac{1}{\hat{p}_j} \sum_{k\neq j} \hat{p}_k \widetilde{A}_k(p).
\end{align}
Let summarize what are the constraints on the modes.
\begin{itemize}
\item $\underline{p}\neq \underline{0}$ and $p_3\neq 0$.\vspace{0.5cm}\\
The weight we use to generate the components is
\begin{align}
\frac{h_p}{2} \hat{p}^2 \left( |\widetilde{A}_1(p)|^2 + |\widetilde{A}_2(p)|^2 \right),
\end{align}
where $h_p$ takes into account the doubling in the action due to the contribution from the complex conjugate partner and it is defined as
\begin{align}
\label{eq:conj_partner}
h_p = \begin{cases}
2 &\text{ when } R(p) \neq p\\
1 &\text{ otherwise }
\end{cases}.
\end{align}
Where the reflection operator in Eq.~\ref{app:reflection_op} is applied to each component of the vector $p$.
The zero component is draw from following distribution
\begin{align}
\frac{h_p}{2} |\underline{\hat{p}}|^2 |\widetilde{A}_0(p)|^2.
\end{align}
\item $\underline{p}\neq \underline{0}$ and $(\hat{p}_1)^2 + (\hat{p}_2)^2\neq 0$.\vspace{0.5cm}\\
In this case we change basis in order to diagonalize the operator. Let us call the new basis $\left(\widetilde{A}_-(p), \widetilde{A}_+(p)\right)$, and the weight is given by
\begin{align}
\frac{h_p}{2} \hat{p}^2 \left( m_-|\widetilde{A}_-(p)|^2 + m_+|\widetilde{A}_+(p)|^2 \right),
\end{align}
where the $m_\mp$ are the following eigenvalues
\begin{align}
\nn
m_- &= \hat{p}^2,\\
\label{eq:massless_eigenvalues}
m_+ &= \hat{p}^2 \left( 1 + \frac{(\hat{p_1})^2 + (\hat{p_2})^2}{(\hat{p_3})^2} \right).
\end{align}
Once we have generated the components in this basis we need to rotate back to the original
\begin{align}
\begin{pmatrix}
\widetilde{A}_1(p)\\
\widetilde{A}_2(p)
\end{pmatrix} = 
 \begin{pmatrix}
 r_2 & r_1\\
 -r_1 & r_2
 \end{pmatrix}
 \begin{pmatrix}
\widetilde{A}_-(p)\\
\widetilde{A}_+(p)
\end{pmatrix}\, ,
\end{align}
where the $r_i$ are given by
\begin{align}
r_j = \frac{\hat{p}_j}{\sqrt{ (\hat{p_1})^2 + (\hat{p_2})^2 }}\quad \text{for } j=1,2.
\end{align}
As in the previous case the zero component is draw from the distribution
\begin{align}
\frac{h_p}{2} |\underline{\hat{p}}|^2 |\widetilde{A}_0(p)|^2.
\end{align}
\item $\underline{p} = \underline{0}$, $p_0\neq 0 $.\vspace{0.5cm}\\
The three spatial components are found independently according to the weight
\begin{align}
\frac{h_p}{2} (\hat{p}_0)^2 \sum_j|\widetilde{A}_j(p)|^2.
\end{align}
The zero component is vanishing, i.e.~$\widetilde{A}_0(p_0, \underline{0}) = 0$.
\end{itemize}

\section{Zero mode fixing equation of motion}
\label{app:ELELC}

The zero mode fixing in Eq.~\ref{eq:zero_mode_fix} 
in momentum space corresponds to the following condition in coordinates space
\begin{align}
\widetilde{A}_\mu(p=0) = c_\mu \Longleftrightarrow \int \de^4 x \, A_\mu(x) = c_\mu\, ,
\end{align}
with $c_\mu\in\mathbb{R}$ constant, for each value of $\mu$ is a non-local condition.

We want to impose the constraint on the zero mode in the partition function, and this reads
\begin{align}
\int \mathcal{D}[A_\mu] \exp{\left[-\text{S}(A)\right]} \prod_{\mu=0}^3\delta\left(\int \de^4 y \, A_\mu(y)-c_\mu\right).
\end{align}
We rewrite the delta function as the limit of a gaussian
\begin{align}
\delta(f) = \lim_{\alpha\rightarrow 0}\, \exp\left(-\frac{f^2}{2\alpha}\right),
\end{align}
then the action becomes an ef\mbox{}fective one written as
\begin{align}
\nn
\text{S}_{\text{ef\mbox{}f}} &= \text{S}(A) + \frac{1}{2\alpha} \sum_{\mu=0}^3 \left(\int \de^4 y \, A_\mu(y)-c_\mu\right)^2\\
&= \int\de^4x\left( F_{\mu\nu}F_{\mu\nu} + j_\mu A_\mu \right) + \frac{1}{2\alpha} \sum_{\mu=0}^3 \left(\int \de^4 y \, A_\mu(y)-c_\mu\right)^2.
\end{align}
The Euler-Lagrange equations, which we recall to be 
\begin{align}
\label{app:euler_lagrange_eq}
\frac{\delta\text{S}}{\delta A_\mu(x)} = \partial_\nu \frac{\delta\text{S}}{\delta \left(\partial_\nu A_\mu(x)\right)},
\end{align}
gives the usual known terms for the local part, and for the new non-local part we get
\begin{align}
\nonumber
\frac{\delta}{\delta A_\mu(x)} \left[ \frac{1}{2\alpha} \sum_{\nu=0}^3 \left(\int \de^4 y \, A_\nu(y)-c_\nu\right)^2\right] &= \frac{1}{2\alpha}\sum_\nu 2\left(\int\de^4y\, A_\nu(y) - c_\nu\right) \int\de^4z\,\delta(z-x)\delta_{\mu\nu}\\
\label{app:b_nu}
&= \frac{1}{\alpha} \left(\int\de^4y\, A_\mu(y) - c_\mu\right) \equiv b_\mu = \text{constant}.
\end{align}
Finally Euler-Lagrange equations are
\begin{align}
\label{app:modified_euler_lagrange}
\sum_\nu \partial_\nu F_{\nu\mu}(x) = j_\mu(x) + b_\mu.
\end{align}

\section{Generation of quenched QED$_{\rm M}$ configurations}

In this section we give the recipe to generate massive QED configurations in Feynman and Coulomb gauge.
The only dif\mbox{}ference with Subsect.~\ref{subsect:feynm_gauge_gen} is the mass term, this reads
\begin{align}
 \frac{m^2_\gamma}{2} \sum_{x,\mu} A_\mu^2 (x+a\hat{\mu}/2).
\end{align}
In momentum space it becomes
\begin{align}
\nn
\frac{m^2_\gamma }{2} \sum_{x,\mu} A_\mu^2(x+a\hat{\mu}/2) &= \frac{m^2_\gamma }{2} \sum_{x,\mu} \frac{1}{V} \sum_{p,p'} \e^{i(p+p')\cdot (x+a\hat{\mu}/2)} \widetilde{A}_\mu(p)\widetilde{A}_\mu(p')\\
\nn
& = \frac{m^2_\gamma }{2} \sum_{\mu,p,p'} \delta(p+p') \e^{i(p+p')_\mu/2} \widetilde{A}_\mu(p)\widetilde{A}_\mu(p')\\
& = \frac{m^2_\gamma }{2} \sum_{\mu,p} \widetilde{A}_\mu(p) \e^{ip_\mu/2} \e^{iR(p)_\mu/2} \widetilde{A}_\mu(p),
\end{align}
where the reflection operator in the first Brillouin zone is given by
Eq.~\ref{app:reflection_op}.
We can identify $\e^{iR(p)_\mu/2} \widetilde{A}_\mu(p') = \left(\e^{ip_\mu/2} \widetilde{A}_\mu(p)\right)^*$ (reality condition of the field $A_\mu$) and obtain
\begin{align}
\frac{m^2_\gamma }{2} \sum_{x,\mu} A_\mu^2(x+a\hat{\mu}/2) &= \frac{m^2_\gamma }{2} \sum_{p,\mu} |\widetilde{A}_\mu(p)|^2.
\end{align}
The Proca action in momentum space is
\begin{align}
\text{S} =  \frac{1}{2}\sum_{p\in\widetilde{\Lambda}} \left[ \sum_{\nu<\mu} |\hat{p}_\mu \widetilde{A}_\nu(p) -  \hat{p}_\nu \widetilde{A}_\mu(p) |^2 + \sum_\mu m^2_\gamma  |\widetilde{A}_\mu(p)|^2\right].
\end{align}

\subsection{Feynman gauge}

Now we need to include the gauge fixing term in Feynman gauge that cancels out the of\mbox{}f diagonal part in $\mu,\nu$ of the field-strength and gives
\begin{align}
\nn
\text{S} = \sum_{x\in\Lambda} \bigg[ \frac{1}{4} \sum_{\mu, \nu} ( \partial^+_\mu A_\nu(x+a\hat{\nu}/2) & - \partial^+_\nu A_\mu(x+a\hat{\mu}/2) )^2 \\
&+ \frac{m^2_\gamma }{2} \sum_{\mu} A_\mu^2(x+a\hat{\mu}/2) + \frac{1}{2}\left(\sum_\mu \partial_\mu A_\mu(x+a\hat{\mu}/2) \right)^2\bigg],
\end{align}
which in Fourier space reads
\begin{align}
\text{S} = \frac{1}{2}\sum_{p\in\widetilde{\Lambda}}\left(\hat{p}^2+m^2_\gamma \right)\sum_\mu|\widetilde{A}_\mu(p)|^2.
\end{align}

\subsection{Coulomb gauge}

We might also consider the Coulomb gauge implemented exactly configuration by configuration. The generation in Coulomb gauge is done by inserting a mass term.\\
In momentum space we can solve the Coulomb gauge fixing constraint with respect the third component, this lead to the following identification
\begin{align}
\label{eq:coulomb_gauge}
\widetilde{A}_3 = -\frac{1}{\hat{p}_3} \left( \hat{p}_1 \widetilde{A}_1 + \hat{p}_2 \widetilde{A}_2 \right),
\end{align}
as long as $\hat{p}_3\neq 0$.
By inserting it in the mass term we get
\begin{align}
\sum_\mu m^2_\gamma  |\widetilde{A}_\mu(p)|^2 = m^2_\gamma  |\widetilde{A}_0(p)|^2 + m^2_\gamma  \left[1+\left(\frac{\hat{p}_1}{\hat{p}_3}\right)^2\right] |\widetilde{A}_1(p)|^2 + m^2_\gamma  \left[1+\left(\frac{\hat{p}_2}{\hat{p}_3}\right)^2\right] |\widetilde{A}_2(p)|^2.
\end{align}
The weight (action) to use to generate the Fourier component is 
\begin{align}
\nn
\sum_{p_3\neq 0} \frac{1}{2} \bigg\{ (\underline{\hat{p}}^2+m^2_\gamma ) |\widetilde{A}_0(p)|^2 & + \hat{p}^2\sum_{k=1}^2 |\widetilde{A}_k(p)|^2 \\
&+ m^2_\gamma \sum_{k=1}^2 \left[1+\left(\frac{\hat{p}_k}{\hat{p}_3}\right)^2\right]|\widetilde{A}_k(p)|^2
 + \frac{m^2_\gamma +\hat{p}^2}{\hat{p}_3^2}\sum_{j,k=1}^2 \hat{p}_j\hat{p}_k \Ree\left(\widetilde{A}_j(p)\widetilde{A}_k(p)\right) \bigg\}.
\end{align}
Let us consider the dif\mbox{}ferent momenta and the corresponding weights.
\begin{itemize}
\item $\underline{p}\neq 0$ and $p_3\neq 0$ ($\hat{p_1}=0=\hat{p_2}$).\\
The action is given by 
\begin{align}
\frac{h_p}{2} (\hat{p}^2+m^2_\gamma ) \left(|\widetilde{A}_1(p)|^2 + |\widetilde{A}_2(p)|^2\right)
\end{align}
where $h_p$ takes into account the doubling in the action due to the contribution from the complex
conjugate partner and it is defined as in Eq.~\ref{eq:conj_partner},
while the third component is fixed through the Coulomb condition Eq.~\ref{eq:coulomb_gauge}.\\
The zero component is draw from the following distribution
\begin{align}
\frac{h_p}{2} (\underline{\hat{p}}^2+m^2_\gamma ) |\widetilde{A}_0(p)|^2.
\end{align}
\item $\underline{p}\neq 0$ and $p_3\neq 0$ ($\hat{p_1}^2+\hat{p_2}^2\neq 0$).\\
In matrix form the action is given by
\begin{align}
\begin{pmatrix}
\widetilde{A}_1^*, & \widetilde{A}_2^*
\end{pmatrix}
\left[\left(\hat{p}^2 + m^2_\gamma \right) \begin{pmatrix}
1+\left(\frac{\hat{p}_1}{\hat{p}_3}\right)^2 & \frac{\hat{p}_1\hat{p}_2}{\hat{p}_3^2}  \\
\frac{\hat{p}_1\hat{p}_2}{\hat{p}_3^2} & 1+\left(\frac{\hat{p}_2}{\hat{p}_3}\right)^2 
\end{pmatrix}\right]
\begin{pmatrix}
\widetilde{A}_1\\
\widetilde{A}_2
\end{pmatrix},
\end{align}
and by diagonalizing it we get the following eigenvalues
\begin{align}
\nn
m_- & \equiv \hat{p}^2 + m^2_\gamma  \xrightarrow{\: m_\gamma \to 0 \: } \hat{p}^2, \\
m_+ & \equiv \underline{\hat{p}}^2\, \frac{\hat{p}^2 + m^2_\gamma }{\hat{p}_3^2} \xrightarrow{\: m_\gamma \to 0 \: } \hat{p}^2\left( 1 + \frac{ \hat{p}_1^2 + \hat{p}_2^2}{\hat{p}_3^2}\right),
\end{align}
that in the massless case become the ones
in Eq.~\ref{eq:massless_eigenvalues}.\\
The weight in the new basis is
\begin{align}
\frac{h_p}{2} (\hat{p}^2+m^2_\gamma ) \left(m_-|\widetilde{A}_-(p)|^2 + m_+|\widetilde{A}_+(p)|^2\right).
\end{align}
Once we have generated the components in this basis we need to rotate back to the original components and this is done through
\begin{align}
\begin{pmatrix}
\widetilde{A}_1\\
\widetilde{A}_2
\end{pmatrix} = \begin{pmatrix}
r_2 & r_1\\
-r_1 & r_2
\end{pmatrix} 
\begin{pmatrix}
\widetilde{A}_-\\
\widetilde{A}_+
\end{pmatrix}\, ,
\end{align}
where the $r_j$ are given by 
\begin{align}
r_j = \frac{\hat{p}_j}{\sqrt{\hat{p}_1^2 + \hat{p}_2^2}}\quad\text{for }j=1,2.
\end{align}
The zero component as the following weight
\begin{align}
\frac{h_p}{2} (\underline{\hat{p}}^2+m^2_\gamma ) |\widetilde{A}_0(p)|^2.
\end{align}
\item $\underline{p}= 0$ and $p_0\neq 0$.\\
The zero component weight is found to be
\begin{align}
\frac{h_p}{2} m^2_\gamma  |\widetilde{A}_0(p)|^2,
\end{align}
while the three spatial components are generated independently according to
\begin{align}
\frac{h_p}{2} \left(m^2_\gamma +\hat{p}_0^2\right) |\widetilde{A}_j(p)|^2.
\end{align}
\item $\underline{p}= 0$ and $p_0= 0$.\\
The zero mode is regularized by the mass term. The four components are integrated independently with the weight
\begin{align}
\frac{h_p}{2} m^2_\gamma  |\widetilde{A}_\mu(p)|^2.
\end{align}
\end{itemize}

\section{Wilson loop in quenched approximation}
\label{app:wils_loop}

In the quenched approximation the expectation value of the Wilson loop is given by \cite{Rothe:1992nt}
\begin{align}
\langle W_C[A] \rangle = 
\frac{\int \mathcal{D}[A] \exp\left[ -\frac{1}{2}\int\de^4 x A_\mu \Omega_{\mu\nu} A_\nu + i eQ\oint \de x_\mu A_\mu \right]}
{\int \mathcal{D}[A] \exp\left[ \frac{1}{2}\int\de^4 x A_\mu \Omega_{\mu\nu} A_\nu \right]}.
\end{align}
In the case of QED in Feynman gauge we know that $\Omega_{\mu\nu} = -\delta_{\mu\nu} \Box$, while in the massive case we simply have $\Omega_{\mu\nu} = \delta_{\mu\nu} (-\Box + m^2_\gamma)$.\\
By completing the square we can compute the Gaussian integral and we find 
\begin{align}
\langle W_C[A] \rangle = 
\exp\left[ -\frac{e^2Q^2}{2} \oint \de z_\mu \oint \de z'_\mu \delta_{\mu\nu} G(z-z') \right],
\end{align}
where $G(x)$ is the Green's function associated to the discretized version of $-\Box$ (in QED) or $-\Box + m^2_\gamma$ in the massive theory for a scalar particle. 
Since the integral is proportional to $\delta_{\mu\nu}$ the integral receives contribution only when $z$ and $z'$ are parallel to each other. We rewrite the expectation value manifestly on the lattice
\begin{align}
\langle W_C[A] \rangle = 
{{4}\choose{2}} \sum_{\mu\neq\nu} \exp\left[ -\frac{e^2Q^2}{2} \sum_{z\in R_{\mu\nu}, z'\in R_{\mu\nu}} G(z-z') \right],
\end{align}
where now $R_{\mu\nu}$ is a rectangle of size $I\times J$ in the plane $(\mu,\nu)$. We introduce the function 
\begin{align}
C_\mu(I, x) = \sum_{i_1, i_2 = 0}^{I-1} G\left[ x + a(i_1-i_2)\hat{\mu}\right],
\end{align}
and by operating the substitution $\tau = i_1 - i_2$ and $\sigma = i_1 + i_2$ we get
\begin{align}
C_\mu(I,xn) = I G(x) + 2\sum_{\tau = 1}^{I-1} (I-\tau) G( x + a\tau\hat{\mu}).
\end{align}
We can now rewrite the Wilson loop in the plane $(\mu,\nu)$ expectation value in its final form
\begin{align}
\langle W_{\mu\nu}[A](I,J) \rangle = 
\exp\left\{ -2e^2Q^2 \left[C_\mu(I, 0) - C_\nu(I,J\hat{\nu}) \right]\right\}.
\end{align}
For Lorentz invariance we can average over $\mu, \nu$ and write directly $w(I,I)$.

\begin{table}[h!t]
\begin{center}
 \begin{tabular}{|c | c c c|} 
  \hline
	$I$ & QED & 
	\multicolumn{2}{|c|}{Massive QED} \\
 \hline
  &  & $m_\gamma = 0.05$  &  $m_\gamma = 5$  \\ [0.5ex] 
 \hline\hline
 1 & 0.500000000(1) &  0.499806547(1) & 0.118374259(1)  \\ 
 \hline
 2 & 1.369311535(1) & 1.368461017(1) & 0.251479961(1) \\
 \hline
 3 & 2.305193057(1) & 2.303260044(1) & 0.381563158(1) \\
 \hline
 4 & 3.261483854(1) & 3.258072049(1) & 0.511425684(1) \\
 \hline
 5 & 4.228829967(1) & 4.223555011(1) & 0.641276934(1) \\
 \hline
 6 & 5.203604529(1) & 5.196090337(1) & 0.771127679(1) \\
 \hline
 7 & 6.183728249(1) & 6.173606728(1) & 0.900978402(1) \\
 \hline
 8 & 7.167859805(1) & 7.154771070(1) & 1.030829125(1) \\
 \hline
 9 & 8.155091868(1) & 8.138684452(1) & 1.160679847(1) \\
 \hline
 10 & 9.144788116(1) & 9.124719122(1) & 1.290530569(1) \\
 \hline
 11 & 10.136487567(1) & 10.112422767(1) & 1.420381292(1) \\
 \hline
 12 & 11.129846532(1) & 11.101460414(1) & 1.550232014(1) \\
 \hline
 13 & 12.124602305(1) & 12.091578088(1) & 1.680082737(1) \\
 \hline
 14 & 13.120549740(1) & 13.082579344(1) & 1.809933459(1) \\
 \hline
 15 & 14.117525653(1) & 14.074309644(1) & 1.939784182(1) \\
 \hline
 16 & 15.115398147(1) & 15.066645656(1) & 2.069634904(1) \\
 \hline
\end{tabular}
 \caption{\emph{Table showing the values of $-\frac{2}{q^2}\ln\left( w(I, I)\right)$ related to the Wilson loops in the infinite lattice in QED and massive QED (Feynman gauge), meant to save PhDs from struggling too much.}}
\end{center}
\end{table}

\section{Coordinate space method}
\label{app:CSM}

The coordinate space method permits us to write the propagator in coordinate space by using a recursion relation.
The only requirement is the knowledge of the Green's function in the corners of the unit hypercube.
In the following we present the methods for the massless and massive propagator, useful for the computation of the
Wilson loops.

\subsection{Luscher-Weisz}

We consider an infinite lattice. The Green's function for a massless particle satisfies
\begin{align}
\label{app:laplacian}
\sum_\mu \partial_\mu^-\partial_\mu^+ G(x) = \delta(x).
\end{align}
We know that in the continuum limit ($x\rightarrow\infty$) $G(x)$ converges to $(4\pi^2x^2)^{-1}$, the Green's function of the continuum laplacian in $\mathbb{R}^4$. We would like to express the propagator through its values close to the origin. The key observation made by Vohwinkel is that
\begin{align}
\label{app:vohw}
\left(\partial_\mu^- + \partial_\mu^+\right) G(x) = x_\mu H(x)\, ,
\end{align}
with the following identifications
\begin{align}
\nn
H(x) &= \int_{-\frac{\pi}{a}}^\frac{\pi}{a} \frac{\de^4 p}{(2\pi)^4} \,\e^{ip\cdot x} \ln(\hat{p}^2),\\
G(x) &= \int_{-\frac{\pi}{a}}^\frac{\pi}{a} \frac{\de^4 p}{(2\pi)^4} \,\frac{\e^{ip\cdot x}}{\hat{p}^2}.
\end{align}
\subsection*{Proof of the formula}
To prove the relation we calculate the l.h.s.~of Eq.~\ref{app:vohw}
\begin{align}
\nn
\partial_\mu^+ G(x) = G(x+a\hat{\mu}) - G(x) &= \int_{-\frac{\pi}{a}}^\frac{\pi}{a} \frac{\de^4 p}{(2\pi)^4}\, \frac{1}{\hat{p}^2}\, \partial_\mu^+\, \e^{ip\cdot x} = \int_{-\frac{\pi}{a}}^\frac{\pi}{a} \frac{\de^4 p}{(2\pi)^4}\, \frac{1}{\hat{p}^2}\, \frac{\e^{ip\cdot x}}{a} \left( \e^{iap_\mu} - 1\right),\\
\nn
\partial_\mu^- G(x) = G(x) - G(x-a\hat{\mu}) &= \int_{-\frac{\pi}{a}}^\frac{\pi}{a} \frac{\de^4 p}{(2\pi)^4}\, \frac{1}{\hat{p}^2}\, \partial_\mu^+\, \e^{ip\cdot x} = \int_{-\frac{\pi}{a}}^\frac{\pi}{a} \frac{\de^4 p}{(2\pi)^4}\, \frac{1}{\hat{p}^2}\, \frac{\e^{ip\cdot x}}{a} \left( 1 - \e^{-iap_\mu}\right),\\
\left(\partial_\mu^- + \partial_\mu^+\right) G(x) &= \int_{-\frac{\pi}{a}}^\frac{\pi}{a} \frac{\de^4 p}{(2\pi)^4}\, \frac{\e^{ip\cdot x}}{\hat{p}^2}\, i\left[\frac{2}{a} \sin\left(ap_\mu\right)\right]. 
\end{align}
We now integrate by parts the r.h.s.~of Eq.~\ref{app:vohw}
\begin{align}
\nn
x_\mu H(x) &= \int_{-\frac{\pi}{a}}^\frac{\pi}{a} \frac{\de^4 p}{(2\pi)^4} \, x_\mu\e^{ip\cdot x} \ln(\hat{p}^2) = \int_{-\frac{\pi}{a}}^\frac{\pi}{a} \frac{\de^4 p}{(2\pi)^4} \, \left(-i\frac{\de}{\de p_\mu}\e^{ip\cdot x}\right) \ln(\hat{p}^2)\\
&= -i \int_{-\frac{\pi}{a}}^\frac{\pi}{a} \frac{\de^4 p}{(2\pi)^4} \, \frac{\de}{\de p_\mu}\left[\e^{ip\cdot x}\ln(\hat{p}^2)\right] + i \int_{-\frac{\pi}{a}}^\frac{\pi}{a} \frac{\de^4 p}{(2\pi)^4} \, \e^{ip\cdot x} \left(\frac{\de}{\de p_\mu}\ln(\hat{p}^2)\right).
\end{align}
The first term vanishes since
\begin{align}
\nn
\int_{-\frac{\pi}{a}}^\frac{\pi}{a} \frac{\de^4 p}{(2\pi)^4} \, \frac{\de}{\de p_\mu}\left[\e^{ip\cdot x}\ln(\hat{p}^2)\right] 
&= \left(   \int_{-\frac{\pi}{a}}^\frac{\pi}{a} \prod_{\nu\neq\mu}\frac{\de p_\nu}{(2\pi^3)} \e^{i p_\nu x_\nu}  \right)  
\left\{\int_{-\frac{\pi}{a}}^\frac{\pi}{a} \frac{\de p_\mu}{2\pi} \, \frac{\de}{\de p_\mu}\left[\e^{ip_\mu x_\mu}\ln(\hat{p}^2)\right]\right\}\\
&= \left(\int_{-\frac{\pi}{a}}^\frac{\pi}{a}\prod_{\nu\neq\mu}\frac{\de p_\nu}{(2\pi^4)}  \e^{i p_\nu x_\nu} \right)
\left[\e^{ip_\mu x_\mu}\ln(\hat{p}^2)\right]_{-\pi/a}^{\pi/a} = 0\, .
\end{align}
The result is
\begin{align}
x_\mu H(x) &= \int_{-\frac{\pi}{a}}^\frac{\pi}{a} \frac{\de^4 p}{(2\pi)^4} \, \frac{\e^{ip\cdot x}}{\hat{p}^2}\, 2i\hat{p}_\mu \cos\left(\frac{ap_\mu}{2}\right) = \int_{-\frac{\pi}{a}}^\frac{\pi}{a} \frac{\de^4 p}{(2\pi)^4} \, \frac{\e^{ip\cdot x}}{\hat{p}^2}\, i\left[\frac{2}{a} \sin\left(ap_\mu\right)\right],
\end{align}
which prove the equation Eq.~\ref{app:vohw}.

\subsection*{Recursion formula}

We can eliminate $H(x)$ in Eq.~\ref{app:vohw} by using Eq.~\ref{app:laplacian}. We start by notice that 
\begin{align}
\sum_{\mu=0}^{3} \partial_\mu^- \partial_\mu^+ = \frac{1}{2} \sum_{\mu=0}^3 \left(\partial_\mu^+ - \partial_\mu^-\right)\, ,
\end{align}
and by taking Eq.~\ref{app:laplacian} we have
\begin{align}
\sum_{\mu=0}^{3} \partial_\mu^- \partial_\mu^+ G(x) = 0 = \frac{1}{2} \sum_{\mu=0}^3 \left(\partial_\mu^+ - \partial_\mu^-\right)G(x) \Longrightarrow \sum_{\mu=0}^3\partial_\mu^+ G(x) = \sum_{\mu=0}^3\partial_\mu^- G(x).
\end{align}
Now we sum over $\mu$ of Eq.~\ref{app:vohw}, and define $\rho = \sum_{\mu}x_\mu$,
\begin{align}
\begin{rcases}
\sum_{\mu} \left(\partial_\mu^- + \partial_\mu^+\right) G(x) = 2\sum_{\mu=0}^3\partial_\mu^- G(x)\\
\sum_{\mu}x_\mu H(x) = \rho H(x)
\end{rcases} H(x) = \frac{2}{\rho} \sum_{\mu=0}^3\partial_\mu^- G(x)\, .
\end{align}
In this way we have eliminated the function $H(x)$ and we can rewrite Eq.~\ref{app:vohw} as
\begin{align}
\left(\partial_\mu^- + \partial_\mu^+\right) G(x) = \frac{2x_\mu}{\rho} \sum_{\mu=0}^3\partial_\mu^- G(x).
\end{align}
We can now relate the value of the propagator at a point $x+a\hat{\mu}$ to the neighbor points
\begin{align}
\label{app:recursion}
G(x+a\hat{\mu}) = G(x-a\hat{\mu}) + \frac{2x_\mu}{\rho} \sum_{\mu=0}^3 \left[ G(x) - G(x-a\hat{\nu}) \right],
\end{align}
for $\rho \neq 0$. 
The propagator is independent of the sign and the order of the coordinates $(x_0, x_1, x_2, x_3)$, from its asymptotic expansion \cite{Luscher:1995zz}. We can restrict the attention to the points $x=an$ with $n_3 \geq n_2 \geq n_1 \geq n_0$ and the recursion relation allows us to express $G(x)$ as a combination of the propagators at the corners of the unit hypercube.\\
Then the 5 initial values are further restricted when we take into account the following relations:
\begin{align}
\nn
& G(0, 0, 0, 0) - G(0, 0, 0, 1) = \frac{1}{8a^2},\\
\nn
& G(0, 0, 0, 0) - 3G(0, 0, 1, 1) - 2G(0, 1, 1, 1) = \frac{1}{a^2\pi^2},\\
& G(0, 0, 0, 0) - 6G(0, 0, 1, 1) - 8G(0, 1, 1, 1) - 3G(1, 1, 1, 1) = 0.
\end{align}
In the end the only initial values we have to put are 
\begin{align}
\nn
a^2 G(0, 0, 0, 0) = 0.154933390231060214084837208107\dots \\
a^2 G(0, 0, 1, 1) = 0.012714703770934215428228758391\dots
\end{align}
The best values of the constants related to $G(0, 0, 0, 0)$ and $G(0, 0, 1, 1)$ are given in \cite{Capitani:2002mp}.

\subsection{Borasoy-Krebs}

The generalization in the case of a massive propagator is done in Ref.~\cite{Borasoy:2005ha}.
The Green's function satisfies the equation
\begin{align}
\left( m^2_\gamma + \sum_\mu \partial_\mu^-\partial_\mu^+ \right) G(x) = \delta(x),
\end{align}
and it still satisfies the Eq.~\ref{app:vohw} with modified functions
\begin{align}
\nn
H(x) &= \int_{-\frac{\pi}{a}}^\frac{\pi}{a} \frac{\de^4 p}{(2\pi)^4} \,\e^{ip\cdot x} \ln(\hat{p}^2 + m^2_\gamma),\\
G(x) &= \int_{-\frac{\pi}{a}}^\frac{\pi}{a} \frac{\de^4 p}{(2\pi)^4} \,\frac{\e^{ip\cdot x}}{\hat{p}^2+m^2_\gamma}.
\end{align}
By following the same lines of arguments in the case of the Luscher-Weisz method we get that the 
recursion formula in the case of a massive propagator is given by
\begin{align}
G(x+a\hat{\mu}) = G(x-a\hat{\mu}) + \frac{2x_\mu}{\rho} \sum_{\mu=0}^3 \left[ \left(1+\frac{m^2_\gamma}{8}\right)G(x) - G(x-a\hat{\nu}) \right].
\end{align}
Now the 5 initial seeds are given by the modified Bessel functions $I_n$,
\begin{align}
G (x_0, x_1, x_2, x_3) = \frac{1}{2} \int_{0}^\infty \de \lambda \exp\left[ -m^2_\gamma\lambda/2 - 4\lambda  \right] I_{x_0}(\lambda) I_{x_1}(\lambda) I_{x_2}(\lambda) I_{x_3}(\lambda).
\end{align}
The seeds used in this work are generated using MATHEMATICA.
Special thanks are given to Langæble, who helped me in writing a code free of raund-of\mbox{}f errors.

\appendix
\setcounter{chapter}{4}
\renewcommand{\chaptername}{Appendix}
\renewcommand{\theequation}{\Alph{chapter}.\arabic{equation}}
\setcounter{equation}{0}

\chapter{Electromagnetic critical mass shift in lattice perturbation theory}
\label{app:em_cr_mass_shift}

In this appendix we calculate from first principle the critical mass shift induced by electromagnetic ef\mbox{}fects, to this purpose we use lattice perturbation theory.
The dif\mbox{}ferent diagrams are calculated, this was done in order to be
able to introduce a mass term for the photon, even though it is not presented here the modification is straightforward.

\section{Feynman rules of the Wilson theory}

The Dirac-Wilson action on the lattice is, with $r=1$,
\begin{align}
\nonumber
\text{S}_L & = \sum_{n\in\Lambda} \psibar_f(n) \left\{ \gamma_\mu \frac{\nabla_\mu^++\nabla_\mu^-}{2}  - \frac{\nabla_\mu^+\nabla_\mu^-}{2} + m_0^f \right\} \psi_f(n)\\
\nonumber
& = \sum_{n\in\Lambda} \bigg\{ \psibar_f(n)\left(m_0^f + 4\right) \psi_f(n)  +\frac{1}{2} \sum_\mu\big[ \psibar_f(n) (\gamma_\mu-1) U_\mu(n) \psi_f(n+\hat{\mu}) \\
\label{eq:dw_act}
& \phantom{= \sum_{n\in\Lambda} \bigg\{ \psibar_f(n)\left(m_0^f + 4\right) \psi_f(n)  + \sum_\mu\big[} - \psibar_f(n) (\gamma_\mu+1) U^\dagger_\mu(n-\hat{\mu}) \psi_f(n-\hat{\mu}) \big] \bigg\}.
\end{align}

\subsection{Photon propagator}

\begin{align}
\begin{tikzpicture}[line width=1.5 pt, scale=2]
	\draw[vector] (0,0) -- (1,0);
	% The labels
	\node at (0.5,0.2){$\gamma$};
	\node at (-0.1,0){$\mu$};
	\node at (1.1,0){$\nu$};
	\node at (2.7,0){$\equiv {\displaystyle \frac{1}{\hat{k}^2} \left[ \delta_{\mu\nu} + (1-\alpha) \frac{\hat{k}_\mu\hat{k}_\nu}{\hat{k}^2} \right] \stackrel{\alpha = 1}{=} \frac{\delta_{\mu\nu}}{\hat{k}^2}}\,.$};
\end{tikzpicture}
\end{align}

\subsection{Fermion propagator}

\begin{align}
\begin{tikzpicture}[line width=1.5 pt, scale=2]
	\draw[fermion] (0,0) -- (1,0);
	% The labels
	\node at (0.5,0.2){$p$};
	\node at (2,0){$\equiv {\displaystyle \frac{ m + \frac{r}{2}\hat{p}^2 - i \slashed{\bar{p}}  }{\left(m + \frac{r}{2}\hat{p}^2\right)^2 + \bar{p}^2 } }\, ,$};
\end{tikzpicture}
\end{align}
where we defined $\bar{p}_\mu = \sin(p_\mu)$.

\subsection{Vertexes}

We want to expand the interaction part of the Dirac-Wilson action in the weak coupling $e$ regime.
We start by expanding the gauge link in $e$ and we drop the index $f$
\begin{align}
U_\mu(n) = \e^{ieQ A_\mu(n+\hat{\mu}/2)} \simeq 1 + ieQ A_\mu(n+\hat{\mu}/2) + \frac{(ieQ)^2}{2}A_\mu(n+\hat{\mu}/2)A_\mu(n+\hat{\mu}/2)+\ord(e^3).
\end{align}
The interaction part is given by
\begin{align}
\nn
\text{S}_L^\text{int} & =  \frac{1}{2} \sum_{n,\mu} \bigg\{ \psibar(n) (\gamma_\mu - 1)\left[ ieQ A_\mu(n+\hat{\mu}/2) - \frac{e^2Q^2}{2}A_\mu(n+\hat{\mu}/2)A_\mu(n+\hat{\mu}/2)\right] \psi(n+\hat{\mu})\\
&\phantom{ = \frac{1}{2} \sum_{n,\mu} \bigg\{  } - \psibar(n) (\gamma_\mu + 1)\left[ - ieQ A_\mu(n-\hat{\mu}/2) - \frac{e^2Q^2}{2}A_\mu(n-\hat{\mu}/2)A_\mu(n-\hat{\mu}/2)\right] \psi(n-\hat{\mu})\bigg\}.
\end{align}
By Fourier transforming we read the Feynman rules directly from the action by recalling that the expansion at the action level is given by $\e^{S} \simeq 1 - S$ 
and we have to take an overall minus sign in the vertexes. 
The photon momentum is always flowing inside the vertex.

\subsection*{Terms in $e$}
\begin{align}
\nn
\text{S}_L^{\text{int}, e} &= \frac{ieQ}{2} \sum_{n,\mu}  \psibar(n) \left[ (\gamma_\mu - 1) A_\mu(n+\hat{\mu}/2) \psi(n+\hat{\mu}) + (\gamma_\mu + 1) A_\mu(n-\hat{\mu}/2)  \psi(n-\hat{\mu}) \right]\\
\nn
& = \frac{ieQ}{2}\frac{1}{V^{3/2}} \sum_{n,\mu}\sum_{\bar{p}, p, k} \widetilde{\psibar}(\bar{p}) \widetilde{A}_\mu(k) \e^{-i\bar{p}\cdot n}\e^{ik\cdot n}\e^{ip\cdot n}  \big[ (\gamma_\mu - 1) \e^{ik_\mu/2} \e^{ip_\mu} \\
&\phantom{ = \frac{ieQ}{2}\frac{1}{V^{3/2}} \sum_{n,\mu}\sum_{\bar{p}, p, k} \widetilde{\psibar}(\bar{p}) \widetilde{A}_\mu(k) \e^{-i\bar{p}\cdot n}\e^{ik\cdot n}\e^{ip\cdot n}  \big[ }
+ (\gamma_\mu + 1) \e^{-ik_\mu/2} \e^{-ip_\mu} \big] \widetilde{\psi}(p),
\end{align}
and recalling that $\delta(p-p') = \frac{1}{V} \sum_n \e^{i(p-p')\cdot n}$,
\begin{align}
\nn
\text{S}_L^{\text{int}, e} & = \frac{ieQ}{2}\frac{1}{\sqrt{V}} \sum_{\mu}\sum_{\bar{p}, p, k} \delta(k+p-\bar{p}) \widetilde{\psibar}(\bar{p}) \widetilde{A}_\mu(k) \left[ \gamma_\mu 2\cos\left(p_\mu+ \frac{k_\mu}{2}\right) -2i\sin\left(p_\mu+ \frac{k_\mu}{2}\right) \right] \widetilde{\psi}(p)\\
& = ieQ\frac{1}{\sqrt{V}} \sum_{\mu}\sum_{p, k} \widetilde{\psibar}(p+k) \widetilde{A}_\mu(k) \left[ \gamma_\mu \cos\left(p_\mu+ \frac{k_\mu}{2}\right) - i\sin\left(p_\mu+ \frac{k_\mu}{2}\right) \right] \widetilde{\psi}(p).\\
&\begin{tikzpicture}[line width=1.5 pt, scale=2]
	\draw[fermion] (0,0.5) -- (0.5,0);
	\draw[fermionbar] (0,-0.5) -- (0.5,0);
	\draw[vector] (0.5,0) -- (1,0);
	% The labels
	\node at (1.2,0){$\gamma$};
	\node at (1.2,-0.2){$k,\mu$};
	\node at (-0.15,0.65){$p$};
	\node at (-0.15,-0.65){$p+k$};
	\node at (3.8,0){$\equiv (V_1)_\mu(p, k) = {\displaystyle -i\frac{eQ}{\sqrt{V}} \left[ \gamma_\mu \cos\left(p_\mu+ \frac{k_\mu}{2}\right) - i r \sin\left(p_\mu+ \frac{k_\mu}{2}\right) \right]}\, .$};
\end{tikzpicture}
\end{align}

\subsection*{Terms in $e^2$}
This is going to be an irrelevant vertex, meaning that in the continuum there is no counterpart
\begin{align}
\nn
\text{S}_L^{\text{int}, e^2} &= -\frac{e^2Q^2}{4} \sum_{n,\mu}  \psibar(n) \big[ (\gamma_\mu - 1) A_\mu(n+\hat{\mu}/2)A_\mu(n+\hat{\mu}/2) \psi(n+\hat{\mu}) \\
\nn
& \phantom{=-\frac{e^2Q^2}{4} \sum_{n,\mu}  \psibar(n) \big[ } - (\gamma_\mu + 1) A_\mu(n-\hat{\mu}/2)A_\mu(n-\hat{\mu}/2)  \psi(n-\hat{\mu})\big] \\
\nn
& = -\frac{e^2Q^2}{4}\frac{1}{V^{2}} \sum_{n,\mu}\sum_{\bar{p}, p, k_1,k_2} \widetilde{\psibar}(\bar{p}) \widetilde{A}_\mu(k_1)\widetilde{A}_\mu(k_2) \e^{-i\bar{p}\cdot n}\e^{ik_1\cdot n}\e^{ik_2\cdot n}\e^{ip\cdot n}  \big[ (\gamma_\mu - 1) \e^{i(k_{1,\mu}+k_{2,\mu})/2} \e^{ip_\mu}  \\
\nn
&  \phantom{=-\frac{e^2Q^2}{4} \sum_{n,\mu}  \psibar(n) \big[}   - (\gamma_\mu + 1) \e^{-i(k_{1,\mu}+k_{2,\mu})/2} \e^{-ip_\mu} \big] \widetilde{\psi}(p)\\
\nn
&= -\frac{e^2Q^2}{4}\frac{1}{V} \sum_{\mu}\sum_{\bar{p}, p, k_1,k_2} \delta\left(p+k_1+k_2-\bar{p}\right) \widetilde{\psibar}(\bar{p}) \widetilde{A}_\mu(k_1)\widetilde{A}_\mu(k_2) \bigg[ \gamma_\mu 2i \sin\left(p_\mu+ \frac{k_{1,\mu}+k_{2,\mu}}{2}\right)  \\
\nn
&  \phantom{=-\frac{e^2Q^2}{4} \sum_{n,\mu}  \psibar(n) \big[}   -2 \cos\left(p_\mu+ \frac{k_{1,\mu}+k_{2,\mu}}{2}\right) \bigg] \widetilde{\psi}(p)\\
\nn
&= -\frac{e^2Q^2}{2}\frac{1}{V} \sum_{\mu}\sum_{p, k_1,k_2} \widetilde{\psibar}(p+k_1+k_2) \widetilde{A}_\mu(k_1)\widetilde{A}_\mu(k_2) \bigg[ \gamma_\mu i \sin\left(p_\mu+ \frac{k_{1,\mu}+k_{2,\mu}}{2}\right)  \\
&  \phantom{=-\frac{e^2Q^2}{4} \sum_{n,\mu}  \psibar(n) \big[}   - \cos\left(p_\mu+ \frac{k_{1,\mu}+k_{2,\mu}}{2}\right) \bigg] \widetilde{\psi}(p).\\
&\begin{tikzpicture}[line width=1.5 pt, scale=2]
	\draw[fermion] (0,0.5) -- (0.5,0);
	\draw[fermionbar] (0,-0.5) -- (0.5,0);
	\draw[vector] (0.5,0) -- (1,0.5);
	\draw[vector] (0.5,0) -- (1,-0.5);
	% The labels
	\node at (1.2,0.5){$\gamma$};
	\node at (1.2,0.3){$k_1, \mu$};
	\node at (1.2,-0.5){$\gamma$};
	\node at (1.2,-0.3){$k_2, \nu$};
	\node at (-0.15,0.65){$p$};
	\node at (-0.15,-0.65){$p+k_1+k_2$};
	\node at (3.5,0){$\equiv {\displaystyle (V_2)_{\mu,\nu}(p, k_1, k_2) = \frac{e^2Q^2}{2V}\delta_{\mu\nu} \bigg[ \gamma_\mu i \sin\left(p_\mu+ \frac{k_{1,\mu}+k_{2,\mu}}{2}\right) }$};
	\node at (4.6,-0.8){${\displaystyle - r \cos\left(p_\mu+ \frac{k_{1,\mu}+k_{2,\mu}}{2}\right) \bigg]}\, .$};
\end{tikzpicture}
\end{align}
 
\subsection{Clover term}

The clover term in the action is given by (dropping the flavor index $f$) 
\begin{align}
\text{S}_\text{clover} & = - \frac{i}{4} r c_{sw} \sum_{n\in\Lambda} \sum_{\mu, \nu} \psibar(n) \sigma_{\mu\nu} \psi(n) \widehat{F}_{\mu\nu},
\end{align}
where $\sigma_{\mu\nu} = \frac{1}{2}\left[\gamma_\mu,\gamma_\nu\right]$ and $\widehat{F}_{\mu\nu} = \frac{1}{8i} \left[ Q_{\mu\nu}(n) - Q_{\nu\mu}(n) \right]$, with the definition
\begin{align}
\nn
Q_{\mu\nu}(n) =& \,U_\mu(n)\, U_\nu(n+\hat{\mu})\, U^\dagger_\mu(n+\hat{\nu})\, U^\dagger_\nu(n)
+ U_\nu(n)\, U^\dagger_\mu(n-\hat{\mu}+\hat{\nu})\, U^\dagger_\nu(n-\hat{\mu})\, U_\mu(n-\hat{\mu})\\
\nn
& + U^\dagger_\mu(n-\hat{\mu})\, U^\dagger_\nu(n-\hat{\mu}-\hat{\nu})\, U_\mu(n-\hat{\mu}-\hat{\nu})\, U_\nu(n-\hat{\nu})\\
\label{eq:clover}
& + U^\dagger_\nu(n-\hat{\nu})\, U_\mu(n-\hat{\nu})\, U_\nu(n+\hat{\mu}-\hat{\nu})\, U^\dagger_\mu(n).
\end{align}
The multiplication of two link is simply the sum of the exponent since we are dealing with an abelian group, i.e.~
\begin{align}
U_\mu(n) U_\nu(n) = \e^{ ieQ \left[ A_\mu(n+\hat{\mu}/2)) + A_\nu(n+\hat{\nu}/2))\right]}.
\end{align}
By expanding Eq.~\ref{eq:clover} linearly in $e$, we obtain for $Q_{\mu\nu}$ 
\begin{align}
\nn
Q_{\mu\nu}^{e} = ieQ \bigg[ &\cancel{A_\mu(n+\hat{\mu}/2)} + A_\nu(n+\hat{\mu}+\hat{\nu}/2) - A_\mu(n+\hat{\nu}+\hat{\mu}/2) - \cancel{A_\nu(n+\hat{\nu}/2)} \\
\nn
& + \cancel{A_\nu(n+\hat{\nu}/2)} - A_\mu(n-\hat{\mu}/2+\hat{\nu}) - A_\nu(n-\hat{\mu}+\hat{\nu}/2) + \cancel{A_\mu(n-\hat{\mu}/2)} \\
\nn
& - \cancel{A_\mu(n-\hat{\mu}/2)} - A_\nu(n-\hat{\mu}-\hat{\nu}/2) + A_\mu(n-\hat{\mu}/2-\hat{\nu}) + \cancel{A_\nu(n-\hat{\nu}/2)} \\
& - \cancel{A_\nu(n-\hat{\nu}/2)} + A_\mu(n+\hat{\mu}/2-\hat{\nu}) + A_\nu(n+\hat{\mu}-\hat{\nu}/2) - \cancel{A_\mu(n+\hat{\mu}/2)} \bigg].
\end{align}
By direct inspection in QED we have that $Q_{\mu\nu}^{e} = -Q_{\nu\mu}^{e}$, i.e.~the $Q^{(e)}$ is antisymmetric and it is contracted, in the action, with $\sigma_{\mu\nu}$ (which is also antisymmetric).
The result is that even powers in $e$ vanish because the second order in $e$ is given by the square of the first order, that is symmetric, $Q_{\mu\nu}^{e^2} = \left(Q_{\mu\nu}^{e}\right)^2 = Q_{\nu\mu}^{e^2}$, contracted with an antisymmetric tensor.
The first order in $e$ of $\widehat{F}$ is then given by
\begin{align}
\nn
\widehat{F}_{\mu\nu}^{(e)} = &\frac{1}{8i} 2 Q_{\mu\nu}^{(e)} = \frac{1}{4i} (ieQ) \bigg[ A_\nu(n+\hat{\mu}+\hat{\nu}/2) - A_\mu(n+\hat{\nu}+\hat{\mu}/2) - A_\mu(n-\hat{\mu}/2+\hat{\nu}) - A_\nu(n-\hat{\mu}+\hat{\nu}/2) \\
& - A_\nu(n-\hat{\mu}-\hat{\nu}/2) + A_\mu(n-\hat{\mu}/2-\hat{\nu}) + A_\mu(n+\hat{\mu}/2-\hat{\nu}) + A_\nu(n+\hat{\mu}-\hat{\nu}/2) \bigg].
\end{align}
Since we want to calculate the extra vertexes coming from the clover term we take clover action at the first order in $e$ and we perform the Fourier transform
\begin{align}
\nn
\text{S}_\text{clover}^{e} & = - \frac{i}{4} \frac{r c_{sw}}{V^{3/2}} \frac{eQ}{4} \sum_{n,\mu\nu} \sum_{\bar{p}, p, k} \widetilde{\psibar}(\bar{p}) \sigma_{\mu\nu} \widetilde{\psi}(p) \e^{i(p+k-\bar{p})\cdot n}\\
\nn
&\phantom{=} \times \bigg[ \e^{i k_\mu} \e^{i k_\nu/2} \widetilde{A}_\nu(k) 
- \e^{i k_\mu/2} \e^{i k_\nu} \widetilde{A}_\mu(k) 
- \e^{-i k_\mu/2} \e^{i k_\nu} \widetilde{A}_\mu(k) 
- \e^{i k_\nu/2} \e^{-i k_\mu} \widetilde{A}_\nu(k)\\
\nn
&\phantom{=\times\bigg[}
- \e^{-i k_\nu/2} \e^{-i k_\mu} \widetilde{A}_\nu(k)
- \e^{-i k_\mu/2} \e^{-i k_\nu} \widetilde{A}_\mu(k) 
+ \e^{i k_\mu/2} \e^{-i k_\nu} \widetilde{A}_\mu(k) 
+ \e^{-i k_\nu/2} \e^{i k_\mu} \widetilde{A}_\nu(k) \bigg]\\
\nn
& = - \frac{ir}{16} \frac{e Q\,c_{sw}}{\sqrt{V}} \sum_{\mu\nu} \sum_{p, k} \widetilde{\psibar}(p+k) \sigma_{\mu\nu} \widetilde{\psi}(p) 
\bigg\{ \widetilde{A}_\nu(k) 2i\sin k_\mu \left[ \e^{ik_\nu/2} + \e^{-ik_\nu/2} \right] \\
\nn
& \phantom{ = - \frac{ir}{16} \frac{e Q\,c_{sw}}{\sqrt{V}} \sum_{\mu\nu} \sum_{p, k} \widetilde{\psibar}(p+k) \sigma_{\mu\nu} \widetilde{\psi}(p) \bigg\{}
+ \widetilde{A}_\mu(k) 2\cos\left(\frac{k_\mu}{2}\right) \left[ \e^{-ik_\nu} - \e^{ik_\nu} \right] \bigg\}\\
& = - \frac{i^2r}{16} \frac{e Q\,c_{sw}}{\sqrt{V}} 4 \sum_{\mu\nu} \sum_{p, k} \widetilde{\psibar}(p+k) \sigma_{\mu\nu} \widetilde{\psi}(p) 
\left\{ \widetilde{A}_\nu(k) \sin k_\mu\cos\left(\frac{k_\nu}{2}\right) - \widetilde{A}_\mu(k) \sin k_\nu\cos\left(\frac{k_\mu}{2}\right) \right\}.
\end{align}
By renaming the first term $\mu \leftrightarrow \nu$ and using that $\sigma_{\mu\nu}$ is antisymmetric we get
\begin{align}
\text{S}_\text{clover}^{e} & =
- \frac{r}{2} \frac{e Q\,c_{sw}}{\sqrt{V}} \sum_{\mu\nu} \sum_{p, k} \widetilde{\psibar}(p+k) \sigma_{\mu\nu} \widetilde{\psi}(p) 
\widetilde{A}_\mu(k) \sin k_\nu\cos\left(\frac{k_\mu}{2}\right),
\end{align}
from which we read the associated vertex
\begin{align}
\begin{tikzpicture}[line width=1.5 pt, scale=2]
	\draw[fermion] (0,0.5) -- (0.5,0);
	\draw[fermionbar] (0,-0.5) -- (0.5,0);
	\draw[vector] (0.5,0) -- (1,0);
	% The labels
	\node at (1.2,0){$\gamma$};
	\node at (1.2,-0.2){$k,\mu$};
	\node at (-0.15,0.65){$p$};
	\node at (-0.15,-0.65){$p+k$};
	\node[circle,fill=black,inner sep=0pt,minimum size=10pt] (a) at (0.5,0) {};
	\node at (3.25,0){$\equiv (V_c)_\mu(p, k) = {\displaystyle \frac{r\,e Q\,c_{sw}}{2\sqrt{V}} \sum_{\nu} \sigma_{\mu\nu} \sin k_\nu\cos\left(\frac{k_\mu}{2}\right)}\, .$};
\end{tikzpicture}
\end{align}

\section{Electromagnetic fermion self-energy}

We want to compute the 1-loop contribution to the quark self-energy, the Feynman diagrams contributing are the following ones.

\subsection*{Tadpole in Feynman gauge}
\begin{align}
\nn
I_T^\text{Feyn} &= \begin{tikzpicture}[line width=1.5 pt, scale=2,baseline=2.25ex]
	\draw[fermion] (0,0) -- (1,0);
	\draw[fermion] (1,0) -- (2,0);;
	\draw[vector] (1.5,0.5) arc (0:360:.5);
	% The labels
	\node at (1.,0.8){$k$};
	\node at (0.2,0.15){$p$};
	\node at (1.8,0.15){$p$};
\end{tikzpicture}
\bigg|_{p=0, m=0}  
=\sum_k \sum_{\mu,\nu} (V_2)_{\mu\nu}(0,-k,k) G_{\mu\nu}(k) \\
&= \sum_k \sum_{\mu,\nu} \delta_{\mu\nu} (-r) \frac{e^2Q^2}{2V} \frac{\delta_{\mu\nu}}{\hat{k}^2}
= -\frac{e^2Q^2}{V} \frac{r}{2} \sum_k \frac{4}{\hat{k}^2}.
\end{align}

\subsection*{Tadpole in Coulomb gauge}
In order to calculate the tadpole diagram in Coulomb gauge it is suf\mbox{}ficient to use the following gauge transformation 
\begin{align}
\widetilde{A}^\text{C}_\mu(p) = \sum_\nu \left[ \delta_{\mu\nu} - \frac{\hat{p}_\mu\cdot \left(0, \underline{\hat{p}}\right)_\nu }{|\underline{\hat{p}}|^2} \right] \widetilde{A}_\nu(p),
\end{align}
where $\widetilde{A}$ is the field generated in Feynman gauge and $\widetilde{A}^\text{C}$ the field in Coulomb gauge.\\
For the tadpole it is suf\mbox{}ficient to calculate the propagator in Coulomb gauge
\begin{align}
\nn
\langle \widetilde{A}^\text{C}_\mu\widetilde{A}^\text{C}_\nu \rangle &=  \sum_{\rho,\sigma} \left[ \delta_{\mu\rho} - \frac{\hat{p}_\mu\cdot \left(0, \underline{\hat{p}}\right)_\rho }{|\underline{\hat{p}}|^2} \right]  \left[ \delta_{\nu\sigma} - \frac{\hat{p}_\nu\cdot \left(0, \underline{\hat{p}}\right)_\sigma }{|\underline{\hat{p}}|^2} \right] \langle\widetilde{A}_\rho(p)\widetilde{A}_\sigma(p)\rangle\\
\nn
&=  \sum_{\rho,\sigma} \left[ \delta_{\mu\rho}\delta_{\nu\sigma} 
- \frac{\delta_{\nu\sigma} \hat{p}_\mu\cdot \left(0, \underline{\hat{p}}\right)_\rho + \delta_{\mu\rho}\hat{p}_\nu\cdot \left(0, \underline{\hat{p}}\right)_\sigma}{|\underline{\hat{p}}|^2} 
+ \frac{\hat{p}_\mu\cdot \left(0, \underline{\hat{p}}\right)_\rho \hat{p}_\nu\cdot \left(0, \underline{\hat{p}}\right)_\sigma }{(|\underline{\hat{p}}|^2)^2} \right] \langle\widetilde{A}_\rho(p)\widetilde{A}_\sigma(p)\rangle\\
\nn
&=   \langle\widetilde{A}_\mu(p)\widetilde{A}_\nu(p)\rangle  - \sum_\rho \frac{\hat{p}_\mu\cdot \left(0, \underline{\hat{p}}\right)_\rho}{|\underline{\hat{p}}|^2} \langle\widetilde{A}_\rho(p)\widetilde{A}_\nu(p)\rangle \\
\nn
& \phantom{=} -\sum_\sigma \frac{ \hat{p}_\nu\cdot \left(0, \underline{\hat{p}}\right)_\sigma}{|\underline{\hat{p}}|^2} \langle\widetilde{A}_\mu(p)\widetilde{A}_\sigma(p)\rangle + \sum_{\rho,\sigma} \frac{\hat{p}_\mu\cdot \left(0, \underline{\hat{p}}\right)_\rho \hat{p}_\nu\cdot \left(0, \underline{\hat{p}}\right)_\sigma }{(|\underline{\hat{p}}|^2)^2} \langle\widetilde{A}_\rho(p)\widetilde{A}_\nu(p)\rangle
\end{align}
\begin{align}
\nn
&=   \frac{\delta_{\mu\nu}}{\hat{p}^2}  - \sum_\rho \frac{\hat{p}_\mu\cdot \left(0, \underline{\hat{p}}\right)_\rho}{|\underline{\hat{p}}|^2} \frac{\delta_{\rho\nu}}{\hat{p}^2}  -\sum_\sigma \frac{ \hat{p}_\nu\cdot \left(0, \underline{\hat{p}}\right)_\sigma}{|\underline{\hat{p}}|^2}  \frac{\delta_{\mu\sigma}}{\hat{p}^2} 
+ \sum_{\rho,\sigma} \frac{\hat{p}_\mu\cdot \left(0, \underline{\hat{p}}\right)_\rho \hat{p}_\nu\cdot \left(0, \underline{\hat{p}}\right)_\sigma }{(|\underline{\hat{p}}|^2)^2} \frac{\delta_{\rho\sigma}}{\hat{p}^2}\\
&=   \frac{\delta_{\mu\nu}}{\hat{p}^2} 
- \frac{\hat{p}_\mu\cdot \left(0, \underline{\hat{p}}\right)_\nu + \hat{p}_\nu\cdot \left(0, \underline{\hat{p}}\right)_\mu}{\hat{p}^2\,|\underline{\hat{p}}|^2} 
+ \sum_{\rho} \frac{\hat{p}_\mu\cdot \left(0, \underline{\hat{p}}\right)_\rho \hat{p}_\nu\cdot \left(0, \underline{\hat{p}}\right)_\rho }{\hat{p}^2\,(|\underline{\hat{p}}|^2)^2}.
\end{align}
Now we can insert the propagator in the tadpole and calculate the value of the integral in Coulomb gauge
\begin{align}
\nn
I_T^\text{Coul} &= -\frac{e^2Q^2}{V} \frac{r}{2} \sum_k \sum_\mu \left[ \frac{1}{\hat{k}^2} 
- 2\frac{\hat{k}_\mu\cdot \left(0, \underline{\hat{k}}\right)_\mu }{\hat{k}^2\,|\underline{\hat{k}}|^2} 
+ \frac{\hat{k}_\mu^2 \sum_j \underline{\hat{k}}^2_j }{\hat{k}^2\,(|\underline{\hat{k}}|^2)^2}  \right] 
= -\frac{e^2Q^2}{V} \frac{r}{2} \sum_k \left[ \frac{4}{\hat{k}^2} 
- 2\frac{\sum_j \hat{k}_j^2}{\hat{k}^2\,|\underline{\hat{k}}|^2} 
+ \frac{\hat{k}^2 }{\hat{k}^2\,|\underline{\hat{k}}|^2}  \right]\\
&= -\frac{e^2Q^2}{V} \frac{r}{2} \sum_k \left[ \frac{4}{\hat{k}^2} 
- \frac{2}{\hat{k}^2} 
+ \frac{1 }{|\underline{\hat{k}}|^2}  \right]
= -\frac{e^2Q^2}{V} \frac{r}{2} \sum_k \left[ \frac{2}{\hat{k}^2} 
+ \frac{1 }{|\underline{\hat{k}}|^2}  \right].
\end{align}
The tadpole result in Coulomb gauge is the same quoted in Ref.~\cite{Duncan:1996xy}.

\subsection*{Sunset in Feynman gauge}
\begin{align}
\nn
I_S^\text{Feyn} &= 	\begin{tikzpicture}[line width=1.5 pt, scale=2,baseline=2.25ex]
	\draw[fermion] (0,0) -- (0.7,0);
	\draw[fermion] (0.7,0) -- (1.4,0);
	\draw[fermion] (1.4,0) -- (2.1,0);
	\draw[vector] (1.5,0) arc (0:180:.5);
	% The labels
	\node at (1.1,0.7){$k$};
	\node at (1.05,0.15){$p-k$};
	\node at (0.1,0.15){$p$};
	\node at (1.8,0.15){$p$};
	\node at (0.5,-0.2){$\mu$};
	\node at (1.5,-0.2){$\nu$};
	\node at (2.45,0){$\bigg|_{p=0, m=0}$};
%	\node at (-0.33,0){$I_S^\text{Feyn} = $};
\end{tikzpicture}  =  \sum_{k} \sum_{\mu,\nu} (V_1)_\nu(-k, 0) S(-k) (V_1)_\mu(0,-k)G_{\mu\nu}(k) \\
\nn
& = -\frac{e^2Q^2}{V} \sum_k \sum_{\mu,\nu,\sigma} \frac{\delta_{\mu\nu}}{\hat{k}^2} \left[ \gamma_\nu \cos\left(\frac{k_\nu}{2}\right) + ir\sin\left(\frac{k_\nu}{2}\right) \right] \frac{ \frac{r}{2}\hat{k}^2_\sigma + i \gamma_\sigma \bar{k}_\sigma  }{\frac{r^2}{4}\left(\hat{k}^2\right)^2 + \bar{k}^2 }  \\
&\phantom{-\frac{e^2Q^2}{V} \sum_k \sum_{\mu,\nu,\sigma} \frac{\delta_{\mu\nu}}{\hat{k}^2}}
\times\left[ \gamma_\mu \cos\left(\frac{k_\mu}{2}\right) + ir\sin\left(\frac{k_\mu}{2}\right) \right].
\end{align}
We retain only the even functions of $k$ since the integration is done in a symmetric interval, this gives us 
\begin{align}
\nn
I_S^\text{Feyn}  =  & -\frac{e^2Q^2}{V} \sum_k \sum_{\mu,\sigma} \frac{1}{\hat{k}^2} \frac{1}{{\frac{r^2}{4}\left(\hat{k}^2\right)^2 + \bar{k}^2 }} \bigg[ \gamma_\mu^2 \frac{r}{2} \hat{k}^2_\sigma \cos^2\left(\frac{k_\mu}{2}\right) + \gamma_\mu\gamma_\sigma (i)^2 r \bar{k}_\sigma \sin\left(\frac{k_\mu}{2}\right)\cos\left(\frac{k_\mu}{2}\right) \\
\nn
& + (ir)^2 \frac{r}{2} \sin^2\left(\frac{k_\mu}{2}\right)\hat{k}_\sigma^2 + \gamma_\sigma\gamma_\mu (i)^2 r \bar{k}_\sigma \sin\left(\frac{k_\mu}{2}\right)\cos\left(\frac{k_\mu}{2}\right) \bigg]\\
\nn
= & -\frac{e^2Q^2}{V} \sum_k \sum_{\mu,\sigma} \frac{1}{\hat{k}^2} \frac{1}{{\frac{r^2}{4}\left(\hat{k}^2\right)^2 + \bar{k}^2 }} \bigg[ \frac{r}{2} \hat{k}^2_\sigma \left(1-\sin^2\left(\frac{k_\mu}{2}\right)\right) - r \bar{k}_\sigma \frac{1}{2}\sin\left(k_\mu\right)\underbrace{\{\gamma_\mu,\gamma_\sigma\}}_{2\delta_{\mu\sigma}}  \\
\nn
& \phantom{-\frac{e^2Q^2}{V} \sum_k \sum_{\mu,\sigma} \frac{1}{\hat{k}^2} \frac{1}{{\frac{r^2}{4}\left(\hat{k}^2\right)^2 + \bar{k}^2 }} \bigg[}
- \frac{r^3}{2} \sin^2\left(\frac{k_\mu}{2}\right)\hat{k}_\sigma^2 \bigg]\\
\nn
= & -\frac{e^2Q^2}{V} \sum_k \sum_{\mu,\sigma} \frac{1}{\hat{k}^2} \frac{1}{{\frac{r^2}{4}\left(\hat{k}^2\right)^2 + \bar{k}^2 }} \bigg[ \frac{r}{2} \hat{k}^2_\sigma \left(1-\frac{\hat{k}^2_\mu}{4}\right) - r \bar{k}_\sigma\bar{k}_\mu \delta_{\mu\sigma}  - \frac{r^3}{2} \frac{\hat{k}_\mu^2}{4}\hat{k}_\sigma^2 \bigg]
\end{align}
\begin{align}
\nn
= & -\frac{e^2Q^2}{V} \sum_k \frac{1}{\hat{k}^2} \frac{r}{{\frac{r^2}{4}\left(\hat{k}^2\right)^2 + \bar{k}^2 }} \bigg[ \frac{\hat{k}^2}{2} \left(1-\frac{\hat{k}^2}{4}\right) - \bar{k}^2   - \frac{r^2}{2} \frac{(\hat{k}^2)^2}{4} \bigg]\\
\nn
= & -\frac{e^2Q^2}{V} r \sum_k \frac{1}{\hat{k}^2} \frac{1}{{\frac{r^2}{4}\left(\hat{k}^2\right)^2 + \bar{k}^2 }} \bigg[ 2\hat{k}^2 - \frac{(\hat{k}^2)^2}{8} - \bar{k}^2 - \frac{r^2}{8} (\hat{k}^2)^2 - \frac{\bar{k}^2}{2} + \frac{\bar{k}^2}{2}\bigg]\\
= & -\frac{e^2Q^2}{V} \frac{r}{2} \sum_k \frac{1}{\hat{k}^2} \left[\frac{\hat{k}^2\left(4 - \hat{k}^2/4\right) - \bar{k}^2}{{\frac{r^2}{4}\left(\hat{k}^2\right)^2 + \bar{k}^2 }} - 1\right].
\end{align}

\subsection*{Clover sunset in Feynman gauge}
\begin{align}
\nn
I_{CS}^\text{Feyn} &= 	\begin{tikzpicture}[line width=1.5 pt, scale=2,baseline=2.25ex]
	\draw[fermion] (0,0) -- (0.7,0);
	\draw[fermion] (0.7,0) -- (1.4,0);
	\draw[fermion] (1.4,0) -- (2.1,0);
	\draw[vector] (1.5,0) arc (0:180:.5);
	% The labels
	\node at (1.1,0.7){$k$};
	\node at (1.05,0.15){$p-k$};
	\node at (0.1,0.15){$p$};
	\node at (1.8,0.15){$p$};
	\node at (0.5,-0.2){$\mu$};
		\node[circle,fill=black,inner sep=0pt,minimum size=10pt] (a) at (0.5,0) {};
	\node at (1.5,-0.2){$\nu$};
		\node[circle,fill=black,inner sep=0pt,minimum size=10pt] (a) at (1.45,0) {};
	\node at (2.45,0){$\bigg|_{p=0, m=0}$};
%	\node at (-0.33,0){$I_{CS}^\text{Feyn} = $};
\end{tikzpicture}  =  \sum_{k} \sum_{\mu,\nu} (V_c)_\nu(-k, 0) S(-k) (V_c)_\mu(0,-k)G_{\mu\nu}(k) \\
& = \frac{r^2 c_{sw}^2 e^2 Q^2}{4V} \sum_k \sum_{\mu,\nu,\rho,\sigma,\gamma} 
\frac{\delta_{\mu\nu}}{\hat{k}^2} 
\left[ -\sigma_{\nu\rho}\sin k_\rho \cos\left(\frac{k_\nu}{2}\right)\right]
\frac{ \frac{r}{2}\hat{k}^2_\gamma + i \gamma_\gamma \bar{k}_\gamma  }{\frac{r^2}{4}\left(\hat{k}^2\right)^2 + \bar{k}^2 }  
\, \sigma_{\mu\sigma}\sin k_\sigma \cos\left(\frac{k_\mu}{2}\right).
\end{align}
As usual we take only the even powers in $k$ and we get
\begin{align}
& I_{CS}^\text{Feyn} = -\frac{r^3 c_{sw}^2 e^2 Q^2}{8V} \sum_{k, \mu,\rho,\sigma} 
\frac{\hat{k}^2}{\hat{k}^2} \frac{\sigma_{\mu\rho}\sigma_{\mu\sigma} \bar{k}_\sigma\bar{k}_\rho \left(1-\hat{k}^2_\mu/4\right)}{\frac{r^2}{4}\left(\hat{k}^2\right)^2 + \bar{k}^2 }.
\end{align}
We need to find what is the product of the two antisymmetric tensors
\begin{align}
4 \sigma_{\mu\rho}\sigma_{\mu\sigma} &= [ \gamma_\mu, \gamma_\rho] [\gamma_\mu, \gamma_\sigma] 
= \left( \gamma_\mu \gamma_\rho - \gamma_\rho \gamma_\mu\right) \left(\gamma_\mu \gamma_\sigma - \gamma_\sigma \gamma_\mu \right)
= 4\delta_{\rho\sigma}\delta_{\mu\sigma} - 2 \gamma_\mu\gamma_\rho\gamma_\sigma\gamma_\mu - 2 \gamma_rho\gamma_\sigma.
\end{align}
The product $\sigma_{\mu\rho}\sigma_{\mu\sigma}$ is contracted with $\bar{k}_\sigma\bar{k}_\rho$ that is symmetric with respect to $\sigma$ and $\rho$, this means we can retain only the symmetric part of the product with respect to $\sigma$ and $\rho$, i.e.~
\begin{align}
\nn
4 \sigma_{\mu\rho}\sigma_{\mu\sigma} \bar{k}_\sigma\bar{k}_\rho 
&= 4\delta_{\rho\sigma}\delta_{\mu\sigma}\bar{k}_\sigma\bar{k}_\rho 
- 2 \gamma_\mu \frac{1}{2}\{\gamma_\rho,\gamma_\sigma\} \gamma_\mu\bar{k}_\sigma\bar{k}_\rho 
- 2 \frac{1}{2}\{\gamma_\rho,\gamma_\sigma\} \bar{k}_\sigma\bar{k}_\rho\\
& = 4\delta_{\rho\sigma}\delta_{\mu\sigma}\bar{k}_\sigma\bar{k}_\rho 
- 2 \delta_{\rho\sigma} \bar{k}_\sigma\bar{k}_\rho 
- 2 \delta_{\rho\sigma} \bar{k}_\sigma\bar{k}_\rho
= 4\bar{k}_\sigma\bar{k}_\rho
\left( \delta_{\rho\sigma}\delta_{\mu\sigma} - \delta_{\rho\sigma} \right).
\end{align}
The substituting back in the original integral we get
\begin{align}
I_{CS}^\text{Feyn} = -\frac{r^3 c_{sw}^2 e^2 Q^2}{8V} \sum_{k, \mu,\rho,\sigma} 
\frac{1-\hat{k}^2_\mu/4}{\frac{r^2}{4}\left(\hat{k}^2\right)^2 + \bar{k}^2 } 
\bar{k}_\sigma\bar{k}_\rho
\left( \delta_{\rho\sigma}\delta_{\mu\sigma} - \delta_{\rho\sigma} \right).
\end{align}
The first term in the parentheses gives
\begin{align}
\nn
\sum_{\mu,\rho,\sigma}\left(1-\hat{k}^2_\mu/4\right) \bar{k}_\sigma\bar{k}_\rho \delta_{\rho\sigma}\delta_{\mu\sigma} 
&= \sum_{\mu,\sigma} \left(1-\hat{k}^2_\mu/4\right) \bar{k}^2_\sigma \delta_{\mu\sigma} 
= \sum_{\mu} \left(1-\hat{k}^2_\mu/4\right) \bar{k}^2_\mu\\
&= \bar{k}^2 - \frac{1}{4} \sum_{\mu} \hat{k}^2_\mu\bar{k}^2_\mu 
= \bar{k}^2 - \frac{1}{4} \hat{k}^4 + \frac{1}{16} \hat{k}^6,
\end{align}
where we used that $\sin(x) = 2 \sin(x/2) \cos(x/2)$. The second term in the parentheses gives
\begin{align}
\sum_{\mu,\rho,\sigma}\left(1-\hat{k}^2_\mu/4\right) \bar{k}_\sigma\bar{k}_\rho \delta_{\rho\sigma}
= \sum_{\mu,\rho}\left(1-\hat{k}^2_\mu/4\right) \bar{k}^2_\rho = \left((4-\hat{k}^2/4\right) \bar{k}^2 
= 4\bar{k}^2-\frac{\bar{k}^2\hat{k}^2}{4}.
\end{align}
By subtracting the second term to the first one and putting everything together we get the final result
\begin{align}
I_{CS}^\text{Feyn} &= \frac{r^3 c_{sw}^2 e^2 Q^2}{8V} \sum_{k} 
\frac{3\bar{k}^2 + \hat{k}^4/4 - \hat{k}^6/16 - \hat{k}^2\bar{k}^2/4 }
{\frac{r^2}{4}\left(\hat{k}^2\right)^2 + \bar{k}^2 } .
\end{align}

\subsection*{Mixed sunset in Feynman gauge}
The first diagram is given by
\begin{align}
\nn
I_{M1}^\text{Feyn} = &	\begin{tikzpicture}[line width=1.5 pt, scale=2,baseline=2.25ex]
	\draw[fermion] (0,0) -- (0.7,0);
	\draw[fermion] (0.7,0) -- (1.4,0);
	\draw[fermion] (1.4,0) -- (2.1,0);
	\draw[vector] (1.5,0) arc (0:180:.5);
	% The labels
	\node at (1.1,0.7){$k$};
	\node at (1.05,0.15){$p-k$};
	\node at (0.1,0.15){$p$};
	\node at (1.8,0.15){$p$};
	\node at (0.5,-0.2){$\mu$};
		\node[circle,fill=black,inner sep=0pt,minimum size=10pt] (a) at (0.5,0) {};
	\node at (1.5,-0.2){$\nu$};
	\node at (2.45,0){$\bigg|_{p=0, m=0}$};
%	\node at (-0.33,0){$I_{M1}^\text{Feyn} = $};
\end{tikzpicture}  
=  \sum_{k} \sum_{\mu,\nu} (V_1)_\nu(-k, 0) S(-k) (V_c)_\mu(0,-k)G_{\mu\nu}(k) \\
\nn
= & -i \left(\frac{e Q}{\sqrt{V}}\right)^2 \frac{rc_{sw}}{2}
\sum_k \sum_{\mu,\nu,\sigma,\rho} 
\frac{\delta_{\mu\nu}}{\hat{k}^2} 
\left[ \gamma_\nu \cos\left(\frac{k_\nu}{2}\right) + i r \sin\left(\frac{k_\nu}{2}\right) \right]
\frac{ \frac{r}{2}\hat{k}^2_\rho + i \gamma_\rho \bar{k}_\rho  }{\frac{r^2}{4}\left(\hat{k}^2\right)^2 + \bar{k}^2 }  \\
\nn
& \phantom{ -i \left(\frac{e Q}{\sqrt{V}}\right)^2 \frac{rc_{sw}}{2} \sum_k \sum_{\mu,\nu,\sigma,\rho} \frac{\delta_{\mu\nu}}{\hat{k}^2} }
\times \sigma_{\mu\sigma}(-\sin k_\sigma) \cos\left(\frac{k_\mu}{2}\right)\\
\stackrel{=}{(\text{even in k})} &  i \left(\frac{e Q}{\sqrt{V}}\right)^2 \frac{rc_{sw}}{2}
\sum_k \sum_{\mu,\sigma,\rho}  
\frac{ ir^2\hat{k}^2_\rho \hat{k}_\mu + i \gamma_\mu\gamma_\rho \bar{k}_\rho \cos\left(\frac{k_\mu}{2}\right) }{\hat{k}^2\left(\frac{r^2}{4}\left(\hat{k}^2\right)^2 + \bar{k}^2\right) }  
\, \sigma_{\mu\sigma} \bar{k}_\sigma \cos\left(\frac{k_\mu}{2}\right).
\end{align}
The second diagrams is
\begin{align}
\nn
I_{M2}^\text{Feyn} = &	\begin{tikzpicture}[line width=1.5 pt, scale=2,baseline=2.25ex]
	\draw[fermion] (0,0) -- (0.7,0);
	\draw[fermion] (0.7,0) -- (1.4,0);
	\draw[fermion] (1.4,0) -- (2.1,0);
	\draw[vector] (1.5,0) arc (0:180:.5);
	% The labels
	\node at (1.1,0.7){$k$};
	\node at (1.05,0.15){$p-k$};
	\node at (0.1,0.15){$p$};
	\node at (1.8,0.15){$p$};
	\node at (0.5,-0.2){$\mu$};
	\node at (1.5,-0.2){$\nu$};
		\node[circle,fill=black,inner sep=0pt,minimum size=10pt] (a) at (1.45,0) {};
	\node at (2.45,0){$\bigg|_{p=0, m=0}$};
%	\node at (-0.33,0){$I_{M2}^\text{Feyn} = $};
\end{tikzpicture}  
=  \sum_{k} \sum_{\mu,\nu} (V_c)_\nu(-k, 0) S(-k) (V_1)_\mu(0,-k)G_{\mu\nu}(k) \\
\nn
= &  -i \left(\frac{e Q}{\sqrt{V}}\right)^2 \frac{rc_{sw}}{2}
\sum_k \sum_{\mu,\nu,\sigma,\rho} 
\frac{\delta_{\mu\nu}}{\hat{k}^2} 
\sigma_{\nu\sigma}\sin k_\sigma \cos\left(\frac{k_\nu}{2}\right)
\frac{ \frac{r}{2}\hat{k}^2_\rho + i \gamma_\rho \bar{k}_\rho  }{\frac{r^2}{4}\left(\hat{k}^2\right)^2 + \bar{k}^2 }  \\
\nn
& \phantom{-i \left(\frac{e Q}{\sqrt{V}}\right)^2 \frac{rc_{sw}}{2}\sum_k \sum_{\mu,\nu,\sigma,\rho} \frac{\delta_{\mu\nu}}{\hat{k}^2} }
\times \left[ \gamma_\mu \cos\left(\frac{k_\mu}{2}\right) + i r \sin\left(\frac{k_\mu}{2}\right) \right]\\
\stackrel{=}{(\text{even in k})} & -i \left(\frac{e Q}{\sqrt{V}}\right)^2 \frac{rc_{sw}}{2}
\sum_k \sum_{\mu,\sigma,\rho}  
\sigma_{\mu\sigma} \bar{k}_\sigma \cos\left(\frac{k_\mu}{2}\right)
\frac{ ir^2\hat{k}^2_\rho \hat{k}_\mu + i \gamma_\rho\gamma_\mu \bar{k}_\rho \cos\left(\frac{k_\mu}{2}\right) }{\hat{k}^2\left(\frac{r^2}{4}\left(\hat{k}^2\right)^2 + \bar{k}^2\right) }  .
\end{align}
By putting the two diagrams together the terms with $\sigma_{\mu\sigma}$ cancel out  because of its antisymmetric nature and we are left with
\begin{align}
\nn
I_{M1}^\text{Feyn} + I_{M2}^\text{Feyn} &= \frac{e^2 Q^2}{V}\,\frac{rc_{sw}}{2} \sum_k
\frac{ 1 }{\hat{k}^2\left(\frac{r^2}{4}\left(\hat{k}^2\right)^2 + \bar{k}^2\right) }\sum_{\mu, \sigma,\rho} \bar{k}_\sigma\bar{k}_\rho \cos^2\left(\frac{k_\mu}{2}\right) \left( \sigma_{\mu\sigma} \gamma_\rho \gamma_\mu - \gamma_\mu \gamma_\rho\sigma_{\mu\sigma} \right)\\
&= \frac{e^2 Q^2}{V}\,\frac{rc_{sw}}{2} \sum_{k, \mu, \sigma,\rho}
\frac{ \bar{k}_\sigma\bar{k}_\rho \left(1-\hat{k}^2_\mu/4\right)  }{\hat{k}^2\left(\frac{r^2}{4}\left(\hat{k}^2\right)^2 + \bar{k}^2\right) } \left( \sigma_{\mu\sigma} \gamma_\rho \gamma_\mu - \gamma_\mu \gamma_\rho\sigma_{\mu\sigma} \right).
\end{align}
The term with the antisymmetric tensor gives
\begin{align}
\left( \sigma_{\mu\sigma} \gamma_\rho \gamma_\mu - \gamma_\mu \gamma_\rho\sigma_{\mu\sigma} \right) = \frac{1}{2} \delta_{\rho\sigma} -\frac{1}{2} \delta_{\sigma\mu}\delta_{\rho\mu},
\end{align}
and collecting everything we get
\begin{align}
\nn
I_{M1}^\text{Feyn} + I_{M2}^\text{Feyn} 
&= \frac{e^2 Q^2}{V}\,\frac{rc_{sw}}{4} \sum_{k, \mu, \sigma,\rho}
\frac{ \bar{k}_\sigma\bar{k}_\rho \left(1-\hat{k}^2_\mu/4\right)  }{\hat{k}^2\left(\frac{r^2}{4}\left(\hat{k}^2\right)^2 + \bar{k}^2\right) } \left(  \delta_{\rho\sigma} - \delta_{\sigma\mu}\delta_{\rho\mu} \right)\\
\nn
&= \frac{e^2 Q^2}{V}\,\frac{rc_{sw}}{4} \sum_{k}
\frac{ 1 }{\hat{k}^2\left(\frac{r^2}{4}\left(\hat{k}^2\right)^2 + \bar{k}^2\right) } \left( \sum_{\mu,\rho}\bar{k}^2_\rho \left(1-\hat{k}^2_\mu/4\right)  - \sum_{\mu}\bar{k}^2_\mu \left(1-\hat{k}^2_\mu/4\right)\right)\\
&= \frac{e^2 Q^2}{V}\,\frac{rc_{sw}}{4} \sum_{k}
\frac{ 3\bar{k}^2 + \hat{k}^4/4   - \hat{k}^6/16 - \hat{k}^2\bar{k}^2/4}{\hat{k}^2\left(\frac{r^2}{4}\left(\hat{k}^2\right)^2 + \bar{k}^2\right) }.
\end{align}

\subsection*{Self-energy resummation}

We symbolically write the self-energy at 1-loop as
{\large
\[
\Sigma^{1}(p; m, e^2) =
\begin{tikzpicture}[line width=1.5 pt, scale=2,baseline=-0.5ex]
	\draw[pattern=north west lines,draw=black, anchor=base,baseline] (2.5,0) arc (0:360:.25);
\end{tikzpicture}
=
\begin{tikzpicture}[line width=1.5 pt, scale=2,baseline=2.25ex]
%%%%%%%%%%%%%%%%%%%%%%%%%%%TADPOLE	
	\draw[vector] (2.5,0.25) arc (0:360:.25);
%%%%%%%%%%%%%%%%%%%%%%%%%%%%%%%%%%%%
\end{tikzpicture}
+
\begin{tikzpicture}[line width=1.5 pt, scale=2,baseline=.7ex]
%%%%%%%%%%%%%%%%%%%%%%%%%%%%%%%%%%%
		\draw[vector] (4.5,0) arc (0:180:.25);
		\draw[fermion, anchor=base] (4,0) -- (4.5,0);
%%%%%%%%%%%%%%%%%%%%%%%%%%%%%%%%%%%%
\end{tikzpicture}\, ,
\]
}
where for simplicity we assume the sunset diagram to be the sum of the ones with dif\mbox{}ferent vertexes insertion, without loss of generality.\\
By summing insertions of self-energy at 1-loop we get that the dressed propagator is given by
{\large
\begin{align*}
\begin{tikzpicture}[line width=1.5 pt, scale=2,baseline=-0.5ex]
	\draw[dressed_fermion, anchor=base,baseline] (0,0) -- (1,0);
\end{tikzpicture}
=&\,
\begin{tikzpicture}[line width=1.5 pt, scale=2]
	\draw[fermion, anchor=base,baseline] (0,0) -- (1,0);
\end{tikzpicture}
+
\begin{tikzpicture}[line width=1.5 pt, scale=2]
		%\node at (1.25,0){$+\,$};
	\draw[fermion, anchor=base] (1.5,0) -- (2.25,0);
%%%%%%%%%%%%%%%%%%%%%%%%%%%TADPOLE	
	\draw[vector] (2.5,0.25) arc (0:360:.25);
%%%%%%%%%%%%%%%%%%%%%%%%%%%%%%%%%%%%	
	\draw[fermion, anchor=base] (2.25,0) -- (3,0);
\end{tikzpicture}
+
\begin{tikzpicture}[line width=1.5 pt, scale=2]
		%\node at (3.25,0){$+\,$};
	\draw[fermion, anchor=base] (3.5,0) -- (4,0);
%%%%%%%%%%%%%%%%%%%%%%%%%%%%%%%%%%%
		\draw[vector] (4.5,0) arc (0:180:.25);
		\draw[fermion, anchor=base] (4,0) -- (4.5,0);
%%%%%%%%%%%%%%%%%%%%%%%%%%%%%%%%%%%%
	\draw[fermion, anchor=base] (4.5,0) -- (5,0);
		%\node at (5.25,0){$+ \dots$};
\end{tikzpicture}\\
&
%\phantom{\begin{tikzpicture}[line width=1.5 pt, scale=2]
%	\draw[dressed_fermion, anchor=base,baseline] (0,0) -- (1,0);
%\end{tikzpicture}
%=}
+
\begin{tikzpicture}[line width=1.5 pt, scale=2]
	\draw[fermion, anchor=base] (1,0) -- (1.75,0);
%%%%%%%%%%%%%%%%%%%%%%%%%%%TADPOLE	
	\draw[vector] (2.,0.25) arc (0:360:.25);
	\draw[fermion, anchor=base] (1.75,0) -- (2.5,0);
	\draw[vector] (2.75,0.25) arc (0:360:.25);
%%%%%%%%%%%%%%%%%%%%%%%%%%%%%%%%%%%%%	
	\draw[fermion, anchor=base] (2.5,0) -- (3,0);
\end{tikzpicture}
+
\begin{tikzpicture}[line width=1.5 pt, scale=2]
	\draw[fermion, anchor=base] (3.5,0) -- (4,0);
%%%%%%%%%%%%%%%%%%%%%%%%%%%%%%%%%%%
		\draw[vector] (4.5,0) arc (0:180:.25);
		\draw[fermion, anchor=base] (4,0) -- (4.5,0);
	\draw[fermion, anchor=base] (4.5,0) -- (5,0);
		\draw[vector] (5.5,0) arc (0:180:.25);
		\draw[fermion, anchor=base] (5,0) -- (5.5,0);
%%%%%%%%%%%%%%%%%%%%%%%%%%%%%%%%%%%%
	\draw[fermion, anchor=base] (5.5,0) -- (6,0);
\end{tikzpicture}
+ \dots\\
%\\
%\vspace{-1cm}
%\phantom{\begin{tikzpicture}[line width=1.5 pt, scale=2,baseline=-0.5ex]
%	\draw[dressed_fermion, anchor=base,baseline] (0,0) -- (1,0);
%\end{tikzpicture}
%}
=&\,
\begin{tikzpicture}[line width=1.5 pt, scale=2,baseline=-0.5ex]
	\draw[fermion, anchor=base,baseline] (0,0) -- (1,0);
\end{tikzpicture}
+
\begin{tikzpicture}[line width=1.5 pt, scale=2,baseline=-0.5ex]
	\draw[fermion, anchor=base] (1.5,0) -- (2,0);
	\draw[pattern=north west lines,draw=black, anchor=base,baseline] (2.5,0) arc (0:360:.25);
	\draw[fermion, anchor=base] (2.5,0) -- (3,0);
\end{tikzpicture}
+
\begin{tikzpicture}[line width=1.5 pt, scale=2,baseline=-0.5ex]
	\draw[fermion, anchor=base] (3.5,0) -- (4,0);
	\draw[pattern=north west lines,draw=black, anchor=base] (4.5,0) arc (0:360:.25);
	\draw[fermion, anchor=base] (4.5,0) -- (5,0);
	\draw[pattern=north west lines,draw=black, anchor=base] (5.5,0) arc (0:360:.25);
	\draw[fermion, anchor=base] (5.5,0) -- (6,0);
\end{tikzpicture}
+\dots\\
%\phantom{\begin{tikzpicture}[line width=1.5 pt, scale=2]
%	\draw[dressed_fermion, anchor=base,baseline] (0,0) -- (1,0);
%\end{tikzpicture}}
= &\,
\begin{tikzpicture}[line width=1.5 pt, scale=2,baseline=-0.5ex]
	\draw[fermion, anchor=base,baseline] (0,0) -- (1,0);
\end{tikzpicture}
\left( 1 +
\begin{tikzpicture}[line width=1.5 pt, scale=2,baseline=-0.5ex]
	\draw[pattern=north west lines,draw=black, anchor=base,baseline] (2.5,0) arc (0:360:.25);
	\draw[fermion, anchor=base] (2.5,0) -- (3,0);
\end{tikzpicture}
+
\begin{tikzpicture}[line width=1.5 pt, scale=2,baseline=-0.5ex]
	\draw[pattern=north west lines,draw=black, anchor=base] (4.5,0) arc (0:360:.25);
	\draw[fermion, anchor=base] (4.5,0) -- (5,0);
	\draw[pattern=north west lines,draw=black, anchor=base] (5.5,0) arc (0:360:.25);
	\draw[fermion, anchor=base] (5.5,0) -- (6,0);
\end{tikzpicture}
+\dots\right)\\
%\phantom{\begin{tikzpicture}[line width=1.5 pt, scale=2]
%	\draw[dressed_fermion, anchor=base,baseline] (0,0) -- (1,0);
%\end{tikzpicture}}
= &\,
\begin{tikzpicture}[line width=1.5 pt, scale=2,baseline=-0.5ex]
	\draw[fermion, anchor=base,baseline] (0,0) -- (1,0);
\end{tikzpicture}
\left( \sum_{k=0}^\infty 
\begin{tikzpicture}[line width=1.5 pt, scale=2,baseline=-0.5ex]
	\draw[pattern=north west lines,draw=black, anchor=base,baseline] (2.5,0) arc (0:360:.25);
	\draw[fermion, anchor=base] (2.5,0) -- (3,0);
\end{tikzpicture}
\right)
=
\frac{\begin{tikzpicture}[line width=1.5 pt, scale=2,baseline=-0.5ex]
	\draw[fermion, anchor=base,baseline] (0,0) -- (1,0);
\end{tikzpicture}}
{1 -
\begin{tikzpicture}[line width=1.5 pt, scale=2,baseline=-0.5ex]
	\draw[pattern=north west lines,draw=black, anchor=base,baseline] (2.5,0) arc (0:360:.25);
	\draw[fermion, anchor=base] (2.5,0) -- (3,0);
\end{tikzpicture}}\\
= &\,
\frac{ 1 }{m + \frac{r}{2}\hat{p}^2 + i \slashed{\bar{p}} - \Sigma^{1}(p; m, e^2) },
\end{align*}
}where we noticed that the series of insertions of self-energy give rise to the geometric series. 
The $\Sigma^{1}(p; m, e^2)$ is a function of the bare parameters. 
Now we add a mass counterterm to the Lagrangian $\delta m \psibar\psi$ that gives the following additional vertex
\begin{align}
\begin{tikzpicture}[line width=1.5 pt, scale=2]
	\draw[fermionnoarrow] (0,0) -- (1,0);
	% The labels
	\node at (0.5,-0.005){$ {\displaystyle\times} $};
%	\node [draw,circle,cross,minimum width=1 cm](B) at (3,0){}; 
%	\node[draw, cross] at (0.5,0) {};
	\node at (1.3,0){$\equiv {\displaystyle \delta m }$};
\end{tikzpicture}.
\end{align}
We interpret the renormalized mass as $m_R = m + \delta m$ and replace in the self-energy calculations $m$ with $m_R$: $\Sigma^{1}(p; m, e^2) \rightarrow \Sigma^{1}(p; m_R, e^2)$. The inverse propagator gets modified as
\begin{align}
m_R + \frac{r}{2}\hat{p}^2 + i \slashed{\bar{p}} - \left(\Sigma^{1}(p; m_R, e^2) + \delta m\right).
\end{align}
We require that the propagator has a pole when the renormalized mass and momentum vanishes, i.e.~$m_R=0$ and $p=0$. By imposing the condition we get the value of $\delta m$
\begin{align}
\nn
& m_R + \frac{r}{2}\hat{p}^2 + i \slashed{\bar{p}} - \left(\Sigma^{1}(p; m_R, e^2) + \delta m\right)\bigg|_{m_R=0, p=0}  = -\Sigma^{1}(p=0; m_R=0, e^2) - \delta m = 0\\
&\Rightarrow \delta m = -\Sigma^{1}(p=0; m_R=0, e^2) 
= - \bigg(
\begin{tikzpicture}[line width=1.5 pt, scale=2,baseline=2.25ex]
	\node at (1.7,0.15){$p$};
	\draw[fermion, anchor=base] (1.5,0) -- (2.25,0);
%%%%%%%%%%%%%%%%%%%%%%%%%%%TADPOLE	
	\draw[vector] (2.5,0.25) arc (0:360:.25);
%%%%%%%%%%%%%%%%%%%%%%%%%%%%%%%%%%%%	
	\node at (2.7,0.15){$p$};
	\draw[fermion, anchor=base] (2.25,0) -- (3,0);
\end{tikzpicture}
+
\begin{tikzpicture}[line width=1.5 pt, scale=2,baseline=2.25ex]
	\node at (3.7,0.15){$p$};
	\draw[fermion, anchor=base] (3.5,0) -- (4,0);
%%%%%%%%%%%%%%%%%%%%%%%%%%%%%%%%%%%
		\draw[vector] (4.5,0) arc (0:180:.25);
			\node at (4.7,0.15){$p$};
		\draw[fermion, anchor=base] (4,0) -- (4.5,0);
%%%%%%%%%%%%%%%%%%%%%%%%%%%%%%%%%%%%
	\draw[fermion, anchor=base] (4.5,0) -- (5,0);
\end{tikzpicture}\bigg)_{m_R=0, p=0}.
\end{align}
The critical mass $m_c$ is the bare mass for which the renormalized mass is vanishing, i.e.~$m_c = -\delta m$, and in terms of that we have the following relation
\begin{align}
m_R = m - m_c = m - \Sigma^{1}(p=0; m_R=0, e^2) = m - \left( I_T + I_S + I_{CS} + I_{M1} + I_{M2} \right).
\end{align}

\section{Tadpole improved lattice perturbation theory}

We improve the perturbation theory result by summing a subclass of tadpole diagrams, which in the case of gluons (QCD) are called cactus diagrams \cite{Constantinou:2006hz}, to all order in perturbation theory.
This procedure is an example of one-sided resummation, meaning that we will resum only tadpole insertions at the same locations. In principle this can ruin the partial cancellation between diagrams that are resummed and the one that are not, furthermore it may spoil the gauge invariance of the result.\\
In QED the gauge invariance is lost by resumming only tadpole diagrams. But the biggest ef\mbox{}fect is given by the tadpole diagrams hence the tadpole resummation is expected to capture most of the cutof\mbox{}f ef\mbox{}fects, see \cite{Duncan:1996xy}.\\
By expanding the interaction part of the action we get that the tadpole diagrams come as
{\large
\[
\text{Impr} = 
\begin{tikzpicture}[line width=1.5 pt, scale=2,baseline=2.25ex]
	\draw[dressed_vector] (2.5,0.25) arc (0:360:.25);
\end{tikzpicture}
=
\begin{tikzpicture}[line width=1.5 pt, scale=2,baseline=2.25ex]
%%%%%%%%%%%%%%%%%%%%%%%%%%%TADPOLE	
	\draw[vector] (2.5,0.25) arc (0:360:.25);
%%%%%%%%%%%%%%%%%%%%%%%%%%%%%%%%%%%%
\end{tikzpicture}
+
\begin{tikzpicture}[line width=1.5 pt, scale=2,baseline=-0.25ex]
%%%%%%%%%%%%%%%%%%%%%%%%%%%TADPOLE	
	\draw[vector, anchor=base] (0.25,0.25) arc (0:360:.25);
	\draw[vector, anchor=base] (0.28,-0.32) arc (0:360:.25);
%%%%%%%%%%%%%%%%%%%%%%%%%%%%%%%%%%%%
\end{tikzpicture}
+\dots
\]
}
\begin{align}
\phantom{\large \text{Impr}} = - \sum_{n=1} \frac{(e^2Q^2)^n}{(2n)!V^n}\left(\sum_k \sum_\mu G_{\mu\mu}(k)\right)^n (2n-1)!!,
\end{align}
where we wrote explicitly the integrals after contractions. There is an overall minus since the vertexes come all from the even powers of the interactions that gives $(ieQ)^{2n}$ and each vertex comes with a factorial $(2n)!$, there is also a symmetry factor in the contractions of the photon legs that gives $(2n-1)!!$. We notice that the factor $(2n)!/(2n-1)!! = (2n)!! = 2^n n!$ and from this we conclude that
\begin{align}
\nn
\text{Impr} &= - \sum_{n=1} \frac{1}{n!} \left(\frac{e^2Q^2}{V}\frac{1}{2} \sum_k \sum_\mu G_{\mu\mu}(k) \right)^n \\
\label{eq:tad_resum}
&= - \sum_{n=1} \frac{1}{n!} \left[ -\Sigma^{1}_T(p=0; m_R=0, e^2) \right]^n = 1-\e^{-\Sigma^{1}_T(0;0,e^2)}.
\end{align}

\subsection*{Electromagnetic critical mass shift}

We imagine to have performed a simulation at a certain value of $\beta = 6/g^2$ for QCD and we want to calculate in lattice perturbation theory the shift in the critical mass in the full theory.
To do so we take the self-energy as the sum of the pure QCD part (from simulations), the pure QED part (from lattice perturbation theory) and the mixed terms in QCD and QED. In the mixed part there are diagrams in which we connect the pure QCD and the pure QED with a massless Dirac-Wilson propagator, and since we are considering zero momentum the only allowed momentum flowing in such propagator is vanishing. The limit of zero momentum of the D-W propagator is
\begin{align}
\lim_{p\rightarrow 0} \frac{m + \frac{r}{2}\hat{p}^2 - i \slashed{\bar{p}}  }{\left(m + \frac{r}{2}\hat{p}^2\right)^2 + \bar{p}^2 }\bigg|_{m=0} \stackrel{r = 1}{=} \frac{1}{8}
\end{align}
In this way we are including all the possible non-$\I$PI diagrams built out of Q(C+E)D. Diagrammatically the self-energy is
{\large
\[
\Sigma(p=0; m=0) =
\begin{tikzpicture}[line width=1.5 pt, scale=2,baseline=-0.5ex]
	\draw[pattern=north west lines,draw=black, anchor=base,baseline] (2.5,0) arc (0:360:.25);
\end{tikzpicture}
\simeq
\begin{tikzpicture}[line width=1.5 pt, scale=2,baseline=2.25ex]
%%%%%%%%%%%%%%%%%%%%%%%%%%%TADPOLE	
	\draw[vector] (2.5,0.25) arc (0:360:.25);
%%%%%%%%%%%%%%%%%%%%%%%%%%%%%%%%%%%%
\end{tikzpicture}
+
\begin{tikzpicture}[line width=1.5 pt, scale=2,baseline=.7ex]
%%%%%%%%%%%%%%%%%%%%%%%%%%%%%%%%%%%SUNSET
		\draw[vector] (4.5,0) arc (0:180:.25);
		\draw[fermion, anchor=base] (4,0) -- (4.5,0);
%%%%%%%%%%%%%%%%%%%%%%%%%%%%%%%%%%%%
\end{tikzpicture}
+
\begin{tikzpicture}[line width=1.5 pt, scale=2,baseline=2.25ex]
%%%%%%%%%%%%%%%%%%%%%%%%%%%TADPOLE	GLUONS
	\draw[gluon] (2.5,0.25) arc (-90:270:.25);
%%%%%%%%%%%%%%%%%%%%%%%%%%%%%%%%%%%%
\end{tikzpicture}
+
\begin{tikzpicture}[line width=1.5 pt, scale=2,baseline=.7ex]
%%%%%%%%%%%%%%%%%%%%%%%%%%%%%%%%%%%SUNSET GLUONS
		\draw[gluon] (4.5,0) arc (0:180:.25);
		\draw[fermion, anchor=base] (4,0) -- (4.5,0);
%%%%%%%%%%%%%%%%%%%%%%%%%%%%%%%%%%%%
\end{tikzpicture}
+
\begin{tikzpicture}[line width=1.5 pt, scale=2,baseline=.7ex]
%%%%%%%%%%%%%%%%%%%%%%%%%%%TADPOLE PHOTONS GLUONS	
	\draw[gluon] (2.3,0.05) arc (-90:270:.25);
	\draw[fermion, anchor=base] (2.25,0) -- (3.,0);
	\draw[vector] (3.25,0.25) arc (0:360:.25);
%%%%%%%%%%%%%%%%%%%%%%%%%%%%%%%%%%%%
\end{tikzpicture}
\]
\vspace{-0.5cm}
\[
\phantom{\Sigma(p=0; m=0) =
\begin{tikzpicture}[line width=1.5 pt, scale=2,baseline=-0.5ex]
	\draw[pattern=north west lines,draw=black, anchor=base,baseline] (2.5,0) arc (0:360:.25);
\end{tikzpicture}=}
+
\begin{tikzpicture}[line width=1.5 pt, scale=2,baseline=.7ex]
%%%%%%%%%%%%%%%%%%%%%%%%%%%TADPOLE	GLUONS + sunset qed
	\draw[gluon] (2.3,0.05) arc (-90:270:.25);
	\draw[fermion, anchor=base] (2.25,0) -- (3.,0);
	\draw[vector] (3.5,0) arc (0:180:.25);
	\draw[fermion, anchor=base] (3,0) -- (3.5,0);
%%%%%%%%%%%%%%%%%%%%%%%%%%%%%%%%%%%%
\end{tikzpicture}
+
\begin{tikzpicture}[line width=1.5 pt, scale=2,baseline=.7ex]
%%%%%%%%%%%%%%%%%%%%%%%%%%%TADPOLE	
	\draw[vector] (2.5,0.25) arc (0:360:.25);
	\draw[fermion, anchor=base] (2.22,0) -- (3.,0);
	\draw[gluon] (3.5,0) arc (0:180:.25);
	\draw[fermion, anchor=base] (3,0) -- (3.5,0);
%%%%%%%%%%%%%%%%%%%%%%%%%%%%%%%%%%%%
\end{tikzpicture}\,\, .
\]
}
The formula reads
\begin{align}
\nn
\Sigma(p=0; m=0) &\simeq \Sigma_e(0; 0) + \Sigma_g(0; 0) + 2\Sigma_e(0; 0)\lim_{k\rightarrow 0}S(k)\Sigma_g(0; 0)\\
&= \Sigma_e(0; 0) + \Sigma_g(0; 0) + \frac{\Sigma_e(0; 0)\Sigma_g(0; 0)}{4},
\end{align}
where we have a factor two in the last addendum because of the symmetry of the diagrams.\\
Recalling that the critical mass (or the self-energy $\Sigma$) is related to the critical $\kappa$ (hopping parameter)
\begin{align}
\kappa_c = \frac{1}{8r + 2\Sigma(0;0)},
\end{align}
we get that
\begin{align}
\nn
\Delta m_c^\text{Q(C+E)D} &\simeq \Sigma(0;0) - \Sigma_g(0;0) = \frac{1}{2\kappa_c} - \frac{1}{2\kappa_c^{e=0}} = \Sigma_e(0;0) + \frac{1}{4} \Sigma_e(0;0)\Sigma_g(0;0)\\
& = \Sigma_e(0;0) + \frac{1}{4} \Sigma_e(0;0)\left( \frac{1}{2\kappa_c^{e=0}} - 4\right) = \frac{\Sigma_e(0;0)}{8\kappa_c^{e=0}}.
\end{align}
We found the electromagnetic shift to the critical mass in the approximation where we sum only the self-energies from QCD and QED connected by a fermion line.\\
We can approximate the EM self-energy with 1-loop result and write
\begin{align}
\label{eq:one-loop-m-c}
\Delta m_c^\text{Q(C+E)D} \simeq \frac{\Sigma_e^{1}(0;0)}{8\kappa_c^{e=0}}.
\end{align}
We can also take as approximate EM self-energy the tadpole resummed one in Eq.~\ref{eq:tad_resum} we get that the shift is given by
\begin{align}
\label{eq:daisy-m-c}
\Delta m_c^\text{Q(C+E)D} \simeq \frac{\Sigma_e^\text{T}(0;0)}{8\kappa_c^{e=0}} = \frac{1}{8\kappa_c^{e=0}} \left( 1 - \e^{-\Sigma^{1}_T(0;0,e^2) } \right)
= \frac{1}{8\kappa_c^{e=0}} \left( 1 - \e^{\delta m_{em} } \right).
\end{align}
\appendix
\setcounter{chapter}{5}
\renewcommand{\chaptername}{Appendix}
\renewcommand{\theequation}{\Alph{chapter}.\arabic{equation}}
\setcounter{equation}{0}

\chapter{Explicit form of the HVP}
\label{app:HVP}

Here we perform the Wick contractions for Eq.~\ref{eq:off_diag}, and we set $r=1$. 
As an example we work out only the following contribution, one out of four,
\begin{align}
\frac{1}{4} \sum_{f,f'} \left[\psibar_f(n+\hat{\mu}) (1+\gamma_\mu) U^\dagger_\mu(n) \psi_f(n)\right]\left[ \psibar_{f'}(\hat{\nu}) (1+\gamma_\nu) U^\dagger_\nu(0) \psi_{f'}(0) \right].
\end{align}
We write all the indexes out, $f, f'$ are the flavor ones, $\alpha,\beta,\gamma,\delta$ the Dirac ones and $\,b,c,d$ the color ones, and we calculate the fermionic part of the expectation value ($\langle\dots\rangle_\text{F} = \prod_{f}^{N_f}\langle\dots\rangle_\text{f}$) that factorizes with respect to flavor
\begin{align}
\nn
I = &\frac{1}{4}  \sum_{f,f'} \bigg\langle \psibar_f^{\,a,\alpha}(n+\hat{\mu}) (1+\gamma_\mu)_{\alpha\beta} \left(U^\dagger_\mu(n)\right)_{ab} \psi_f^{b,\beta}(n)  \psibar_{f'}^{\,c,\gamma} (\hat{\nu}) (1+\gamma_\nu)_{\gamma \delta} \left(U^\dagger_\nu(0)\right)_{cd} \psi_{f'}^{d,\gamma}(0) \bigg\rangle_{\text{F}} \\
= &  \sum_{f,f'}  \frac{1}{4}(1+\gamma_\mu)_{\alpha\beta} \left(U^\dagger_\mu(n)\right)_{ab} (1+\gamma_\nu)_{\gamma \delta} \left(U^\dagger_\nu(0)\right)_{cd}
\bigg\langle \psibar_f^{\,a,\alpha}(n+\hat{\mu})  \psi_f^{b,\beta}(n)  \psibar_{f'}^{\,c,\gamma} (\hat{\nu}) \psi_{f'}^{d,\gamma}(0) \bigg\rangle_{\text{F}}.
\end{align}
We recall that Wick contractions give the propagator,
\begin{align}
\langle \psi^{a, \alpha}(n) \psibar^{\,b, \beta}(m) \rangle = \left(D^{-1} (n,m)\right)_{ab}^{\alpha \beta},
\end{align}
and can be used to obtain
\begin{align}
\nn
I & = \frac{(-1)^2}{4}(1+\gamma_\mu)_{\alpha\beta} \left(U^\dagger_\mu(n)\right)_{ab} (1+\gamma_\nu)_{\gamma \delta} \left(U^\dagger_\nu(0)\right)_{cd}
\bigg\{  \sum_{f\neq f'} \bigg\langle \psi_f^{b,\beta}(n) \psibar_f^{\,a,\alpha}(n+\hat{\mu})  \bigg\rangle_{f}  \\
\nn
&\phantom{\frac{1}{4}(1+\gamma_\mu)_{\alpha\beta}}
\times \bigg\langle \psi_{f'}^{d,\gamma}(0) \psibar_{f'}^{\,c,\gamma} (\hat{\nu})\bigg\rangle_{f'} 
+ \sum_{f} \bigg\langle \psi_{f}^{d,\gamma}(0) \psibar_f^{\,a,\alpha}(n+\hat{\mu})  \bigg\rangle_{f}  \bigg\langle \psi_f^{b,\beta}(n) \psibar_{f}^{\,c,\gamma} (\hat{\nu})\bigg\rangle_{f'} \bigg\}\\
\nn
& = \frac{1}{4}(1+\gamma_\mu)_{\alpha\beta} \left(U^\dagger_\mu(n)\right)_{ab}  (1+\gamma_\nu)_{\gamma \delta} \left(U^\dagger_\nu(0)\right)_{cd}  
\bigg\{ \sum_{f\neq f'} \left(D^{-1}_f (n+\hat{\mu},n)\right)_{ab}^{\alpha \beta} \left(D^{-1}_{f'} (\hat{\nu},0)\right)_{dc}^{\delta \gamma} \\
& \phantom{= \frac{1}{4}(1+\gamma_\mu)_{\alpha\beta} \left(U^\dagger_\mu(n)\right)_{ab}  (1+\gamma_\nu)_{\gamma \delta} \left(U^\dagger_\nu(0)\right)_{cd}  \bigg\{} 
+ \sum_{f} \left(D^{-1}_f (n+\hat{\mu},0)\right)_{ab}^{\alpha \beta} \left(D^{-1}_{f} (\hat{\nu},n)\right)_{dc}^{\delta \gamma} \bigg\}.
\end{align}
In the first step we used the factorization of the fermionic expectation value with respect to flavor and we divided the case in which $f=f'$ and $f\neq f'$, recalling that the fermionic variables are Grassmann variables and they anticommute. In the second step we performed the Wick contraction and displayed the propagator.\\
By collecting all the indexes together we get 
\begin{align}
\nn
I = & \underbrace{\frac{1}{4} \left(\sum_f \tr\left\{ D^{-1}_f (n+\hat{\mu},n) U_\mu^\dagger(n) (1+\gamma_\mu) \right\}\right)
\left(\sum_{f'} \tr\left\{ D^{-1}_f (\hat{\nu},0) U_\nu^\dagger(0) (1+\gamma_\nu) \right\}\right)}_{\text{disconnected piece}}\\
& + \underbrace{\frac{1}{4} \sum_{f} \tr\left\{ (1+\gamma_\nu) U_\nu^\dagger(0) D^{-1}_f (n+\hat{\mu},0) (1+\gamma_\mu) U_\mu^\dagger(n) D^{-1}_f (\hat{\nu},n) \right\}}_{\text{connected piece}}.
\end{align}
By using $\gamma_5$-hermiticity of the Dirac-Wilson operator ($D^{-1} = \gamma_5 \left(D^\dagger\right)^{-1} \gamma_5$) we obtain
\begin{align}
I_\text{connected} = \frac{1}{4} \sum_{f} \tr\left\{ (1+\gamma_\nu) U_\nu^\dagger(0) \gamma_5 
\left[\left(D^\dagger\right)^{-1}_f (0, n+\hat{\mu})\right]  \gamma_5
(1+\gamma_\mu)
 U_\mu^\dagger(n) D^{-1}_f (\hat{\nu},n) \right\}.
\end{align}
Analogously one can calculate the other pieces of the product and eventually obtain, for the connected part,
\begin{align}
\nonumber
\Pi_{\mu\nu}^{\text{connected}} (n) = \frac{1}{4} \sum_f \bigg(  \tr & \bigg\{ (1+\gamma_\nu) U_\nu^\dagger(0) \gamma_5 
\left[\left(D^\dagger\right)^{-1}_f (0, n+\hat{\mu})\right] \gamma_5 (1+\gamma_\mu)
  U_\mu^\dagger(n) D^{-1}_f (\hat{\nu},n)\\
\nonumber
  - & (1-\gamma_\nu) U_\nu(0) \gamma_5 
\left[\left(D^\dagger\right)^{-1}_f (\hat{\nu}, n+\hat{\mu})\right] \gamma_5 (1+\gamma_\mu)
  U_\mu^\dagger(n) D^{-1}_f (0,n)\\
\nonumber
  - & (1+\gamma_\nu) U_\nu^\dagger(0) \gamma_5 
\left[\left(D^\dagger\right)^{-1}_f (0, n)\right] \gamma_5 (1-\gamma_\mu)
  U_\mu(n) D^{-1}_f (\hat{\nu},n+\hat{\mu})\\
\label{eq:conn}
  + & (1-\gamma_\nu) U_\nu(0) \gamma_5 
\left[\left(D^\dagger\right)^{-1}_f (\hat{\nu},n)\right] \gamma_5 (1-\gamma_\mu)
  U_\mu(n) D^{-1}_f (0,n+\hat{\mu})\bigg\}\bigg).
\end{align}
The contraction for Eq.~\ref{eq:diag}, the contact term, are done in the same way
\begin{align}
\nn
\Pi_{\mu\nu}^{\text{ct}}(0) & = - \frac{\delta_{\mu\nu}}{2} \sum_f \bigg\langle  \psibar_f(0)U_\mu(0)(1-\gamma_\mu)\psi_f(\hat{\mu}) + \psibar_f(\hat{\mu})U^\dagger_\mu(0)(\gamma_\mu+1)\psi_f(0) \bigg\rangle_\text{F}\\
\nn
& = - \frac{\delta_{\mu\nu}}{2} \sum_f \bigg\langle  \psibar_f^{\, a,\alpha}(0)  \left(U_\mu(0)\right)_{ab} (1-\gamma_\mu)_{\alpha\beta} \psi_f^{b,\beta} (\hat{\mu}) + \psibar_f^{\, a,\alpha}(\hat{\mu})  \left(U^\dagger_\mu(0)\right)_{ab} (\gamma_\mu+1)_{\alpha\beta}  \psi_f^{b,\beta}(0) \bigg\rangle_f\\
\nn
& = (-)^2 \frac{\delta_{\mu\nu}}{2} \sum_f \bigg[ \left(U_\mu(0)\right)_{ab} (1-\gamma_\mu)_{\alpha\beta} \bigg\langle \psi_f^{b,\beta} (\hat{\mu}) \psibar_f^{\, a,\alpha}(0)   \bigg\rangle_f\\
\nn
&\phantom{= (-)^2 \frac{\delta_{\mu\nu}}{2} \sum_f \bigg[ }
+ \left(U^\dagger_\mu(0)\right)_{ab} (\gamma_\mu+1)_{\alpha\beta}  \bigg\langle   \psi_f^{b,\beta}(0)   \psibar_f^{\, a,\alpha}(\hat{\mu})     \bigg\rangle_f\bigg]
\end{align}
\begin{align}
\nn
& = \frac{\delta_{\mu\nu}}{2} \sum_f \left[ \left(U_\mu(0)\right)_{ab} (1-\gamma_\mu)_{\alpha\beta} \left(D^{-1}_{f} (0,\hat{\mu})\right)_{ba}^{\beta\alpha} + \left(U^\dagger_\mu(0)\right)_{ab} (\gamma_\mu+1)_{\alpha\beta}  \left(D^{-1}_{f} (\hat{\mu},0)\right)_{ba}^{\beta\alpha}\right]\\
\nn
& = \frac{\delta_{\mu\nu}}{2} \sum_f \left( \tr \left\{ (1-\gamma_\mu) U_\mu(0) \left[D^{-1}_{f} (0,\hat{\mu})\right] + (1+ \gamma_\mu) U^\dagger_\mu(0) \left[D^{-1}_{f} (\hat{\mu},0)\right]\right\}\right)\\
& = \frac{\delta_{\mu\nu}}{2} \sum_f \left( \tr \left\{ (1-\gamma_\mu) U_\mu(0) \gamma_5 \left[\left(D^\dagger\right)^{-1}_{f}  (\hat{\mu},0)\right] \gamma_5 + (1+ \gamma_\mu) U^\dagger_\mu(0) \left[D^{-1}_{f} (\hat{\mu},0)\right]\right\}\right),
\end{align}
and by recalling that $\{\gamma_5,\gamma_\mu\} = 0 \Rightarrow \gamma_5\gamma_\mu\gamma_5 = -\gamma_\mu$ and $\gamma_5^2=\I$, we find
\begin{align}
\label{eq:ct}
\Pi_{\mu\nu}^{\text{ct}}(0) & = \frac{\delta_{\mu\nu}}{2} \sum_f \left( \tr \left\{ (1+\gamma_\mu) U_\mu(0) \left[\left(D^\dagger\right)^{-1}_{f} (\hat{\mu},0)\right] + (1+ \gamma_\mu) U^\dagger_\mu(0) \left[D^{-1}_{f} (\hat{\mu},0)\right]\right\}\right).
\end{align}
By looking at Eq.~\ref{eq:conn} and Eq.~\ref{eq:ct} we see that to evaluate $\Pi_{\mu\nu}$ we need to do five inversions on each gauge field configuration, which make the calculations quite expensive in computer time.\\
The identification with the results in the appendix of \cite{Gockeler:2003cw} is easily made.

\bibliography{citation}{}

\end{document}